\documentclass[aip,jap,preprint,
showkeys,
amsmath,amssymb,amsfonts,
floatfix,
a4paper]{revtex4-1}

\usepackage{graphicx}
\usepackage{dcolumn}
\usepackage{bm}

\usepackage[utf8]{inputenc}
\usepackage[T1]{fontenc}

\usepackage[english]{babel}
\usepackage{csquotes}
\usepackage{mathptmx}
\usepackage{etoolbox}
\usepackage{amstext}
\usepackage{placeins}
\usepackage{tabularray}
\usepackage{breakcites}

\usepackage{color}
\usepackage[table,xcdraw,dvipsnames]{xcolor}
\usepackage[breaklinks]{hyperref}
\usepackage{miller}
\usepackage[capitalize]{cleveref}
\usepackage{upgreek}
\usepackage{textcomp}
\usepackage{siunitx}
\usepackage{threeparttable}
\usepackage[inline]{enumitem}
\usepackage{comment}
\usepackage{mytikz}
\usepackage{orcidlink}

\DeclareUnicodeCharacter{2212}{\dash}
\DeclareRobustCommand\dash{%
  \unskip\nobreak\thinspace\textemdash\allowbreak\thinspace\ignorespaces}

\DeclareUnicodeCharacter{0305}{} 

\DeclareSIUnit\angstrom{\text {Å}}

\pgfdeclarelayer{bg}    
\pgfsetlayers{bg,main}

\DeclareTblrTemplate{contfoot-text}{default}{}
\DeclareTblrTemplate{conthead-text}{default}{}
\DeclareTblrTemplate{caption-tag}{default}{}
\DeclareTblrTemplate{caption-sep}{default}{}

\renewcommand{\selectlanguage}[1]{}
\renewcommand{\rightarrow}{{\ \tikz[baseline=-3.7pt,>=stealth',->] \draw[->] (0,0) -- (0.4,0);\ }}
\newcommand{\revision}[1]{#1}

\hypersetup{
  colorlinks=True,
  linkbordercolor=RoyalBlue,
  urlbordercolor=RoyalBlue,
  urlcolor=RoyalBlue,
  linkcolor=green!50!black,
  citecolor=orange!70!black,
  breaklinks=True
}

\newcommand*{\reftext}[1]{{\textrm{[#1]}}}
\newcommand*{\kb}{k_\mathrm{B}}

\newcommand*{\bkgCol}[1]{
\begin{tikzpicture}[baseline]\fill[#1, draw=#1] (0,-2pt) rectangle ++(1em,1em);\end{tikzpicture}
}

\newcommand*{\appE}[2]{\noindent\reftext{#1}~\cite{#2}:}
\newcommand*{\glosHeading}[1]{\textbf{#1}}

\newenvironment{appendixTable}
    {
    \begingroup
    \centering
    \setlength{\tabcolsep}{4pt}
    \renewcommand{\arraystretch}{1.2}
  \begin{longtblr}[caption={},label=none, postsep=4.5\bigskipamount]{width=\linewidth, colspec={Q[l,0.15\linewidth] Q[l,0.8\linewidth]}}
    }
    {
    \end{longtblr}
    \endgroup
    }

\newenvironment{glossaryTableLong}
{
  \begin{tblr}[long, caption={},label=none, presep=1.5\bigskipamount,
    postsep=2.5\bigskipamount]{width=\linewidth, rowsep=-3pt, colspec={Q[r,0.2\linewidth]Q[l,0.05\linewidth]Q[l,0.75\linewidth]}}
}
    {
    \end{tblr}
    }

\newcommand*{\er}{\varepsilon_\mathrm{r}}
\newcommand*{\ezero}{\varepsilon_0}
\newcommand*{\erpe}{\varepsilon_\mathrm{r}^\perp}
\newcommand*{\erpa}{\varepsilon_\mathrm{r}^\parallel}

\newcommand*{\es}{\varepsilon_\mathrm{s}}
\newcommand*{\espe}{\varepsilon_\mathrm{s}^\perp}
\newcommand*{\espa}{\varepsilon_\mathrm{s}^\parallel}
\newcommand*{\ei}{\varepsilon_\infty}
\newcommand*{\eipe}{\varepsilon_\infty^\perp}
\newcommand*{\eipa}{\varepsilon_\infty^\parallel}

\newcommand*{\omlo}{\omega_{\mathrm{LO}}}
\newcommand*{\omlope}{\omega_{\mathrm{LO}}^\perp}
\newcommand*{\omlopa}{\omega_{\mathrm{LO}}^\parallel}
\newcommand*{\omto}{\omega_{\mathrm{TO}}}
\newcommand*{\omtope}{\omega_{\mathrm{TO}}^\perp}
\newcommand*{\omtopa}{\omega_{\mathrm{TO}}^\parallel}

\newcommand{\memg}{m^*_{\mathrm{M}\Gamma}}
\newcommand{\memk}{m^*_{\mathrm{MK}}}
\newcommand{\meml}{m^*_{\mathrm{ML}}}
\newcommand{\mdeperp}{m^*_{\mathrm{de}\perp}}
\newcommand{\mdepara}{m^*_{\mathrm{de}\parallel}}
\newcommand{\mceperp}{m^*_{\mathrm{ce}\perp}}
\newcommand{\mcepara}{m^*_{\mathrm{ce}\parallel}}

\newcommand{\mcond}{m^*_\mathrm{c}}
\newcommand{\mde}{m^*_\mathrm{de}}
\newcommand{\mce}{m^*_\mathrm{ce}}
\newcommand{\mdos}{m^*_\mathrm{d}}

\newcommand{\mhgm}{m^*_{\Gamma \mathrm{M}}}
\newcommand{\mhgk}{m^*_{\Gamma \mathrm{K}}}
\newcommand{\mhga}{m^*_{\Gamma \mathrm{A}}}
\newcommand{\mdhperp}{m^*_{\mathrm{dh}\perp}}
\newcommand{\mdhpara}{m^*_{\mathrm{dh}\parallel}}
\newcommand{\mchperp}{m^*_{\mathrm{ch}\perp}}
\newcommand{\mchpara}{m^*_{\mathrm{ch}\parallel}}
\newcommand{\mh}{m^*_\mathrm{h}}
\newcommand{\mdh}{m^*_{\mathrm{dh}}}
\newcommand{\mch}{m^*_{\mathrm{ch}}}
\newcommand{\mhhh}{m^*_{\mathrm{hh}}}
\newcommand{\mhlh}{m^*_{\mathrm{lh}}}

\newcommand{\md}{m^*_\mathrm{d}}
\newcommand{\mc}{M_\mathrm{C}}
\newcommand{\mpol}{m_\mathrm{p}}

\newcommand*{\eg}{E_{\mathrm{g}}}
\newcommand*{\egx}{E_{\mathrm{gx}}}
\newcommand*{\ex}{E_{\mathrm{x}}}
\newcommand*{\egn}{E_{\mathrm{gn}}}
\newcommand*{\egp}{E_{\mathrm{gp}}}

\newcommand*{\tg}{T_{\mathrm{g}}}

\newcommand*{\anc}{A_{\mathrm{nc}}}
\newcommand*{\anv}{A_{\mathrm{nv}}}
\newcommand*{\apc}{A_{\mathrm{pc}}}
\newcommand*{\apv}{A_{\mathrm{pv}}}

\newcommand*{\bnc}{B_{\mathrm{nc}}}
\newcommand*{\bnv}{B_{\mathrm{nv}}}
\newcommand*{\bpc}{B_{\mathrm{pc}}}
\newcommand*{\bpv}{B_{\mathrm{pv}}}

\newcommand*{\cpv}{C_{\mathrm{pv}}}

\newcommand*{\enc}{E_{\mathrm{nc}}}
\newcommand*{\env}{E_{\mathrm{nv}}}
\newcommand*{\epc}{E_{\mathrm{pc}}}
\newcommand*{\epv}{E_{\mathrm{pv}}}

\newcommand*{\rs}{r_{\mathrm{s}}}

\newcommand*{\ad}{a_\mathrm{D}}
\newcommand*{\fd}{f_\mathrm{D}}
\newcommand*{\debyeT}{\Theta_\mathrm{D}}

\newcommand*{\ecb}{E_{\mathrm{CB}}}
\newcommand*{\evb}{E_{\mathrm{VB}}}

\newcommand*{\ekbt}{E_{\kb T}}

\newcommand*{\jn}{J_{\mathrm{n}}}
\newcommand*{\jp}{J_{\mathrm{p}}}

\newcommand*{\vn}{v_{n}}
\newcommand*{\vp}{v_{p}}

\newcommand*{\eion}{E_{\mathrm{i}}}
\newcommand*{\ephonon}{E_{\mathrm{p}}}
\newcommand*{\tl}{T_{\mathrm{L}}}
\newcommand*{\omop}{\omega_{\mathrm{OP}}}

\newcommand*{\aelec}{a_\mathrm{n}}
\newcommand*{\belec}{b_\mathrm{n}}
\newcommand*{\celec}{c_\mathrm{n}}
\newcommand*{\delec}{d_\mathrm{n}}
\newcommand*{\ahole}{a_\mathrm{p}}
\newcommand*{\bhole}{b_\mathrm{p}}
\newcommand*{\chole}{c_\mathrm{p}}
\newcommand*{\dhole}{d_\mathrm{p}}

\newcommand*{\apara}{a_\parallel}
\newcommand*{\aperp}{a_\perp}
\newcommand*{\bpara}{b_\parallel}
\newcommand*{\bperp}{b_\perp}

\newcommand*{\alphapara}{\alpha_\parallel}
\newcommand*{\alphaperp}{\alpha_\perp}
\newcommand*{\betapara}{\beta_\parallel}
\newcommand*{\betaperp}{\beta_\perp}

\newcommand*{\DN}{\Delta_N}
\newcommand*{\excessn}{\Delta_{n}}
\newcommand*{\excessp}{\Delta_{p}}

\newcommand*{\RSRH}{R_\mathrm{SRH}}
\newcommand*{\RSRHb}{R_\mathrm{SRH}^\mathrm{b}}
\newcommand*{\RSRHs}{R_\mathrm{SRH}^\mathrm{s}}
\newcommand*{\Rbim}{R_\mathrm{bim}}
\newcommand*{\RAuger}{R_\mathrm{Auger}}

\newcommand*{\tauSRH}{\tau_\mathrm{SRH}}
\newcommand*{\taubim}{\tau_\mathrm{bim}}
\newcommand*{\tauAuger}{\tau_\mathrm{Auger}}
\newcommand*{\taur}{\tau_\mathrm{r}}
\newcommand*{\taup}{\tau_\mathrm{p}}
\newcommand*{\taun}{\tau_\mathrm{n}}
\newcommand*{\taunp}{\tau_\mathrm{n,p}}
\newcommand*{\taumax}{\tau_\mathrm{max}}

\newcommand*{\nintr}{n_\mathrm{i}}
\newcommand*{\eintr}{E_\mathrm{i}}

\newcommand*{\etrap}{E_\mathrm{t}}
\newcommand*{\gtrap}{g_\mathrm{t}}
\newcommand*{\ntrap}{N_\mathrm{t}}
\newcommand*{\nitrap}{N_\mathrm{it}}

\newcommand*{\sigmanp}{\sigma_\mathrm{n,p}}
\newcommand*{\vthermal}{v_\mathrm{th}}

\newcommand*{\eact}{E_\mathrm{act}}
\newcommand*{\eactsurf}{E_\mathrm{bb}}

\newcommand*{\surfn}{s_\mathrm{n}}
\newcommand*{\surfp}{s_\mathrm{p}}
\newcommand*{\nsurf}{n_\mathrm{s}}
\newcommand*{\psurf}{p_\mathrm{s}}

\newcommand*{\cn}{C_\mathrm{n}}
\newcommand*{\cp}{C_\mathrm{p}}

\newcommand*{\na}{N_\mathrm{A}}

\newcommand*{\nam}{N_\mathrm{A}^-}
\newcommand*{\nd}{N_\mathrm{D}}

\newcommand*{\ndp}{N_\mathrm{D}^+}
\newcommand*{\nk}{N_\mathrm{K}}

\newcommand*{\ga}{g_\mathrm{A}}
\newcommand*{\gd}{g_\mathrm{D}}

\newcommand*{\ef}{E_\mathrm{F}}
\newcommand*{\efn}{E_\mathrm{F,n}}
\newcommand*{\efp}{E_\mathrm{F,p}}

\newcommand*{\ed}{E_\mathrm{D}}
\newcommand*{\ded}{\Delta \ed}
\newcommand*{\edh}{E_\mathrm{Dh}}
\newcommand*{\edk}{E_\mathrm{Dc}}
\newcommand*{\ea}{E_\mathrm{A}}
\newcommand*{\dea}{\Delta \ea}
\newcommand*{\eah}{E_\mathrm{Ah}}
\newcommand*{\eak}{E_\mathrm{Ac}}
\newcommand*{\ec}{E_\mathrm{C}}
\newcommand*{\ev}{E_\mathrm{V}}

\newcommand*{\nc}{N_\mathrm{C}}
\newcommand*{\nv}{N_\mathrm{V}}

\newcommand{\mumin}{\mu_\mathrm{min}}
\newcommand{\mumax}{\mu_\mathrm{max}}
\newcommand{\mulow}{\mu_\mathrm{low}}
\newcommand{\nref}{N_\mathrm{ref}}
\newcommand{\munperp}{\mu_\mathrm{n}^\perp}
\newcommand{\munpara}{\mu_\mathrm{n}^\parallel}
\newcommand{\mupperp}{\mu_\mathrm{p}^\perp}
\newcommand{\muppara}{\mu_\mathrm{p}^\parallel}

\newcommand{\mun}{\mu_\mathrm{n}}
\newcommand{\mup}{\mu_\mathrm{p}}
\newcommand{\muc}{\mu_\mathrm{c}}
\newcommand{\muccs}{\mu_\mathrm{ccs}}

\newcommand{\gmin}{\gamma_\mathrm{min}}
\newcommand{\gmax}{\gamma_\mathrm{max}}
\newcommand{\gref}{\gamma_\mathrm{ref}}
\newcommand{\gdelta}{\gamma_\mathrm{\delta}}
\newcommand{\gNNref}{\gamma_\mathrm{NNref}}
\newcommand{\gI}{\gamma_\mathrm{I}}
\newcommand{\gsat}{\gamma_\mathrm{sat}}
\newcommand{\gbeta}{\gamma_\mathrm{\beta}}

\newcommand{\bp}{\beta_\mathrm{p}}
\newcommand{\bmin}{\beta_\mathrm{min}}
\newcommand{\bmax}{\beta_\mathrm{max}}

\newcommand{\Np}{N_\mathrm{p}}

\newcommand{\muH}{\mu_\mathrm{H}}
\newcommand{\rH}{r_\mathrm{H}}
\newcommand{\RH}{R_\mathrm{H}}
\newcommand{\nH}{n_\mathrm{H}}

\newcommand{\vsat}{v_\mathrm{sat}}
\newcommand{\vmax}{v_\mathrm{max}}
\newcommand{\vs}{v_\mathrm{sat}}
\newcommand{\vsperp}{v_\mathrm{sat}^\perp}
\newcommand{\vspara}{v_\mathrm{sat}^\parallel}

\newcommand{\Tref}{T_\mathrm{ref}}

\makeatletter
\def\@email#1#2{%
 \endgroup
 \patchcmd{\titleblock@produce}
  {\frontmatter@RRAPformat}
  {\frontmatter@RRAPformat{\produce@RRAP{*#1\href{mailto:#2}{#2}}}\frontmatter@RRAPformat}
  {}{}
}%

\def\CT@@do@color{%
  \global\let\CT@do@color\relax
        \@tempdima\wd\z@
        \advance\@tempdima\@tempdimb
        \advance\@tempdima\@tempdimc
\advance\@tempdimb\tabcolsep
\advance\@tempdimc\tabcolsep
\advance\@tempdima2\tabcolsep
        \kern-\@tempdimb
        \leaders\vrule
                \hskip\@tempdima\@plus  1fill
        \kern-\@tempdimc
        \hskip-\wd\z@ \@plus -1fill }

\makeatother
\begin{document}


\title{TCAD Parameters for 4H-SiC: A Review}
\author{J{\"u}rgen \surname{Burin}\,\orcidlink{0000-0002-0965-5746}}%
\author{Philipp \surname{Gaggl}\,\orcidlink{0000-0002-4154-1837}}%
\author{Simon \surname{Waid}}%
\author{Andreas \surname{Gsponer}\,\orcidlink{0000-0002-5012-7371}}%
\author{Thomas \surname{Bergauer}\,\orcidlink{0000-0002-5786-0293}}%
\email[e-mail: ]{thomas.bergauer@oeaw.ac.at}
\affiliation{ 
Institute of High Energy Physics, Austrian Academy of Sciences, Nikolsdorfer Gasse 18, 1050 Wien
}%

\date{\today}






\begin{abstract}
  In this literature review we investigate the permittivity, density-of-state mass, band gap, impact ionization, charge carrier recombination, incomplete ionization and mobility in 4H silicon carbide. We provide a comprehensive overview over characterization methods, models and parameters to lower the entrance barrier for newcomers and allow a critical evaluation of common material property descriptions. We further highlight areas for future research by identifying gaps in the current knowledge base.

  For each investigated property we found a large amount of models and parameter sets based on measurements, calculations or fittings. With literal and/or graphical comparisons we reveal qualitative good agreement but also flawed data values, misinterpretations of research results and inconsistencies among multiple investigations, even those directly referencing each other. We identify parameter variations, e.g., due to temperature, with high impact that are rarely considered in 4H-SiC analyses and common values that are based on old research of deviating materials or properties. We further show the slow accommodation of recent research results within the scientific community and reveal missing characterization data but also insufficient models in state-of-the-art technology computer aided design (TCAD) tools. Overall, our review enables scientifically based decisions on 4H-SiC material parameters and unravels the demand for further investigations to validate commonly used values, confirm hypothesis and cover additional dependencies.
\end{abstract}
\keywords{4H-SiC, silicon carbide, material properties, TCAD simulations,
  simulation parameters, permittivity, density-of-states mass, band gap, impact
  ionization, charge carrier recombination, incomplete ionization, mobility, review}

\maketitle
\tableofcontents
\newpage


\section*{\label{sec:glossary}Glossary}

\begin{glossaryTableLong}
  \SetCell[r=1,c=3]{l} {\glosHeading{General}} && \\
  && \\
  $T$ && temperature in \si{\K} \\
  $e$ && elementary charge (\SI{1.602e-19}{\coulomb}) \\
  $\hbar$ && reduced Planck constant (\SI{6.582e-16}{\electronvolt\s}) \\
  $\kb$ && Boltzmann constant (\SI{8.617e-5}{\electronvolt\per\K}) \\
  $t$ && time \\     
  $E$ && energy \\
  $F$ && electric field\\     
  $m_0$ && electron rest mass \\
  $n,p$ && electron/hole concentration \\
  && \\
  && \\
  \SetCell[r=1,c=3]{l} {\glosHeading{Permittivity}} && \\
  && \\
  $\ezero$ && vacuum permittivity (\SI{8.854e-12}{\farad\per\m}) \\
  $\varepsilon',\varepsilon''$ && real/imaginary part of the complex
  permittivity \\
  $\es,\ei$ && static/optical relative permittivity \\
  $\varepsilon^{\parallel},\varepsilon^{\perp}$ && permittivity
  parallel/perpendicular to c-axis \\
  $hf$ && photon energy \\
  $\omlo,\omto$ && longitudinal/transversal optical phonon frequency \\
  && \\
  && \\
  \SetCell[r=1,c=3]{l} {\glosHeading{Density-of-States Mass}} && \\
  && \\
  $\nc,\nv$ && conduction/valence band density-of-states \\
  $\memg,\memk,\meml$ && effective electron masses in the spatial directions $\perp,\perp,\parallel$ to the c-axis \\
  $\mhgm,\mhgk,\mhga$ && effective hole masses in the spatial directions $\perp,\perp,\parallel$ to the c-axis\\
  $\md,\mde,\mdh$ && effective/electron/hole density-of-state mass \\
  $\mcond$ && effective conductivity mass \\
  $\mhhh, \mhlh$ && heavy hole/light hole relative mass \\
  $\mpol$ && polaron mass \\
  $\alpha$ && Fr{\"o}hlich constant \\
  $\mc$ && number of conduction band minima \\
  $z_x, n_x, \eta$ && temperature dependency parameters\\
  && \\
  && \\ \pagebreak
  \SetCell[r=1,c=3]{l} {\glosHeading{Band Gap}} && \\
  && \\
  $\eg$ && band gap energy \\
  $\egx$ && exciton band gap energy \\
  $\ex$ && free exciton binding energy \\
  $\Delta E_\mathrm{th},\Delta E_\mathrm{ph}$ && bandgap variation due to
  thermal expansion/electron-phonon interaction \\ 
  $\ecb,\evb$ && doping induced variation of conduction/valence band \\
  $\enc,\env$ && donor induced variation of conduction/valence band \\
  $\epc,\epv$ && acceptor induced variation of conduction/valence band \\
  $\egn,\egp$ && donor/acceptor induced variation of the band gap \\
  $\ndp,\nam$ && ionized donor/acceptor concentration \\
  $\alpha,\beta,\Theta,p$ && temperature dependency
  parameters \\
  $\Delta$ && phonon dispersion \\
  $A,B, C$ && doping dependency parameters\\
  && \\
  && \\
  \SetCell[r=1,c=3]{l} {\glosHeading{Impact Ionization}} && \\
  && \\
  $\nd,\na$ && donor/acceptor concentration \\
  $\ndp,\nam$ && ionized donor/acceptor doping concentration \\
  $\alpha,\beta$ && electron/hole impaction ionization coefficient \\
  $\ephonon$ && optical phonon energy \\
  $\eion$ && electron-hole pair ionization energy \\
  $\langle \eion \rangle$ && effective ionization threshold \\
  $\lambda$ && mean free path \\
  $a$ && impact ionization coefficient at $F=\infty$ \\
  $b,m$ && field dependency parameters in \cref{eq:II_OC} \\
  $c,d$ && temperature dependency parameters in \cref{eq:II_OC}\\
  $\gamma$ && temperate scaling parameter \\
  $\omop$ && optical phonon energy\\
  && \\
  && \\
  \SetCell[r=1,c=3]{l} {\glosHeading{Charge Carrier Recombination}} && \\
  && \\
  $R,G$ && recombination/generation rate \\
  $\RSRH,\Rbim,\RAuger$ && SRH/Bimolecular/Auger recombination rate \\
  $B$ && bimolecular recombination coefficient \\ \pagebreak
  $\cn,\cp$ && electron/hole Auger recombination coefficients \\
  $\taun,\taup$ && electron/hole life time\\
  $\tauSRH,\taubim,\tauAuger$ && Shockley-Read-Hall/Bimolecular/Auger recombination lifetime \\
  $\tau_{T0}, \tau_0, \tau_\infty$ && lifetime at
  $T=T_0$, $T=\SI{0}{\K}$, $T=\infty$ \\
  $\tau^\mathrm{ll},\tau^\mathrm{hl}$ && lifetime under low/high level injection \\
  $n_0,p_0$ && equilibrium electron/hole concentration \\
  $\DN,\nintr$ && excess/intrinsic charge carrier concentration \\
  $\etrap$ && trap energy level \\
  $\eintr$ && effective Fermi level \\
  $\eg$ && band gap energy \\
  $\ec,\ev$ && conduction/valence band energy \\
  $\eact$ && activation energy in temperature scaling \\
  $\vthermal$ && thermal charge carrier velocity \\
  $\surfn,\surfp$ && electron/hole surface recombination velocity \\
  $\gtrap$ && trap degeneracy \\
  $\sigma$ && capture cross section \\
  $\alpha,C$ && temperature dependency parameter \\
  $\gamma$ && doping dependency parameter \\
  && \\
  && \\
  \SetCell[r=1,c=3]{l} {\glosHeading{Incomplete Ionization}} && \\
  && \\
  $\ef$ && intrinsic Fermi energy \\
  $\ec,\ev$ && conduction/valence band energy \\
  $\efn, \efp$ && quasi Fermi energy for electrons/holes \\
  $\ded,\dea$ && donor/acceptor ionization energy \\
  $\Delta \edh, \Delta \edk$ && donor ionization energy at hexagonal/cubic
  lattice sites \\
  $\Delta \eah, \Delta \eak$ && acceptor ionization energy at hexagonal/cubic
  lattice sites \\
  $\Delta E_0$ && ionization energy at zero doping \\
  $\gd,\ga$ && donor/acceptor degeneracy \\
  $\nintr$ && intrinsic carrier concentration \\
  $n_0,p_0$ && equilibrium electron/hole concentration \\
  $\nd,\na$ && donor/acceptor doping concentration \\
  $\ndp,\nam$ && ionized donor/acceptor doping concentration \\
  $\nk$ && compensation carrier concentration \\
  $\alpha,N_\mathrm{E},c$ && doping dependency parameter \\
  $\sigma$ && capture cross section\\
  && \\
  && \\ \pagebreak
  \SetCell[r=1,c=3]{l} {\glosHeading{Mobility}} && \\
  && \\
  $\sigma$ && conductivity \\
  $\rho$ && resistivity \\
  $\mun,\mup$ && electron/hole majority carrier mobility \\
  $\muccs$ && carrier-carrier scattering mobility \\
  $\mumin,\mumax$ && mobility at a doping concentration of $\infty$/zero\\
  $\muH$ && Hall mobility\\
  $v$ && charge carrier velocity \\
  $\vsat$ && saturation carrier velocity \\
  $\nref,N_\mathrm{ref2},\delta,\kappa$ && doping dependency paramters \\
  $\gamma$ && temperature dependency parameter \\
  $\beta$ && scaling factor for high-field mobility \\
  $a,b,c$ && parameters for temperature scaling of $\beta$ \\
  $\rH,\RH$ && Hall scattering factor/coefficient \\
  $\nH$ && Hall charge carrier count
\end{glossaryTableLong}

\FloatBarrier

\newpage



\section{\label{sec:intro}Introduction}

Silicon carbide (SiC) is a wide band gap semiconductor that enables the design of devices for high-voltage applications with low power losses and high power densities due to its favorable properties of high breakdown voltage, low leakage current, and fast switching speed~\cite{jiya2020}. The inherent radiation hardness makes SiC also attractive for spacecrafts, nuclear-medicine medical devices, and high-energy physics~\cite{denapoli2022}. To optimize the design process for these areas of application and decrease the number of design iterations, the materials parameters have to be known accurately. These commonly serve as input for \textit{technology computer-aided design} (TCAD) simulations that predict the behavior of devices at an early development stage and allow insights into physical processes taking place inside the semiconductor, which is impossible with measurements. The accuracy and reliability of these simulations depend on the employed physical models and the provided material parameters.

For 4H-SiC, the most popular polytype in the industry, the main challenge when parameterizing a TCAD model is not the lack of parameters in the literature but the overwhelming amount. In the last $70$ years, the physical properties of 4H-SiC were extensively measured and described by various approaches. This led to the essential problem of proper parameter selection: Finding, analyzing, and comparing the multitude of parameters is time-consuming, while assessing the \revision{reliability} of the available data is challenging. Consequently, parameters are often taken from secondary literature, passing them on in unverified reference chains. Unavoidable rounding and typographic mistakes in this process led to substantial parameter deviations over time. The resulting contradictions make it difficult to choose one option over the other, especially for newcomers.

To tackle these issues, overview papers for the 4H-SiC properties permittivity~\cite{bechstedt1997}, density-of-states mass~\cite{bechstedt1998,chen1997a,harris1995,pensl2005,schadt1997,tairov1977,troffer1998,takahashi2007,shur2006}, band gap energy~\cite{devaty1997,janzen2008,stefanakis2014,tian2020,neudeck2013,beyer2011,neuberger1971}, impact ionization~\cite{feng2004a,levinshtein2001,neilainglesias2012,raynaud2009,stefanakis2021,sankin2002}, charge carrier recombination~\cite{klein2009,marinella2010}, incomplete ionization~\cite{wang2023,pensl2005,levinshtein2001b,heera2001,harris1995,bluet2000,atabaev2018,scaburri2011a,schoner1994} and mobility~\cite{adachi2005,beyer2011,harris1995,ioffe2023,lebedev2017,levinshtein2001,neilainglesias2012,roschke2001,tian2020,stefanakis2014} were published. These, however, solely list values and parameters from an incomplete set of sources without addressing the following questions: How where the presented values characterized? What are the most commonly used values within the community? Are there additional descriptions/parameter sets and how do they compare against each other? Exist discrepancies among different models and can these be explained or is further research required? Where do we still lack information for proper approximations?

With this review, we aim to answer all these question by providing an overview of the existing literature on the material parameters permittivity, density-of-states mass, band gap, impact ionization, charge carrier recombination, incomplete ionization and mobility. For each topic we
\begin{enumerate*}[label=(\roman*)]
\item introduce the utilized mathematical models,
\item list characterization methods,
\item identify fundamental investigations, show the achieved model
  parameters and analyze inconsistencies over time,
\item discuss the agreement among the models and comment on discrepancies,
\item depict the abundance of specific values in scientific publications and
\item highlight shortcomings and propose areas for future research.
\end{enumerate*}
These analyses provide newcomers a starting point on 4H-SiC, experienced users the possibility to evaluate their models/parameter sets and the community to plan future research goals.

\revision{Despite our detailed analyses we are not able to provide recommendations for single parameter values or typical applications due to the uniqueness of physical entities. Each proposed parameter set matched the behavior of a specific device, making it impossible for us to rate one result over the other. In fact, simulation parameters found in literature may only serve as starting point to approach the measurements of an actual device.}

This work is organized in the following fashion: In \cref{sec:methods}, we introduce the general methods utilized in this review, followed by a description of (4H) silicon carbide in \cref{sec:sic}. We then review the permittivity (\cref{sec:permittivity}), density-of-states mass (\cref{sec:dos}), band gap (\cref{sec:bandgap}), impact ionization (\cref{sec:II}), charge carrier recombination (\cref{sec:regen}), incomplete ionization (\cref{sec:incompIon}) and mobility (\cref{sec:mobility}) before we conclude the paper in \cref{sec:conclusion}.



\section{\label{sec:methods}Methodology}

Before we started this research we tried to gather the desired information with the help of artificial intelligence tools. These, however, only provided a very narrow range of publications for each topic and failed to properly extract the correct information. Consequently, we conducted the literature review by executing the tasks described in the sequel.  The data, evaluation scripts and figures of this study are openly available in the repository \href{https://gitlab.com/dd-hephy/HiBPM/review_4HSiC}{Data of 4H SiC TCAD Parameter Review}~\cite{burin2025}.

\textit{\underline{Literature Review:} } To identify and acquire a comprehensive set of investigations we utilized scientific search engines and followed references respectively citations in suitable contributions. In total, we collected and analyzed $1087$ scientific publications, but only included in the manuscript those who either
\begin{enumerate*}[label=(\roman*)]
\item provided a theoretical description,
\item conducted measurements,
\item developed models,
\item provided model parameters,
\item referenced previous results or
\item simply named parameter values.
\end{enumerate*}
The latter two were necessary to show the evolution of parameters when being referenced and which values are most commonly used.  In a two step process the single contributions were first evaluated in isolation before we aggregated the extracted information by topic.

\textit{\underline{Data Extraction:} } Our initial belief that transferring parameter values is a straightforward task was quickly falsified. Understanding what the authors intended to say was often challenging and required careful considerations. One common problem was that the data were presented in many shapes and formalisms and had to be converted to a common one. Accompanying information, such as temperature, doping concentration/type, or spatial direction, were difficult to grasp and demanded an analysis of the text. Thus, we concluded that an automated procedure was not feasible in this regard and transferred the data by hand, although this method is error prone.

\textit{\underline{Theoretical Background:} }
For each topic we first compared the theoretical explanations, equations, and models described in the reviewed publications. This was sometimes surprisingly challenging because
\begin{enumerate*}[label=(\roman*)]
\item fundamentally different values were denoted with the same symbol (e.g.,
  the free and bound exciton binding energy in \cref{sec:bandgap}), 
\item parameters were declared ambiguously in various notations (see charge carrier recombination in \cref{sec:regen}) and
\item models were denoted by multiple names while being subsets of each other (see impact ionization in \cref{sec:II}).
\end{enumerate*}

In this review we summarize at the beginning of each section the relevant theoretical concepts in a concise introduction. For this purpose we classified the differences among the models and merged, wherever possible, various approaches into a common description. For the latter, we aligned with state-of-the-art TCAD simulation tools.

We intend these summaries to result in a steep learning curve for newcomers, enabling them to easier comprehend the dedicated publications outside this review. Experienced readers might find the collection of the most important information in one place beneficial as a reference and to refresh their memory.

\textit{\underline{Symbol Definitions:} } Symbols are used with different meanings in the literature depending on the context. For example, $\alpha$ and $\beta$ are commonly used to denote the impact ionization coefficients for electrons and holes (cp. \cref{sec:II}), are parameters for the temperature-dependent band gap (cp. \cref{sec:bandgap}) and denote the 3C ($\beta$) and hexagonal (including 4H) ($\alpha$) polytypes of SiC. To prevent ambiguities while sticking to the common literature, we defined the symbols for each topic separately in the glossary. Since the sections of this review are self-contained, we think that the risk of mix-ups is acceptable.

\textit{\underline{Characterization Methods:} } \revision{To compare TCAD parameters}, it is essential to know how they were determined, i.e., the utilized characterization method. In this review, we encountered calculations, measurements, and fittings to existing data. A good mix among these is preferable to prohibit incorrect assumptions in the calculations or disturbances in the measurement setup that falsify the results. In this work, we simply list the found methods; for further details, we refer the interested reader to the specific publications or dedicated literature~\cite{hess1991,friedrichs2009}.

\textit{\underline{Data Analysis:} } As a first step of the analysis, we identified fundamental investigations, i.e., those that proposed parameter values for the first time. We documented the characterization method and further details, such as the value range used for fitting, spatial orientation, or doping concentration/type. For non-fundamental publications, we assumed that the provided values were referenced. We say ``assume'' because parameters were commonly stated without any further comment. Wherever possible, we inferred a relationship based on the used values and marked these as ``guess'' in the data.

A major challenge was to verify the referenced data, i.e., to compare citing and referenced publications. In the case of a mismatch, we noted the discrepancy in alphabetic order (see Appendix~\ref{sec:appendix}) and tried to find explanations. These included confusion of polytypes/spatial directions, erroneous theoretical concepts, rounding, and typographical errors. Instances where we were unsuccessful to resolve a mismatch are discussed in the text of the respective topic.

\textit{\underline{Origin Tracing:} } Unfortunately, fundamental investigations were not always referenced directly, resulting in long chains of citations, which we call ``reference chains''. We wrote a Python script that explored these chains automatically and displayed the utilized values, which allowed us to investigate the changes introduced along the path.

Missing references could sometimes be inferred from the parameter values, but often, a unique mapping was not possible. We even encountered cases where the utilized values could not be traced back to any scientific publication, leaving the exact origin unknown. One possible explanation in such cases is that these value served as the default ones in simulation tools and were assigned within the respective company.

\textit{\underline{Presentation:} } For improved readability, we transformed the data into more comprehensible formats such as tables and figures, where we ordered the contributions in regard to their publication year to visualize changes over time. We also varied colors and line styles but solely for the purpose of increased readability. Self-written \textit{Python} scripts automatized this task to reduce the chances of typographical errors.

For each topic, we summarized in a table the fundamental investigations together with their parameters. We chose this representation over a figure because the exact values decrease the effort to implement the respective model in the reader's simulation. Complementary to the references provided by \LaTeX, we labelled each paper with a unique badge of the form $[XxxxYYZ]$, where $Xxxx$ denotes the first four letters of the main author's surname, $YY$, the last two digits of the publication year and $Z$ an optional letter to ensure the uniqueness of the label within this review. This scheme supports the reader in connecting the results of a single publication, reconstructing important relationships, and searching for a contribution throughout the document.

In addition to the parameter table, we showed the specific aspects of the models, e.g., temperature, doping, or field dependency graphically. All available parameter sets were combined in these plots to identify outliers and tendencies at the cost of readability. To identify every detail, we recommend the reader to zoom into the figures using the digital version of this review. Plotting each model ourselves had the side effect that, in rare cases, we identified incorrect parameters in the fundamental investigation and were able to suggest replacements. For further convenience \begin{tikzpicture}[scale=0.6] \node (elSign) at (0,0) {}; \drawElecSign;\end{tikzpicture} denotes plots for electrons and \begin{tikzpicture}[scale=0.6] \node (holeSign) at (0,0) {}; \drawHoleSign;\end{tikzpicture} those for holes.

The identified reference chains were also converted to a figure featuring the referenced parameter values. Since not all values in the literature were properly referenced and, thus, do not show up in the reference chain plot, we added a figure depicting the abundance of specific values. We allowed multiple publications from the same author to preserve a quantitative relation among values.

\textit{\underline{Discussion:} } Last but not least, we commented on trends and shortcomings in the data to highlight future avenues of research, which, hopefully, trigger a constructive reevaluation within the scientific community.

\revision{We did not provide suggestions on which parameter set to use, because this decision depends on the investigated device structure. Instead the reader is faced with the non-trivial task to pick suitable values. Unfortunately, we did not find dedicated literature or methods on parameter selection in the literature. A common approach is to simulate various models, compare the results to measurements and pick the parameters that fit best~\cite{tian2020}. For dedicated simulations the parameters shown in this review should anyhow just be considered the starting point for additional optimizations to match the unique device at hand.

  For predictive simulations the selection is even more cumbersome. In this case we recommend to pick parameters that were determined on devices that are equal or at least similar to the desired one. This is important, because TCAD parameters are fitted to measurements and thus are always device specific. Consequently, applying the same parameters in differing occasions can lead to varying accuracies. Many parameters were even characterized in simple 4H-SiC epilayers and can not be attached to any device. For these reasons, we are going to discuss the characterization methods and the investigated devices in each respective section.  }

\section{\label{sec:sic}Silicon Carbide (SiC)}

This section provides a historical introduction of silicon carbide in general and 4H in detail. We are also going to define the characteristics of this material, which we require to model specific material properties.

\subsection{\label{sec:history}Brief History}

The first-ever reference of silicon carbide (SiC) dates back to a report of a synthesized compound material containing silicon-carbon bonds by Berzelius in 1824~\cite{berzelius1824}. Despite the initial interest in the material due to its diamond-like hardness, this makes it one of the earliest investigated semiconductor in history~\cite{kimoto2014a, tsao2018, feng2023, svarna2024}.  Through mass-production, enabled by the Acheson process in 1892~\cite{acheson1895}, it soon became a leading abrasive material~\cite{ayalew2004, kimoto2014a}. The poor crystalline quality of the byproducts, however, only allowed rudimentary insights into the outstanding physical and electrical properties of SiC, such as its electroluminescent nature \cite{round1907, kimoto2014a}. In 1905, Moissan reported the first natural occurrence of SiC, naming the mineral Moissanite~\cite{moissan1905}.

After the invention of the Lely process in 1955, which enabled the synthesis of relatively pure single-crystalline SiC (mostly the polytype 6H)~\cite{lely1955, lely1958}, scientific interest re-emerged. The semiconductor properties were heavily investigated in the subsequent decades~\cite{choyke1969, kimoto2014a, dipaoloemilio2024}, mainly focusing on its potential as blue LED and within high-temperature environments\cite{oconnor1960, marshall1974}. In the 1970s, silicon based technologies surpassed SiC, which reduced research targeted at the latter to a minimum~\cite{powell1993, kimoto2014a, dipaoloemilio2024}. Only in the late 1990s, following significant improvements in manufacturing by Tairov and Tsvektov (modified Lely method in 1981)~\cite{tairov1978, tairov1981}, Barret (1991)~\cite{barrett1991}, and Davis (1995)~\cite{davis1995}, SiC became relevant again for the semiconductor industry~\cite{powell2002, kimoto2014a, dipaoloemilio2024}.

Although today's electronics market is still dominated by Si, SiC has become increasingly interesting due to its higher band gap (compared to Si), which enables the development of devices with less power consumption, higher peak voltage, and higher temperature stability~\cite{kimoto2014a, jiya2020, dipaoloemilio2024}. After basic research into Schottky barrier diodes (SBDs), MOSFETs, and JFETs in the early 2000s and their respective first mass production only a decade later, SiC has rapidly evolved into a cornerstone of the high-power electronics industry \cite{jiya2020, dipaoloemilio2024}.  The following cost reduction and increasing accessibility of high-quality material quickly rekindled the interest of the high-energy physics (HEP) community in SiC (especially the 4H-SiC polytype) for particle detection.  The primary goal has been to develop devices that fulfill the ever-growing demands for operation in high-luminosity and high-energy environments~\cite{nava2008, denapoli2022}.

\subsection{\label{sec:polytypes}Crystal Structure \& Polytypes}

SiC is a compound semiconductor formed by Si and C in equal parts, which both possess four valence electrons. Within a crystalline lattice, each Si-atom bonds with exactly four C-atoms and vice versa. The SiC crystal can arrange, at least theoretically, in infinitely many constellations~\cite{fisher1990, powell1993, kimoto2014a}. While the occurrence of a given compound in more than one crystal structure is defined as polymorphism, SiC itself is a prime example of a specialized form, called polytypism, where all stable phases (called polytypes) differ from each other in a specific way~\cite{powell1993}.

Each SiC polytype is constructed by stacking sheets of equal Si-C bilayers in a varying fashion upon each other. Thus, SiC polytypes are identical in two dimensions and only differ within the dimension/axis perpendicular to the bilayer sheets (often referred to as the basal plane). This axis is known as the principal axis, c-axis, and [0001]-axis in literature~\cite{fisher1990, powell1993, bechstedt1997, ayalew2004, nava2008, kimoto2014a}.

\begin{figure}[t]
\centering
\includegraphics[width=0.6\linewidth]{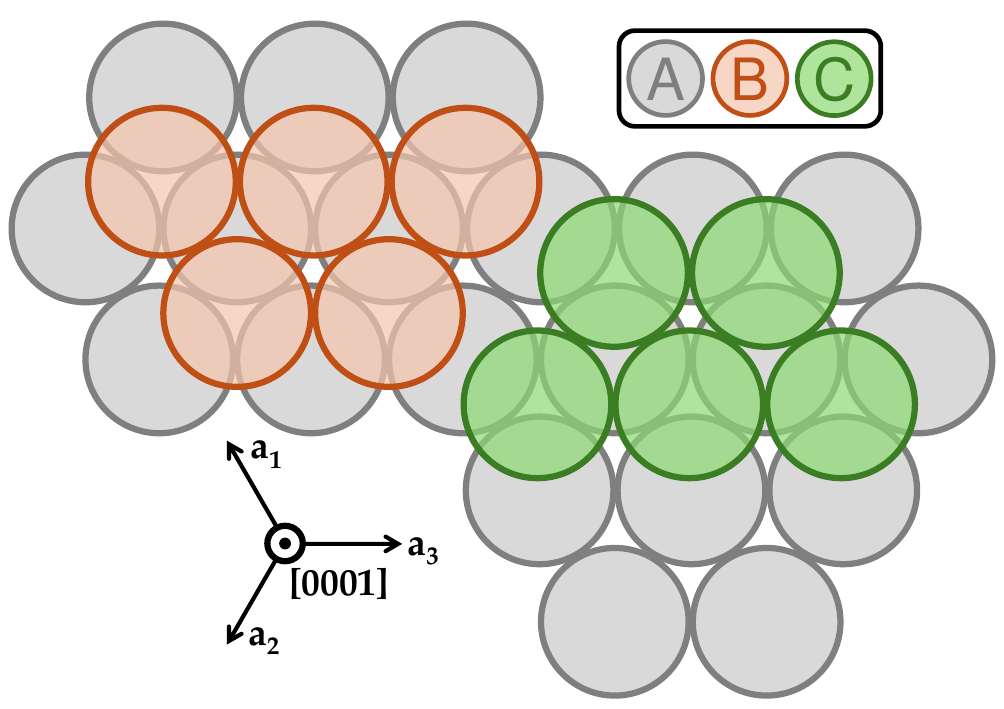}
\caption{\label{fig:crystal_structure}The three possible occupation sites (\textit{A}, \textit{B}, and \textit{C}) in a hexagonal close-packed system of Si-C bilayers stacked along the [0001] axis. \textit{a\textsubscript{1}}, \textit{a\textsubscript{2}}, and \textit{a\textsubscript{3}} form base vectors with an angle of $120^\circ$ that are consistent with the \textit{Miller-Bravais notation}~\cite{powell1993}. Reproduced with permission from Fundamentals of Silicon Carbide Technology~\cite[chap. 2]{kimoto2014a}. Copyright 2014, John Wiley \& Sons Singapore Pte. Ltd.}
\end{figure}

\begin{figure}[t]
\centering
\includegraphics[width=0.95\textwidth]{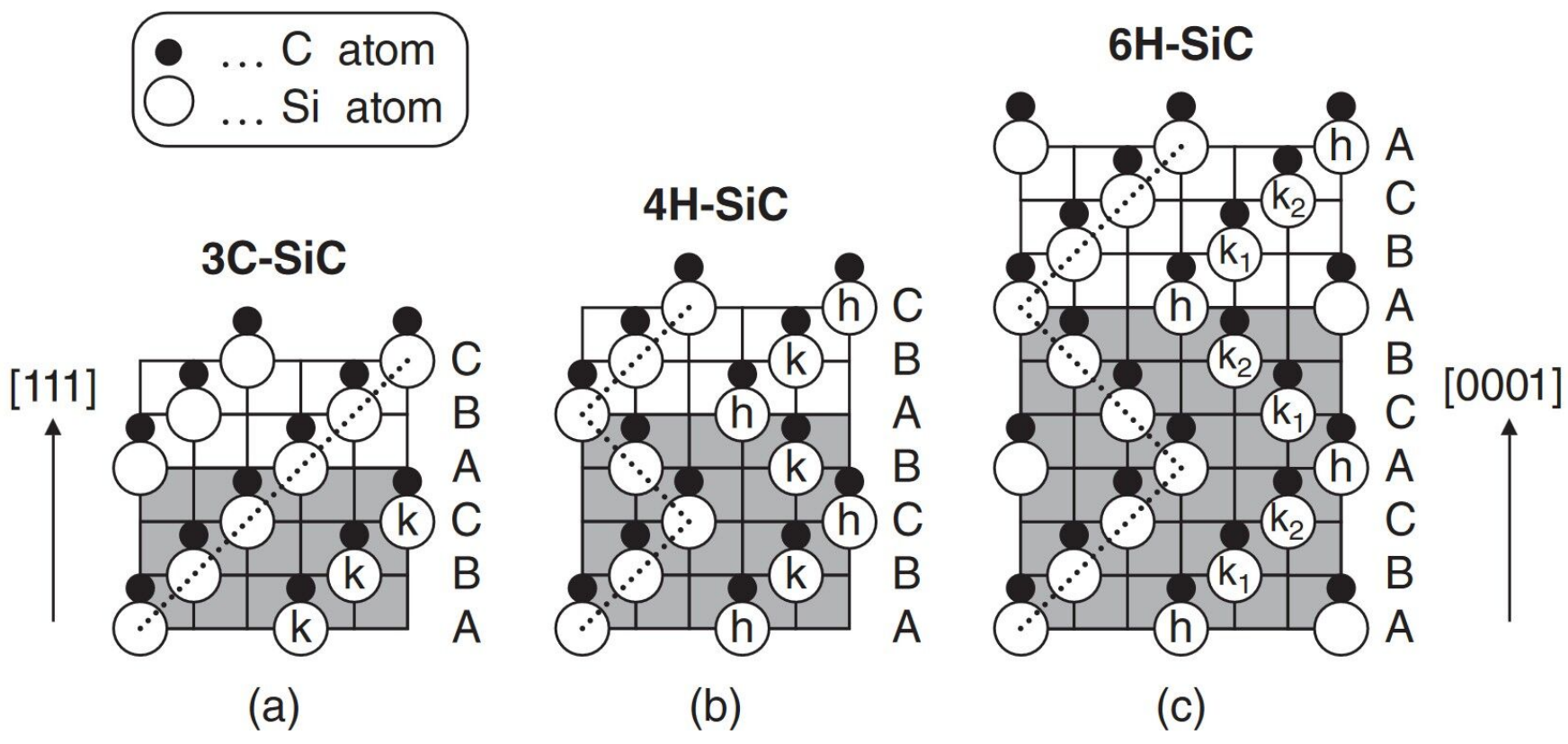}
\caption{\label{fig:polytypes_stackings}Stacking structures of (a) 3C-SiC, (b) 4H-SiC, and (c) 6H-SiC along the principal axis. \textit{h} and \textit{k} denote whether the structure surrounding the lattice site is of cubic or hexagonal form. Gray layers highlight the name-giving number of stacks until the structure is repeated. Reproduced with permission from Fundamentals of Silicon Carbide Technology~\cite[chap. 2]{kimoto2014a}. Copyright 2014, John Wiley \& Sons Singapore Pte. Ltd.}
\end{figure}

In a hexagonal close-packed system, subsequent layers can not occupy the same lattice sites but have to alternate (see \cref{fig:crystal_structure}). The resulting stacking order of these three positional sites then defines the SiC polytype~\cite{powell1993}, which is named after two distinctive features: The number of layers until the respective stacking order repeats itself, and whether the emerging structure shows a cubic (C), hexagonal (H), or rhombohedral (R) symmetry~\cite{powell1993, nava2008, kimoto2014a}. Besides the polytype 4H-SiC investigated in this review, the most prominent ones are 3C-SiC and 6H-SiC, whose stacking order and structure only differ slightly (see \cref{fig:polytypes_stackings}).

\subsection{\label{anisotropy}Anisotropy of 4H-SiC}

Despite the infinite number of stacking variations, only 3C-SiC is isotropic in nature (cubic symmetry). Every other polytype features a certain degree of anisotropy, which denotes a directional dependence of material and electrical properties~\cite{powell1993, ayalew2004, kimoto2014a}. Though this anisotropic nature can be simplified into two components, parallel and perpendicular to the c-axis, a very basic introduction regarding the complexity of the crystallographic structure of 4H-SiC is attempted. For more detailed information, the reader is referred to the dedicated literature~\cite{kaeckell1996, ayalew2004, buono2012}.

\begin{figure}[t]
\centering
\includegraphics[width=0.65\textwidth]{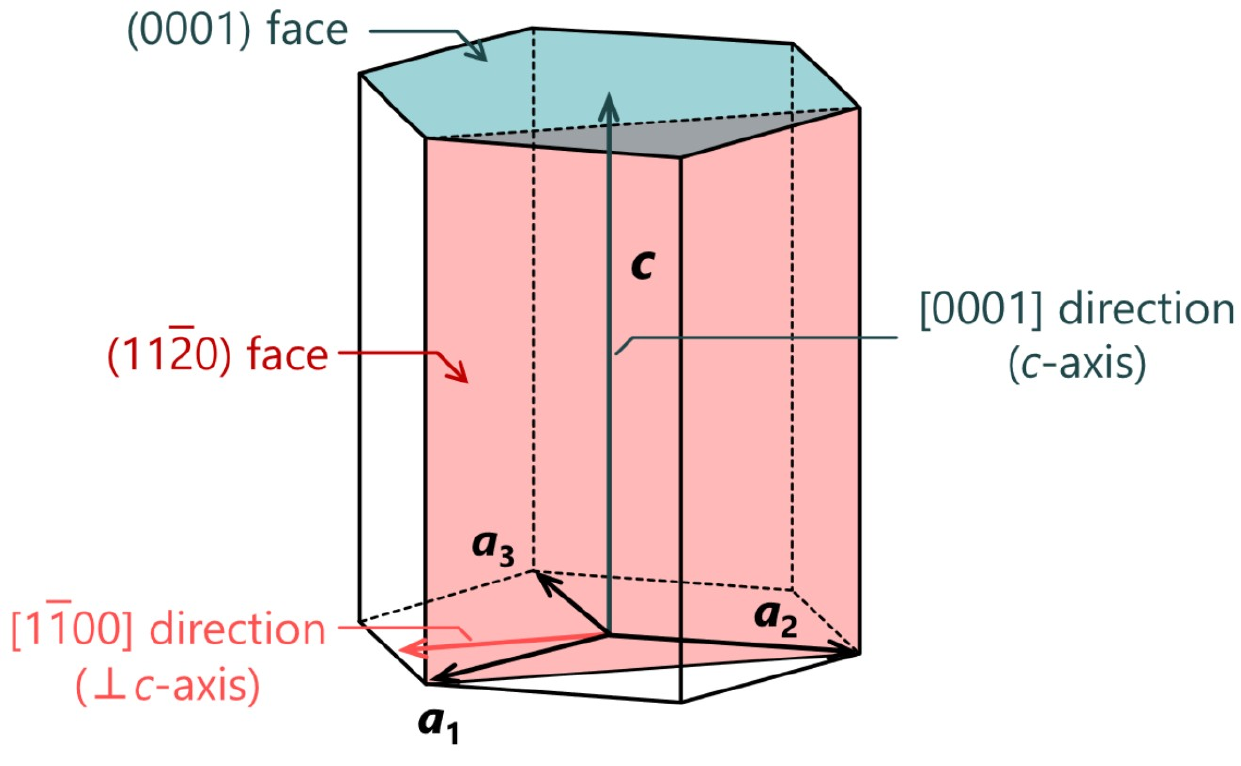}
\caption{\label{fig:unit_cell}4H-SiC unit cell. \textit{$a_1$}, \textit{$a_2$}, \textit{$a_3$}, and \textit{$c$} represent the base vectors of a Bravais lattice, while the \hkl[0001] and \hkl[1-100] direction, as well as the \hkl(0001) and \hkl(11-20) faces, are given in Miller-Bravais notation~\cite{west2012}. Not shown, but often encountered within the literature, is the \hkl[11-20] direction, perpendicular to both the c-axis and \hkl[1-100] direction. Reproduced with permission from Physica Status Solidi (b) 260, 10 (2023)~\cite{ishikawa2023}. Copyright 2023, Wiley-VCH GmbH.}
\end{figure}

In the field of crystallography, structures are usually defined via a unit cell. Such a formation, often also called a primitive cell, represents the smallest repeating unit that still represents the full symmetry of the overlying crystal structure\cite{west2012}. In the case of 4H-SiC, the unit cell is of hexagonal type and consists of four Si and C atoms each. The given base vectors \textit{$a_1$}, \textit{$a_2$}, \textit{$a_3$} (forming a $120^\circ$ angle between each other), and \textit{$c$} form a Bravais lattice, while the so-called Miller-Bravais-notation is used to describe other important directions/axes and faces via a complex linear superposition (see \cref{fig:unit_cell})~\cite{west2012}. For example, the \hkl(0001)-face and the opposite \hkl(000-1)-face are also often called the Si and C face~\cite{kimoto2019}, while the former is also known as the basal plane~\cite{ayalew2004}. Considering common SiC-manufacturing methods, the \hkl(0001)-face (Si-face) is usually parallel to a wafer surface, implying that the c-axis (\hkl[0001] direction) is perpendicular to it~\cite{kaeckell1996, powell1993, ayalew2004, kimoto2014a, ishikawa2023}.

\begin{figure}[t]
\centering
\includegraphics[width=0.4\textwidth]{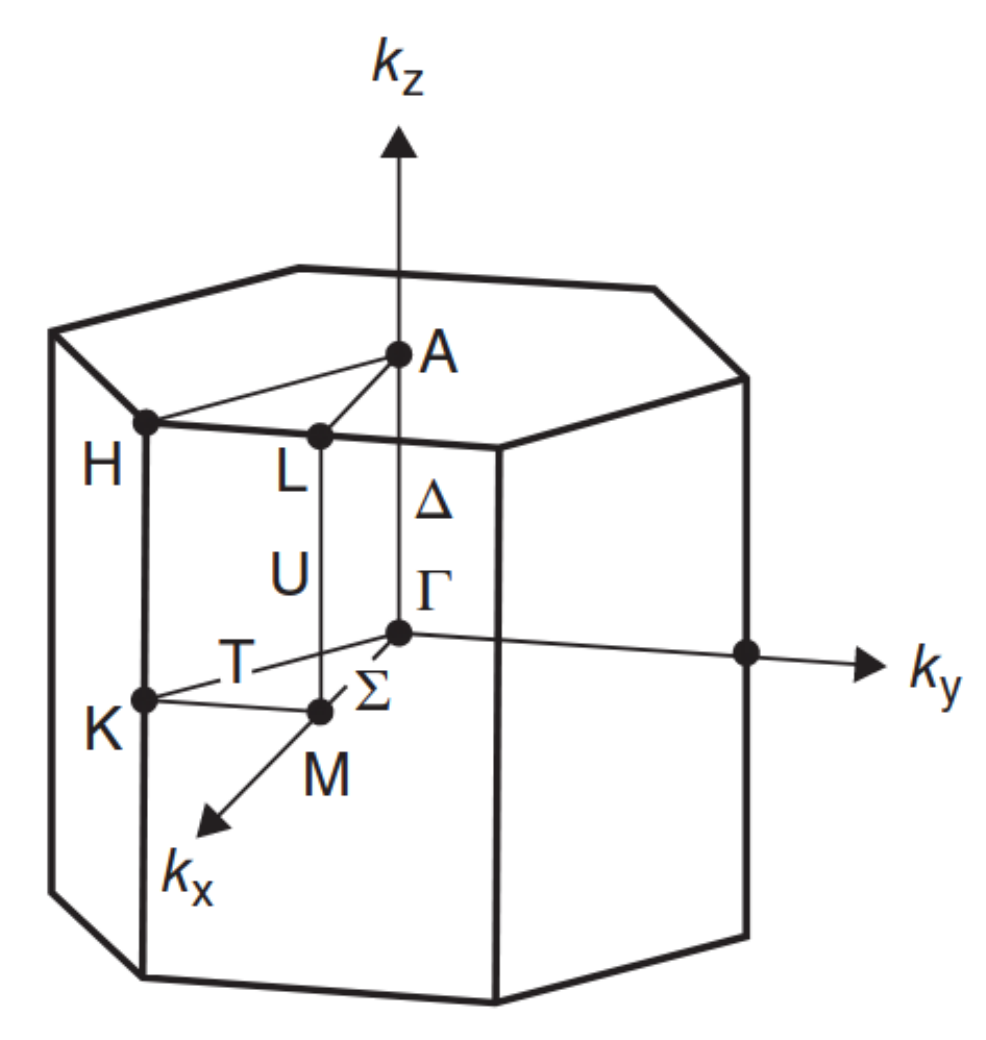}
\caption{\label{fig:brillouin_zone}Characteristic lattice locations of 4H-SiC
  within the unit cell. The included \textit{$k_x$}, \textit{$k_y$}, and \textit{$k_z$} directions are according to Miller-Bravais notation. Reproduced with permission from Fundamentals of Silicon Carbide Technology~\cite[chap. 2]{kimoto2014a}. Copyright 2014, John Wiley \& Sons Singapore Pte. Ltd.}
\end{figure}

Within the unit cell the characteristic symmetry points $\Gamma$, M, K and A were defined
(see \cref{fig:brillouin_zone})~\cite{tanaka2018, wang2020a}, resulting in the following directions:
\begin{itemize}
    \item $\Gamma-$A~\rightarrow~\hkl[0001] direction
    \item $\Gamma-$M~\rightarrow~\hkl[1-100] direction
    \item $\Gamma-$K\rightarrow~\hkl[11-20] direction
\end{itemize}

\begin{figure}[t]
\centering
\includegraphics[width=0.69\textwidth]{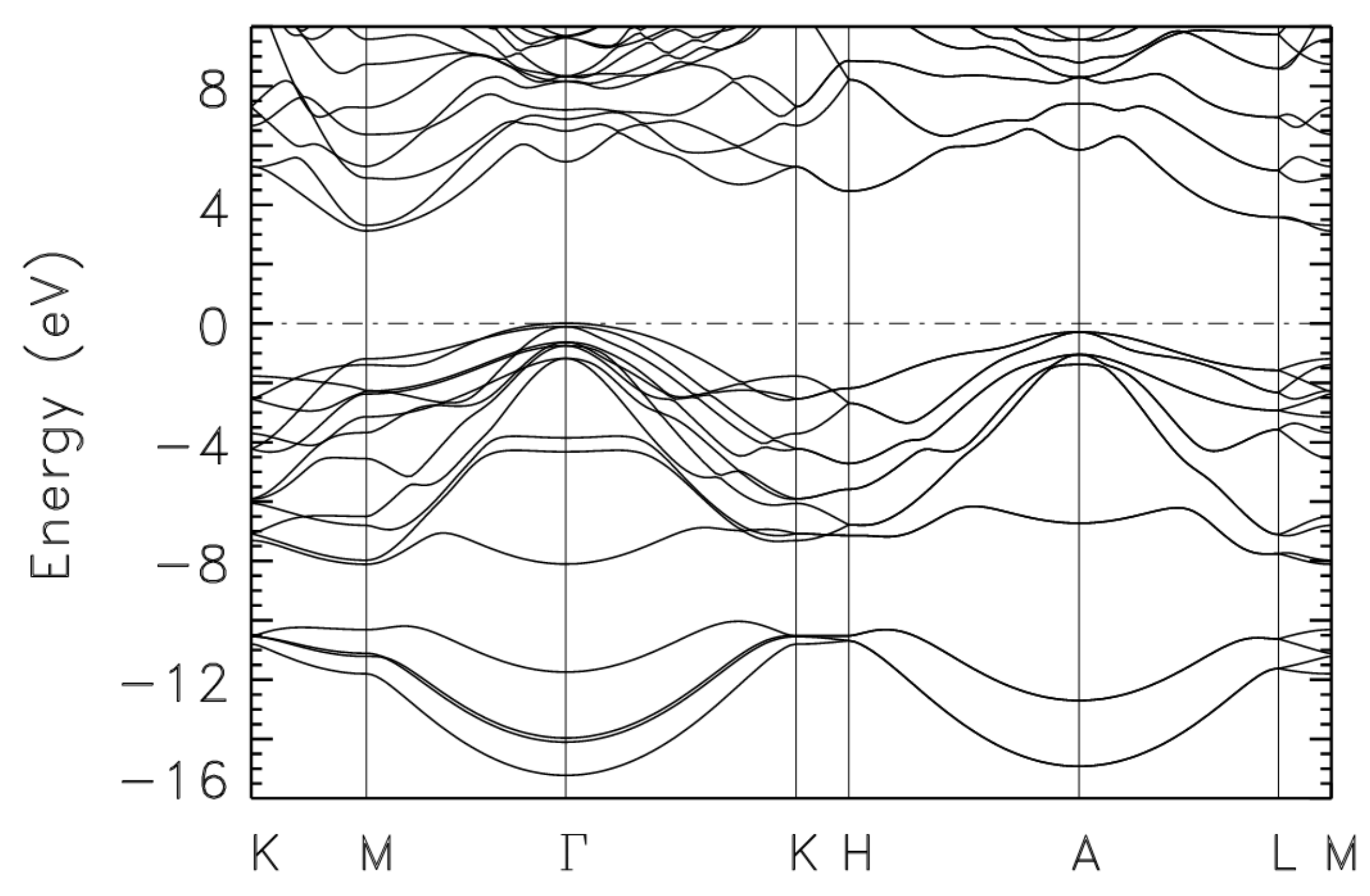}
\caption{\label{fig:band_structure}Electronic energy band structure of 4H-SiC
  calculated utilizing the \textit{ab initio} linear combination of atomic
  orbital (LCAO) method (reprinted from \citet{zhao2000a} under open access license).}
\end{figure}

\revision{The calculated electronic band structure in between these symmetry
  points~\cite{bellotti2000,kackell1994,wellenhofer1997,bellotti1999,ching2006,zhao2000a,chen1997}
  reveals that 4H-SiC is an indirect semiconductor (see
  \cref{fig:band_structure}). In detail, the conduction band is located at the M
  point with two conduction bands within \SI{200}{\milli\eV} while the valence
  band maximum resides at the $\Gamma$ point, featuring three valence bands separated by less than
  \SI{100}{\milli\eV}.}

The anisotropic nature of 4H-SiC in simulations is usually handled in a much simpler way than the above definitions may indicate. In general, we differentiate two cases when discussing material and electrical parameters of 4H-SiC: A parallel and a perpendicular component with respect to the c-axis (or principal axis and [0001] direction)~\cite{powell1993, ayalew2004, nava2008, buono2012, kimoto2014a, kimoto2019, denapoli2022, dipaoloemilio2024}. These components will be denoted by $\parallel$ and $\perp$ throughout this work.


\section{\label{sec:permittivity}Permittivity}

The permittivity $\varepsilon$ describes charge movement in response to internal/external potentials, how the latter are screened within a material~\cite{bellotti1999}, the dielectric properties that influence electromagnetic wave propagation as well as their reflections on interfaces~\cite{komarov2005}. In TCAD simulations, the main application of the permittivity is the Poisson equation~\cite{sze2012}, which describes the electric field induced by a charge. This is important, for example, to
\begin{enumerate*}[label=(\roman*)]
    \item calculate capacitances within a device,
    \item determine the doping concentrations of measured samples and
    \item investigate the impact of defects and traps.
\end{enumerate*}

All TCAD tools require a value input of the relative permittivity $\er=\varepsilon/\varepsilon_0$, i.e., the ratio of $\varepsilon$ to the vacuum permittivity $\ezero = \SI{8.854e-12}{\farad\per\m}$. Accurate values are important for two reasons:
\begin{enumerate*}[label=\arabic*)]
\item Basic characteristics like the capacitance in response to an applied external voltage are often the foundation for more advanced investigations. Small errors in the beginning multiply, causing larger deviations down the line.
\item Depending on material and investigation, the amount of charge carriers can
  be very high (\SI{e18}{\per\cubic\centi\m} and more), which amplifies any initial inaccuracy.
\end{enumerate*}

Our review shows that although fundamental investigations are published up to the present day, they rarely were referenced in the literature. Instead, the mainly deployed values trace back to a publication from the year 1970, which is itself based on measurements of the 6H-SiC polytype from 1944.

\subsection{Introduction}

The relative permittivity is a complex function of the frequency $\omega$ (see \cref{eq:permittivity_er}). The real part $\varepsilon'$ represents the energy stored in the material when exposed to an electric field, and the imaginary part $\varepsilon''$ the losses, e.g., absorption and attenuation~\cite{komarov2005}.
\begin{equation}
    \label{eq:permittivity_er}
   \er^*(\omega) = \varepsilon'(\omega) + \mathrm{i} \varepsilon''(\omega)
\end{equation}
$\varepsilon'$ and $\varepsilon''$ are tightly interconnected via the Kramers-Kronig (KK) relation~\cite{adachi2005} (see \cref{eq:permittivity_kk}). In TCAD simulations of semiconductor devices, $\varepsilon''$ is of low importance, while $\varepsilon'$ is featured in the Poisson equation. Consequently, we will focus on $\varepsilon'$ in the sequel.
\begin{subequations}
\label{eq:permittivity_kk}
\begin{eqnarray}
    \varepsilon'(\omega) &=& 1 + \frac{2}{\mathrm{\uppi}} \int_0^\infty \frac{\omega' \varepsilon''(\omega')}{\omega'^2-\omega^2} \mathrm{d}\omega' \\ 
     \varepsilon''(\omega) &=& - \frac{2\omega}{\uppi} \int_0^\infty \frac{ \varepsilon'(\omega')}{\omega'^2-\omega^2} \mathrm{d}\omega'
\end{eqnarray}    
\end{subequations}

\subsubsection{Static and High-Frequency Relative Permittivity}

Although investigations on the frequency-dependent $\varepsilon'$ exist~\cite{naftaly2016,tarekegne2019,li2023,hartnett2011}, we focus here on the static ($\es$) and high-frequency resp. optical ($\ei$) relative permittivity. The former is declared as $\es=\varepsilon'(\omega \to 0)$, while a definition for the latter is more complicated. $\ei$ denotes $\varepsilon'$ at the end of the reststrahlen range towards higher frequencies, where the real part of the refractive index is null~\cite{olivares2013}. In a publication that focused on optical high-frequency analysis~\cite{ching2006}, $\ei$ was even denoted as $\varepsilon'(0)$, which must not be confused with $\es$. These inconsistencies were explained by \citet{patrick1970} in the following fashion: ``We shall use $\ei$ to denote the extrapolation $\hdots$ to zero frequency. This somewhat contradictory notation arose because $\ei$, the "optical" dielectric constant, was often set $\hdots$ at a frequency much higher than the lattice frequency but low compared with electronic transition frequencies. In many substances no suitable frequency exists, and it is preferable to extrapolate optical data to zero frequency $\hdots$''

Thus, $\ei$ is measured at frequencies well above the long-wavelength longitudinal optical (LO) phonon frequency $\omlo$~\cite{adachi2005}. The latter is used in conjunction with the transversal optical (TO) phonon frequency $\omto$ in the Lyddane-Sachs-Teller (LST) relationship~\cite{lyddane1941} to translate between $\es$ and $\ei$ (see \cref{eq:permittivity_lst}).
\begin{equation}
\label{eq:permittivity_lst}
\frac{\es}{\ei} = \left( \frac{\omega_{LO}}{\omega_{TO}}\right)^2
\end{equation}

\subsubsection{Characterization Methods}

\revision{One possibility to determine the relative permittivity are
  calculations, which include density functional theory (DFT) based local density approximations
(LDA)~\cite{adolph1997,peng-shou2004}
using the HSE06 functional (HSE06)~\cite{coutinho2017}, plane-wave
pseudopotential method (PWP)~\cite{karch1996,wellenhofer1996},
full-potential linear muffin-tin-orbital method (FPLMTO)~\cite{ahuja2002},
orthogonalized linear combination of atomic orbitals method
(OLCAO)~\cite{ching2006}, QP model (LDA+QP)~\cite{persson2005}
or the effective mass theory~\cite{ivanov2006}. These approaches determine the band
structure and, thus, $\varepsilon''$ before transforming it into $\varepsilon'$
using the Kramers-Kronig relations (\cref{eq:permittivity_kk}).} Measurements use
either some form of resonator (RES)~\cite{hartnett2011,jones2011,li2023},
spectroscopy (SPEC)~\cite{naftaly2016,tarekegne2019,gao2022a}, spectroscopic ellipsometry (SE)~\cite{ninomiya1994,cheng2022}, infrared ellipsometry (IRE)~\cite{mainali2024}, the refractive index (RI)~\cite{patrick1970,ikeda1980,harima1995} and a fitting to $\ln(J/F)$~\cite{yang2022a}, with $J$ the current density and $F$ the field in a pin-diode. In two instances the transversal and longitudinal optical phonon frequencies were used in the Lyddane-Sachs-Teller relationship (\cref{eq:permittivity_lst})~\cite{cheng2022,ninomiya1994}.

If $\varepsilon''$ is negligible, the permittivity can be computed from its tight relationship with the complex refractive index $n^*$, i.e., $n^*(E) = \sqrt{\varepsilon_r(E)}$~\cite{adachi2005}, with $E=\hbar \omega$. In four publications~\cite{patrick1970,ikeda1980,tarekegne2019,harima1995} $\ei$ was fitted by the representation shown in \cref{eq:permittivity_refr}, with $hf$ being the photon energy, $\eg$ an "average band gap" and $\varepsilon_\mathrm{g}$ proportional to the oscillator strength~\cite{patrick1970}. Six publications investigated the relative permittivity at \si{\milli\meter}-wave frequencies (MM) (\SI{10}{\GHz} to \SI{10}{\THz})~\cite{dutta2006,chen2008,jones2011,li2023a,li2024b,afsar2005}, whose results got summarized by \citet{li2023} and \citet{yanagimoto2024}.
\begin{equation}
    \label{eq:permittivity_refr}
    n^2 = 1 + \frac{\varepsilon_g}{1-(hf/\eg)^2} \approx \ei + \varepsilon_\mathrm{g} \left( \frac{hf}{\eg} \right)
\end{equation}

\subsection{Results \& Discussion}


Publications had to be discarded for this review when no polytype was specified~\cite{schoner2004, polaert2017,hofman1957,agarwal2004} or powder material was investigated~\cite{baeraky2002,yang2013}. \citet{ivanov2006} improved upon the value $\es=9.95$~\cite{ivanov2003,ivanov2003a} causing us to discard the outdated publications. For measurements at \si{\milli\meter}-wave frequencies, the data fluctuated, or solely single data points were provided. Since this makes an interpolation to zero frequency ($=\es$) impossible we did not consider the respective data. We were unable to confirm some of the values referenced in the overview of high-frequency permittivities by \citet{bechstedt1997} because we could not acquire the book by \citet{bimberg1982}. Out of $206$ reviewed publications, only $179$ provided a comprehensive set of data and were included in our analysis.

\subsubsection{Fundamental Values}

\begin{table*}[p!]
    \centering
        \setlength{\tabcolsep}{9pt}
        \renewcommand{\arraystretch}{1.2}
        \caption{\label{tab:permittivity}Fundamental investigations on
          permittivity. \bkgCol{no4HColor} mark publications focused on 6H-SiC and \bkgCol{calcColor} calculations.}
    \resizebox{0.98\linewidth}{!}{%
        \begin{tabular}{l *{6}{c} *{3}{c}}
ref. & $\es$ & $\espa$ & $\espe$ & $\ei$ & $\eipa$ & $\eipe$ & method & SiC & doping \\ \hline 
\rowcolor{no4HColor} \reftext{Patr70}\cite{patrick1970}\footnotemark[1] & \num{9.78}\footnotemark[8] & \num{10.03} & \num{9.66} & \num{6.58}\footnotemark[8] & \num{6.7} & \num{6.52} & RI & 6H & -- \\ 
\reftext{Iked80}\cite{ikeda1980}\footnotemark[2] & \num{9.94}\footnotemark[8] & \num{10.32} & \num{9.76} & -- & -- & -- & RI & 4H & -- \\ 
\rowcolor{calcColor} \reftext{Chen94}\cite{chen1994} & -- & -- & -- & \num{7.04}\footnotemark[8] & \num{7.20} & \num{6.96} & DFT-LDA & 4H & -- \\ 
\rowcolor{no4HColor} \reftext{Nino94}\cite{ninomiya1994} & \num{9.83}\footnotemark[8] & \num{9.98} & \num{9.76} & \num{6.62}\footnotemark[8] & \num{6.67} & \num{6.59} & SE & 6H & -- \\ 
\reftext{Hari95}\cite{harima1995}\footnotemark[3] & -- & -- & -- & \num{6.63}\footnotemark[8] & \num{6.78}\footnotemark[4] & \num{6.56}\footnotemark[4] & RI & 4H & -- \\ 
\rowcolor{calcColor} \reftext{Karc96}\cite{karch1996} & \num{10.53}\footnotemark[8] & \num{10.9} & \num{10.352} & \num{7.02}\footnotemark[8] & \num{7.169} & \num{6.946} & PWP & 4H & -- \\ 
\rowcolor{calcColor} \reftext{Well96}\cite{wellenhofer1996} & -- & -- & -- & \num{7.02}\footnotemark[8] & \num{7.17} & \num{6.95} & PWP & 4H & -- \\ 
\rowcolor{calcColor} \reftext{Adol97}\cite{adolph1997} & -- & -- & -- & \num{7.56}\footnotemark[8] & \num{7.61} & \num{7.54} & DFT-LDA & 4H & -- \\ 
\rowcolor{calcColor} \reftext{Ahuj02}\cite{ahuja2002} & -- & -- & -- & \num{7.11}\footnotemark[8] & \num{7.47} & \num{6.94} & FPLMTO & 4H & n-type \\ 
\rowcolor{calcColor} \reftext{Peng04}\cite{peng-shou2004} & -- & -- & -- & \num{6.31}\footnotemark[8] & \num{6.44} & \num{6.25} & DFT-LDA & 4H & -- \\ 
\rowcolor{calcColor} \reftext{Pers05}\cite{persson2005} & \num{9.73}\footnotemark[8] & \num{9.94} & \num{9.63} & \num{6.47}\footnotemark[8] & \num{6.62} & \num{6.40} & LDA+QP & 4H & intrinsic \\ 
\rowcolor{calcColor} \reftext{Chin06}\cite{ching2006} & -- & -- & -- & \num{6.81} & -- & -- & OLCAO & 4H & intrinsic \\ 
\reftext{Dutt06}\cite{dutta2006} & \num{9.97+-0.02}\footnotemark[5] & -- & -- & -- & -- & -- & MM & 4H & high purity \\ 
\reftext{Ivan06}\cite{ivanov2006} & \num{9.93+-0.01} & -- & -- & -- & -- & -- & EMT & 4H & intrinsic \\ 
\reftext{Hart11}\cite{hartnett2011} & \num{9.77}\footnotemark[7] & -- & -- & -- & -- & -- & RES & 4H & high purity \\ 
\reftext{Jone11}\cite{jones2011} & \num{9.60}\footnotemark[6] & -- & -- & -- & -- & -- & RES & 4H & high purity \\ 
\reftext{Naft16}\cite{naftaly2016} & \num{10.11}\footnotemark[8] & \num{10.53}\footnotemark[7] & \num{9.91}\footnotemark[7] & -- & -- & -- & SPEC & 4H & undoped \\ 
\rowcolor{calcColor} \reftext{Cout17}\cite{coutinho2017} & \num{10.13}\footnotemark[8] & \num{10.65} & \num{9.88} & -- & -- & -- & HSE06 & 4H & -- \\ 
\reftext{Tare19}\cite{tarekegne2019} & -- & -- & -- & \num{6.587+-0.003} & -- & -- & SPEC & 4H & -- \\ 
\reftext{Chen22}\cite{cheng2022} & -- & -- & \num{9.97} & -- & -- & -- & SE & 4H & -- \\ 
\reftext{Gao22a}\cite{gao2022a} & -- & -- & -- & \num{6.51} & -- & -- & SPEC & 4H & -- \\ 
\reftext{Yang22a}\cite{yang2022a} & \num{10.21} & -- & -- & -- & -- & -- & $\ln(J/F)$ & 4H & p-type \\ 
\reftext{Li23}\cite{li2023} & -- & \num{10.27+-0.03} & -- & -- & -- & -- & RES & 4H & high purity \\ 
\reftext{Li24b}\cite{li2024b} & \num{9.91}\footnotemark[8] & \num{10.2+-0.05} & \num{9.77+-0.01} & -- & -- & -- & RES & 4H & high purity \\ 
\reftext{Main24}\cite{mainali2024} & -- & -- & -- & -- & -- & \num{6.40+-0.2} & IRE & 4H & -- \\ 
\end{tabular}
        \footnotetext[1]{fitted to refractive index values by \citet{thibault1944}}
        \footnotetext[2]{fitted to refractive index values by \citet{shaffer1971} using 6H phonon frequencies}
        \footnotetext[3]{fitted to refractive index values by \citet{shaffer1971}}
        \footnotetext[4]{referenced by \citet{bechstedt1997} from \reftext{Bimb82}~\cite{bimberg1982}}
        \footnotetext[5]{frequency range \SIrange{131}{145}{\GHz}, for lower resp. higher frequencies $\es=9.74$ was achieved}
        \footnotetext[6]{calculated from refractive index $n$ as $\es=n^2$}
        \footnotetext[7]{wavelength and/or temperature dependent}
        \footnotetext[8]{$\er = ((\erpe)^2 \erpa)^{1/3}$}        
    }
\end{table*}

We found static relative permittivity ($\es$) values within the range \numrange{9.6}{10.65} and high-frequency relative permittivity ($\ei$) within \numrange{6.25}{7.61} (see \cref{tab:permittivity}), whereas the direction parallel to the c-axis achieves larger values than the perpendicular ones. Wherever necessary, we calculated the relative permittivity according to $\es=(\espa\,(\espe)^2)^{\frac{1}{3}}$ resp.  $\ei=(\eipa\,(\eipe)^2)^{\frac{1}{3}}$~\cite{lindefelt1998,pernot2001,pernot2005,koizumi2009,tanaka2018}, which was more often used than $\es=(\espa \espe)^{\frac{1}{2}}$ proposed by~\citet{ivanov2006}.

Early investigations~\cite{patrick1970,ikeda1980,harima1995} utilized \cref{eq:permittivity_refr} to extract the permittivity from varying refractive index data, some measured on 6H-SiC samples. After the year 1996 dominated calculations of $\ei$ and past 2006 measurements of $\es$. Despite these different phases, we were unable to identify any trends over time (see \cref{fig:permittivity_fund}). On the contrary: The 6H-SiC permittivity values are similar to 4H. This is underlined by the statements of \citet{zollner1999} who stated that for energies below \SI{3}{\eV} their results are comparable to 6H values~\cite{choyke1997a}.

\revision{The high frequency permittivity was mainly determined by calculations, which deliver higher values than the measurements (with two exceptions~\cite{persson2005,peng-shou2004}) and the statistical mean of $\ei=6.79$. We made the same observation for the static permittivity with a mean of $\es=9.96$, with the difference that in this case measurements were more often deployed than calculations. \citet{karch1996} proposed remarkably high values using the plane-wave pseudopotential method, while \citet{jones2011} extracted very low ones from the refractive index.}

\begin{figure}[t!]
    \centering
    \resizebox{0.99\textwidth}{!}{%
    \input{figures/permittivity_fund}}
    \caption{\label{fig:permittivity_fund}Fundamental investigations on permittivity in chronological order.}
\end{figure}

\subsubsection{Origin of Parameters}


The primary source for 4H-SiC permittivity in literature is the investigation by \citet{patrick1970} (for a detailed analysis see \cref{fig:permittivity_ref_chain} in Appendix~\ref{sec:refChainPermittivity}) based on 6H measurements published in 1944~\cite{thibault1944}. The usage of 6H values was justified in multiple publications~\cite{harris1995,ladesmartin2000,pernot2001,persson1999,son2004,hatakeyama2013,bakowski1997} by the unavailability of 4H specific parameters. The fact that this claim is repeated up to the year 2022~\cite{cheng2022} shows the high demand for a systematic parameter analysis. In the majority of cases, 6H values were, however, used for 4H analyses without further remark, which left the false impression of specific 4H parameters.

This unawareness of dedicated 4H-SiC permittivities is surprising, especially because \citet{ikeda1980} published corresponding values already in 1980. Based on measurements by \citet{shaffer1971} but using the phonon frequency of 6H, direction-dependent values for $\es$ were, easy to overlook, added to a comment in the reference list. Interestingly, we encountered the proposed values on multiple occasions~\cite{kimoto2014a,kimoto2019,pensl2005,reshanov2005,schadt1997,troffer1998} but never found a direction citation of that particular paper. Not even cross-references among the citing publications exist.

The negligence of 4H-SiC permittivity investigations seems to be deeply rooted within the scientific community. Only $5$ out of the $23$ fundamental research articles were cited within the boundaries of this review. The fact that six got published after the year 2021 provides just a partial explanation. Although long reference chains (cp. \cref{fig:permittivity_ref_chain} in Appendix~\ref{sec:refChainPermittivity}) hide the fact that these are 6H-based values, we also found papers within the last five years referencing directly to \citet{patrick1970}. Another interpretation of these insights is that the permittivity has little impact in TCAD simulations, such that somewhat accurate 6H values are often sufficient. The difference is anyhow well within \SI{5}{\percent}. Nevertheless, we highly encourage the scientific community to also adopt more recent permittivity evaluations on 4H-SiC in future publications, or at least clearly specify and comment on the usage of 6H parameter values.

\subsubsection{Literature Values}

\begin{figure*}[p!]
    \centering
    \resizebox{\textwidth}{!}{%
    \input{figures/permittivity_eps}
    }
    \caption{\label{fig:permittivity_eps}Published values for the static permittivity. The size of values indicates the abundance in literature and connections that at least one direct reference was found. Brighter colors highlight the initial values. \bkgCol{no4HColor} publications are not focused on 4H-SiC and \bkgCol{fundColor} are novel analyses on 4H-SiC.}
\end{figure*}

We found evidence that the 6H values by \citet{patrick1970} were used in $41$ publication, either referenced directly or via intermediate publications. Thereby, the values were transformed by rounding, e.g., $9.66 \rightarrow 9.7 \rightarrow 10$~\cite{neudeck2001,arpatzanis2006} (cp. \cref{fig:permittivity_ref_chain} in Appendix~\ref{sec:refChainPermittivity}), defining directional permittivities as effective ones, e.g., $\espe=9.66 \rightarrow \es=9.7$ or $\espa=10.03 \rightarrow \es=10$, and mere typographical errors, e.g., turning \num{9.66} into \num{9.67}~\cite{kovalchuk2020} (see Appendix~\ref{sec:perm_appendix} for a comprehensive analysis of all encountered inconsistencies). Over the years, these changes expanded the initial data set to such an extent that the majority of all values in literature are covered (see \cref{fig:permittivity_eps,fig:permittivity_hf}). Due to missing references, we cannot say for sure that all authors who picked the same value made their selection based on the same data. For example, values like $\es=9.7$ could be derived from both \citet{patrick1970} and \citet{ikeda1980}. In total, we found a direct connection of these two papers alone to the values deployed in \SI{80}{\percent} of the investigated publications, which confirms again the negligence of later 4H-SiC permittivity investigations.

\begin{figure*}[t!]
    \centering
    \resizebox{\textwidth}{!}{%
    \input{figures/permittivity_hf}
    }
    \caption{\label{fig:permittivity_hf}Published values for the static permittivity. The size of values indicates the abundance in literature and connections that at least one direct reference was found. Brighter colors highlight the initial values. \bkgCol{no4HColor} publications are not focused on 4H-SiC and \bkgCol{fundColor} are novel analyses on 4H-SiC.}
\end{figure*}

In two cases, we had a hard time extracting and verifying permittivity results. \citet{chow1993} specified the permittivity as a multiple of the permittivity in Si. For a comparison we picked $\es{}_{\mathrm{Si}}=11.7$~\cite{ioffe2023}.
Challenging is $\es=8.5584$~\cite{acharyya2017,banerjee2021}, according to the authors computed from $\espe=9.66$ and $\espa=10.03$~\cite{ioffe2023}. We were not able to calculate this value analytically since the result is lower than both constituents. Only by adding the high-frequency relative permittivity to the mix we achieved a somewhat close value of $8$.

\subsubsection{Temperature Dependency and Phonon Frequencies}

The static permittivity is temperature-dependent. Although TCAD tools do not support temperate-dependent permittivities by default, it is possible to cover these changes by custom code in the simulations. For a frequency around \SI{40}{\GHz}, \citet{hartnett2011} provided a polynomial approximation up to degree four (see Eq.~\ref{eq:permittivity_hartnett}). \citet{cheng2022} approximated $\espe=9.82 + \num{4.87e-4}\,T$ and \citet{li2024b} $\espe = 9.77\,(1 + \num{6e-5} (T-\SI{300}{\K}))$ and $\espa = 10.2\,(1 + \num{1e-4} (T-\SI{300}{\K}))$. All these models show an increase of the permittivity with rising temperature (\cref{fig:permittivity_temp}).
\begin{equation}
    \begin{split}   
    \label{eq:permittivity_hartnett}
    \es(T) =  9.7445 &+ \num{3.1862e-5}\,T - \num{6.3026e-7}\,T^2 \\
    &+ \num{5.9848e-9}\,T^3 - \num{8.2821e-12}\,T^4
    \end{split}
\end{equation}

\begin{figure}
    \centering
    \resizebox{0.7\textwidth}{!}{%
    \input{figures/permittivity_temp}
    }
    \caption{Permittivity vs. temperature. The models are only shown in the interval used for characterization.}
    \label{fig:permittivity_temp}
\end{figure}

While we investigated the permittivity, we encountered various values of the longitudinal ($\omlo$) and transversal optical ($\omto$) phonon frequencies in 4H-SiC (see \cref{tab:permittivity_phonon}). For $\omlo$, we found values of \SIrange{104}{120}{\milli\electronvolt} and \SIrange[per-mode=power]{964}{992}{\per\centi\m}. The two ranges do not fully overlap, as an energy of \SI{104}{\milli\electronvolt} translates to \SI[per-mode=power]{838}{\per\centi\m}. For $\omto$ the values of \SIrange{95}{100}{\milli\electronvolt} and \SIrange[per-mode=power]{776}{798}{\per\centi\m} match better. Due to the fact that this was not the result of a targeted search we do not claim completeness.

\begin{table}[th!]
    \setlength{\tabcolsep}{7pt}
    \centering
    \caption{\label{tab:permittivity_phonon}Phonon frequencies in 4H-SiC. No claim for completeness.}
    \resizebox{0.95\linewidth}{!}{%
    \begin{tabular}{c| *{8}{c}}
         ref & $\omlo$ & $\omlo$ & $\omlopa$ & $\omlope$ & $\omto$ & $\omto$ & $\omtopa$ & $\omtope$  \\
         & [\si{\milli\electronvolt}] & [\si{\per\centi\meter}] & [\si{\per\centi\meter}]& [\si{\per\centi\meter}] & [\si{\milli\electronvolt}] & [\si{\per\centi\meter}]& [\si{\per\centi\meter}]& [\si{\per\centi\meter}] \\ \hline
         &&&&&&&& \\[-7pt] 
         \reftext{Feld68}\cite{feldman1968} &--& \numrange{967}{971} &--&--&--&--&--&-- \\ %
         \reftext{Neub71}\cite{neuberger1971} &  \num{120}\footnotemark[1] &--&--&--& \num{100} &--&--&--\\
         \reftext{Hari95}\cite{harima1995} &--&--& \num{964.2} & \num{966.4} &--&--& \num{783} & \num{798} \\
         \reftext{Frei95}\cite{freitas1995} & \num{104.1+-0.2} &--&--&--& \num{95} &--&--&--\\
         \reftext{Hari98}\cite{harima1998} &--& \num{964} &--&--&--& \num{783} &--&-- \\
         \reftext{Tiwa99}\cite{tiwald1999} &--&--& \num{967+-3} & \num{971} &--&--& \num{782} & \num{797} \\
         \reftext{Levi01}\cite{levinshtein2001} & \num{104.2}\footnotemark[2] &--&--&--&--&--&--&--\\
         \reftext{Sun11a}\cite{sun2011a} &--& \num{984+-21} &--&--&--& \num{776.4+-1} &--&-- \\
         \reftext{Arvi17}\cite{arvinte2017} &--& \num{974} &--&--&--& \num{793} &--&-- \\
         \reftext{Zhen19}\cite{zheng2019} &--& \num{964} &--&--&--& -- &--&-- \\
         \reftext{Main24}\cite{mainali2024} &--&--&--& \num{992.1+-0.2} &--&--&--& \num{797.7+-0.3}
    \end{tabular}
    \footnotetext[1]{corresponds to $\SI{29}{\THz}$, same value proposed in \cite{ahuja2002,iwata2001,hjelm2003,mickevicius1998a,persson1997,bellotti1999}}
    \footnotetext[2]{denoted as 6H by \citet{neuberger1971}}
    }
\end{table}


\section{Density-of-States Mass}
\label{sec:dos}

TCAD tools use effective masses to simplify the description of specific effects. Examples are tunneling masses, quantum well masses, effective mass at the contact in a channel, thermionic relative masses, conductivity masses, and density-of-states masses. Detailed information on each can be found in the respective simulation framework manual. In this section we review the density-of-states (DOS) mass~\cite{adachi2003}, which is used in TCAD simulations, e.g., to calculate the charge carrier concentration or impact and incomplete ionization.

Overall, our review reveals that the majority of investigations on the density-of-states masses in 4H-SiC were conducted in a single decade between $1994$ and $2004$. The earliest studies in the 1970s were focused on measurements, while nowadays calculations are utilized predominantly. This led to the concerning circumstance that we were only able to identify two measurements for the hole DOS mass. The demand for a more thorough characterization is also fueled by the significant temperature dependency (the hole mass more than doubles between zero and \SI{300}{K}) that was predicted by calculations but is rarely considered in literature.


\subsection{Introduction}

\revision{The density-of-states masses are calculated from the direction-dependent effective masses of electrons and holes. The latter are used to approximate the energy bands in the conduction band minimum respectively valence band maximum in various directions~\cite{adachi2005,kackell1994} (cp. \cref{fig:band_structure} in \cref{sec:sic}). In a parabolic approximation (see \cref{eq:dos_parabolic})~\cite{adachi2005} the effective mass $m$, which is denoted for better readability relative to the free electron mass $m_0$, i.e., $m^* = m/m_{0}$~\cite{son2000}, denotes the reciprocal of the curvature. Consequently, rapid changes in the energy bands result in low masses while slow changes indicate high masses.}
\begin{equation}
  \label{eq:dos_parabolic}
  E = E_0 \pm \frac{(\hbar k)^2}{2\,m^*}
\end{equation}

We start our analysis with a discussion of the direction-dependent masses and
then show how the DOS (and also the conductivity) masses are derived. For
further information, the interested reader is referred to the dedicated
literature~\cite{adachi2005,harris1995,chen1997a,bechstedt1998,persson2005,schadt1997,son2004,wellenhofer1997,pensl2005}.

\subsubsection{Effective Masses along Principal Directions}

For electrons, the masses are specified in the directions starting in the conduction band minimum at the M point~\cite{persson1996,ayalew2004,chen1997a,karch1995,dong2004} (cp. \cref{sec:sic}) towards the $\Gamma$, K and L point, denoted in the sequel as $\memg$, $\memk$ and $\meml$. The first two are perpendicular to the c-axis, the latter one parallel~\cite{ayalew2004,bakowski1997,iwata2003,nakashima1997}. In the M point, two conduction bands are very close together, such that both can influence the effective mass~\cite{dong2004,schadt1997}. \citet{zhao2000} even used three conduction bands. If not stated otherwise, we will focus solely on the lowest one within this paper.

The valence band maximum is in the $\Gamma$ point~\cite{ayalew2004,bakowski1997,troffer1998}. Consequently, the three relative hole masses are termed $\mhgm$, $\mhgk$ (perpendicular), and $\mhga$ (parallel), indicating the directions towards the M, K, and A point~\cite{kuroiwa2019,dong2004}. Three bands are very close at the $\Gamma$ point~\cite{persson1996,bakowski1997,son2004,kuroiwa2019,lambrecht1997}. The two topmost are called heavy-hole (hl) and light-hole (ll), and the third one crystal split off (so)~\cite{son2004,lambrecht1997}. Although all may influence the effective mass, often only a subset is used.


\subsubsection{Density-of-States (DOS) Mass}

The effective density-of-states is, for example, used to evaluate the electron and hole concentration. It is defined separately for conduction ($\nc$) and valence ($\nv$) band~\cite{scaburri2011a,blakemore1962,zippelius2011,ladesmartin2000,gotz1993,suttrop1991} (see \cref{eq:dos_dos}) with $\mc$ equal the number of conduction band minima in the first Brillouin zone~\cite{gotz1993}.
\begin{equation}
    \label{eq:dos_dos}
\begin{split}  
 \nc &= 2\;\mc \left( \frac{2\pi \mde \kb T}{h^2}\right) ^{3/2} \\
 \nv &= 2 \left( \frac{2\pi \mdh \kb T}{h^2}\right) ^{3/2}
 \end{split}
\end{equation}
$\mde$ and $\mdh$ denote the effective density-of-states masses, which are defined as the multiplicative average of the direction-dependent masses (see \cref{eq:dos_masses})~\cite{wellenhofer1997,pensl2005,egilsson1999,bakowski1997,suttrop1991,kohlscheen2003,son2004}.
\begin{equation}
    \label{eq:dos_masses}
    \begin{split}  
    \mde &= ((\mdeperp)^2\,\mdepara)^{1/3} = (\memg \memk \meml)^{1/3} \\
    \mdh &= ((\mdhperp)^2\,\mdhpara)^{1/3} = (\mhgm \mhgk \mhga)^{1/3}\\
    \mdeperp &= \sqrt{\memg \memk} \ , \qquad \mdepara =\meml \\
    \mdhperp &= \sqrt{\mhgm \mhgk} \ , \qquad \mdhpara = \mhga
    \end{split}
\end{equation}

In this case $\mde$ is called the \textit{single valley} DOS electron effective mass~\cite{pensl2005,ayalew2004,egilsson1999,hatakeyama2013}, as the factor $\mc$ is not considered. It is common, e.g., in some TCAD tools, to add $\mc$ to the effective mass~\cite{adachi2005,albanese2010,bakowski1997,hemmingsson1997,scajev2013a} (see \cref{eq:dos_MC}). In this review, we will only present the single valley values and highlight all publications where we found expressions including $\mc$.
\begin{equation}
    \label{eq:dos_MC}
    \mde = (\mc^2\,(\mdeperp)^2\,\mdepara)^{1/3}
\end{equation}

In addition to the direction, the effective mass of the holes also has to combine multiple bands, i.e., the masses of heavy ($\mhhh$) and light ($\mhlh$) holes (see \cref{eq:dos_mdh})~\cite{lindefelt1998,sze2007,pernot2005}.
\begin{equation}
    \label{eq:dos_mdh}
    \mdh = \left(\mhhh {}^{3/2} + \mhlh {}^{3/2} \right)^{2/3}
\end{equation}
This expression is already a simplification because the energy differences among the bands also have to be considered (see \cref{eq:dos_mdh_temperature})~\cite{bakowski1997,raynaud2010}. $\Delta E_2$ and $\Delta E_3$ denote the energy difference to the highest band, which were determined as \SI{9}{\milli\eV}~\cite{bakowski1997} resp. \SI{8.6}{\milli\eV}~\cite{persson1997} for the second band and \SI{73}{\milli\eV}~\cite{bakowski1997}, \SI{77}{\milli\eV}~\cite{persson1998a} resp. \SI{72}{\milli\eV}~\cite{raynaud2010} for the third band.
%
\begin{equation}
    \label{eq:dos_mdh_temperature}
    \mh(T) = \left[m_{\mathrm{h}1}^* {}^{3/2} + m_{\mathrm{h}2}^* {}^{3/2}
      \exp\left(-\frac{\Delta E_2}{\kb T}\right)+  m_{\mathrm{h}3}^* {}^{3/2} \exp\left(-\frac{\Delta E_3}{\kb T}\right)\right]^{2/3}
\end{equation}

The changing amount of charge carriers with temperature can be compactly modeled
by using a \textit{thermal DOS effective
  mass}~\cite{wellenhofer1997,khanna2023}. There is no explicit form available
but it was calculated by \citet{wellenhofer1997} (electrons and holes
separately) and \citet{tanaka2018}(average effective mass). Later
\citet{schadt1997,hatakeyama2013} fitted the results from
\citet{wellenhofer1997} with an equation used for silicon (see
\cref{eq:dos_temperature})~\cite{lang1983}.
\begin{equation}
    \label{eq:dos_temperature}
    m^*(T) = \left(\frac{z_0 + z_1 T+ z_2 T^2 + z_3 T^3 + z_4 T^4}{1+n_1 T + n_2 T^2 + n_3 T^3 + n_4 T^4}\right)^\eta   
\end{equation}

A temperature-dependent change according to \cref{eq:dos_temperature} can be included in some of the modern TCAD tools, but we found no possibility of implementing \cref{eq:dos_mdh_temperature}. Not supported are also doping-dependent DOS masses~\cite{adachi2005}, which were already shown for 6H-SiC~\cite{persson1999a}, but not yet for 4H.


\subsubsection{Effective Conductivity Mass}

We regularly encountered the conductivity mass in our review. Initially, we found it very challenging to distinguish it from the DOS mass, because both are calculated from the direction-dependent effective masses. Therefore, we decided to explicitly highlight the differences here.

The conductivity mass $\mcond$ is a simplification to model the mobility $\mu$
(see \cref{sec:mobility}) in
4H-SiC~\cite{ishikawa2021,ishikawa2024,huang1998,iwata2000,iwata2000a,iwata2001,iwata2003a,kinoshita1998,kordina1995,schadt1997,tanaka2018}
(see \cref{eq:dos_cond})~\cite{chen1997a}, with $\tau$ the charge carrier
lifetime (see \cref{sec:regen}).
\begin{equation}
    \label{eq:dos_cond}
    \mu = \frac{e \tau}{\mcond\,m_0},
\end{equation}
The differences to the DOS mass are visible in the definition of the conductivity mass, which is reciprocal (see \cref{eq:dos_cond_def})~\cite{adachi2005,camassel2008,ishikawa2021,ishikawa2024}.
\begin{equation}
    \label{eq:dos_cond_def}
    \begin{split}
    \mce &= \frac{3\mceperp\mcepara}{\mceperp + 2\mcepara}\\
    \mch &= \frac{3\mchperp\mchpara}{\mchperp + 2\mchpara} \\
    \frac{2}{\mceperp} &= \frac{1}{\memk} + \frac{1}{\memg} \, , \qquad \mcepara = \meml\\
    \frac{2}{\mchperp} &= \frac{1}{\mhgm} + \frac{1}{\mhgk} \,, \qquad \mchpara = \mhga
    \end{split}
\end{equation}
Combining these definitions leads to condensed expressions for the conductivity mass we encountered regularly in literature (see \cref{eq:dos_cond_merged})~\cite{huang1998}.
\begin{equation}
    \label{eq:dos_cond_merged}
    \begin{split}
    \frac{3}{\mce} = \frac{1}{\memg} + \frac{1}{\memk} + \frac{1}{\meml} \\
    \frac{3}{\mch} = \frac{1}{\mhgm} + \frac{1}{\mhgk} + \frac{1}{\mhga}
    \end{split}    
\end{equation}


\subsubsection{Polaron Mass}

In the Si-C bond of silicon carbide, carbon atoms are more electronegative than silicon ones, resulting in a partly ionic crystal~\cite{son2004}. \revision{A moving charge carrier can spend some of its energy to further polarize the crystal, reducing its own energy and the energy of the whole system~\cite{landau1948}. In equilibrium a quasiparticle, which is called a polaron, is formed~\cite{devreese2003,landau1948}. \citet{landau1948} pointed out that the band states of electrons are actually unstable and that the current carrier is actually the polaron.}

The process mentioned above was also described by longitudinal optical vibrations that generate an electric field along the direction of the vibration which interacts with the charge carriers~\cite{son2004,son2000}. Effectively, the carrier is ``dressed'' by a charge leading to a deviating effective mass~\cite{persson1997}. The adapted mass $\mpol$, called \textit{polaron mass}, is slightly higher than the bare mass $m$ and can be calculated according to Eq.~(\ref{eq:dos_polaron})~\cite{son2004,persson1997,persson2005}, where $\alpha$ is called the Fr{\"o}hlich constant (see \cref{eq:dos_polaron_frohlich}). If the CGS unit system is used~\cite{mahan1990,devreese2003,frohlich1954} the term $1/4\pi\es$ of Eq.~(\ref{eq:dos_polaron_frohlich}) has to be removed.
\begin{eqnarray}
    \label{eq:dos_polaron}
    \mpol &= m \frac{1 - \num{8e-4}\alpha^2}{1- \alpha/6 + \num{3.4e-3} \alpha^2} \approx m \left(1-\frac{\alpha}{6}\right)^{-1} \\
    \label{eq:dos_polaron_frohlich}
    \alpha &= \frac{1}{2} \left( \frac{1}{\ei} - \frac{1}{\es}\right) \frac{e^2}{\hbar\omlo} \left(\frac{2m\omlo}{\hbar}\right)^{1/2}\ \frac{1}{4\pi\es}
\end{eqnarray}


\subsubsection{Characterization Methods}

\revision{Effective masses were mainly determined by band structure calculations. The most commonly used approach was the density functional theory (DFT)~\cite{akturk2009} based local density approximation (LDA)~\cite{dong2004,iwata2003,kaeckell1996,nilsson1996,persson1998a,wellenhofer1997} using the projector augmented wave method (PAW)~\cite{kuroiwa2019}, (full-potential) linearized augmented plane wave method ((FP)LAPW)~\cite{persson1996,persson1997,persson1999b,persson2005,son2004}, (orthogonalized) linear combination of atomic orbital ((O)LCAO)~\cite{ching2006,zhao2000a}, full-potential linear muffin-tin orbital method (FPLMTO)~\cite{lambrecht1995}, GW approximation (GW)~\cite{wenzien1995}, hybrid pseudo-potential and tight-binding (HPT)~\cite{chen1997} or the HSE functional (HSE)~\cite{lu2021}. Additional calculations feature empirical pseudo potentials (EPM)~\cite{bellotti2000,pennington2001,kackell1994,karch1995,nilsson1996} and RSP Hamiltonians (RSPH)~\cite{lambrecht1997}.}

Published values also served as starting point for Monte Carlo (MC)~\cite{akturk2009,nilsson1996} simulations and genetic algorithm fittings (GAF)~\cite{ng2010,ng2011} that condense multiple investigations into new parameter sets. For example, the data by \citet{son1995} were refined by \citet{nilsson1996}, whose results served as a starting point for a fitting by \citet{mickevicius1998}. Similarly, \citet{mikami2024} calculated the hole mass as the second derivative of the E-k dispersion by \citet{persson1996}.

The calculations are complemented by measurements of the infrared spectroscopic ellipsometry (IRSE)~\cite{tiwald1999}, photoluminescence~\cite{egilsson1999}, infrared absorption (IR)~\cite{gotz1993}, Raman scattering~\cite{harima1995} and Hall effect~\cite{iwata2003a,koizumi2009,lomakina1973,lomakina1974}. Also prominent are investigations based on optically detected cyclotron resonance (ODCR)~\cite{kordina1995,son1995,son2000,son2000a,volm1996}, but in these cases, polaron masses were achieved.

   
\subsection{Results \& Discussion}

In this section, we present and discuss the DOS mass values found in the literature. We discarded conductivity masses~\cite{mikami2024}, publications that did not clearly specify the SiC polytype~\cite{bosch1966}, the results by \citet{son1995} because \citet{son2004} stated that the smaller values are due to ``errors caused by a broad and asymmetric ODCR line shape with the peak position slightly shifted to lower magnetic fields'' and the values by \citet{bellotti1999}, which were later published again~\cite{bellotti2000}. Overall, $138$ out of $153$ collected publications were included in our analyses.

\subsubsection{Effective Mass along Principal Directions}

\begin{table*}[p!]
    \centering
    \setlength{\tabcolsep}{10pt}
    \renewcommand{\arraystretch}{1.1}
    \caption{\label{tab:dos_principal}Effective masses in principal directions. Multiple values for the same band are calculated by differing algorithms. \bkgCol{measColor} denotes measurements.}
    \resizebox{0.95\linewidth}{!}{%
    \begin{tabular}{l|*{4}{c}|*{4}{c}|cc}
 & \multicolumn{4}{|c}{electron} & \multicolumn{4}{|c|}{hole} && \\ 
ref. & $\memg$ & $\memk$ & $\meml$ & band & $\mhgm$ & $\mhgk$ & $\mhga$ & band & method & polaron \\ 
 & [$1$] & [$1$] & [$1$] && [$1$] & [$1$] & [$1$] & && \\ \hline 
 &&&&&&&&&& \\[-7pt] 
\reftext{Kack94}\cite{kackell1994}& \num{0.62} & \num{0.13} & \num{0.39} & -- & \num{4.23} & \num{2.41} & \num{1.73} & hh & EPM & --  \\ 
& -- & -- & -- & -- & \num{0.45} & \num{0.77} & \num{1.73} & lh & EPM & --  \\ 
& -- & -- & -- & -- & \num{0.74} & \num{0.51} & \num{0.21} & so & EPM & --  \\ 
\reftext{Karc95}\cite{karch1995}& \num{0.66} & \num{0.31} & \num{0.3} & -- & -- & -- & -- & -- & EPM & --  \\ 
\reftext{Lamb95}\cite{lambrecht1995}& \num{0.58} & \num{0.28} & \num{0.31} & -- & -- & -- & -- & -- & FPLMTO & --  \\ 
\reftext{Wenz95}\cite{wenzien1995}& \num{0.6} & \num{0.28} & \num{0.19} & -- & -- & -- & -- & -- & GW & --  \\ 
\reftext{Kaec96}\cite{kaeckell1996}& \num{0.57} & \num{0.32} & \num{0.32} & -- & -- & -- & -- & -- & DFT-LDA & --  \\ 
\reftext{Nils96}\cite{nilsson1996}& \num{0.43} & \num{0.43} & \num{0.28} & 1 & -- & -- & -- & -- & DFT-LDA/EPM & --  \\ 
& \num{0.52} & \num{0.21} & \num{0.45} & 2 & -- & -- & -- & -- & DFT-LDA/EPM & --  \\ 
\reftext{Pers96}\cite{persson1996}& \num{0.57} & \num{0.28} & \num{0.31} & -- & -- & -- & -- & -- & LAPW & --  \\ 
\rowcolor{measColor}\reftext{Volm96}\cite{volm1996}& \num{0.58+-0.01} & \num{0.31+-0.01} & \num{0.33+-0.01} & -- & -- & -- & -- & -- & ODCR & y  \\ 
\reftext{Chen97}\cite{chen1997}& \num{1.2} & \num{0.19} & \num{0.33} & -- & -- & -- & -- & -- & HPT & --  \\ 
\reftext{Pers97}\cite{persson1997}& \num{0.57} & \num{0.28} & \num{0.31} & 1 & -- & -- & -- & -- & LAPW & --  \\ 
& \num{0.59} & \num{0.31} & \num{0.34} & 1 & -- & -- & -- & -- & LAPW & --  \\ 
& \num{0.61} & \num{0.29} & \num{0.33} & 1 & -- & -- & -- & -- & LAPW & y  \\ 
& \num{0.78} & \num{0.16} & \num{0.71} & 2 & -- & -- & -- & -- & LAPW & --  \\ 
& \num{0.8} & \num{0.18} & \num{0.75} & 2 & -- & -- & -- & -- & LAPW & --  \\ 
& \num{0.85} & \num{0.17} & \num{0.77} & 2 & -- & -- & -- & -- & LAPW & y  \\ 
\reftext{Pers98a}\cite{persson1998a}\footnotemark[1]& -- & -- & -- & -- & \num{0.7} & \num{3.04} & \num{1.64} & 1 & DFT-LDA & --  \\ 
& -- & -- & -- & -- & \num{0.6} & \num{0.34} & \num{1.64} & 2 & DFT-LDA & --  \\ 
\reftext{Bell00}\cite{bellotti2000}& \num{0.57} & \num{0.23} & \num{0.27} & -- & -- & -- & -- & -- & EPM & --  \\ 
\reftext{Zhao00a}\cite{zhao2000a}& \num{0.62+-0.03} & \num{0.27+-0.02} & \num{0.31+-0.02} & -- & -- & -- & -- & -- & LCAO & --  \\ 
\reftext{Penn01}\cite{pennington2001}& \num{0.6+-0.05} & \num{0.2+-0.02} & \num{0.36+-0.02} & -- & -- & -- & -- & -- & EPM & --  \\ 
\reftext{Iwat03}\cite{iwata2003}& \num{0.59} & \num{0.29} & -- & -- & -- & -- & -- & -- & DFT-LDA & --  \\ 
\rowcolor{measColor}\reftext{Iwat03a}\cite{iwata2003a}& \num{0.58} & \num{0.3} & -- & -- & -- & -- & -- & -- & Hall & --  \\ 
\reftext{Dong04}\cite{dong2004}& \num{0.53} & \num{0.27} & \num{0.3} & -- & \num{0.86} & \num{0.95} & \num{1.58} & 1 & DFT-LDA & --  \\ 
& -- & -- & -- & -- & \num{0.55} & \num{0.52} & \num{1.32} & 2 & DFT-LDA & --  \\ 
& -- & -- & -- & -- & \num{1.13} & \num{1.30} & \num{0.21} & 3 & DFT-LDA & --  \\ 
\reftext{Chin06}\cite{ching2006}& \num{0.47} & -- & \num{0.38} & -- & -- & -- & -- & -- & OLCAO & --  \\ 
\reftext{Ng10}\cite{ng2010}& \num{0.66} & \num{0.31} & \num{0.34} & -- & -- & -- & -- & -- & GAF & --  \\ 
\reftext{Kuro19}\cite{kuroiwa2019}& \num{0.54} & \num{0.28} & \num{0.31} & -- & \num{0.54} & \num{0.54} & \num{1.48} & -- & PAW & --  \\ 
\reftext{Lu21}\cite{lu2021}& \num{0.54} & \num{0.3} & -- & -- & \num{2.77} & \num{1.82} & \num{1.52} & -- & HSE & --  \\ 
\end{tabular}
        \footnotetext[1]{fitted to~\cite{persson1996,persson1997}}
    }
\end{table*} 

The relative masses in the principal directions (see \cref{tab:dos_principal}) are predominantly determined by calculations. Out of $21$ publications, only two~\cite{volm1996,iwata2003a} conducted measurements, and both focused solely on electrons. Consequently, calculations are, at the moment, the only source for hole masses. In literature, the hole bands are either denoted as heavy-hole (hh), light-hole (lh), and crystal split-off (so) or simply as 1, 2, 3. By comparison of the respective values, we are confident to say that 1 $\equiv$ hh, 2 $\equiv$ lh, and 3 $\equiv$ so.

We did not include the results by \citet{pennington2004}, because 
\begin{enumerate*}[label=(\roman*)]
    \item we were unable to retrace the stated effective masses of the second conduction band in the provided reference~\cite{pennington2001} and
    \item two out of three values were identical to band $1$, which contradicted other investigations of band $2$.
\end{enumerate*}
We also excluded the values $\memg=0.58,\memk=0.31,\meml=0.35$~\cite{bellotti2000} because we were unable to acquire the referenced publication by \citet{goano1999} to confirm them.

\begin{figure}[t!]
    \centering
    \resizebox{0.9\linewidth}{!}{%
    \input{figures/dos_mass_stat}
    }
    \caption{\label{fig:dos_mass_stat}Statistical analysis of masses in principal directions. We only considered band $1$ and values without further specifications. Shown are the 0th, 25th, 50th, 75th and 100th quartile. The mean value is added in numerical form.}
\end{figure}

According to our statistical analysis (see \cref{fig:dos_mass_stat}) the electron masses in literature are clustered closely around the mean values of $0.58$ ($\memg$), $0.28$ ($\memk$) and $0.31$ ($\meml$) with few outliers, for example $\memg=1.2$~\cite{chen1997}, $\memg=0.47$~\cite{ching2006}, $\memk<0.2$~\cite{kackell1994,chen1997} or $\meml=0.19$~\cite{wenzien1995}. Polaron masses deviate not distinguishable. For meaningful statement about the second band more information would be required, which is the reason why we do not show any statistical analysis on band $2$ here.

Although we only found five parameter sets for holes, the agreement among them is bad. Solely for $\mhga$ an acceptable spread around the mean of $1.58$ was achieved. The oldest~\cite{kackell1994} and newest~\cite{lu2021} investigation report exceptional high values. We considered methodical or temperature differences to explain these discrepancies but were not successful in doing that.  \revision{The value $\mhgk=3.04$ derived by \citet{persson1998a} might be a typographical error because the same value reappeared in the original table and the mass $0.34$ of the second band is only a single number flip apart. Nevertheless, the presented band diagram shows a steep decrease, rendering also the higher value possible.}

Despite these discrepancies, the DOS masses are a relatively inactive field of research. From the $21$ fundamental investigations $17$ were conducted in the years $1994$ to $2004$. In contrast, we found only three papers on this topic within the last $15$ years.

\subsubsection{Density-of-States Mass}

\begin{table}[p!]
    \setlength{\tabcolsep}{10pt}
    \caption{\label{tab:dos_effMass}DOS masses[1/2]. Multiple values for the same band are calculated by differing algorithms. \bkgCol{measColor} denotes measurements.}
    \resizebox{0.95\linewidth}{!}{%
    \begin{tabular}{l|*{4}{c}|*{4}{c}|cc}
 & \multicolumn{4}{|c}{electron} & \multicolumn{4}{|c|}{hole} && \\ 
ref. & $\mde$ & $\mdeperp$ & $\mdepara$ & band & $\mdh$ & $\mdhperp$ & $\mdhpara$ & band & method & polaron \\ 
 & [$1$] & [$1$] & [$1$] && [$1$] & [$1$] & [$1$] & && \\ \hline 
 &&&&&&&&&& \\[-7pt] 
\rowcolor{measColor}\reftext{Loma73}\cite{lomakina1973}& \num{0.20}\footnotemark[1] & \num{0.21} & \num{0.19} & -- & -- & -- & -- & -- & Hall & --  \\ 
\rowcolor{measColor}\reftext{Loma74}\cite{lomakina1974}& \num{0.20}\footnotemark[1] & \num{0.21} & \num{0.19} & -- & -- & -- & -- & -- & Hall & --  \\ 
\rowcolor{measColor}\reftext{Gotz93}\cite{gotz1993}& \num{0.19} & \num{0.176} & \num{0.224} & -- & -- & -- & -- & -- & IR & --  \\ 
\reftext{Kack94}\cite{kackell1994}& \num{0.31}\footnotemark[1] & \num{0.28}\footnotemark[2] & \num{0.39}\footnotemark[3] & -- & \num{2.60}\footnotemark[4] & \num{3.19}\footnotemark[5] & \num{1.73}\footnotemark[6] & hh & EPM & --  \\ 
& -- & -- & -- & -- & \num{0.84}\footnotemark[4] & \num{0.59}\footnotemark[5] & \num{1.73}\footnotemark[6] & lh & EPM & --  \\ 
& -- & -- & -- & -- & \num{0.43}\footnotemark[4] & \num{0.61}\footnotemark[5] & \num{0.21}\footnotemark[6] & so & EPM & --  \\ 
\rowcolor{measColor}\reftext{Hari95}\cite{harima1995}& \num{0.35}\footnotemark[1] & \num{0.3+-0.07} & \num{0.48+-0.12} & -- & -- & -- & -- & -- & Raman & --  \\ 
\reftext{Karc95}\cite{karch1995}& \num{0.39}\footnotemark[1] & \num{0.45}\footnotemark[2] & \num{0.30}\footnotemark[3] & -- & -- & -- & -- & -- & EPM & --  \\ 
\rowcolor{measColor}\reftext{Kord95}\cite{kordina1995}& -- & \num{0.42} & -- & -- & -- & -- & -- & -- & ODCR & y  \\ 
\reftext{Lamb95}\cite{lambrecht1995}& \num{0.35}\footnotemark[1] & \num{0.4} & \num{0.27} & -- & -- & -- & -- & -- & FPLMTO & --  \\ 
\rowcolor{measColor}\reftext{Son95}\cite{son1995}& \num{0.37}\footnotemark[1] & \num{0.42} & \num{0.29+-0.03} & -- & -- & -- & -- & -- & ODCR & y  \\ 
\reftext{Wenz95}\cite{wenzien1995}& \num{0.32}\footnotemark[1] & \num{0.41}\footnotemark[2] & \num{0.19}\footnotemark[3] & -- & -- & -- & -- & -- & GW & --  \\ 
\reftext{Kaec96}\cite{kaeckell1996}& \num{0.39}\footnotemark[1] & \num{0.43}\footnotemark[2] & \num{0.32}\footnotemark[3] & -- & -- & -- & -- & -- & DFT-LDA & --  \\ 
\reftext{Nils96}\cite{nilsson1996}& \num{0.37}\footnotemark[1] & \num{0.43}\footnotemark[2] & \num{0.28}\footnotemark[3] & 1 & -- & -- & -- & -- & DFT-LDA/EPM & --  \\ 
& \num{0.37}\footnotemark[1] & \num{0.33}\footnotemark[2] & \num{0.45}\footnotemark[3] & 2 & -- & -- & -- & -- & DFT-LDA/EPM & --  \\ 
\reftext{Pers96}\cite{persson1996}& \num{0.37}\footnotemark[1] & \num{0.40}\footnotemark[2] & \num{0.31}\footnotemark[3] & -- & -- & -- & -- & -- & LAPW & --  \\ 
\rowcolor{measColor}\reftext{Volm96}\cite{volm1996}& \num{0.39}\footnotemark[1] & \num{0.42}\footnotemark[2] & \num{0.33}\footnotemark[3] & -- & -- & -- & -- & -- & ODCR & y  \\ 
\reftext{Bako97}\cite{bakowski1997}& -- & -- & -- & -- & \num{0.84} & -- & -- & 1 & -- & --  \\ 
& -- & -- & -- & -- & \num{0.79} & -- & -- & 2 & -- & --  \\ 
& -- & -- & -- & -- & \num{0.78} & -- & -- & 3 & -- & --  \\ 
\reftext{Chen97}\cite{chen1997}& \num{0.42}\footnotemark[1] & \num{0.48}\footnotemark[2] & \num{0.33}\footnotemark[3] & -- & -- & -- & -- & -- & HPT & --  \\ 
\rowcolor{measColor}\reftext{Hemm97}\cite{hemmingsson1997}& -- & -- & -- & -- & \num{1}\footnotemark[7] & -- & -- & -- & FIT & --  \\ 
\reftext{Lamb97}\cite{lambrecht1997}& -- & -- & -- & -- & \num{0.85}\footnotemark[4] & \num{0.62} & \num{1.6} & hh & RSPH & --  \\ 
& -- & -- & -- & -- & \num{0.84}\footnotemark[4] & \num{0.62} & \num{1.55} & lh & RSPH & --  \\ 
& -- & -- & -- & -- & \num{0.81}\footnotemark[4] & \num{1.58} & \num{0.21} & so & RSPH & --  \\ 
\reftext{Pers97}\cite{persson1997}& \num{0.37}\footnotemark[1] & \num{0.40}\footnotemark[2] & \num{0.31}\footnotemark[3] & 1 & \num{0.82}\footnotemark[4] & \num{0.59} & \num{1.56} & 1 & LAPW & --  \\ 
& \num{0.40}\footnotemark[1] & \num{0.43}\footnotemark[2] & \num{0.34}\footnotemark[3] & 1 & \num{0.82}\footnotemark[4] & \num{0.59} & \num{1.6} & 1 & LAPW & --  \\ 
& \num{0.39}\footnotemark[1] & \num{0.42}\footnotemark[2] & \num{0.33}\footnotemark[3] & 1 & \num{0.82}\footnotemark[4] & \num{0.59} & \num{1.56} & 2 & LAPW & y  \\ 
& \num{0.44}\footnotemark[1] & \num{0.35}\footnotemark[2] & \num{0.71}\footnotemark[3] & 2 & \num{0.82}\footnotemark[4] & \num{0.59} & \num{1.6} & 2 & LAPW & --  \\ 
& \num{0.48}\footnotemark[1] & \num{0.38}\footnotemark[2] & \num{0.75}\footnotemark[3] & 2 & \num{0.78}\footnotemark[4] & \num{1.49} & \num{0.21} & 3 & LAPW & --  \\ 
& \num{0.48}\footnotemark[1] & \num{0.38}\footnotemark[2] & \num{0.77}\footnotemark[3] & 2 & \num{0.79}\footnotemark[4] & \num{1.49} & \num{0.22} & 3 & LAPW & y  \\ 
\end{tabular}
        \footnotetext[1]{$\mde = (\mdeperp {}^2 \mdepara)^{1/3}$}
        \footnotetext[2]{$\mdeperp = \sqrt{\memg \memk}$}
        \footnotetext[3]{$\mdepara = \meml$}
        \footnotetext[4]{$\mdh = (\mdhperp {}^2 \mdhpara)^{1/3}$}
        \footnotetext[5]{$\mdhperp = \sqrt{\mhgm \mhgk}$}
        \footnotetext[6]{$\mdhpara = \mhga$}
        \footnotetext[7]{fitted to \reftext{Pers96}\cite{persson1996}}
    }
\end{table} 

\begin{table}[p]
    \setlength{\tabcolsep}{8pt}
    \caption{\label{tab:dos_effMassII}DOS masses[2/2]. Multiple values for the same band are calculated by differing algorithms. \bkgCol{measColor} denotes measurements.}
    \resizebox{0.95\linewidth}{!}{%
    \begin{tabular}{l|*{4}{c}|*{4}{c}|cc}
 & \multicolumn{4}{|c}{electron} & \multicolumn{4}{|c|}{hole} && \\ 
ref. & $\mde$ & $\mdeperp$ & $\mdepara$ & band & $\mdh$ & $\mdhperp$ & $\mdhpara$ & band & method & polaron \\ 
 & [$1$] & [$1$] & [$1$] && [$1$] & [$1$] & [$1$] & && \\ \hline 
 &&&&&&&&&& \\[-7pt] 
\reftext{Well97}\cite{wellenhofer1997}& \num{0.394} & -- & -- & -- & -- & -- & -- & -- & DFT-LDA & --  \\ 
\reftext{Lind98}\cite{lindefelt1998}& -- & -- & -- & -- & \num{1.7} & -- & -- & hh & LAPW & --  \\ 
& -- & -- & -- & -- & \num{0.48} & -- & -- & lh & LAPW & --  \\ 
\reftext{Pers98a}\cite{persson1998a}& -- & -- & -- & -- & \num{0.94} & \num{1.46}\footnotemark[5] & \num{1.64}\footnotemark[6] & 1 & DFT-LDA & --  \\ 
& -- & -- & -- & -- & \num{0.84} & \num{0.45}\footnotemark[5] & \num{1.64}\footnotemark[6] & 2 & DFT-LDA & --  \\ 
& -- & -- & -- & -- & \num{0.88} & -- & -- & 3 & DFT-LDA & --  \\ 
\rowcolor{measColor}\reftext{Egil99}\cite{egilsson1999}& \num{0.37} & -- & -- & -- & -- & -- & -- & -- & PL & --  \\ 
\reftext{Pers99b}\cite{persson1999b}& -- & -- & -- & -- & \num{0.85}\footnotemark[4] & \num{0.62} & \num{1.61} & 1 & LAPW & --  \\ 
& -- & -- & -- & -- & \num{0.84}\footnotemark[4] & \num{0.61} & \num{1.62} & 1 & LAPW & --  \\ 
& -- & -- & -- & -- & \num{0.78}\footnotemark[4] & \num{0.56} & \num{1.52} & 2 & LAPW & --  \\ 
& -- & -- & -- & -- & \num{0.78}\footnotemark[4] & \num{0.58} & \num{1.42} & 2 & LAPW & --  \\ 
& -- & -- & -- & -- & \num{0.78}\footnotemark[4] & \num{1.5} & \num{0.21} & 3 & LAPW & --  \\ 
& -- & -- & -- & -- & \num{0.76}\footnotemark[4] & \num{1.46} & \num{0.21} & 3 & LAPW & --  \\ 
\rowcolor{measColor}\reftext{Tiwa99}\cite{tiwald1999}& \num{0.36}\footnotemark[1] & \num{0.36} & \num{0.36} & -- & -- & -- & -- & -- & IRSE & --  \\ 
\reftext{Bell00}\cite{bellotti2000}& \num{0.33}\footnotemark[1] & \num{0.36}\footnotemark[2] & \num{0.27}\footnotemark[3] & -- & -- & -- & -- & -- & EPM & --  \\ 
\rowcolor{measColor}\reftext{Son00}\cite{son2000}& \num{0.39}\footnotemark[1] & \num{0.45+-0.02} & \num{0.30+-0.02} & -- & \num{0.91}\footnotemark[4] & \num{0.66+-0.02} & \num{1.75+-0.02} & -- & ODCR & y  \\ 
\reftext{Zhao00}\cite{zhao2000}\footnotemark[7]& \num{0.37}\footnotemark[1] & \num{0.42} & \num{0.28} & 1 & -- & -- & -- & -- & FIT & --  \\ 
& \num{0.44}\footnotemark[1] & \num{0.35} & \num{0.71} & 2 & -- & -- & -- & -- & FIT & --  \\ 
& \num{0.40}\footnotemark[1] & \num{0.66} & \num{0.15} & 3 & -- & -- & -- & -- & FIT & --  \\ 
\reftext{Zhao00a}\cite{zhao2000a}& \num{0.37}\footnotemark[1] & \num{0.41+-0.02} & \num{0.31}\footnotemark[3] & -- & -- & -- & -- & -- & LCAO & --  \\ 
\reftext{Penn01}\cite{pennington2001}& \num{0.34}\footnotemark[1] & \num{0.35+-0.02} & \num{0.31+-0.05} & -- & -- & -- & -- & -- & EPM & --  \\ 
\reftext{Iwat03}\cite{iwata2003}& -- & \num{0.41}\footnotemark[2] & -- & -- & -- & -- & -- & -- & DFT-LDA & --  \\ 
\rowcolor{measColor}\reftext{Iwat03a}\cite{iwata2003a}& -- & \num{0.42}\footnotemark[2] & -- & -- & -- & -- & -- & -- & Hall & --  \\ 
\reftext{Dong04}\cite{dong2004}& \num{0.35}\footnotemark[1] & \num{0.38}\footnotemark[2] & \num{0.30}\footnotemark[3] & -- & \num{1.09}\footnotemark[4] & \num{0.90}\footnotemark[5] & \num{1.58}\footnotemark[6] & 1 & DFT-LDA & --  \\ 
& -- & -- & -- & -- & \num{0.72}\footnotemark[4] & \num{0.53}\footnotemark[5] & \num{1.32}\footnotemark[6] & 2 & DFT-LDA & --  \\ 
& -- & -- & -- & -- & \num{0.67}\footnotemark[4] & \num{1.21}\footnotemark[5] & \num{0.21}\footnotemark[6] & 3 & DFT-LDA & --  \\ 
\reftext{Chin06}\cite{ching2006}& -- & -- & \num{0.38}\footnotemark[3] & -- & -- & -- & -- & -- & OLCAO & --  \\ 
\reftext{Aktu09}\cite{akturk2009}& \num{0.40} & -- & -- & -- & -- & -- & -- & -- & DFT & --  \\ 
\rowcolor{measColor}\reftext{Koiz09}\cite{koizumi2009}& -- & -- & -- & -- & \num{0.5} & -- & -- & -- & Hall & --  \\ 
\reftext{Ng10}\cite{ng2010}& \num{0.41}\footnotemark[1] & \num{0.45}\footnotemark[2] & \num{0.34}\footnotemark[3] & -- & -- & -- & -- & -- & GAF & --  \\ 
\reftext{Kuro19}\cite{kuroiwa2019}& \num{0.36}\footnotemark[1] & \num{0.39}\footnotemark[2] & \num{0.31}\footnotemark[3] & -- & \num{0.76}\footnotemark[4] & \num{0.54}\footnotemark[5] & \num{1.48}\footnotemark[6] & -- & PAW & --  \\ 
\reftext{Lu21}\cite{lu2021}& -- & \num{0.40}\footnotemark[2] & -- & -- & \num{1.97}\footnotemark[4] & \num{2.25}\footnotemark[5] & \num{1.52}\footnotemark[6] & -- & HSE & --  \\ 
\end{tabular}
        \footnotetext[1]{$\mde = (\mdeperp {}^2\,\, \mdepara)^{1/3}$}
        \footnotetext[2]{$\mdeperp = \sqrt{\memg \memk}$}
        \footnotetext[3]{$\mdepara = \meml$}
        \footnotetext[4]{$\mdh = (\mdhperp {}^2\,\, \mdhpara)^{1/3}$}
        \footnotetext[5]{$\mdhperp = \sqrt{\mhgm \mhgk}$}
        \footnotetext[6]{$\mdhpara = \mhga$}
        \footnotetext[7]{fitted to \reftext{Pers97}\cite{persson1997}}
    }
\end{table} 

For the DOS masses we complemented values found in literature by those computed from the effective masses along the principal directions presented in the previous section using Eq.~(\ref{eq:dos_masses}) (see \cref{tab:dos_effMass} and \cref{tab:dos_effMassII}). We also show parameters for a third conduction band by \citet{zhao2000}, which is \SI{2}{\electronvolt} above band $1$ in the M-point. Again, the majority ($31$ out of $40$) of the investigations were carried out in the years between $1993$ and $2006$, versus only five in the last $15$ years. Calculations clearly dominate, but the amount of measurements increased significantly: We found eleven~\cite{lomakina1973,lomakina1974,harima1995,kordina1995,son1995,volm1996,egilsson1999,tiwald1999,son2000,iwata2003a} measurements for electrons but still only two~\cite{son2000,koizumi2009} for holes.

\begin{figure}[t!]
    \centering
    \resizebox{0.9\linewidth}{!}{%
    \input{figures/dos_dosmass_stat}
    }
    \caption{\label{fig:dos_dosmass_stat}Statistical analysis of DOS masses. For electrons, we display band $1$, and for holes, bands $1$, $2$, and $3$. Shown are the 0th, 25th, 50th, 75th and 100th quartile. The mean value is added in numerical form.}
\end{figure}

The statistical analysis shows a comparable distribution for electrons, but a higher variance for holes (see \cref{fig:dos_dosmass_stat}). For $\mdh$ and $\mdhperp$ the majority of scientific analyses deliver values $<1$ and only few publications, which we already discussed for the direction-dependent masses, values bigger than $2$. An exceptional agreement was reached for $\mdhpara=\num{0.21+-0.01}$ of the third band among five publications.

The presented values adhere to some restrictions. \citet{schadt1997} stated that they are only valid close to the band minimum/maximum as they are calculated there or at very low temperatures. \citet{son2004} claimed that if the polaron effect is added to the results from \citet{persson1996,persson1999b}
the values $\mdhperp=0.66$ and $\mdhpara = 1.76$ fit well to the results from ODCR measurements.

The value of $\mc$ changed over the years. In early publications we encountered rather high values of $\mc=12$~\cite{patrick1965,schadt1994a,gotz1993} or $\mc=6$~\cite{pensl1993,tairov1977,itoh1995}, but more recent publications agree upon $\mc=3$~\cite{hemmingsson1997,hatakeyama2013,bakowski1997,ioffe2023,ivanov2000,iwata2001,khanna2023,ladesmartin2000,levinshtein2001,lindefelt1998,maximenko2023,nipoti2016,pensl2005,pernot2000,persson1998a,scaburri2011a,schadt1997,zippelius2011}.

In rare cases, the mathematical models to calculate the DOS masses (cp. Eq.~(\ref{eq:dos_masses})) differed. \citet{pennington2004} used $\mde=\sqrt{\memg \memk}$, which \citet{tilak2007} reused in combination with $\mdeperp$ calculated according to Eq.~(\ref{eq:dos_masses}). This resulted in $\mde \approx \mdeperp$. \citet{persson1998a} included only one perpendicular mass into the calculation of the effective mass, i.e., $\mde = ((m^*_{de\perp 1})^2\,\, \mdepara)^{1/3}$ and  \citet{gao2022a} stated $\mde=\mdepara$.
\citet{ivanov2003a} mixed parallel and perpendicular directions, i.e., $\mdeperp = \sqrt{\meml \memk}=0.32$ resp. $\mdepara = \meml = 0.58$. The authors referenced these equations~\cite{faulkner1969} but we were unable to locate them there. Transformed to the set of equations used in this paper, we got $\mdeperp = 0.42$ and $\mdepara=0.33$. A more comprehensive list of all inconsistencies we identified is presented in Appendix~\ref{sec:dos_appendix}.

\subsubsection{Temperature Dependency}

\citet{wellenhofer1997} and \citet{tanaka2018} calculated the temperature dependency of the DOS masses (see \cref{fig:dos_temperature}). Their results showed that the electron mass stays almost constant around $0.4$ but predicted for $\mdh$ an increase of more than \SI{100}{\percent} in the range \SIrange{0}{200}{\K}. This implies that parameters extracted from calculations at \SI{0}{\kelvin} are quite inaccurate for simulations at \SI{300}{\kelvin}. Despite these large variations, temperature is only occasionally considered in literature~\cite{troffer1998,zippelius2011}. \citet{sozzi2019} ($\mde=0.4$, $\mdh = 2.6$) and \citet{luo2020} ($\mde=0.4$, $\mdh = 2.64$) extracted explicitly mass values at \SI{300}{\kelvin} from the calculations. \citet{rakheja2020} claimed that the values they used were measured at \SI{300}{\kelvin}, although the referenced investigation~\cite{son2000} was carried out at \SI{4.4}{\kelvin}.

\begin{figure}[t]
    \centering
    \resizebox{0.95\linewidth}{!}{%
    \input{figures/dos_temperature}
    }
    \caption{\label{fig:dos_temperature}Temperature dependency of the DOS mass. Dashed lines indicate fittings to the calculations by \citet{wellenhofer1997}.}
\end{figure}

The result of the calculations by \citet{wellenhofer1997} can not be written in an implicit form, such that two fittings using \cref{eq:dos_temperature} were proposed~\cite{schadt1997,hatakeyama2013}. The corresponding parameters are shown in \cref{tab:dos_temperature}. In \cref{eq:dos_mdh_temperature} we introduced an approach proposed by \citet{bakowski1997} to combine the three valence bands in the $\Gamma$ point, which also contained the temperature. Compared to the calculations by \citet{wellenhofer1997} $\mdh$ increases only moderately for the coefficients $m_{\mathrm{h}1}=0.84$, $m_{\mathrm{h}2}=0.79$, $m_{\mathrm{h}3}=0.78$, $\Delta E_2=\SI{9}{\milli\electronvolt}$ and $\Delta E_3=\SI{73}{\milli\electronvolt}$ (see \cref{fig:dos_temperature}). We plotted the model only up to \SI{100}{\K}, because the authors remarked that the presented parameters are only valid for low temperatures.  Almost the same values, i.e., $m_{\mathrm{h}1}=0.82$, $m_{\mathrm{h}2}=0.82$, $m_{\mathrm{h}3}=0.78$, $\Delta E_2=\SI{8.6}{\milli\electronvolt}$ and $\Delta E_3=\SI{72}{\milli\electronvolt}$ were reported by \citet{persson1997}.

\begin{table*}[t]
    \centering
    \setlength{\tabcolsep}{2pt}
    \renewcommand{\arraystretch}{1.1}
    \sisetup{zero-decimal-to-integer,
    }
    \caption{\label{tab:dos_temperature}Fitting parameters in \cref{eq:dos_temperature} to calculations by \citet{wellenhofer1997}.}
    \resizebox{1\linewidth}{!}{%
    \begin{tabular}{cc|*{5}{c}|*{4}{c}|{c}}
         ref. & mass& $z_0$ & $z_1$ & $z_2$ & $z_3$ & $z_4$ & $n_1$ & $n_2$ & $n_3$ & $n_4$ & $\eta$ \\
         && [1] & [\si{\K\tothe{-1}}] & [\si{\K\tothe{-2}}] & [\si{\K\tothe{-3}}] & [\si{\K\tothe{-4}}] &
         [\si{\K\tothe{-1}}] & [\si{\K\tothe{-2}}] & [\si{\K\tothe{-3}}] &[\si{\K\tothe{-4}}] &[1] \\ \hline
         &&&&&&&&&&& \\[-7pt] 
         \reftext{Scha97}~\cite{schadt1997} & $\mde$ & \num{0.3944} & \num{-6.822e-4} & \num{1.335e-6} & \num{3.597e-10} & \num{0} & \num{-1.776e-3} & \num{3.65e-6} & \num{0} & \num{0} & \num{1} \\
         \reftext{Scha97}~\cite{schadt1997} & $\mdh$ & \num{1.104} & \num{1.578e-2}  & \num{3.087e-3} & \num{-7.635e-8} & \num{0} & \num{1.387e-2} & \num{1.126e-3} & \num{0} & \num{0} & \num{1} \\         
         \reftext{Hata13}~\cite{hatakeyama2013} & $\mde$ & \num{0.394} & \num{0} & \num{3.09e-8} & \num{2.23e-10} & \num{-1.65e-13} & \num{0} & \num{0} & \num{0} & \num{0} & \num{1} \\
         \reftext{Hata13}~\cite{hatakeyama2013} & $\mdh$ & \num{1} & \num{6.92e-2} & \num{0} & \num{0} & \num{1.88e-6} & \num{0} & \num{6.58e-4} & \num{0} & \num{4.32e-7} & 2/3
    \end{tabular}
    }
\end{table*}

\subsubsection{Origin of Parameters}


For electrons we identified the investigations by \citet{gotz1993}, \citet{son1995}, \citet{persson1996}, \citet{volm1996}, \citet{persson1997} and \citet{wellenhofer1997} as the most influential ones (a graphical representation with all details is available in \cref{fig:dos_ref_chain_e} in Appendix~\ref{sec:refChainDOS}). Upon citation the values were transferred accurately: Solely the parameters proposed by \citet{gotz1993} got rounded to two digits. Values were properly referenced, forcing us only occasionally to predict dependencies based on the utilized values and reference chains were found only to a level of two (one intermediate publication). The same statements can be made about hole masses (see \cref{fig:dos_ref_chain_h} in Appendix~\ref{sec:refChainDOS}), with the difference that only two publications, i.e., \citet{persson1997} and \citet{son2000}, have more than four citations.


\subsubsection{Literature Values}

For a comprehensive overview of values utilized for $\mde$ and $\mdh$ we extended those we found in literature with computations based on fundamental values using Eq.~(\ref{eq:dos_masses}) (see \cref{fig:dos_TCAD}). For $\mde$ we used only the first conduction band but for holes we merged the first two bands using \cref{eq:dos_mdh}.

\begin{figure}[t]
    \centering
    \resizebox{0.94\linewidth}{!}{%
    \input{figures/dos_TCAD}
    }
    \caption{\label{fig:dos_TCAD}Effective DOS masses. Size of values indicates abundance in literature. \bkgCol{calcColor}~denotes values that we calculated from fundamental masses using Eqs.~(\ref{eq:dos_masses}) and (\ref{eq:dos_mdh}).}
\end{figure}

For $\mde$ we found dominantly values of \num{0.37+-0.05}, which matches the mean value of the fundamental studies. For holes the value spread is more significant. Although the majority of fundamental investigations (cp. \cref{fig:dos_dosmass_stat}) agree on values $<1$, e.g., $0.82$ and $0.91$, we found many values $\ge 1$. Possible explanations are high masses from calculations (colored background in the figure), confusions between direction-dependent and effective masses as well as temperature considerations ($\mdh=2.64$ and $2.6$).

The value $\mdh=1.2$ has multiple origins: While \citet{bakowski1997} derived it from the temperature dependent model at \SI{300}{\K}~\cite{huang1998,shah1998,lee2002,raynaud2010} some referenced it from \citet{persson1997}, where it was stated as $\meml$ in 6H~\cite{biondo2012}. A third possibility is that, based on the parameters by \citet{son2000} (cp. \cref{fig:dos_ref_chain_h} in Appendix~\ref{sec:refChainDOS}), the expression $\mdh=(\mdhperp (\mdhpara)^2)^{1/3}={1.26}$~\cite{reshanov2005} was used instead of $\mdh=((\mdhperp)^2 \mdhpara)^{1/3}=0.91$, which then may got rounded to $\mdh=1.2$~\cite{albanese2010,yang2019}.  The prominent value of $\mdh=1$ was often provided without reference or explanation. Only \citet{hemmingsson1997} fitted it to the calculations by \citet{persson1996}. \citet{pensl2003} used the results by \citet{vandaal1963}, who investigated 6H-SiC.

\section{\label{sec:bandgap}Band Gap}

The conductivity of a material depends, among others, on the number of free charge carriers, which possess a direction and a momentum-based energy. The physically allowed energy values are described in the \textit{band diagram}: free electrons are located in the \textit{conduction band} and free holes in the \textit{valence band} (cp. \cref{sec:sic}). When an electron is excited from the valence to the conduction band, e.g., thermally or optically, it leaves a hole behind, such that in each band, a free charge carrier is created. If both bands touch, this generation process is simple, resulting in good conductivity, i.e., a conductor. A large forbidden energy region between the bands, called the \textit{band gap}, makes transitions unlikely, leading to lower conductance, i.e., an insulator. Materials with intermediate band gaps are called \textit{semi}conductors.

TCAD simulations use the band gap energy $\eg$, i.e., the difference between the
minimum conduction band energy and the maximum valence band energy, to determine
\begin{enumerate*}[label=(\roman*)]
    \item the local carrier concentration,
    \item the electric currents and
    \item the barrier heights in Schottky contacts.
\end{enumerate*}    
In these cases, $\eg$ is often used in exponential functions, which demands high accuracy to achieve a realistic description. In this section, we will therefore focus on band gap energies by extending previous analyses~\cite{neudeck2013,beyer2011,neuberger1971}.

Our research revealed that the majority of the currently used values are, with high confidence, based on measurements conducted in the year 1964~\cite{choyke1964, choyke1964a, zanmarchi1964}. However, these studies focused on the exciton band gap energy, which differs slightly from the band gap, at low temperatures. Coincidentally, recent measurements suggest that these values are reasonable for $\eg$ at room temperature, although the uncertainty in the data is still big. Therefore, further investigations are required for confirmation.


\subsection{Introduction}

The band gap in 4H-SiC is measured between the top of the valence band in the $\Gamma$-point (see \cref{sec:sic}) and the bottom of the conduction band in the M-point~\cite{ching2006,ioffe2023,son2004}, making 4H-SiC an \textit{indirect} semiconductor. Consequently, some transfer of moment, e.g., by a phonon, is required to raise an electron from the valence band maximum to the conduction band minimum. To create electron-hole pairs directly in the $\Gamma$-point, more than the band gap energy, i.e., the ionization energy, is required. \citet{gsponer2024a} provided an overview of ionization energies, whereat the authors extracted a value of \SI{7.83\pm 0.02}{\electronvolt} from their own measurements.

The statement that electrons are lifted from the valence to the conduction is not completely accurate. Actually, the electron first forms an exciton by Coulomb interactions with the hole it left behind~\cite{kwasnicki2014,schroder2005,wolfe1984,elliott1957}. This setup can be compared to the hydrogen atom but with larger radii due to the lower effective masses (cp. \cref{sec:dos}). Additional energy has to be provided to inject the electron into the conduction band. Consequently, the band gap energy $\eg$ is the sum of the energy required to generate the exciton, i.e., the \textit{exciton band gap energy} $\egx$, and the energy to free the electron from the exciton, i.e., the \textit{free exciton binding energy} $\ex$ (see \cref{eq:bandgap_eg})~\cite{albanese2010}.
\begin{equation}
    \label{eq:bandgap_eg}
    \eg = \egx + \ex
\end{equation}
An exciton is neutral and thus roams easily within the lattice, but it can achieve an energetically more favorable configuration by attaching itself to an impurity~\cite{kwasnicki2014}. The energy reduction relative to $\egx$ is denoted as \textit{bound} exciton binding energy~\cite{choyke1969,haberstroh1994,gotz1993,itoh1994,itoh1996,patrick1965,ikeda1980,ikeda1980b,dubrovskii1975,kordina1995,klahold2020,kwasnicki2014}, which depends on the impurity atom, the lattice site and the charge state~\cite{choyke1964a}. \revision{In contrast, the \textit{free} exciton binding energy denotes the additional energy required to free the electron from the exciton, i.e., that it can move freely.}

In our review, we found it quite challenging to distinguish the free and bound exciton binding energy, although they describe different processes and must not be confused. The main reasons were:
\begin{enumerate*}[label=(\arabic*)]
    \item Both are denoted by the same symbol $\ex$.
    \item They share similar values in the range of a few \si{\milli\electronvolt}.
    \item Publications often use only the term exciton binding energy.
\end{enumerate*}

The band gap energy is not constant. First and foremost, it varies among the polytypes of silicon carbide. It was empirically shown that the band gap scales linearly with the degree of hexagonality, i.e., the ratio of hexagonal to cubic lattice sites~\cite{choyke1964,zanmarchi1964,choyke2004,kimoto2014a} (4H-SiC has a hexagonality of 50 \%~\cite{freer1990}).
In addition, the band gap shows a pressure~\cite{adachi2005}, strain~\cite{kuroiwa2019}, temperature and doping dependency. It is important to take these influences into account because shifting band edges can create potential barriers that then influence the carrier transport~\cite{habib2011}.

In this review, we describe changes in the band gap by \cref{eq:bandgap_TN}. $\eg(T)$ denotes the temperature dependent variations and $\Delta \eg (\ndp,\nam)$ the doping dependent ones, with $\ndp$/$\nam$ the concentration of ionized donors/acceptors (see also \cref{sec:incompIon}).
\begin{equation}
    \label{eq:bandgap_TN}
    \eg (T,\ndp,\nam) = \eg(T) - \Delta \eg (\ndp,\nam)
\end{equation}


\subsubsection{Temperature Dependency}

The temperature is equivalent to the amount of lattice vibrations in a material, which can cause a shift of the band energies and the electron-lattice interaction energy~\cite{fan1951}. Since these changes are not uniform for all bands, the effective band gap is altered\cite{das2015}.

The temperature dependent band gap (see \cref{eq:bandgap_temp})~\cite{adachi2005} contains effects from thermal expansion ($\Delta E_\mathrm{th}(T)$) and electron-phonon interaction ($\Delta E_\mathrm{ph}(T)$). These effects are hard to separate in experiments~\cite{adachi2005}; we only found one example~\cite{cheng2022} in literature where such a distinction was made. In general, the models for $\Delta E_\mathrm{ph}(T)$ are used to describe both with reasonable accuracy, because $\Delta E_\mathrm{th}(T)$ has a much weaker impact~\cite{passler1997} and the scaling of both contributions is comparable~\cite{passler1998}. \citet{arvanitopoulos2017} extended \cref{eq:bandgap_temp} by adding an additive term $\Delta_\mathrm{g}^{\mathrm{Fermi}}$ to account for carrier statistics. This is, however, only necessary for devices whose size is close to the de-Broglie wavelength.
\begin{equation}
    \label{eq:bandgap_temp}
    \eg(T) = \eg(0) - \Delta E_{\mathrm{th}}(T) - \Delta E_{\mathrm{ph}}(T)
\end{equation}

At high temperatures the band gap decreases linearly~\cite{galeckas2002,passler2002,ladesmartin2000} while for low temperatures non-linear, i.e., quadratic~\cite{passler2002} or plateau-like~\cite{odonnell1991}, behavior was observed. An empirical law to describe the temperature dependent band gap is the Varshni relation (see \cref{eq:bandgap_varshni})~\cite{varshni1967}, with $\tg$ as arbitrary characterization temperature. In this case, $\eg \propto T^2$ at low temperatures.
\begin{equation}
    \label{eq:bandgap_varshni}
    \eg(T) = \eg(\tg) + \alpha \left( \frac{\tg^2}{\tg+\beta} - \frac{T^2}{T+\beta} \right)
\end{equation}

Two shortcomings of the Varshni relation were reported~\cite{passler1997,passler1998,odonnell1991}: First, available data sets can not be approximated simultaneously for low and high temperatures~\cite{grivickas2007,passler1999}. For example, \citet{passler1999} tried to match the values from \citet{choyke1964} but achieved only unrealistic results. Second, it is impossible to correlate the parameters of the Varshni relation with physical mechanism-specific quantities. For example, $\beta$ was said to be approximately the Debye temperature, but we even found negative values in the literature~\cite{balachandran2005,nawaz2010,usman2014}. For these reasons, \citet{scajev2009} proposed a description that explicitly used the Debye temperature $\debyeT$ (see Eq.~(\ref{eq:bandgap_debyeT})).
\begin{equation}
    \label{eq:bandgap_debyeT}
    \begin{split}
        \eg(T) &= \eg(0) - \ad\, \fd(\debyeT,T) \\
        \fd(\debyeT,T) &= \frac{18\,\kb\,T^4}{\debyeT^3} \int_0^{\debyeT/T} \left( \exp(x) -1 \right)^{-1}\, x^3\, \mathrm{d}x
    \end{split}
\end{equation}

Plateau-like behavior at cryogenic temperatures is achieved by physics-based models of Bose-Einstein type (see \cref{eq:bandgap_bose_einstein})~\cite{odonnell1991,vina1984} with $\Theta_\mathrm{B}$ as the mean frequency of the involved phonons and $\alpha_\mathrm{B}$ as the strength of the electron-phonon interaction~\cite{adachi2005}.
\begin{equation}
    \label{eq:bandgap_bose_einstein}
    \eg(T) = E_\mathrm{B} - \alpha_\mathrm{B} \left( 1 + \frac{2}{\mathrm{e}^{\Theta_\mathrm{B}/T} -1 } \right)
\end{equation}

The rate of change of the band gap at low temperatures depends on the phonon dispersion $\Delta$ of a material~\cite{passler1999}. \cref{eq:bandgap_varshni} is most accurate for $\Delta > 1$~\cite{passler2002,passler2003} and \cref{eq:bandgap_bose_einstein} represents the lower limit ($\Delta \to 0$), most suitable for $\Delta < \frac{1}{3}$~\cite{passler1999}. Since the dispersion of most semiconductors is in the range of \numrange{0.3}{0.6}~\cite{passler1999} \citet{passler1997,passler1999} proposed the model shown in \cref{eq:bandgap_passler} with $\varepsilon$ the entropy and $\Theta_\mathrm{p}$ as the average phonon temperature.
\begin{equation}
\label{eq:bandgap_passler}
    \begin{split}
        \eg(T) =& \eg(0) - \frac{\varepsilon \Theta_\mathrm{p}}{2} \left[\sqrt[p]{1+\left(\frac{2T}{\Theta_\mathrm{p}}\right)^p}-1\right]\\
        p \approx& \sqrt{\frac{1}{\Delta^2}+1}
    \end{split}
\end{equation}
This model was subsequently extended by \citet{passler2002} to cover a wider phonon dispersion ($\Delta$) range and, thus, promises to bridge the gap between \cref{eq:bandgap_varshni} and \cref{eq:bandgap_bose_einstein}. Because this advanced description has not been adopted by the community yet, we will employ the more commonly used model described in \cref{eq:bandgap_passler} in this review.


\subsubsection{Doping Dependency}

Doping-dependent changes of the band gap are caused by many-body effects of free carriers, i.e., their interactions with dopants and with each other. These become dominant for small carrier-to-carrier distances~\cite{schubert1993}. Possible are interactions within a band, across bands, and with ionized dopants~\cite{schubert1993,jain1991,lindefelt1998}. These effects were investigated extensively on their own~\cite{schubert1993} and were, for 4H-SiC, eventually merged by \citet{lindefelt1998} (see \cref{eq:bandgap_lindefelt}) for ionized charge carrier concentrations (see \cref{sec:incompIon}) above a few \SI{e18}{\per\cubic\centi\meter} based on a similar analysis for silicon by \citet{jain1991}. $E_\mathrm{(n/p)(c/v)}$ denotes the changes to the conduction (c) resp. valence (v) band due to n- or p-type doping. With increasing doping concentrations, the conduction band energy level drops, i.e., $\Delta \enc<0$ and $\Delta \epc < 0$, and the valence band energy level increases, i.e., $\Delta \env>0$ and $\Delta \epv > 0$. To correctly account for the overall reduction of the band gap, also called \textit{band gap narrowing}, in \cref{eq:bandgap_TN}, the factors concerning the conduction band are subtracted in the first line of \cref{eq:bandgap_lindefelt}. The term $\cpv$ was added later in an extension proposed by \citet{persson1999}.
\begin{equation}
\label{eq:bandgap_lindefelt}
\begin{split}  
    \Delta \eg (\ndp,\nam) &= \Delta \evb - \Delta \ecb \\
    &=\Delta E_{\mathrm{nv}} (\ndp) + \Delta E_{\mathrm{pv}} (\nam) - \Delta E_{\mathrm{nc}} (\ndp) - \Delta E_{\mathrm{pc}} (\nam) \\
    \Delta \enc (\ndp) &= \anc \left( \frac{\ndp}{\num{e18}} \right)^{1/3} + \bnc \left( \frac{\ndp}{\num{e18}} \right)^{1/2} < 0\\
    \Delta \env (\ndp) &= \anv \left( \frac{\ndp}{\num{e18}} \right)^{1/4} + \bnv \left( \frac{\ndp}{\num{e18}} \right)^{1/2} > 0\\
    \Delta \epc (\nam) &= \apc \left( \frac{\nam}{\num{e18}} \right)^{1/4} + \bpc \left( \frac{\nam}{\num{e18}} \right)^{1/2} < 0 \\
    \Delta \epv (\nam) &= \apv \left( \frac{\nam}{\num{e18}} \right)^{1/3} + \bpv \left( \frac{\nam}{\num{e18}} \right)^{1/2} + \cpv \left( \frac{\nam}{\num{e18}} \right)^{1/4}\ > 0
\end{split}
\end{equation}

We found publications that grouped the contributions to the band gap narrowing by doping type and, occasionally, also merged the parameters and the denominator \num{e18}. This description is often referred to as Jain-Roulston model~\cite{jain1991} (see \cref{eq:bandgap_jain}). The correlations among the parameters of \cref{eq:bandgap_lindefelt,eq:bandgap_jain} result to $A_{x\mathrm{c}}^\mathrm{j}/A_{x\mathrm{c}} = \num{e-6}$, $A_{x\mathrm{c}}^\mathrm{j}/A_{x\mathrm{v}} \approx \num{3.162e-5}$ and $B_{xy}^\mathrm{j}/B_{xy}=\num{e-9}$ with $x\in \{\mathrm{n,p}\}, y \in \{\mathrm{c,v}\}$. For better comparison, we transferred all parameters of \cref{eq:bandgap_jain} to their respective counterparts in \cref{eq:bandgap_lindefelt}, except the sum of $B_{xy}$, which could not be separated.
\begin{equation}
\label{eq:bandgap_jain}
\begin{split}  
    \Delta \eg (\ndp,\nam) &= \Delta \egp (\nam) + \Delta \egn (\ndp) \\
    \Delta \egn (\ndp) &= - \anc^\mathrm{j} \left(\ndp \right)^{1/3} + (\bnv^\mathrm{j} - \bnc^\mathrm{j}) \left(\ndp \right)^{1/2} + \anv^\mathrm{j} \left(\ndp\right)^{1/4} \\
    \Delta \egp (\nam) &= - \apc^\mathrm{j} \left(\nam \right)^{1/3} + (\bpv^\mathrm{j}-\bpc^\mathrm{j}) \left(\nam \right)^{1/2} + \apv^\mathrm{j} \left(\nam\right)^{1/4}
\end{split}
\end{equation}

An alternative approach to describe the doping dependency is the Slotboom model (see \cref{eq:bandgap_slotboom}), which was originally developed for Si. $N$ denotes the sum of all dopants and $N_\mathrm{n,p}$ a reference concentration. \citet{ruff1994} proposed parameters for 6H before \citet{ladesmartin2000} fitted \cref{eq:bandgap_slotboom} to the results of \cref{eq:bandgap_lindefelt}.
\begin{equation}
    \label{eq:bandgap_slotboom}
    \Delta \eg = C_{\mathrm{n,p}} \left( \ln\left( \frac{N}{N_\mathrm{n,p}} \right) + \sqrt{\left( \ln\left(\frac{N}{N_\mathrm{n,p}}\right) \right)^2 + G} \right)
\end{equation}

Finally, the band gap reduction can also be interpreted as renormalization due to electron-electron interactions alone, which \citet{schubert1993} described by \cref{eq:bandgap_schubert}. 
\begin{equation}
    \label{eq:bandgap_schubert}
    \Delta \eg = \frac{e^2}{4 \uppi \varepsilon \rs}
\end{equation}
%
The screening radius $\rs$ depends on the charge carrier concentration $n$ and is given by the Debye and Thomas-Fermi radii for the non-degenerate (see Eq.~(\ref{eq:bandgap_debye})) and degenerate (see Eq.~(\ref{eq:bandgap_thomas_fermi})) case~\cite{schubert1993}.
\begin{eqnarray}
    \Delta \eg &= \frac{e^3 \sqrt{n}}{4 \uppi \varepsilon^{3/2} \sqrt{\kb T}} &\qquad \text{(Debye, non-degenerate)} \label{eq:bandgap_debye} \\
    \Delta \eg &= \frac{e^3 \sqrt{\mde (3n)^{1/3}}}{4 \uppi^{5/3} \varepsilon^{3/2} \hbar} &\qquad \text{(Thomas-Fermi, degenerate)} \label{eq:bandgap_thomas_fermi}
\end{eqnarray}

State-of-the-art TCAD tools only support \cref{eq:bandgap_varshni,eq:bandgap_lindefelt,eq:bandgap_slotboom} out of the box, although some feature the possibility to write custom code for band gap narrowing. \cref{eq:bandgap_jain} is only partially supported, which can cause problems if solely the sum $(\bnv^\mathrm{j} - \bnc^\mathrm{j})$ respectively $(\bpv^\mathrm{j}-\bpc^\mathrm{j})$ is provided in the literature.


\subsubsection{Methods}

Appropriate methods to determine the band gap were partially discussed by \citet{denapoli2022} and \citet{nava2008}. One possibility are measurements, e.g., transmission spectroscopy (TS)~\cite{ahuja2002, cheng2022}, spectroscopic ellipsometry~\cite{mainali2024}, photo absorption (PA)~\cite{choyke1957,choyke1964,choyke1964a,grivickas2007}, optical admittance (OA)~\cite{evwaraye1996}, exciton electroabsorption (EE)~\cite{dubrovskii1975}, free carrier absorption (FCA)~\cite{galeckas2002}, free exciton luminescence (FEL)~\cite{ikeda1980b}, photoluminescence (PL)~\cite{itoh1996,ivanov1998,kordina1995}, photoconductivity (PC)~\cite{ivanov2002}, transient absorption spectrum (TAS)~\cite{fang2018}, optical absorption spectrum (OAS)~\cite{miller2000} and wavelength-modulated absorption (WMA)~\cite{klahold2020,sridhara2000}. Sometimes, the results of multiple measurements are combined to improve the accuracy~\cite{choyke1969}. \citet{stefanakis2014} stated that the free carrier absorption method overestimates the band gap while optical absorption studies deliver more accurate results.

\revision{Calculations are a common alternative and include empirical pseudopotential methods (EPM)~\cite{bellotti2000,junginger1970,vanhaeringen1997}, density functional theory based local density approximation (DFT-LDA)~\cite{dong2004,kaeckell1996} using the orthogonalized linear combination of atomic orbitals method (OLCAO)~\cite{ching2006}, first-principles self-consistent linear muffin-tin orbital method (LMTO-ASA)~\cite{gavrilenko1990}, generalized gradient approximation (GGA)~\cite{huang2022b,nuruzzaman2015}, GW approximation (GW)~\cite{sinelnik2021,wenzien1995,ummels1998,son2004}, HSE06 approximation (HSE06)~\cite{yan2020,kuroiwa2019,lu2021,torres2022}, hybrid pseudo-potential tight-binding method (HPT)~\cite{chen1997}, full-potential linearized augmented plane wave method (LAPW)~\cite{persson1997}, modified Becke-Johnson potential (mBJ)~\cite{yamaguchi2018} and BZW procedure (BZW)~\cite{zhao2000a}. Further calculations were based on rectangular barriers of finite height (RB)~\cite{dubrovskii1977} and estimations based on the crystal hexagonality (HEX)~\cite{zanmarchi1964}. \citet{backes1994} considered the lattice of polytypes as natural superlattices consisting of pure 3C SiC segments (NASU). Available results in literature were approximated by fitting (FIT)~\cite{hatakeyama2013,kimoto2014a,ladesmartin2000,levinshtein2001,lechner2021} and genetic algorithm fitting (GAF)~\cite{ng2010}.}

\subsection{Results \& Discussion}

In the time span from the early 1960s until today, we identified $51$ fundamental investigations. Out of those were $21$ measurements, $20$ calculations, $6$ fittings to existing data and $4$ models whose origin could not be determined (see \cref{tab:bandgap_models,tab:bandgap_modelsII}). Seven out of the $51$ fundamental investigations were published within the last five years, showing that research on the band gap is still very active.

\begin{table}[p!]
    \setlength{\tabcolsep}{9pt}
    \caption{\label{tab:bandgap_models}Band gap energies and temperature dependency fittings according to \cref{eq:bandgap_varshni} [1/2]. \bkgCol{no4HColor} indicates research not focused on 4H-SiC and \bkgCol{calcColor} calculations.}
    \resizebox{0.99\linewidth}{!}{%
    \begin{tabular}{l | *{4}{c} | *{2}{c} |  cc}
& \multicolumn{4}{c}{band gap} & \multicolumn{2}{|c|}{temperature dep.} && \\ref. & $\eg$ & $\egx$ & $\ex$ & $\tg$ & $\alpha$ & $\beta$ & range & method \\ 
 & [\si{\eV}] & [\si{\eV}] & [\si{\meV}] & [\si{\K}] & [\si{\eV\per\K}] & [\si{\K}] & [\si{\K}] & \\ \hline 
 &&&&&&&& \\[-7pt] 
\rowcolor{no4HColor} \reftext{Choy57}\cite{choyke1957}\,\footnotemark[1]& --& --& --& --& \num{3.3e-4}& -- & \numrange{300}{710} & PA \\ 
\reftext{Choy64}\cite{choyke1964}& --& \num{3.263+-0.003}& --& \num{4}& -- & --  & -- & PA \\ 
\reftext{Choy64a}\cite{choyke1964a}\,\footnotemark[2]& --& \num{3.265}& --& \num{4.7}& -- & --  & -- & PA \\ 
\reftext{Zanm64}\cite{zanmarchi1964}& --& \num{3.23}& --& \num{300}& -- & --  & -- & HEX \\ 
\rowcolor{calcColor} \reftext{Jung70}\cite{junginger1970}& \num{2.8}& --& --& \num{0}&--&--&--& EPM\\ 
\reftext{Dubr75}\cite{dubrovskii1975}& --& --& \num{20+-1.5}& \num{2}& -- & --  & -- & EE \\ 
\rowcolor{calcColor} \reftext{Dubr77}\cite{dubrovskii1977}& \num{3.2}& --& --& \num{0}&--&--&--& RB\\ 
\reftext{Iked80b}\cite{ikeda1980b}& --& \num{3.2639}& \num{10}& \num{77}& -- & --  & -- & FEL \\ 
\rowcolor{calcColor} \reftext{Gavr90}\cite{gavrilenko1990}& \num{2.89}& --& --& \num{0}&--&--&--& LMTO-ASA\\ 
\rowcolor{calcColor} \reftext{Back94}\cite{backes1994}& \num{3.28}& --& --& \num{0}&--&--&--& NASU\\ 
\rowcolor{calcColor} \reftext{Park94}\cite{park1994}& \num{2.14}& --& --& \num{0}&--&--&--& EPM\\ 
\reftext{Kord95}\cite{kordina1995}& --& \num{3.265}& --& \num{4.2}& -- & --  & -- & PL \\ 
\rowcolor{calcColor} \reftext{Wenz95}\cite{wenzien1995}& \num{3.56}& --& --& \num{0}&--&--&--& GW\\ 
\reftext{Evwa96}\cite{evwaraye1996}& \num{3.41+-0.03}& --& --& \num{40}& -- & --  & -- & OA \\ 
\reftext{Itoh96}\cite{itoh1996}\,\footnotemark[2]& --& 3.265& --& 4.25& -- & --  & -- & PL \\ 
\rowcolor{calcColor} \reftext{Kaec96}\cite{kaeckell1996}& \num{2.18}& --& --& \num{0}&--&--&--& DFT-LDA\\ 
\rowcolor{calcColor} \reftext{Chen97}\cite{chen1997}& \num{3.27}& --& --& \num{0}&--&--&--& HPT\\ 
\rowcolor{calcColor} \reftext{Pers97}\cite{persson1997}& \num{2.9}& --& --& \num{0}&--&--&--& DFT-LDA\\ 
\reftext{Ivan98}\cite{ivanov1998}& --& \num{3.266}& --& \num{2}& -- & --  & -- & PL \\ 
\rowcolor{calcColor} \reftext{Umme98}\cite{ummels1998}& \num{3.35}& --& --& \num{0}&--&--&--& GW\\ 
\rowcolor{calcColor} \reftext{Bell00}\cite{bellotti2000}& \num{3.05}& --& --& \num{0}&--&--&--& EPM\\ 
\reftext{Lade00}\cite{ladesmartin2000}\,\footnotemark[4]& --& 3.265& \num{40}& 0& \num{3.3e-4} & \num{1050}& \numrange{4}{200} & FIT\\ 
& -- & 3.265 & \num{40}& 0 & \num{3.3e-2} & \num{1e5}& \numrange{4}{600} & FIT\\ 
& -- & 3.342 & \num{40}& 0 & \num{3.3e-4} & \num{0}& \numrange{300}{700} & FIT\\ 
\reftext{Mill00}\cite{miller2000}& \num{3.378+-0.001}& --& --& \num{15}& -- & --  & -- & OAS \\ 
\reftext{Srid00}\cite{sridhara2000}& --& \num{3.267}& --& \num{2}& -- & --  & -- & WMA \\ 
\rowcolor{calcColor} \reftext{Zhao00a}\cite{zhao2000a}& \num{3.11}& --& --& \num{0}&--&--&--& BZW\\ 
\reftext{Levi01}\cite{levinshtein2001}\,\footnotemark[3]& \num{3.23}& --& --& \num{300}& \num{6.5e-4}& \num{1300} & \numrange{0}{800} & FIT \\ 
\reftext{Ahuj02}\cite{ahuja2002}& \num{3.26+-0.098}& --& --& \num{300}& -- & --  & -- & TS \\ 

\end{tabular}
        \footnotetext[1] {according to the shown band gap, the data are for 21R
          but others~\cite{bakowski1997} denote them as 6H}
        \footnotetext[2] {solely measurement for lowest temperature shown}
        \footnotetext[3] {model fitted to results by \citet{choyke1969}}
        \footnotetext[4] {model fitted to 6H-SiC results by \citet{choyke1962}}
    }
\end{table}

\begin{table}[p!]
    \setlength{\tabcolsep}{9pt}
    \caption{\label{tab:bandgap_modelsII}Band gap energies and temperature dependency fittings according to \cref{eq:bandgap_varshni} [2/2]. \bkgCol{no4HColor} indicates research not focused on 4H-SiC and \bkgCol{calcColor} calculations.}
    \resizebox{0.99\linewidth}{!}{%
    \begin{tabular}{l | *{4}{c} | *{2}{c} |  cc}
& \multicolumn{4}{c}{band gap} & \multicolumn{2}{|c|}{temperature dep.} && \\ref. & $\eg$ & $\egx$ & $\ex$ & $\tg$ & $\alpha$ & $\beta$ & range & method \\ 
 & [\si{\eV}] & [\si{\eV}] & [\si{\meV}] & [\si{\K}] & [\si{\eV\per\K}] & [\si{\K}] & [\si{\K}] & \\ \hline 
 &&&&&&&& \\[-7pt] 
\reftext{Gale02}\cite{galeckas2002}& 3.285& --& --& 0& \num{3.5e-4} & \num{1100}& \numrange{0}{650} & FCA\\ 
& 3.2625 & -- & --& 300 & \num{2.4e-4} & \num{0}& \numrange{300}{650} & FCA\\ 
\reftext{Ivan02}\cite{ivanov2002}& \num{3.285}& --& \num{20.5+-1}& \num{2}& -- & --  & -- & PC \\ 
\reftext{Shal02}\cite{shalish2002}& \num{3.18}& --& --& \num{300}& -- & --  & -- & PL \\ 
\rowcolor{calcColor} \reftext{Dong04}\cite{dong2004}& \num{2.194}& --& --& \num{0}&--&--&--& DFT-LDA\\ 
\rowcolor{calcColor} \reftext{Son04}\cite{son2004}& \num{3.35}& --& --& \num{0}&--&--&--& GW\\ 
\reftext{Bala05}\cite{balachandran2005}\,\footnotemark[1]& \num{3.26}& --& --& \num{300}& \num{4.15e-4}& \num{-131} & -- & -- \\ 
\rowcolor{calcColor} \reftext{Chin06}\cite{ching2006}& \num{2.433}& --& --& \num{0}&--&--&--& OLCAO\\ 
\reftext{Griv07}\cite{grivickas2007}& --& 3.267& \num{30+-10}& 0& -- & --  & \numrange{0}{500} & PA \\ 
\reftext{Tama08a}\cite{tamaki2008a}\,\footnotemark[1]& \num{3.23}& --& --& \num{300}& \num{7.036e-4}& \num{1509} & -- & -- \\ 
\reftext{Ng10}\cite{ng2010}& \num{3.28}& --& --& \num{0}& -- & --  & -- & GAF \\ 
\reftext{Khal12}\cite{khalid2012}\,\footnotemark[1]& \num{3.285}& --& --& \num{0}& \num{2.206e-2}\footnotemark[4] & \num{1e5} & -- & -- \\ 
\reftext{Hata13}\cite{hatakeyama2013}\,\footnotemark[3]& \num{3.285}& \num{3.265}& --& \num{0}& \num{9.06e-4}& \num{2030} & \numrange{0}{800} & FIT \\ 
\reftext{Kimo14a}\cite{kimoto2014a}\,\footnotemark[3]& --& \num{3.265}& --& \num{2}& \num{8.2e-4}& \num{1800} & \numrange{0}{800} & FIT \\ 
\reftext{Stef14}\cite{stefanakis2014}\,\footnotemark[2]& \num{3.285}& --& --& \num{0}& \num{3.3e-4}& \num{240} & -- & -- \\ 
\rowcolor{calcColor} \reftext{Nuru15}\cite{nuruzzaman2015}& \num{2.45}& --& --& \num{0}&--&--&--& GGA\\ 
\reftext{Fang18}\cite{fang2018}& \num{3.22}& --& --& \num{300}& -- & --  & -- & TAS \\ 
\rowcolor{calcColor} \reftext{Yama18}\cite{yamaguchi2018}& \num{3.12}& --& --& \num{0}&--&--&--& mBJ\\ 
\rowcolor{calcColor} \reftext{Kuro19}\cite{kuroiwa2019}& \num{3.15}& --& --& \num{0}&--&--&--& HSE06\\ 
\reftext{Klah20}\cite{klahold2020}& --& \num{3.2659}& \num{40}& \num{1.4}& -- & --  & -- & WMA \\ 
\rowcolor{calcColor} \reftext{Yan20}\cite{yan2020}& \num{3.19}& --& --& \num{0}&--&--&--& HSE06\\ 
\reftext{Lech21}\cite{lechner2021}\,\footnotemark[1]& \num{3.265}& --& --& \num{0}& \num{10.988e-3}& \num{32744.3} & -- & FIT \\ 
\rowcolor{calcColor} \reftext{Lu21}\cite{lu2021}& \num{3.17}& --& --& \num{0}&--&--&--& HSE06\\ 
\rowcolor{calcColor} \reftext{Sine21}\cite{sinelnik2021}& \num{3.26}& --& --& \num{0}&--&--&--& GW\\ 
\reftext{Chen22}\cite{cheng2022}& \num{3.28}& --& --& \num{300}& \num{5.27e-4}& \num{0} & \numrange{300}{620} & TS \\ 
\rowcolor{calcColor} \reftext{Huan22b}\cite{huang2022b}& \num{3.18}& --& --& \num{0}&--&--&--& GGA\\ 
\rowcolor{calcColor} \reftext{Torr22}\cite{torres2022}& \num{3.17}& --& --& \num{0}&--&--&--& HSE06\\ 
\reftext{Main24}\cite{mainali2024}& \num{3.3+-0.02}& --& --& \num{300}& -- & --  & -- & SE 
\end{tabular}
        \footnotetext[1] {origin of Varshni parameters unknown}    
        \footnotetext[2] {values referenced from a TCAD tool manual, no corresponding scientific publication found}
        \footnotetext[3] {model fitted to results by \citet{choyke1969}}
        \footnotetext[4] {this parameter is negative in the paper, leading to an increasing band gap with temperature}
        \footnotetext[5] {model fitted to results by \citet{miller2000}}
    }
\end{table}

\citet{ivanov2002} conducted the only measurement of the band gap energy $\eg$ at temperatures below $\SI{5}{\K}$, who extracted a value of $\approx \SI{3.285}{\eV}$. This is consistent with the analysis by \citet{galeckas2002} who measured the absorption coefficient down to $\SI{50}{\K}$ and then added measurement values taken at $\SI{2}{\K}$ by \citet{sridhara1998} to extrapolate $\eg(0)=\SI{3.285}{\eV}$. At slightly elevated temperatures \citet{miller2000} ($\SI{3.378+-0.001}{\eV}$ at $\SI{15}{\K}$) and \citet{evwaraye1996} ($\SI{3.41+-0.03}{\eV}$ at $\SI{40}{\K}$) achieved higher values for the band gap.

An explanation for this lack of data on $\eg$ is the fact the exciton band gap energy $\egx$ was mainly determined for low temperatures. We found seven investigations that agreed remarkably well on $\egx(0)= \SI{3.265}{\eV}$ (see \cref{fig:bandgap_fund_stat} for a statistical representation)\revision{, although different measurement techniques were used. Solely optical admittance (OA) and optical absorption spectrum (OAS) delivered values that are $\approx\SI{100}{\milli\eV}$ higher.} \citet{grivickas2007} predicted $\egx(0)= \SI{3.267}{\eV}$ by combining measurements down to $\SI{100}{\K}$ with the low-temperature results by \citet{itoh1996}. The authors thereby shifted the latter, which predicted $\egx(4.25)= \SI{3.265}{\eV}$~\cite{itoh1996}, by $\SI{2}{\milli\eV}$ to achieve a better match with their own results.

The free exciton binding energy $\ex$ was extracted either directly from measurements~\cite{dubrovskii1975,ikeda1980b,ivanov2002,grivickas2007} or calculated by using the reduced mass of the electron-hole pair $m_{\mathrm{red}}$ and the permittivity $\varepsilon$ in the hydrogen model shown in \cref{eq:bandgap_hydrogen}~\cite{klahold2020,elliott1957}.
\begin{equation}
    \label{eq:bandgap_hydrogen}
    \ex = \frac{m_{\mathrm{red}} e^4}{2 \hbar^2 \varepsilon^2}
\end{equation}
The values of $\ex$ range from \SIrange{10}{40}{\milli\electronvolt}, whereas \citet{devaty1997} argued that \SI{10}{\milli\electronvolt}~\cite{ikeda1980b}, which denote the activation energy for thermal quenching of free excitons, is too low to be the free exciton binding energy. Two out of five investigations propose an average value of $\SI{20}{\milli\eV}$, which would match the observation of $\egx=\SI{3.265}{\electronvolt}$ and $\eg=\SI{3.285}{\electronvolt}$ below five Kelvin.

At room temperature, measurements were only conducted for $\eg$. The spread of values is, compared to $\egx$ at low temperatures, big (cp. \cref{fig:bandgap_fund_stat}). Single measurements, such as the investigation by \citet{ahuja2002} with an uncertainty of \SI{98}{\milli\electronvolt}, showed variations in the range of a few \%. Even the latest three results published in the last seven years propose values of \SI{3.26+-0.04}{\electronvolt}. A possible explanation are doping-dependent band gap variations. Therefore, we ordered the results by the utilized n-type doping concentration in \cref{tab:bandgap_doping}. However, no clear tendency could be identified, meaning that future research is required to converge on a common value or find a suitable explanation for the deviations.

\begin{figure}[t]
    \centering
    \resizebox{0.85\linewidth}{!}{%
    \input{figures/bandgap_fund_stat}
    }
    \caption{\label{fig:bandgap_fund_stat}Statistical analysis of band gap measurements. Shown are the 0th, 25th, 50th, 75th and 100th quartile. The mean value is added in numerical form.}
\end{figure}

\begin{table}[ht]
    \centering
    \caption{\label{tab:bandgap_doping}Measurements of $\eg$ at \SI{300}{\K} ordered by the utilized n-type doping concentration.}
    \setlength{\tabcolsep}{9pt}
    \resizebox{0.5\linewidth}{!}{%
    \begin{tabular}{l ccc}
         ref. & dopant & doping conc. & band gap \\
         && [\si{\per\cubic\cm}] & [\si{\eV}] \\ \hline
         \reftext{Chen22}\cite{cheng2022} & -- & -- & \num{3.28} \\         
         \reftext{Gale02}\cite{galeckas2002} & N & \numrange{7e14}{4e15} & \num{3.2625} \\
         \reftext{Shal02}\cite{shalish2002} & N & \num{1e16} & \num{3.18} \\
         \reftext{Fang18}\cite{fang2018} & -- & \num{1.1e18}  & \num{3.22} \\ 
         \reftext{Main24}\cite{mainali2024} & -- & \num{3.7e18} & \num{3.30+-0.02} \\
         \reftext{Ahuj02}\cite{ahuja2002} & -- & \num{7e18} & \num{3.26+-0.098}
    \end{tabular}
    }
\end{table}

\revision{The results achieved by calculations are in less agreement. In our analyses we discovered a dependency on the utilized computational method. The closest results to measurements were achieved using the hybrid pseudo-potential tight-binding method (HPT) (\SI{3.27}{\eV}) and considering 4H as superlattice of pure 3C SiC segments (NASU) (\SI{3.28}{\eV}). The GW approximation resulted in close (\SI{3.26}{\eV}\cite{sinelnik2021}) but also higher values of \SI{3.35}{\eV}\cite{son2004,ummels1998} and \SI{3.56}{\eV}~\cite{wenzien1995}. The HSE06 approximation achieved consistent but lower values in the range \SIrange{3.15}{3.19}{\eV} (see \cref{tab:bandgap_calculation}).}

\begin{table}[ht]
  \centering
  \setlength{\tabcolsep}{12pt}
  \renewcommand{\arraystretch}{0.8}
  \caption{\label{tab:bandgap_calculation}Band gap results grouped by the utilized
    calculation method.}
  \resizebox{0.75\linewidth}{!}{%
    \begin{tabular}{l cl}
method & $\eg$ & publications \\ 
 & [\si{\eV}] & \\ \hline 
 && \\[-7pt] 
BZW & \num{3.11} & \reftext{Zhao00a}\cite{zhao2000a}\\ 
DFT-LDA & \numrange{2.18}{2.90} & \reftext{Kaec96}\cite{kaeckell1996}, \reftext{Pers97}\cite{persson1997}, \reftext{Dong04}\cite{dong2004}\\ 
EPM & \numrange{2.14}{3.05} & \reftext{Jung70}\cite{junginger1970}, \reftext{Park94}\cite{park1994}, \reftext{Bell00}\cite{bellotti2000}\\ 
GGA & \numrange{2.45}{3.18} & \reftext{Nuru15}\cite{nuruzzaman2015}, \reftext{Huan22b}\cite{huang2022b}\\ 
GW & \numrange{3.26}{3.56} & \reftext{Wenz95}\cite{wenzien1995}, \reftext{Umme98}\cite{ummels1998}, \reftext{Son04}\cite{son2004}, \reftext{Sine21}\cite{sinelnik2021}\\ 
HPT & \num{3.27} & \reftext{Chen97}\cite{chen1997}\\ 
HSE06 & \numrange{3.15}{3.19} & \reftext{Kuro19}\cite{kuroiwa2019}, \reftext{Yan20}\cite{yan2020}, \reftext{Lu21}\cite{lu2021}, \reftext{Torr22}\cite{torres2022}\\ 
LMTO-ASA & \num{2.89} & \reftext{Gavr90}\cite{gavrilenko1990}\\ 
NASU & \num{3.28} & \reftext{Back94}\cite{backes1994}\\ 
OLCAO & \num{2.43} & \reftext{Chin06}\cite{ching2006}\\ 
RB & \num{3.20} & \reftext{Dubr77}\cite{dubrovskii1977}\\ 
mBJ & \num{3.12} & \reftext{Yama18}\cite{yamaguchi2018}\\ 

\end{tabular}
  }
\end{table}    

We only found one publication that described the anisotropy~\cite{grivickas2007} of the band gap. The results revealed uniform exciton (band gap) energies for fields parallel resp. perpendicular to the c-axis.

\subsubsection{Temperature Dependency}

We found four measurement campaigns on the temperature dependency of the band gap in 4H-SiC~\cite{choyke1964a,cheng2022,galeckas2002,grivickas2007}. Additional models~\cite{hatakeyama2013,kimoto2014a,levinshtein2001,ladesmartin2000} were developed by fitting to data published in the 1960s~\cite{choyke1964a,choyke1964,choyke1969} (see \cref{tab:bandgap_models,tab:bandgap_modelsII}). Since the development of dedicated temperature-dependent models only started in the 2000s, earlier publications were forced to rely on models developed for 6H-SiC~\cite{bakowski1997,wang1999}. However, in 2004~\cite{ayalew2004} and even in 2015~\cite{megherbi2015}, 6H based models were still in use for 4H-SiC, showing the importance of a comprehensive literature overview on this topic.

Some of the proposed model parameters~\cite{balachandran2005,tamaki2008a,khalid2012,stefanakis2014,lechner2021} could not be related to any scientific publication, but in some cases to default values of prominent TCAD simulation suites~\cite{stefanakis2014}. Although \citet{tamaki2008a} provided a reference~\cite{casady1996}, we were unable to reproduce their parameters.  The model by \citet{khalid2012} is extraordinary, as it has a positive temperature dependency, i.e., an increase of the band gap with temperature is predicted. In our opinion, the chances are high that this was a typographical error, and we corrected the value for this review.

In the analyzed publications, the model in \cref{eq:bandgap_varshni} is predominantly used to describe the temperature dependency. The respective parameters are shown in \cref{tab:bandgap_models,tab:bandgap_modelsII}. We did not include $\alpha=\SI{3.2e-4}{\eV\per\K}$ and $\beta=\SI{565}{\K}$ presented by \citet{grivickas2007}, because the sole purpose of these values was to highlight the bad fit to the experimental data. Nevertheless, they still got referenced by \citet{levcenco2011}. The model introduced in \cref{eq:bandgap_passler} was utilized only once by \citet{grivickas2007} with the parameters shown in Eq.~(\ref{eq:bandgap_passler_params}). The model proposed by \citet{scajev2009} (see Eq.~(\ref{eq:bandgap_debyeT})) used the parameters shown in Eq.~(\ref{eq:bandgap_debyeT_params}), which were fitted to data by \citet{miller2000}.
\begin{eqnarray}
    \label{eq:bandgap_passler_params}
    \Theta_\mathrm{p} = \SI{450}{\K}\ , &\varepsilon =
                                                \SI{3e-4}{\electronvolt\per\K}\ , &p = \num{2.9} \\
    \label{eq:bandgap_debyeT_params}
    \eg(0)=\SI{3.3762}{\eV}\ , &\ad=\num{1.31}\ ,&\debyeT=\SI{1000+-50}{\K}
\end{eqnarray}

\begin{figure}[p]
    \centering
    \resizebox{0.99\linewidth}{!}{%
    \input{figures/bandgap_models}
    }
    \caption{\label{fig:bandgap_models}Band gap measurements and models. The latter are only shown in the interval used during characterization, with their corresponding equations referenced at the end of each entry.}
\end{figure}

For an easier comparison, we plotted all the measurements and models presented earlier (see \cref{fig:bandgap_models}). The fittings~\cite{levinshtein2001,hatakeyama2003,kimoto2014a} to the data by \citet{choyke1969} are all very similar, so we only show two out of the three explicitly. Clearly observable is the similarity of these models with the one proposed by \citet{tamaki2008a}, suggesting that also the latter was fitted to \citet{choyke1969}.  Similarly, the parameters provided by \citet{lechner2021} reproduce the fitting from \citet{ladesmartin2000} (\SIrange{4}{600}{\K}), causing us to not show the former. The results by \citet{scajev2009} are not shown as they are identical to those by \citet{miller2000}.

We identified some confusion between $\eg$ and $\egx$ in the literature. For example, some authors~\cite{levinshtein2001,tamaki2008a,stefanakis2014,lechner2021} stated that they describe $\eg$ but the values agree more to $\egx$. The model by \citet{khanna2023} matches $\eg$ at low temperatures, but at around $\SI{200}{\K}$, it is equal to $\egx$ and follows that value from there onward. The fitting by \citet{balachandran2005} is only feasible for $T \geq \SI{300}{\K}$ as it has a singularity at \SI{131}{\K}. Newer models, e.g., the one by \citet{cheng2022}, suggested a steeper slope, actually crossing the traces of $\egx$ with $\eg$.

The plateau described by \cref{eq:bandgap_passler} is barely visible, showing that the deviations are only subtle. \citet{ladesmartin2000} approximated the shape with the Varshni model by just fitting it in a very narrow temperature range. The phonon dispersion of 4H-SiC $\Delta=0.29$~\cite{grivickas2007} is rather low, which would actually indicate that \cref{eq:bandgap_bose_einstein} is the most suitable; however, \cref{eq:bandgap_passler} seems to be accurate as well. Even more, \citet{stefanakis2014} compared the single models and identified \cref{eq:bandgap_varshni} as the most fitting one.

\citet{miller2000} is the only publication we found that used \cref{eq:bandgap_bose_einstein}. The authors also proposed a second fitting (see \cref{eq:bandgap_miller_II}) with the difference that the exponent $-1$ was missing. Due to unreasonable results, we consider this a typographical error that we corrected in this review. The values achieved with $\eg(0)=\SI{3.378+-0.001}{\eV}$, $\kappa=\num{0.345+-0.031}$ and $\theta_\mathrm{E}=\num{613+-23}$ were equal to those by \cref{eq:bandgap_bose_einstein} so we do not show them explicitly.

\begin{equation}
    \label{eq:bandgap_miller_II}
    \eg(T) = \eg(0) - \kappa \left( \exp(\theta_\mathrm{E}/T) -1 \right)^{-1}
\end{equation}

In the figure, the tremendous spread of measurement values for $\eg(300)$ is visible. Clearly, no conclusive statements can be made based on these data, highlighting the need for further detailed analyses. It would also be possible to use $\eg=\egx+\ex$ for this purpose. However, we did not find any investigations of $\ex$ with increasing temperature.

\subsubsection{Doping Dependency}

We encountered two fittings for the doping-dependent narrowing model in \cref{eq:bandgap_lindefelt} (see \cref{tab:bandgap_lindefelt}), which are solely based on calculations. Measurement results are available~\cite{weingartner2002,grivickas2007}, but due to their sparsity (see \cref{fig:bandgap_lindefelt}), they are not suitable to verify the calculations. The models predict an increasing rate of change with doping concentration, whereas the narrowing induced by n-type doping is bigger than the one by p-type doping. The additional parameters proposed by \citet{persson1999} only led to small differences.

\begin{table*}[t]
    \centering
    \setlength{\tabcolsep}{7pt}
    \renewcommand{\arraystretch}{0.8}
    \sisetup{
    scientific-notation = fixed, fixed-exponent = 0,
    round-mode=places, round-precision=2}
    \caption{\label{tab:bandgap_lindefelt}Parameters for ionized dopants induced band gap narrowing in \cref{eq:bandgap_lindefelt}.}
    \resizebox{0.95\linewidth}{!}{%
        \begin{tabular}{ll*{4}{c}|*{5}{c}}
&&\multicolumn{4}{c|}{n-type} & \multicolumn{5}{c}{p-type} \\ 
ref. & method & $\anc$ & $\bnc$ & $\anv$ & $\bnv$ & $\apc$ & $\bpc$ & $\apv$ & $\bpv$ & $\cpv$\\ 
 && [\si{\meV}] & [\si{\meV}] & [\si{\meV}] & [\si{\meV}] & [\si{\meV}] & [\si{\meV}] & [\si{\meV}] & [\si{\meV}] & [\si{\meV}] \\ \hline 
 &&&&&&&&&& \\[-11pt] 
 \reftext{Lind98}\cite{lindefelt1998} & calculation  & \num{-1.500000e+01} & \num{-2.930000e+00} & \num{1.900000e+01} & \num{8.740000e+00} & \num{-1.570000e+01} & \num{-3.870000e-01} & \num{1.300000e+01} & \num{1.150000e+00} & -\\ 
 \reftext{Pers99}\cite{persson1999} & calculation  & \num{-1.791000e+01} & \num{-2.200000e+00} & \num{2.823000e+01} & \num{6.240000e+00} & \num{-1.615000e+01} & \num{-1.070000e+00} & \num{-3.507000e+01} & \num{6.740000e+00} & \num{5.696000e+01}
\end{tabular}
    }
\end{table*}    

\citet{lindefelt1998} highlighted that the valence band displacement is larger than the one of the conduction band (see \cref{fig:bandgap_lindefelt_cv}). In their model, this is consistent for both doping types. Other publications distributed the band gap narrowing equally across valence and conduction band~\cite{arvanitopoulos2017} or chose a contribution of $\mid\Delta E_\mathrm{c}\mid/\mid\Delta \eg\mid=0.7$~\cite{donnarumma2012}. The model predictions are right in the middle of these two cases, i.e., around a value of $0.6$.

\begin{figure*}[t!]
    \centering
    \resizebox{0.95\linewidth}{!}{%
        \input{figures/bandgap_lindefelt}
    }
    \caption{\label{fig:bandgap_lindefelt}Doping induced band gap narrowing due to n- ($\egn$) and p-type ($\egp$) doping. Black dots denote measurements. }
\end{figure*}

For the Slotboom model, that is used by various publications~\cite{ladesmartin2000,ayalew2004,lechner2021}, we found one set of parameters~\cite{ladesmartin2000} (see \cref{tab:bandgap_slotboom}). In contrast to the other models, the band gap narrowing is linear in the semi-logarithmic plot (see \cref{fig:bandgap_lindefelt}). Up to a doping concentration of \SI{e20}{\per\cubic\centi\m}, the predictions for p-type doping-induced narrowing are comparable to the earlier model, but for n-type doping, they are up to \SI{100}{\percent} higher.

\begin{table}[t]
    \centering
    \setlength{\tabcolsep}{8pt}
    \renewcommand{\arraystretch}{0.8}
    \caption{\label{tab:bandgap_slotboom}Parameters for ionized dopants induced band gap narrowing Slotboom model.}
    \resizebox{0.45\linewidth}{!}{%
    \begin{tabular}{l*{4}{c}}
    ref. & type & $C$ & $N$ & $G$ \\ 
    && [\si{\electronvolt}] & [\si{\per\cubic\centi\meter}] & [1] \\ \hline 
    &&&& \\[-11pt] 
    \reftext{Lade00}\cite{ladesmartin2000} & n & \num{2e-2} & \num{e17} & \num{0.5} \\
     & p & \num{9e-3} & \num{e17} & \num{0.5}
    \end{tabular}
    }
\end{table} 

\begin{figure*}[t!]
    \centering
    \resizebox{0.99\linewidth}{!}{%
        \input{figures/bandgap_lindefelt_cv}
    }
    \caption{\label{fig:bandgap_lindefelt_cv}Doping induced band gap narrowing. The changes to the conduction and valence band are shown for both dopants.}
\end{figure*}

The band gap narrowing according to Eq.~(\ref{eq:bandgap_thomas_fermi}) is also shown because it is occasionally used in literature~\cite{stefanakis2014,donnarumma2012}. \citet{stefanakis2014} used a slightly different form of the Thomas-Fermi radius shown in \cref{eq:bandgap_stefanakis} with $n_0$ as the equilibrium carrier density.
\begin{equation}
    \label{eq:bandgap_stefanakis}
    \Delta \eg = \frac{e^2}{4 \uppi \ezero \es} \left(\frac{3 n_0 e^2}{2 \ezero \ef}\right)^{1/2}
\end{equation}
Unfortunately we were unable to reproduce the results achieved by this formalism in the mentioned publications so we extracted the curves from \citet{donnarumma2012}. \citet{johannesson2016} used the Debye radius from Eq.~(\ref{eq:bandgap_debye}) scaled by $3/4$, based on the calculations by \citet{lanyon1979}.

\citet{miller2000} analyzed semi-insulating and doped 4H-SiC (without providing further details such as the doping concentration) with the surprising result that the band gap for the latter is bigger. \citet{fang2018} only registered a band gap narrowing of \SI{10}{\milli\eV}, which is explainable by the employed doping concentrations of \SI{1.1e18}{\per\cubic\cm} and \SI{9.1e18}{\per\cubic\cm}.

\subsubsection{Origin of Parameters}

Tracing parameters back to their origin was a challenging task because frequently references were missing. From the ones that were stated and some that we inferred based on the used values, we concluded that the most influential publications are from 1964 by \citet{choyke1964} and \citet{choyke1964a} (for a detailed graphical representation see \cref{fig:bandgap_ref_chain_bandgap} in Appendix~\ref{sec:refChainBandgap}). The outcome of more recent measurements was not yet adopted in the scientific community.


Through reference chains, the values previously found in low-temperature measurements were altered. Already in 1970 \citet{junginger1970} rounded the values of $\egx=\SI{3.263} {\electronvolt}$ to $\eg=\SI{3.26}{\electronvolt}$, i.e., changing both value and meaning. This was not the only time that we found such a change in literature~\cite{bellotti1999,kohlscheen2003,choi2005,tanner2007,zippelius2011,biondo2012,uhnevionak2015,rescher2018,berens2021,capan2022,denapoli2022,johannesson2019}, sometimes even extended by a change of the temperature to \SI{300}{\K}. We also found evidence that $\eg=\SI{3.26}{\electronvolt}$ was turned into $\eg=\SI{3.3}{\electronvolt}$~\cite{dhanaraj2010} or $\eg=\SI{3.2}{\electronvolt}$~\cite{lebedev1999}. The latter was also derived from $\eg=\SI{3.25}{\electronvolt}$~\cite{soler2019} and proposed by \citet{dubrovskii1977} based on measurements, although some references~\cite{madelung1991,madelung1996,ioffe2023} interpreted it as exciton band gap energy. The value $\egx=\SI{3.23}{\electronvolt}$ was first introduced by \citet{zanmarchi1964}, but we were unable to find a direct connection to $\eg=\SI{3.23}{\electronvolt}$. \citet{rakheja2020} used this value based on the models proposed by \citet{ladesmartin2000}, but we were unable to reproduce it.  We only achieved $\eg=\SI{3.23}{\electronvolt}$ by using the 2H band gap from \citet{persson1997} and scaling it with temperature. $\eg=\SI{3.25}{\electronvolt}$ was derived once from $\egx=\SI{3.265}{\electronvolt}$~\cite{hudgins2003} and once from $\egx=\SI{3.263}{\electronvolt}$~\cite{pearton2023}, but no temporal coherent connections could be found. We thus summarized publications featuring this value in the figure under the reference \textit{unknown}.

Sometimes it is beneficial to look at the missing data to reason about the validity of certain values: We did not find a single measurement that proposed a band gap energy at room temperature of either $\eg=\SI{3.2}{\electronvolt}$, $\SI{3.23}{\electronvolt}$ or $\eg=\SI{3.25}{\electronvolt}$. These values were exclusively extracted from models that were characterized by low-temperature measurements of $\eg$ and sometimes even $\egx$, as we showed in the previous section. The only exception is $\eg=\SI{3.26}{\electronvolt}$, which was derived by \citet{ahuja2002}. However, the uncertainty of $\SI{98}{\milli\electronvolt}$ stated by the authors is so big that it also contains all other energy values mentioned so far.

The main source for the free exciton binding energy is the analysis by \citet{dubrovskii1975} from 1975 (see \cref{fig:bandgap_ref_chain_Ex} in Appendix~\ref{sec:refChainBandgap}). Results based on 6H are available~\cite{nedzvetskii1969,sankin1975} but were rarely utilized. The overall amount of references in regard to $\ex$ is very limited: the latest ones we found were from the year 2014.

The most commonly used temperature scaling factor $\alpha=\SI{3.3e-4}{\electronvolt\per\K}$ (see \cref{fig:bandgap_ref_chain_temp} in Appendix~\ref{sec:refChainBandgap}) was determined by \citet{choyke1957} in 1957 for an undefined polytype of SiC. \citet{bakowski1997} and \citet{yoshida1995} argued that it was 6H but the shown absolute band gap energy ($\approx \SI{2.9}{\eV}$ at \SI{0}{\K}) indicates a different polytype. Despite that, a wide range of 4H-SiC fittings up to this day utilize this value, implying that 4H and 6H share the same temperature dependency~\cite{zhao2003}. The also popular $\alpha=\SI{6.5e-4}{\electronvolt\per\K}$ was proposed by \citet{levinshtein2001} based on a fitting to the low-temperature measurements by \citet{choyke1969} (basically the same data as in \citet{choyke1964a}). Therefore, most of the parameters for the temperature dependency are based on data from the 50s and 60s.


In contrast to the band gap energy, more recent models for the temperature dependency were also utilized in the literature. In some occasions, we were again either unable to identify the measurements used for the model fitting~\cite{balachandran2005,tamaki2008a} or the models were based on the same old data~\cite{kimoto2014a,hatakeyama2013}.

In regard to variations caused by doping, the publication by \citet{lindefelt1998} is the most referenced (see \cref{fig:bandgap_ref_chain_doping} in Appendix~\ref{sec:refChainBandgap}) one. Its values were transferred in most cases without changes, although the usage of \cref{eq:bandgap_jain} made a direct comparison of the parameters challenging. For example, $\bnc$ and $\bnv$ resp. $\bpc$ and $\bpv$ got merged~\cite{albanese2010,pezzimenti2013,zeghdar2019,zeghdar2020} but also $\apc$ and $\cpv$~\cite{lophitis2018}. Occasionally, we identified typographical mistakes, which we documented together with all other detected inconsistencies in Appendix~\ref{sec:bandgap_appendix}.


\subsubsection{Literature Values}

In the previous section, we discussed the changes to measurement results along reference chains. Together with models that utilize various combinations of energy values and temperature scaling factors, these lead to many (exciton) band gap energy values in literature (see \cref{fig:bandgap_values}).

\begin{figure*}[p]
    \centering
    \resizebox{1\linewidth}{!}{%
    \input{figures/bandgap_values}
    }
    \caption{Values for band gaps at varying temperatures. \bkgCol{calcColor} values correspond to calculations, \bkgCol{fundColor} ones to measurements and \bkgCol{guessColor} ones are values calculated from models.}
    \label{fig:bandgap_values}
\end{figure*}

The majority of calculations for $\eg$ at $\SI{0}{\K}$ reported rather low values. In our opinion, the origin of this discrepancy is not well understood because recent investigations also predicted values of only \SI{3.15+-0.03}{\eV}. Models and measurements $<\SI{5}{\K}$ derived higher values, starting at $\SI{3.25}{\electronvolt}$ up to $>\SI{4}{\electronvolt}$. These extreme values often indicate that the respective parameters were fitted to $\SI{300}{\K}$ because the model in \cref{eq:bandgap_varshni} is not able to simultaneously describe the band gap at low and high temperatures. We identified value clusters for $\eg(0)$ at \SI{3.26}{\electronvolt}, \SI{3.263}{\electronvolt}, \SI{3.265}{\electronvolt} and \SI{3.285}{\electronvolt}, but also at \SI{3.359}{\electronvolt}. The latter is achieved for $\eg(300)=\SI{3.26}{\electronvolt}$ and $\alpha=\SI{3.3e-4}{\electronvolt\per\K}$. The other values overlap, in large parts, with measurements of $\egx$.

At room temperature the most commonly used values for $\eg$ include \SI{3.2}{\electronvolt}, \SI{3.23}{\electronvolt} and \SI{3.26}{\electronvolt}. We already showed direct connections between these values and the exciton band gap at low temperatures, which means that a large share of currently used values denote a different quantity at a deviating temperature. 

The investigated publications largely agree upon $\ex \approx \SI{20}{\milli\eV}$, but we found only a single usage of this value in the last decade. The same statement can be made about the exciton band gap energy, which is seemingly pushed out of focus of the community. For a precise discussion of band gap parameters, the knowledge about the existence and the influence of excitons is indispensable and needs to be reintroduced.

\section{Impact Ionization}
\label{sec:II}

An electric field accelerates free charge carriers. When their kinetic energy exceeds the \textit{ionization energy} of the material, the charge carrier is able to generate an excessive electron-hole pair. This process is called \textit{impact ionization} and is sometimes deliberately used, e.g., in avalanche diodes, to increase the responsiveness of a detection device~\cite{adachi2005}. However, in most devices, it is an undesired effect causing breakdown and destruction. Consequently, impact ionization simulations are crucial to identify and subsequently guard the unsafe operation regions of a device.

The amount of generated electron-hole pairs via impact ionization differs between holes and electrons and depends on conditions such as field strength and the spatial direction. Of special importance is the temperature dependency, as a fraction of the deposited energy during impact ionization is transferred to the lattice, increasing its temperature. If the impact ionization is enhanced with rising temperature, a self-amplifying process can be started, eventually destroying the device thermally (thermal runaway).

Our review shows that most of the existing models agree upon the relevant parameters and also, that the impact ionization decreases with temperature. However, many small inaccuracies were discovered (mostly typographical errors and confusion of units), causing big differences in the results. Past research primarily focused on the direction parallel to the c-axis, although the parameters for the electron indicate an anisotropy that has to be considered.


\subsection{Introduction}

TCAD tools describe impact ionization via a charge carrier generation rate (see \cref{eq:II_G})~\cite{bakowski1997,buono2012} with $\jn$/$\jp$ and $\vn$/$\vp$ being current and velocity respectively, while \textit{n} and \textit{p} denote parameters representing either electrons or holes.
\begin{equation}
    \label{eq:II_G}
    G_{II} = \frac{1}{q} \left(\alpha \jn + \beta \jp\right) = \frac{1}{q}\left(\alpha n \vn + \beta p \vp\right)
\end{equation}
The impact ionization coefficients for electrons ($\alpha$) and holes ($\beta$) denounce the number of electron-hole pairs a single charge carrier can generate per unit length when moving in an electric field $F$ (see~\cref{eq:II_coeff})~\cite{adachi2005}. In the sequel, we will introduce some of the models to describe $\alpha$ and $\beta$.
For further details, the interested reader is deferred to the dedicated literature~\cite{capasso1985,selberherr1984,adachi2005,feng2004a,neilainglesias2012}.
\begin{equation}
    \label{eq:II_coeff}
    \alpha = \frac{1}{n} \frac{\mathrm{d}n}{\mathrm{d}x} \si{\per\cm}\ \ \ ,\ \ \ \beta = \frac{1}{p} \frac{\mathrm{d}p}{\mathrm{d}x} \si{\per\cm}
\end{equation}

The still very popular empirical Chynoweth's law (see \cref{eq:II_chynoweth})~\cite{chynoweth1958,chynoweth1960}, also called Van Overstraeten-de Man model~\cite{vanoverstraeten1970}, was published in the late 1950s and was based upon the same formalism used to model impact ionization in gases.
\begin{equation}\label{eq:II_chynoweth}
   \alpha, \beta(F) = a\ \exp\left[-\frac{b}{F}\right]
\end{equation}
At the time of publication, this empirical fitting was the only possibility to describe the charge carrier multiplication at low electric fields. A physical explanation was only available for high field strengths (see \cref{eq:II_wolf})~\cite{wolff1954}, which utilizes the optical phonon energy $\ephonon$, the mean free path $\lambda$, and the ionization energy $\eion$.
\begin{equation}
    \label{eq:II_wolf}
   \alpha,\beta(F) = \frac{eF}{\eion} \exp\left[-\frac{3\ephonon\eion}{(eF\lambda)^2}\right]
\end{equation}
In 1961 \citet{shockley1961} modeled the low-field impact ionization coefficient according to \cref{eq:II_shockley}. The term preceding the exponential is equal to \cref{eq:II_wolf} and denotes how often a carrier's kinetic energy can reach the ionization energy $\eion$ (generate an excessive electron-hole pair) per unit length. The exponential scales this value by the chance of an uninterrupted acceleration, which is characterized by the mean free path $\lambda$. This model is therefore called Shockley's ``lucky electron''~\cite{konstantinov1997}.
\begin{equation}\label{eq:II_shockley}
    \alpha,\beta(F) = \frac{eF}{\eion} \exp\left[-\frac{\eion}{eF\lambda}\right]
\end{equation}
The close relationship between \cref{eq:II_chynoweth} and \cref{eq:II_shockley} ($a=eF/E_i$, $b=E_i/e\lambda$) finally enabled a physically based calculation of parameters $a$ and $b$~\cite{lackner1991}.

The models in \cref{eq:II_wolf} and \cref{eq:II_shockley} were combined in Baraff's theory~\cite{baraff1962} in 1962, which resulted in  $\alpha,\beta \propto \exp[-b/F]$ for low fields and $\alpha,\beta \propto \exp[-c/F^2]$ for high ones. It was later extended by \citet{thornber1981} to the expression shown in \cref{eq:II_TH}~\cite{capasso1985} with $\langle \eion \rangle$ the effective ionization threshold~\cite{capasso1985} and $\ekbt$ the energy due to thermal effects~\cite{nouketcha2020}. For the ionization energy $\eion$ originally a value of $3/2\,\eg$\cite{nouketcha2020} $\approx \SI{4.89}{\eV}$~\cite{galeckas2002} at \SI{300}{\K} ($\eg$ the band gap; see~\cref{sec:bandgap}) was assumed. Recent measurements, however, revealed for 4H-SiC values in the range of \SIrange{7.28}{8.6}{\eV}~\cite{gsponer2024a,chaudhuri2013,garcia2013,chandrashekhar2006,logiudice2005,phlips2005,lebedev2004,ivanov2004a,bertuccio2003}.
\begin{equation}
    \label{eq:II_TH}
    \alpha,\beta(F) = \frac{eF}{\langle \eion \rangle} \exp\left[-\frac{\langle \eion\rangle}{[(eF\lambda)^2/3\ephonon] + eF\lambda + E_{\kb T}}\right]
\end{equation}

We also encountered variations of \cref{eq:II_TH}: \citet{konstantinov1997} removed everything but the high-field part for $\alpha$, while \citet{kyuregyan2016} proposed a transformation that used only three condensed parameters. As Baraff's theory failed to satisfy all demands~\cite{okuto1972}, \citet{okuto1975} extended \cref{eq:II_chynoweth} by adding the electric field as a multiplicative factor, an exponential parameter $m$ and a temperature dependency via $c$ and $d$ (see \cref{eq:II_OC}). The simplified version with $c=d=n=0$ is often referred to as Selberherr model~\cite{selberherr1984}. All investigated publications use $n=0$ (\citet{biondo2012} stated $n=1$, but the result did only match when we set $n=0$), so we will not consider this parameter in the sequel.
\begin{equation}\label{eq:II_OC}
   \alpha, \beta(F) = a\{1+c(T-300)\}\ F^n\exp\left[-\left(\frac{b\{1+d(T-300)\}}{F}\right)^m\right]
\end{equation}

To enable calculations we found models for the impact ionization coefficients using a power law, i.e., $\alpha,\beta \propto F^A$~~\cite{bartsch2010,stum2014,lutz2011}. We also encountered descriptions based on multi-stage~\cite{acharyya2014} and inelastic collision events~\cite{acharyya2017}, which are not yet supported by TCAD tools. \citet{banerjee2021} proposed a slightly deviation equation (see \cref{eq:II_banerjee}), whose predictions deviated from the remaining results. Therefore we did not include it in our analysis.
\begin{equation}
    \begin{split}
    \label{eq:II_banerjee}
    \alpha =& \left( \frac{F}{\aelec} \right) \exp \left( - \frac{\belec}{F^2} \right) \\
    \beta =& \left( \frac{F}{7} \right) \exp \left[ -\frac{1}{\aelec\,F^2 + \belec\,F} \right]
    \end{split}
\end{equation}

TCAD tools also support more advanced models, including a sophisticated temperature dependency in \cref{eq:II_OC}~\cite{valdinoci1999} and the consideration of the initial location of the impact carrier (Lackner model)~\cite{lackner1991}. We found, however, no application to 4H-SiC so far.

\subsubsection{Anisotropy}

Impact ionization is anisotropic in 4H-SiC. The breakdown field perpendicular to the c-axis, i.e., in  \hkl[11-20] direction, was reported to be \SIrange{80}{85}{\percent}~\cite{kitawaki2024}, \SI{75}{\percent}~\cite{nakamura2002} respectively \SIrange{50}{60}{\percent}~\cite{hatakeyama2004} to that parallel to the c-axis, i.e., in \hkl[0001] direction. Nevertheless, \citet{bhargav2024} pointed out that this discrepancy becomes less pronounced at high fields due to the dominant hole coefficient along the c-axis.

Multiple approaches to deal with direction-dependent impact ionization were developed:
\begin{enumerate*}[label=(\roman*)]
\item \citet{hatakeyama2009} developed a formalism (already available in modern TCAD tools) that combines measurements along the principal axis to predict the impact ionization coefficient in any desired direction.
    \item \citet{jin2024} reused available parameters but introduced a new approach to calculate the ``driving force'' by considering the field direction for constant carrier temperature.
    \item \citet{nida2019} adapted the field strength to an effective $F^* = (\md/m_F)^{1/2}F$, with $m_F$ the effective masses along the direction of $F$ and $\md$ the density-of-states effective mass (see \cref{sec:dos}).
\end{enumerate*}

\subsubsection{Temperature Dependency}

To depict the changing behavior with temperature, the dedicated parameters in \cref{eq:II_OC} were commonly used, whereas \citet{niwa2014} extended their capabilities by providing a second-degree polynomial in $T$ for $\ahole$. In contrast, \citet{hamad2015} extracted parameter values for eight temperatures independently.

A different approach applied the multiplicative factor $\gamma$ shown in \cref{eq:II_gamma}~\cite{crowell1966} to the parameters $a$ and $b$ of \cref{eq:II_chynoweth}~\cite{hatakeyama2009,ladesmartin2000,ayalew2004,nallet1999,nallet2000}, with $T_0$ being a reference temperature (usually \SI{300}{\kelvin}), $\tl$ the lattice temperature and $\omop$ the optical phonon energy. We are confident that $\omop$ corresponds to the longitudinal optical phonon energy $\omlo$ (see \cref{sec:permittivity}) due to matching values. \citet{nida2019} scaled the mean free path in \cref{eq:II_TH} by $\sqrt{\gamma}$ and the ionization energy by the ratio of the band gap at the lattice temperature $\tl$ and \SI{300}{\kelvin}.
\begin{equation}
    \gamma = \frac{\tanh\left(\frac{\hbar\omop}{2\,\kb\,T_0}\right)}{\tanh\left(\frac{\hbar\omop}{2\,\kb\,\tl}\right)} 
    \label{eq:II_gamma}
\end{equation}


\subsubsection{Methods}


To measure the impact ionization coefficients, an equal amount of charge carriers is generated in a space charge region, either by (pulsed) electron (electron beam induced current (EBIC))~\cite{raghunathan1999} or optical beams (optical beam induced current (OBIC))~\cite{konstantinov1997,konstantinov1998,loh2008,ng2003,nguyen2011,nguyen2012,niwa2014,stefanakis2020,zhao2019,hamad2015}. Defects have a significant impact on the measured coefficients, such that EBIC is used to extract parameters at defect-free regions~\cite{baliga2019}.

The charge carrier generation is executed at varying field strengths. Recording the respective terminal currents enables a comparison against the no-field current and, thus, the determination of the effective amplification. The readout of the current can be executed in DC mode~\cite{konstantinov1997,konstantinov1998}, which complicates the elimination of leakage current~\cite{raghunathan1999}, AC mode~\cite{raghunathan1999,nguyen2011,nguyen2012}, or both combined~\cite{loh2008,ng2003}. Additional challenges are the selection of suitable test structures, (e.g., p-n/n-p diodes or pnp/npn transistors) and the proper separation of electron and hole multiplication phenomena. For further information, the interested reader is referred to the dedicated literature~\cite{konstantinov1997,kyuregyan2016,stefanakis2020,nguyen2011,green2012,niwa2014}.

Monte Carlo simulations were also used to investigate impact ionization. Some authors extracted the impact coefficients as the reciprocal of the average scattering distance~\cite{cheang2021,sun2012} while others solely presented simulated values without fitting them to any of the earlier presented models~\cite{bellotti2000,akturk2009,brennan2000,fujita2017a,hjelm2003,mickevicius1998, nilsson2000,sun2011,tanaka2022,zhao2000,zhu2006}. Further investigations focused on non-localized models~\cite{concannon1995,katayama1989} and the impact of defects~\cite{raghunathan1998} in the presence of a magnetic field~\cite{mukherjee2020,mukherjee2017}.

We found multiple fittings of \cref{eq:II_OC} to the OBIC results by \citet{konstantinov1997} (FOBIC)~\cite{sheridan2000,morisette2001,zhao2003,brosselard2004,choi2005,bellone2014,baliga2019,banerjee2021}, who originally used \cref{eq:II_TH}, and to the Monte Carlo results by \citet{nilsson2000} (FMC)~\cite{bertilsson2000,biondo2012}.

\subsection{Results \& Discussion}

In the sequel, we present the results of our research on impact ionization. We did not include the parameters provided by \citet{banc2002}, as the authors focused on transistor channels. We also discarded publications that did not clearly specify the used polytype of SiC~\cite{kyuregyan1989,cheong2003,cheong2015,kimoto2021}, as well as those solely investigating the deviations of the breakdown voltage~\cite{konstantinov1998a,palmour1994,neudeck1999,wei2011,ng2004}.

A common figure of merit for impact ionization in literature is the critical electric field strength. We found values in the range of \SIrange{2}{3}{\mega\V\per\cm}~\cite{neudeck1998,neudeck2001,zetterling2002,werner2001,yoder1996,pensl2005,pearton2018,elasser2002,han2003}, and dependencies on the doping concentration~\cite{baliga2019,hefner2001,kimoto2018,kimoto2019,konstantinov1997,lutz2011,nakamura2002,niwa2015,palmour1997,peters2000,raghunathan1999,shur2006,singh2002,zhao2019}. These values are, most commonly, determined for uniformly doped non-punch-through diodes using power law approximations of the impact coefficients~\cite{raghunathan1999}. Consequently, such values have to be corrected according to the actual structure and doping level~\cite{slobodyan2022,niwa2015}. For example, over short distances, a higher field is required to achieve breakdown and vice versa. Therefore, the product of distance and field is a more important parameter for impact ionization than the critical electric field alone. In TCAD simulations the latter is automatically achieved based on the provided impact ionization coefficients.

\subsubsection{Impact Ionization Coefficients}

The earliest investigations on 4H-SiC impact ionization were published by \citet{konstantinov1997} and \citet{raghunathan1999}. The proposed values, however, differed by almost one order of magnitude, forcing authors to consciously use 6H based values~\cite{wang1999}. Manifold explanations were provided for this discrepancy: \citet{bellotti1999} reported that \citet{konstantinov1997} used non-defect-free material, explaining the higher coefficients through defect-assisted ionization. \citet{nilsson2000} stated that the techniques used by \citet{raghunathan1999} are considered more accurate and \citet{bertilsson2000} argued that the deviations could be explained by the anisotropy of the impact ionization coefficients, which the author later revoked~\cite{bertilsson2004}. \citet{feng2004a} concluded that the results by \citet{raghunathan1999} deviated from practical results because a large share of the intrinsic defects was simply missed due to the focused beam that was used in the analysis.

\begin{table*}[p]
    \centering
        \setlength{\tabcolsep}{7pt}
            \renewcommand{\arraystretch}{1.1}
    \caption{\label{tab:II_parameter_OC} Impact ionization parameters for \cref{eq:II_OC}. Single values indicate that the crystal direction was not specified. A dash represents $1$ for $m$ and $0$ otherwise. \bkgCol{konsColor} marks fittings to the OBIC results by \citet{konstantinov1997} and \bkgCol{nilsColor} fittings to the MC investigation by \citet{nilsson2000}.}
    \resizebox{1\linewidth}{!}{%
    \begin{tabular}{lrlrl*{4}{c}|rlrl*{4}{c}|c}
& \multicolumn{8}{c|}{electron} & \multicolumn{8}{c|}{hole} & \\ 
ref. & $\apara$ & $\aperp$ & $\bpara$ & $\bperp$ & $c$ & $d$ & $m$ & $F$ region & $\apara$ & $\aperp$ & $\bpara$ & $\bperp$ & $c$ & $d$ & $m$ & $F$ region &  method\\ 
 & \multicolumn{2}{c}{[\SI{e6}{\per\centi\m}]} & \multicolumn{2}{c}{[\si{\mega\V\per\cm}]} & [\SI{e-3}{\per\K}] & [\SI{e-3}{\per\K}] & [1] & [\si{\mega\V\per\cm}] & \multicolumn{2}{c}{[\SI{e6}{\per\centi\m}]} & \multicolumn{2}{c}{[\si{\mega\V\per\cm}]} & [\SI{e-3}{\per\K}] & [\SI{e-3}{\per\K}] & [1] & [\si{\mega\V\per\cm}] & \\ \hline 
\reftext{Ragh99}\cite{raghunathan1999} & -- & -- & -- & -- & -- & -- & -- & -- & \multicolumn{2}{c}{\num{3.09}} & \multicolumn{2}{c}{\num{17.9+-0.4}} & \num{-3.46} & -- & -- & \numrange{2.5}{3.2} & EBIC\\ 
\rowcolor{nilsColor}\reftext{Bert00}\cite{bertilsson2000} & \num{0.4} & \num{48.0} & \multicolumn{2}{c}{\num{15.0}} & -- & -- & \num{1.15} & \numrange{2.0}{4.0} & \num{1.8} & \num{45.0} & \multicolumn{2}{c}{\num{15.0}} & -- & -- & -- & \numrange{2.0}{4.0} & FMC\\ 
\rowcolor{konsColor}\reftext{Sher00}\cite{sheridan2000}\footnotemark[1] & -- & -- & -- & -- & -- & -- & -- & -- & \num{5.18} & -- & \num{14.0} & -- & -- & -- & -- & \numrange{1.5}{10.0} & FOBIC\\ 
\rowcolor{konsColor}\reftext{Mori01}\cite{morisette2001} & \num{1.69} & -- & \num{9.69} & -- & -- & -- & \num{1.6} & \numrange{2.5}{10.0} & \num{3.32} & -- & \num{10.7} & -- & -- & -- & \num{1.1} & \numrange{1.5}{10.0} & FOBIC\\ 
\reftext{Ng03}\cite{ng2003} & \multicolumn{2}{c}{\num{1.98}} & \multicolumn{2}{c}{\num{9.46}} & \num{-2.02}\footnotemark[2] & -- & \num{1.42} & \numrange{1.8}{4.0} & \multicolumn{2}{c}{\num{4.38}} & \multicolumn{2}{c}{\num{11.4}} & \num{-0.91}\footnotemark[2] & -- & \num{1.06} & \numrange{1.8}{4.0} & OBIC\\ 
\rowcolor{konsColor}\reftext{Zhao03}\cite{zhao2003} & \num{7.26} & -- & \num{23.4} & -- & -- & -- & -- & \numrange{2.5}{10.0} & \num{6.85} & -- & \num{14.1} & -- & -- & -- & -- & \numrange{1.5}{10.0} & FOBIC\\ 
\rowcolor{konsColor}\reftext{Bros04}\cite{brosselard2004} & \num{0.408} & -- & \num{16.7}\footnotemark[3] & -- & -- & -- & -- & \numrange{1.25}{3.0} & \num{16.3} & -- & \num{16.7} & -- & -- & -- & -- & \numrange{1.0}{3.0} & FOBIC\\ 
\reftext{Hata04}\cite{hatakeyama2004} & \num{176.0} & \num{21.0} & \num{33.0} & \num{17.0} & -- & -- & -- & \numrange{2.0}{5.0} & \num{341.0} & \num{29.6} & \num{25.0} & \num{16.0} & -- & -- & -- & \numrange{2.0}{5.0} & OBIC\\ 
\rowcolor{konsColor}\reftext{Choi05}\cite{choi2005} & \num{16.5} & -- & \num{25.8} & -- & -- & -- & -- & \numrange{2.7}{5.0} & \num{5.5} & -- & \num{13.5} & -- & -- & -- & -- & \numrange{1.5}{4.0} & FOBIC\\ 
\reftext{Loh08}\cite{loh2008} & \multicolumn{2}{c}{\num{2.78}} & \multicolumn{2}{c}{\num{10.5}} & -- & -- & \num{1.37} & \numrange{1.25}{5.0} & \multicolumn{2}{c}{\num{3.51}} & \multicolumn{2}{c}{\num{10.3}} & -- & -- & \num{1.09} & \numrange{1.0}{5.0} & OBIC\\ 
\reftext{Loh09}\cite{loh2009} & -- & -- & -- & -- & -- & -- & -- & -- & \multicolumn{2}{c}{\num{3.321}} & \multicolumn{2}{c}{\num{10.385}} & \num{-2.78} & \num{0.48}\footnotemark[4] & \num{1.09} & \numrange{1.33}{2.0} & OBIC\\ 
\reftext{Nguy11}\cite{nguyen2011} & \multicolumn{2}{c}{\num{0.46}} & \multicolumn{2}{c}{\num{17.8}} & -- & -- & -- & \numrange{2.0}{2.7} & \multicolumn{2}{c}{\num{15.6}} & \multicolumn{2}{c}{\num{17.2}} & -- & -- & -- & \numrange{1.5}{2.7} & OBIC\\ 
\rowcolor{nilsColor}\reftext{Bion12}\cite{biondo2012}\footnotemark[5] & \multicolumn{2}{c}{\num{0.325}} & \multicolumn{2}{c}{\num{17.1}} & \num{-32.9} & -- & -- & \numrange{2.0}{4.0} & \multicolumn{2}{c}{\num{3.25}} & \multicolumn{2}{c}{\num{17.1}} & \num{-32.9} & -- & -- & \numrange{2.0}{4.0} & FMC\\ 
\reftext{Gree12}\cite{green2012} & \multicolumn{2}{c}{\num{0.019}} & \multicolumn{2}{c}{\num{2.888}} & -- & -- & \num{4.828} & \numrange{1.8}{2.5} & \multicolumn{2}{c}{\num{6.0}\footnotemark[6]} & \multicolumn{2}{c}{\num{13.87}\footnotemark[6]} & -- & -- & \num{0.96} & \numrange{1.6}{4.0} & OBIC\\ 
\reftext{Nguy12}\cite{nguyen2012} & \multicolumn{2}{c}{\num{3.36}} & \multicolumn{2}{c}{\num{22.6}} & -- & -- & -- & \numrange{1.5}{4.8} & \multicolumn{2}{c}{\num{8.5}} & \multicolumn{2}{c}{\num{15.97}} & -- & -- & -- & \numrange{1.5}{4.8} & OBIC\\ 
\reftext{Song12}\cite{song2012}\footnotemark[7] & \multicolumn{2}{c}{\num{3.78}} & \multicolumn{2}{c}{\num{10.5}} & \num{-1.47} & -- & \num{1.37} & \numrange{1.25}{5.0} & \multicolumn{2}{c}{\num{4.51}} & \multicolumn{2}{c}{\num{10.5}} & \num{-1.56} & -- & \num{1.1} & \numrange{1.0}{5.0} & FIT\\ 
\reftext{Sun12}\cite{sun2012} & \multicolumn{2}{c}{\num{1.803}} & \multicolumn{2}{c}{\num{13.52}} & -- & -- & \num{1.2} & \numrange{2.0}{5.0} & \multicolumn{2}{c}{\num{1.861}} & \multicolumn{2}{c}{\num{9.986}} & -- & -- & \num{1.11} & \numrange{1.5}{4.0} & MC\\ 
\rowcolor{konsColor}\reftext{Bell14}\cite{bellone2014} & \num{0.407} & -- & \num{16.7} & -- & -- & -- & -- & \numrange{2.5}{10.0} & \num{16.3} & -- & \num{16.7} & -- & -- & -- & -- & \numrange{1.5}{10.0} & FOBIC\\ 
\reftext{Niwa14}\cite{niwa2014} & \multicolumn{2}{c}{\num{8190.0}} & \multicolumn{2}{c}{\num{39.4}} & -- & -- & -- & \numrange{2.2}{2.7} & \multicolumn{2}{c}{\num{4.513}} & \multicolumn{2}{c}{\num{12.82}} & x\footnotemark[8] & \num{1.38} & -- & \numrange{1.4}{2.5} & OBIC\\ 
\reftext{Hama15}\cite{hamad2015} & \multicolumn{2}{c}{\num{0.99}} & \multicolumn{2}{c}{\num{12.9}} & -- & -- & -- & \numrange{2.5}{7.0} & \multicolumn{2}{c}{\num{1.61}} & \multicolumn{2}{c}{\num{11.5}} & -- & -- & -- & \numrange{2.5}{7.0} & OBIC\\ 
\reftext{Niwa15}\cite{niwa2015} & \num{0.143} & -- & \num{4.93} & -- & --\footnotemark[9] & --\footnotemark[9] & \num{2.37} & \numrange{2.2}{3.3} & \num{3.14} & -- & \num{11.8} & -- & \num{6.3}\footnotemark[9] & \num{1.23}\footnotemark[9] & \num{1.02} & \numrange{1.0}{2.35} & OBIC\\ 
\reftext{Shar15}\cite{sharma2015} & \num{186.0} & -- & \num{28.0} & -- & -- & -- & -- & -- & \num{301.0} & -- & \num{20.5} & -- & -- & -- & -- & -- & DIV\\ 
\reftext{Kyur16}\cite{kyuregyan2016}\footnotemark[10] & \multicolumn{2}{c}{\num{38.6+-15.0}} & \multicolumn{2}{c}{\num{25.6+-0.1}} & -- & -- & -- & \numrange{1.2}{5.0} & \multicolumn{2}{c}{\num{5.31+-0.3}} & \multicolumn{2}{c}{\num{13.1+-0.01}} & -- & -- & -- & \numrange{1.0}{5.0} & FIT\\ 
\reftext{Zhan18}\cite{zhang2018} & \multicolumn{2}{c}{\num{1.31}} & \multicolumn{2}{c}{\num{13.0}} & \num{-1.47} & -- & -- & -- & \multicolumn{2}{c}{\num{2.98}} & \multicolumn{2}{c}{\num{13.0}} & \num{-1.56} & -- & -- & -- & --\\ 
\rowcolor{konsColor}\reftext{Bali19}\cite{baliga2019} & \num{313.0} & -- & \num{34.5} & -- & -- & -- & -- & \numrange{2.0}{5.0} & \num{8.07} & -- & \num{15.0} & -- & -- & -- & -- & \numrange{1.1}{4.0} & FOBIC\\ 
\reftext{Zhao19}\cite{zhao2019} & \num{0.339} & -- & \num{5.15} & -- & -- & -- & \num{2.37} & \numrange{2.4}{3.2} & \num{3.56} & -- & \num{11.7} & -- & \num{6.19} & \num{1.15} & \num{1.02} & \numrange{1.0}{2.5} & OBIC\\ 
\reftext{Stef20}\cite{stefanakis2020} & -- & \num{6.4} & -- & \num{12.5} & -- & -- & -- & \numrange{1.6}{2.0} & -- & \num{6.0} & -- & \num{13.3} & -- & -- & -- & \numrange{1.5}{2.0} & OBIC\\ 
\reftext{Chea21}\cite{cheang2021}\footnotemark[11] & \multicolumn{2}{c}{\num{0.932}} & \multicolumn{2}{c}{\num{7.19}} & -- & -- & \num{1.95} & \numrange{2.0}{10.0} & \multicolumn{2}{c}{\num{1.75}} & \multicolumn{2}{c}{\num{6.56}} & -- & -- & \num{1.45} & \numrange{1.6}{10.0} & MC\\ 
\reftext{Stef21}\cite{stefanakis2021}\footnotemark[12] & \num{2.8} & -- & \num{20.7} & -- & \num{-1.0}\footnotemark[13] & \num{-0.29}\footnotemark[13] & -- & \numrange{1.25}{10.0} & \num{2.5} & -- & \num{12.1} & -- & \num{1.74}\footnotemark[13] & \num{0.59}\footnotemark[13] & -- & \numrange{1.25}{10.0} & FIT\\ 
\reftext{Kita24}\cite{kitawaki2024} & -- & \num{6.91} & -- & \num{14.4} & \num{3.75} & \num{0.657} & -- & \numrange{1.6}{2.1} & -- & \num{34.6} & -- & \num{17.5} & \num{4.39} & \num{0.604} & -- & \numrange{1.6}{2.1} & OBIC\\ 
\end{tabular} 
        {\scriptsize
        \footnotetext[1]{same values achieved as 6H investigation by \citet{ruff1994}}
        \footnotetext[2]{provided by~\citet{cha2008a}} 
        \footnotetext[3]{changed value $\bhole=\SI{1.67e5}{\V\per\centi\m}$ to $\bhole=\SI{1.67e7}{\V\per\centi\m}$ as the results were too high}
        \footnotetext[4]{we changed $\bhole=\num{8.9e6} - \num{4.95e3}\, T$ to $\num{8.9e6} + \num{4.95e3}\, T$ to match the plots in the paper}
        \footnotetext[5]{we changed $m$ from $2$ to $1$ and $n$ from $1$ to $0$ to match the specified reference~\cite{nilsson2000}}
        \footnotetext[6]{we changed $\ahole=\SI{6e4}{\per\cm}$ to $\SI{6e6}{\per\cm}$ and $\bhole=\SI{1.387e6}{\V\per\cm}$ to \SI{1.387e7}{\V\per\cm} to fit plots} 
        \footnotetext[7]{fitted to results by \citet{loh2008}}
        \footnotetext[8]{$\ahole = \num{3.94e6} - \num{1.96e4}\, T + 71.7\, T^2$}
        \footnotetext[9]{different values suggested by~\citet{steinmann2023}} 
        \footnotetext[10]{values achieved by averaging of~\cite{konstantinov1997,konstantinov1998,ng2003,ng2004,loh2008,loh2009,green2012,niwa2014}}
        \footnotetext[11]{$a$ and $b$ presumably stated in \si{\per\meter} and \si{\mega\volt\per\meter} in paper, converted to \si{\per\centi\meter} and \si{\mega\volt\per\centi\meter}}
        \footnotetext[12]{values fitted to~\cite{nguyen2012,loh2008,konstantinov1997,raghunathan1997,zhao2019,nilsson2002,bellotti2000,ng2003,ruff1994}}
        \footnotetext[13]{provided by~\citet{steinmann2023}} 
        }
}
\end{table*}

While early Monte Carlo simulations~\cite{bertilsson2000,bellotti2000,ruden2000,nilsson2000} supported the results by \citet{raghunathan1999}, more advanced simulation models~\cite{sun2012} and additional measurements showed a better agreement with \citet{konstantinov1997}. The results of the latter were, in hindsight, more commonly accepted within the community, which is also highlighted by seven different fittings (see \cref{tab:II_parameter_OC} and \cref{tab:II_parameter_TH}).

\begin{table*}[ht]
    \centering
        \setlength{\tabcolsep}{7pt}
        \renewcommand{\arraystretch}{1.1}
    \caption{\label{tab:II_parameter_TH} Impact ionization parameters for \cref{eq:II_TH}.}
    \resizebox{0.95\linewidth}{!}{%
    \begin{tabular}{lc|*{5}{c}|*{5}{c}|c}
&& \multicolumn{5}{c|}{electron} & \multicolumn{5}{c|}{hole} & \\ 
ref. & dir & $\langle \eion \rangle$ & $\lambda$ & $\ephonon$ & $E_{\kb T}$ & $F$ region & $\langle \eion \rangle$ & $\lambda$ & $\ephonon$ & $E_{\kb T}$ & $F$ region & method\\ 
 && [\si{\eV}] & [\si{\angstrom}] & [\si{\milli\eV}] & [\si{\milli\eV}] & [\si{\mega\V\per\cm}] & [\si{\eV}] & [\si{\angstrom}] & [\si{\milli\eV}] & [\si{\milli\eV}] & [\si{\mega\V\per\cm}] & \\ \hline 
\reftext{Kons97}\cite{konstantinov1997}\footnotemark[1] & $\parallel$ & \num{10.0} & \num{29.9} & \num{120.0} & \num{0.0} & \numrange{2.5}{10.0} & \num{7.0} & \num{32.5} & \num{120.0} & \num{0.0} & \numrange{1.5}{10.0} & OBIC\\ 
\reftext{Nida19}\cite{nida2019}\footnotemark[2] & $\parallel$ & \num{10.61} & \num{27.0} & \num{120.0} & $\kb T$ & \numrange{1.0}{10.0} & \num{10.87} & \num{39.49} & \num{85.0} & $\kb T$ & \numrange{1.0}{10.0} & FIT\\ 
\reftext{Nouk20}\cite{nouketcha2020}\footnotemark[3] & -- & \num{7.5} & \num{10.0} & \num{92.5} & \num{14.4} & \numrange{0.9}{10.0} & \num{6.62} & \num{4.8} & \num{9.0} & \num{102.0} & \numrange{0.9}{10.0} & FIT\\ 
\reftext{Stei23}\cite{steinmann2023}\footnotemark[4] & -- & \num{10.6} & \num{27.0} & \num{95.0} & $\kb T$ & \numrange{1.0}{10.0} & \num{10.9} & \num{62.0} & \num{95.0} & $\kb T$ & \numrange{1.0}{10.0} & FETIV\\ 
 &  & \num{10.6} & \num{27.0} & \num{87.0} & $\kb T$ & \numrange{1.0}{10.0} & \num{10.9} & \num{80.0} & \num{87.0} & $\kb T$ & \numrange{1.0}{10.0} & FETIV\\ 
\end{tabular}
        \footnotetext[1]{only high field asymptotics (without linear term in denominator) used for $\alpha$}
        \footnotetext[2]{fitting to \cite{konstantinov1997,hatakeyama2009,raghunathan1999,niwa2015,loh2008}} 
        \footnotetext[3]{$3/2\, \eg$ instead of $\langle \eion \rangle$ used in the exponential, fitting to~\cite{ng2003,loh2008,konstantinov1998,niwa2015,nida2019,raghunathan1997}} 
        \footnotetext[4]{parameters by \citet{nida2019} used as starting point}
    }
\end{table*}

We discovered multiple issues regarding these fittings:
\begin{enumerate*}[label=(\roman*)]
\item The commonly used $m=1$ in \cref{eq:II_OC} is unable to match the curvature of the original data. Therefore, proper model selection in regard to the expected field strengths is key to avoid unacceptable discrepancies.
\item \citet{brosselard2004}, \citet{bellone2014}, and \citet{raynaud2010} derived the same parameters, but their prediction for electrons is based on a faulty assumption. \citet{raynaud2010} described the procedure as fitting the holes and then using $\beta/\alpha=40$, but \citet{konstantinov1997} used this value only in an intermediate calculations, whereas the final results revealed a non-constant relationship of $\beta/\alpha$. For this reason, the predictions of $\alpha$ by the stated authors are too low.
\item Although \citet{baliga2019} stated to use the results by \citet{konstantinov1997}, the model approximates the measurements by \citet{hatakeyama2004} much better.
\end{enumerate*}

In general, the literature is dominated by measurements (see \cref{tab:II_parameter_OC} and \cref{tab:II_parameter_TH}),\revision{ which focused on diodes with beveled edges but we also found planar diodes~\cite{nguyen2011}, Schottky diodes~\cite{raghunathan1999}, merged pin-schottky (MPS) diodes~\cite{sharma2015} and field-effect transistors (FETs)~\cite{zhao2003}}. Since 2008, multiple studies on the impact ionization of 4H-SiC, twelve of them within the last decade, were published, demonstrating an active area of research.  By comparing the models with the plots in the respective papers, we identified inaccuracies in five out of $33$ investigations, including errors in presented parameter values~\cite{brosselard2004,green2012,cheang2021} and equations~\cite{loh2009,biondo2012} (see footnotes of \cref{tab:II_parameter_OC}). Appendix~\ref{sec:II_appendix} presents a more comprehensive listing of encountered inaccuracies.

The measurements are complemented by the investigation of \citet{stefanakis2021}, who provided fittings to Monte Carlo simulations~\cite{nilsson2002, bellotti2000} and to existing 4H-SiC models~\cite{nguyen2012,loh2008,konstantinov1997,raghunathan1997,zhao2019,ng2003} (as well as 6H~\cite{ruff1994}), by enforcing $m=1$ in \cref{eq:II_OC}. Based on these fittings the authors achieved a ``global fit'' model. \revision{\citet{nida2019} fitted their model to multiple sources~\cite{konstantinov1997,hatakeyama2009,raghunathan1999,niwa2015,loh2008} and \citet{nouketcha2020} reused these data, extended by further investigations~\cite{ng2003,loh2008,konstantinov1998,niwa2015,nida2019,raghunathan1997}, in a genetic algorithm}. \citet{kyuregyan2016} calculated the average of available parameter values without conducting any fitting. According to the authors, this is supposed to remove statistical inaccuracies and uncertainties introduced by the characterization methods. \citet{song2012} fitted to \citet{loh2008}, who in turn fitted high-field values from \citet{ng2003} and low-field ones of own measurements.



\begin{figure*}
    \centering
    \resizebox{0.95\linewidth}{!}{%
    \input{figures/II_elec}
    }
    \caption{\label{fig:II_elec}Impact ionization coefficient for electrons. Each model is limited to the interval used for characterization. To distinguish the models proposed by \reftext{Stei23}~\cite{steinmann2023}, we added ($\ephonon$).}
\end{figure*}

Vastly deviating parameter values in \cref{eq:II_OC} can lead to similar results, because $a$ and $b$ compensate each other. \revision{We encountered for parameter $a$ of electrons values in the range \SIrange{0.019e16}{8190e16}{\per\centi\m}. This implies, that a direct comparison of the parameters is not reasonable but instead the models need to be plotted for an efficient comparison.} The impact ionization coefficients for electrons (see \cref{fig:II_elec}) show a spread of more than one order of magnitude, whereas deviations increase for low fields. \citet{zhang2018} resp. \citet{sharma2015} predicted higher values than most other models and \citet{bellone2014,bertilsson2000,biondo2012} lower ones. \citet{hatakeyama2004} modeled the increase of $\alpha$ with the electric field steeper than the majority in literature. \revision{Since there are no extreme outliers the often utilized optical beam induced current method (OBIC) can be considered reliable.}

\begin{figure*}
    \centering
    \resizebox{0.95\linewidth}{!}{%
    \input{figures/II_hole}
    }
    \caption{\label{fig:II_hole}Impact ionization coefficient for holes. Each model is limited to the interval used for characterization. To distinguish the models proposed by \reftext{Stei23}~\cite{steinmann2023}, we added ($\ephonon$).}
\end{figure*}

For holes (see \cref{fig:II_hole}) the spread in values is much smaller, especially close to a field strength around $\SI{2}{\mega\volt\per\centi\meter}$. Nevertheless, the results of \citet{steinmann2023} at low fields are several orders of magnitude higher than the average value in literature, while the results by \citet{raghunathan1999}, \citet{biondo2012} and \citet{bertilsson2000} are around one order of magnitude lower. In fact, the latter match the impact ionization coefficient of electrons so well that they even got interpreted as $\alpha$~\cite{loh2008}. Similar to electrons, \citet{hatakeyama2004} predicted a much steeper increase with field strength, leading to higher than average values (comparable to \citet{sharma2015}).

In contrast to Silicon, 4H-SiC shows higher hole than electron current amplification (i.e., $\beta>\alpha$~\cite{ng2002a,adachi2005,chow2000a}), which is attributed to discontinuities in the electron spectrum~\cite{nilsson1996}. Consequently, the Shockley approximation of the ``lucky electron'' can only be applied to holes, as the electron's energy is prohibited to continuously increase~\cite{konstantinov1997}. For this reason, \citet{konstantinov1997} only used the high-field part for $\alpha$.

\subsubsection{Anisotropy}

As outlined earlier, the breakdown field in \hkl[11-20] is lower than in \hkl[0001] direction, meaning that the respective impact ionization coefficient values ought to be higher. Already in the year 2000, \citet{nilsson2000} proposed a non-constant relationship between the single directions, though early simulations/calculations did not manage to depict this correctly~\cite{bertilsson2000}, using constant factors instead. We found $\betaperp/\betapara=25$ and $\alphaperp/\alphapara=120$~\cite{bertilsson2000}, as well as $\alphaperp/\alphapara=1/3.5$~\cite{bakowski1997,ladesmartin2000}, whereas the latter was most likely derived for 6H.

In \cref{tab:II_parameter_OC} we denote the directions only in clearly distinguishable cases, though it is safe to assume that all investigations concentrated on the direction parallel to the c-axis~\cite{bhargav2024}. We found only five exceptions~\cite{bertilsson2000,hatakeyama2004,stefanakis2020,kitawaki2024,tanaka2024}: \citet{tanaka2024} conducted Monte Carlo simulations to show that the impact ionization coefficient for holes perpendicular to the c-axis is only slightly higher than parallel to it (see \cref{fig:II_hole}). The results by \citet{stefanakis2020} even shows an overlap of the coefficients perpendicular and parallel to the c-axis.

For electrons the perpendicular coefficients turned out to be approximately one order of magnitude higher than the parallel ones. In fact, the values of $\alphaperp$ are comparable to $\betapara$~\cite{tanaka2024}, implying that electrons have to be carefully considered in breakdown analyses as well. The low value of $\alpha$ parallel to the c-axis was explained by the discontinuities in the conduction band~\cite{konstantinov1997,nakamura2002,tanaka2024}, making it difficult for electrons to gain enough energy~\cite{tanaka2024}. Nevertheless, more information is required for definite statements.

\subsubsection{Temperature Dependency}


Temperature analyses are rare in the literature. Some of the available data were even proposed by other authors years after the original publication, e.g., by \citet{cha2008}, or by \citet{steinmann2023}, who fitted the linear and quadratic temperature coefficients of the breakdown voltage.

\citet{nida2019} presented the high-temperature evolution of selected models, though well outside of their experimental temperature range. The proposed calculations require a band gap model, whereas we picked the one from \citet{galeckas2002} (see \cref{sec:bandgap}). \citet{hatakeyama2009} used the temperature scaling shown in \cref{eq:II_gamma} (multiplication with $\gamma$), which can also be found in additional publications~\cite{ladesmartin2000,schroeder2006,nallet2000}. We only show this temperature scaling for electrons, as it was not specified for holes explicitly. \citet{hatakeyama2009} noted that for a good fit $\omop=\SI{190}{\meV}$ had to be used, thereby contradicting experimental results of $\omlo=\SI{120}{\meV}$. \citet{nallet2000} used $\hbar\omop=\SI{90}{\meV}$ with $T_0=\SI{600}{\kelvin}$.

All models we found predicted a decreasing hole impact ionization coefficient with increasing temperature (see \cref{fig:II_hole_temp}), matching reports of a direct relationship between breakdown voltage and temperature~\cite{cha2008,guo2005}. Consequently, thermal runaway is prevented. \citet{steinmann2023} provided a fitting to their own measurements for the model by \citet{niwa2015} with the parameters shown in \cref{eq:II_steinmann_niwa_h}. We do not explicitly plot this fitting, as it is very close to the original model. Extraordinary is the model by \citet{biondo2012}, who proposed a value of $\chole=\SI{-32.9e-3}{\per\K}$ (comparable to \citet{loh2009}), leading to $\beta=0$ at $\approx \SI{330}{\K}$. We are unsure about the origin of the underlying data, as the reference used for fitting~\cite{nilsson2000} did not provide a temperature analysis. Similarly \citet{song2012} denoted \citet{loh2008} as source, who also did not provide the respective data. \citet{hamad2015} fitted the model shown in \cref{eq:II_OC} at various temperatures, which explains the weird shape of the plot.
\begin{equation}
    \label{eq:II_steinmann_niwa_h}
    \chole = \SI{2.05e-3}{\per\K}\qquad , \qquad \dhole = \SI{0.65e-3}{\per\K}
\end{equation}

An exception to the general statement of decreasing impact coefficient with temperature is the model by \citet{zhao2019}. However, the indicated increase within the range from \SIrange{150}{300}{\K} is not supported by the plots shown in the very same publication.

\begin{figure*}
    \centering
    \resizebox{0.5\linewidth}{!}{%
    \input{figures/II_hole_temp}
    }
    \caption{\label{fig:II_hole_temp}Temperature dependence of the hole impact ionization coefficient at $1/F = \SI{0.4}{\centi\m\per\mega\V}$. The references in bracket denote the used models.}
\end{figure*}

The results for electrons (see \cref{fig:II_elec_temp}) are comparable to those for holes, in that $\alpha$ approximately halves between \SI{100}{\K} and \SI{600}{\K}. \citet{kimoto2018} assigned the discontinuities in the conduction band, called ``minigaps'' in the paper, as the reason for the low-temperature dependence of $\alpha$. 

The only exception are the fittings of \citet{steinmann2023} to \citet{niwa2015} and \citet{stefanakis2021}, who reported an increase of $\alpha$. The model for the former only delivered reasonable results after we changed $\delec=\SI{-0.72e-3}{\per\K}$ to $\delec=\SI{-0.72e-6}{\per\K}$. Calculations by \citet{tanaka2024} also predicted an increase of $\alpha$ due to a small Brillouin zone width and narrow bandwidth of the $E-k$ dispersion in the conduction band. Comparable to holes, \citet{biondo2012} proposed the same high value for $\celec$, leading to $\alpha=0$ at $\approx \SI{330}{\K}$. In the plot, we also added the temperature modeling by the $\gamma$ factor (see \cref{eq:II_gamma}), showing a decline with increasing temperature. \citet{niwa2015} did not detect any changes of $\alpha$ with temperature.

\begin{figure*}
    \centering
    \resizebox{0.5\linewidth}{!}{%
    \input{figures/II_elec_temp}
    }
    \caption{\label{fig:II_elec_temp}Temperature dependence of the electron impact ionization coefficient at $1/F = \SI{0.3}{\centi\m\per\mega\V}$. The references in bracket denote the used models.}
\end{figure*}

\subsubsection{Origin of Parameters}

The majority of the currently utilized impact ionization coefficients are based on 4H measurements (for a detailed graphical representation see \cref{fig:II_ref_chain} in Appendix~\ref{sec:refChainII}). Care has to be taken, especially for publications prior to the year 2000, as those are often based on 6H-SiC~\cite{bakowski1997,kyuregyan1989,ruff1994,trew1991}. The main reason is that 4H values were not available or simply~\cite{wang1999} because available fittings from \citet{raghunathan1997,konstantinov1998} deviated, resulting in an inconclusive picture.  Nevertheless, the respective values often found their way in later 4H-SiC publications~\cite{ladesmartin2000,neilainglesias2012,wright1996,wright1998,bhatnagar2005}.


The most influential publication was published by \citet{hatakeyama2004} but also the one by \citet{konstantinov1997} was extensively used for fittings. In total, twelve fundamental investigations were referenced at least once in literature.

During our analyses, we discovered some inconsistencies: The results from \citet{bakowski1997} were changed by \citet{ladesmartin2000} who calculated the parameters at \SI{273}{\K} and introduced a typographical error for the hole coefficient $a$, which was stated as \SI{3.24e6}{\per\cm} instead of \SI{2.24e6}{\per\cm}. \citet{nallet1999} used 6H values of $\beta$ proposed by \citet{raghunathan1997} for both electrons and holes. Surprisingly, multiple authors~\cite{pezzimenti2013,nallet1999,akturk2008} followed that example. \cref{fig:II_ref_chain} in Appendix~\ref{sec:refChainII} indicates incorrect values for $\bhole$ in publications referencing \reftext{Ragh99}~\cite{raghunathan1999}. This is not the case, because \citet{raghunathan1999} proposed a temperature dependent value of $\bhole=\SI{3.09}{\mega\V\per\cm}$, which we adopted in this review, and $\bhole=\SI{3.25+-0.3e6}{\mega\V\per\cm}$, which was also used in literature. \citet{khalid2012} correctly referenced $m=2$ from \citet{biondo2012} but, as our analyses showed, this value is unreasonable. Appendix~\ref{sec:II_appendix} presents a more comprehensive listing of encountered inaccuracies.

\section{Charge Carrier Recombination}
\label{sec:regen}

In a semiconductor electron-hole pairs are continuously created, for example due to thermal processes, and annihilated by recombination. In thermal equilibrium these process are balanced, resulting in the electron ($n_0$) and hole ($p_0$) equilibrium carrier concentrations that satisfy the condition $n_0\,p_0 = \nintr^2$, with $\nintr$ being the intrinsic carrier concentration.

In TCAD tools the equilibrium state is not explicitly modeled, only how the semiconductor returns to it. The deviation of the charge carriers to $n_0$ and $p_0$ is denoted as excess carrier, non-equilibrium, generated~\cite{hayashi2011,hayashi2011a} or solely carrier concentration~\cite{neimontas2006} $\DN$ and its rate of change includes diffusion ($D$ equals the ambipolar diffusion coefficient), recombination ($R$) and generation ($G$) (see \cref{eq:regen_dN})~\cite{scajev2013,fang2018,mao2023}. The latter can be caused for example by impact ionization, which we already investigated in \cref{sec:II}, or optical generation.
\begin{equation}
    \label{eq:regen_dN}
    \frac{\mathrm{d} \DN}{\mathrm{d}t} = D \frac{\mathrm{d}^2 \DN}{\mathrm{d}\mathbf{x}^2} - R + G
\end{equation}

In this section we are going to review charge carrier recombination. Accurate models enable high-resolution charge carrier concentrations over time, which influence the conductivity and internal electric fields. We focused our investigation on minority charge carriers, i.e., electrons in p-type and holes in n-type material, because these are the most common ones found in literature.

Our analyses revealed that the lifetime, i.e., the average time between two recombination events, depends on $\DN$, the temperature and the doping concentration. We found measurements that proposed values in a range of four orders of magnitude, but we were unable to identify a clear trend with time or measurement technique. Therefore, we conclude a strong dependency on the quality of the device. Less and even contradicting data is available for temperature and doping related changes of the recombination rate, calling for further investigations in the future.

\subsection{Introduction}

The decline of charge carriers towards their equilibrium values $n_0$ resp. $p_0$ is described by the recombination rate $R$~\cite{schroder1982}, which includes the trap-assisted Shockley-Read-Hall ($\RSRH$), the bimolecular ($\Rbim$) and the Auger ($\RAuger$) recombination rate~\cite{galeckas1997} (see Eq.~(\ref{eq:regen_recomb}))~\cite{arafat2012,scajev2010,fang2018,nagaya2020,yang2019}. An alternative representation is to use lifetimes $\tau_\mathrm{x}$, which denote the average time between two recombination events (see Eq.~(\ref{eq:regen_recomb_lifetime}))~\cite{albanese2010}.
\begin{eqnarray}
    R &=& \RSRH + \Rbim + \RAuger \label{eq:regen_recomb} \\ 
    &=& \frac{\DN}{\tauSRH} + \frac{\DN}{\taubim} + \frac{\DN}{\tauAuger} = \frac{\DN}{\taur} \label{eq:regen_recomb_lifetime}
\end{eqnarray}

The individual contributions to the recombination will be investigated separately in the sequel. For a  comprehensive description the interested reader is referred to the dedicated literature~\cite{arafat2012,landsberg1992,sze2007,schroder2005}.

\subsubsection{Shockley-Read-Hall Recombination}
\label{sec:srh}


The term $\RSRH$ denotes the successive capturing of a hole and an electron in a trap level with energy $E_\mathrm{t}$ inside the band gap~\cite{arafat2012,hangleiter1988}, sometimes also called monomolecular recombination~\cite{galeckas1997}. The following expressions were introduced to describe this process (see also \cref{fig:regen_capture_emission})~\cite{shockley1952,booker2015,degraaff1990}: \textit{electron capture} denotes the transition of an electron from the conduction band into the trap and \textit{electron emission} the reverse process. Similarly, during \textit{hole capture} a hole rises from the valence band to the trap level, i.e., an electron drops from the trap into the valence band, and \textit{hole emission} denotes the reverse case. In any capture event the electron, thus, looses energy, while in any emission event it gains some. The electron/hole \textit{capture cross sections} quantify the possibility to capture an electron/hole~\cite{abakumov1978}, but they can also describe the thermal emission rate $e$ (see \cref{eq:regen_emission})~\cite{suproniuk2020} with $E_\mathrm{a}$ the defect activation energy.
\begin{equation}
    \label{eq:regen_emission}
    e = \sigma\, \gamma\, T^2\, \exp\left( - \frac{E_\mathrm{a}}{\kb\,T} \right)
\end{equation}

\begin{figure}[t]
    \centering
    \resizebox{0.5\linewidth}{!}{%
    \input{figures/regen_capture_emission}
    }
    \caption{\label{fig:regen_capture_emission}Capture and emission of charge carriers described by $\RSRH$. $E_\mathrm{t}$ denotes the trap level energy.}
\end{figure}

The energy released/consumed during these transitions is exchanged with lattice vibrations (phonons)~\cite{arafat2012}, whereat different phonon interactions are distinguished, e.g., multi- or cascade-phonon interaction~\cite{fossum1982,landsberg1992}. Among these the capture cross section and their respective temperature dependencies differ~\cite{booker2015}.

The recombination rate $R$ depends on the material quality~\cite{arafat2012}: Structural defect positions~\cite{mori2005} can cause a decrease of $R$ but certain impurities or damages in the lattice, which are called effective ``lifetime killers''~\cite{ayedh2017}, lead to a steep increase. The recombination even varies across a single wafer. Typically it is smallest in the middle where the best growth conditions are available~\cite{cui2019,erlekampf2019,erlekampf2021,galeckas1997a,galeckas1999a,galeckas2001,kato2007,klein2008,kordina1996,lilja2013,hassan2009,tsuchida2018,zhang2023a,murata2019}, but also in thick 4H-SiC layers variations were reported~\cite{mahadik2017}. Lots of effort was undertaken to refine growth conditions in order to achieve cleaner samples~\cite{jenny2006,grivickas2001,lilja2013,hayashi2012,ichikawa2012,kaji2015,kawahara2012,kimoto2008,kimoto2010,miyazawa2013,okuda2013a,okuda2014,storasta2007,storasta2008,tawara2016,tsuchida2018,wang2020,yang2019,zhang2023a,hiyoshi2009,yan2024,khemka1999}.

An accurate description of $\RSRH$ requires detailed information about energy
level, type (acceptor or donor) and cross section of the defects in the
device. We encountered a large amount of investigations targeting this
topic~\cite{aberg1999,afanasev2000,albanese2010,alfieri2005,alfieri2020,anikin1985,arvanitopoulos2017,arvinte2017,ayalew2004,ayalew2005,ayedh2016,bakowski1997,balachandran2005,baliga2006,bathen2019,bellone2011,bergman1997,bhatnagar2005,bluet2000,booker2016,brodar2020,buono2012,capan2018,capan2018a}\newline
{}~\cite{capano2000,castaldini2004,chaudhuri2021,chen2000,chen1997,choyke1997d,cochrane2007,cochrane2009,coutinho2017,dalibor1997,dalibor2003,danno2007,danno2006,darmody2019,david2004,denapoli2022,devaty1997,doyle1997,egilsson1999,evwaraye1996,fan2014,feng2004a,gerhardt2011,gotz1993,greulich-weber1997,haberstroh1994,hagen1973,harris1995,hatakeyama2013}\newline
{}~\cite{hazdra2017,hazdra2021,hemmingsson1998,hemmingsson1997,henisch2013,hornos2011,huang2022,huang2022a,huh2006,ikeda1980,ioffe2023,itoh1994,itoh1998,ivanov2005,ivanov2003,iwata2001,izzo2008,janzen2008,johannesson2019,kagamihara2004,kaji2015,karsthof2020,kasamakova-kolaklieva2004,kawahara2012,kimoto2018,kimoto1995,kimoto2014a,kimoto2015,kimoto2019,kimoto2008}\newline
{}~\cite{kimoto2003,kimoto1997,klein2006,klein2009,kleppinger2021,koizumi2009,kuznetsov1995,kwasnicki2014,ladesmartin2000,laube2004,lavia2005,lebedev1999,levinshtein2001,li2003,lichtenwalner2023,liu2015,lomakina1973,lomakina1974,lophitis2018,lu2021,lutz2011,madelung1991,mandal2020,martinez2002,matsunami1974,matsuura2004,matsuura1999,megherbi2018,mickevicius1998}\newline
{}~\cite{murata2019,muzykov2012,nawaz2010,negoro2003,neudeck2001,neudeck2006,nguyen2015,pankove2014,parisini2013,pastuovic2017,pearton2023,pensl1993,pensl2005,pernot2005,pernot2001,persson1999a,persson1998,persson2005,pezzimenti2013,rakheja2020,rao1999,reshanov2005,roschke1998,schadt1997,schmid2004,sharma2015,shur2006,smith1999,son2004a,son2012}\newline
{}~\cite{sozzi2019,storasta2004,storasta2002,storasta2007,suzuki1977,suzuki1973,tanaka2000,tawara2004,terziyska2003,torres2022,tripathi2019,troffer1997,troffer1998,uhnevionak2015,wang2023,wijesundara2011,xu2023,yan2020,yoshioka2018,zetterling2002,zhang2003,zhu2008,zippelius2011,zippelius2012,suproniuk2020,torpo2001}, which makes a comprehensive analysis within this review infeasible. Instead, we refer the interested reader to the report by~\citet{gaggl2025}, who investigated defects in 4H-SiC and developed a TCAD model for it. In regard to recombination, the most important ones are called $\mathrm{Z}_{1/2}$ and $\mathrm{EH}_{6/7}$~\cite{tawara2004,yamashita2020,zhang2003,marinella2010} and, presumably, denote different charge states of a carbon vacancy~\cite{booker2016}.

The Shockley-Read-Hall recombination includes a bulk ($\RSRHb$) and surface ($\RSRHs$) contribution. Both occur simultaneously and are therefore often hard to separate~\cite{arafat2012,kato2024}. Nevertheless, we will describe them in isolation in the sequel.

\paragraph{Bulk Recombination Rate}

\citet{shockley1952,hall1952} first described the recombination rate of the bulk mathematically in 1952 (see Eq.~(\ref{eq:regen_SRH_bulk}))~\cite{shockley1952,hall1952,arvanitopoulos2017,degraaff1990,lutz2011,selberherr1984,schroder2005}.  $\taunp$ denotes the electron resp. hole lifetime, $\sigmanp$ the electron resp. hole cross section, $\ntrap$ the trap concentration, $\vthermal=\sqrt{3 \kb T/\md}$~~\cite{degraaff1990,suttrop1991} the thermal velocity and $\gtrap$ the trap degeneracy factor. The latter is often neglected as it is usually one~\cite{buono2012}. \citet{shockley1952} stated that the trap energy level ($\etrap$) is an effective one that they calculated from the trap level and the degeneracies of the empty ($w_\mathrm{p}$) and full ($w$) trap, i.e., $\etrap = \etrap(\text{true}) + \kb T \ln(w_\mathrm{p}/w)$. TCAD tools accept either the lifetimes $\tau_\mathrm{n,p}$ or the cross sections $\sigma_\mathrm{n,p}$, but they do not consider the degeneracy factor yet.
\begin{eqnarray}
    \RSRHb &=& \frac{np - \nintr^2}{\taup(n+n_1) + \taun(p+p_1)} \label{eq:regen_SRH_bulk} \\
    n_1 &=& \frac{1}{\gtrap} \nc \exp\left(-\frac{\ec - \etrap}{\kb T}\right) \label{eq:regen_n1}\\
    p_1 &=& \gtrap\, \nv\, \exp\left(-\frac{\etrap - \ev}{\kb T}\right)\label{eq:regen_p1}\\
    \nintr &=& \sqrt{n_1 p_1} = \sqrt{\nc\nv} \exp\left(-\frac{\ec - \ev}{2\kb T}\right) = \sqrt{\nc\nv} \exp\left(-\frac{\eg}{2\kb T}\right) \\
    \taunp &=& (\sigmanp \vthermal \ntrap)^{-1} \label{eq:regen_taunp}
\end{eqnarray}

$p_1$ and $n_1$ can also be described by the intrinsic carrier concentration and an effective Fermi level $\eintr$ (see \cref{eq:regen_np1_intr})~\cite{lee2002,albanese2010,fossum1982}. $\eintr$, which was defined in one case as the mid gap energy level~\cite{biondo2012}, differs from the intrinsic Fermi level $\ef$ that defines the actual carrier concentration $n = \nc \exp [ (\ef-\ec)/\kb T]$ and $p = \nv \exp [ (\ev-\ef)/\kb T]$.
%
\begin{equation}
\begin{split}
    \label{eq:regen_np1_intr}
    n_1 = \frac{1}{\gtrap} \nintr \exp \left( -\frac{\etrap - \eintr}{\kb T}\right) \\
    p_1 = \gtrap \nintr \exp \left( -\frac{\eintr - \etrap}{\kb T}\right)
    \end{split}
\end{equation}


The SRH carrier lifetime changes with the doping concentration $N_\mathrm{A,D}$~\cite{lilja2017}. On the one hand dopants are defects (see \cref{sec:incompIon}), i.e., recombination centers, whose impact on the recombination is described with Eq.~(\ref{eq:regen_SRH_bulk})~\cite{rakheja2020}. On the other hand the doping process introduces additional damage~\cite{selberherr1984} that lowers the lifetime according to the empirical Scharfetter relation (see \cref{eq:regen_scharfetter})~\cite{scharfetter1967,arvanitopoulos2017,fossum1982,roulston1982,law1991,bellone2011,selberherr1984,tyagi1983,liaugaudas2015,ruff1994}. \citet{shao2024} recently provided a detailed analysis on the change of the lifetime for various relations of donor and acceptor concentrations.
\begin{equation}
    \label{eq:regen_scharfetter}
    \tau = \frac{\taumax}{1+\left( \frac{\na + \nd}{\nref}\right)^\gamma}
\end{equation}

The defect induced lifetime reduction was determined in more simplistic fashions as well, e.g., $\tau(\si{\us}) = \num{1.5e13} / N_{\mathrm{VC}}$~~\cite{yamashita2020,sapienza2020} and $\tau(\si{\us}) = \num{1.8e13} / N_{\mathrm{VC}}$~~\cite{kimoto2018} for carbon vacancies, and $\tau(\si{\us}) = \num{1.6e13} / N_{\mathrm{Z1/2}}$~~\cite{kimoto2014a,meyers2017} and $\tau(\si{\us}) = \num{2e13} / N_{\mathrm{Z1/2}}$~~\cite{kimoto2010} for the $\mathrm{Z}_{1/2}$ defect, with $N_x$ the respective concentration in \si{\per\cubic\centi\m}~\,\cite{klein2010}.

At high temperatures the energetic charge carrier has to approach the center of the defect more closely to be captured~\cite{tyagi1983}. Consequently, the lifetime increases with temperature~\cite{tawara2016,kordina1996,levinshtein2005,okuda2014,agarwal2001,ivanov1999,ivanov2006a}, which was analyzed by \citet{udal2007}. The cross section changes in the order of $T^{-x}$~~\cite{wu1982}, which leads, in conjunction with the change of the thermal velocity according to $\sqrt{T}$, to a power law description~\cite{sapienza2020}. TCAD tools use the model shown in \cref{eq:regen_powerT0}~\cite{goebel1992,klaassen1992a,schenk1992}, whereat $\tau_{T0}$ denotes the lifetime at some reference temperature $T_0$.
\begin{equation}
    \label{eq:regen_powerT0}
    \taumax = \tau_{T0} \left( \frac{T}{T_0} \right)^{\alpha}
\end{equation}
Because this approximation predicts $\taumax \to 0$ for $T\to 0$ a modified
approach shown in \cref{eq:regen_power0} was proposed~\cite{rakheja2020}. By
comparison we find $\tau_0 = \tau_{T0}/2$ at $T=T_0$ and $\taumax=\tau_0$ at $T=\SI{0}{\K}$.
\begin{align}
    \label{eq:regen_power0}
    \taumax = \tau_{0} \left( 1+ \left(\frac{T}{T_0}\right)^{\alpha} \right)
\end{align}

We also found an exponential description (see \cref{eq:regen_actEnergy})~\cite{udal2007} for the temperature dependency of $\tau$, with $\eact$ being an activation energy and $\tau_\infty$ the lifetime for $T \to \infty$.
\begin{align}
    \label{eq:regen_actEnergy}
    \taumax = \tau_{\infty} \exp \left( - \frac{\eact}{\kb T} \right)
\end{align}
Since the meaning of $\tau_\infty$ is hard to grasp, it can be replaced by $\tau_{T0} \exp (\eact/\kb T_0)$~\cite{tamaki2008,kimoto2014a}, with $\tau_{T0}$ the lifetime at $T=T_0$. For $\eact = \SI{0.105}{\eV}$ and $T_0 = \SI{300}{\K}$ we get $\exp (\eact/\kb T_0)=57.9$, which was sometimes stated explicitly~\cite{tamaki2008,kimoto2014a}. A similar model shown in \cref{eq:regen_actEnergy51}~\cite{scajev2013} approaches a value of $51 \tau_0$ for $T \to \infty$.
\begin{align}
    \label{eq:regen_actEnergy51}
    \taumax = \tau_{0} \left(1 + \frac{100}{1+\exp\left(\eact/\kb T\right)} \right)
\end{align}

The lifetimes in \cref{eq:regen_actEnergy} and \cref{eq:regen_actEnergy51} stall for high temperatures, which makes them only suitable for temperatures between \SIrange{300}{500}{\K}~\cite{udal2007}. The model shown in \cref{eq:regen_expSimple}~\cite{lechner2021,nallet1999} circumvents this problem.
\begin{align}
    \label{eq:regen_expSimple}
    \taumax = \tau_{T0} \exp^{C\left(\frac{T}{T_0}-1 \right)}
\end{align}

A recent investigation by \citet{lechner2021} identified an increase of the
lifetime up to a temperature between \SIrange{600}{700}{\K} followed by a
decrease at higher ones. Based on the research by \citet{schenk1992} the author
proposed the fitting shown in \cref{eq:regen_expUpDown}~\cite{lechner2021}.
\begin{align}
    \label{eq:regen_expUpDown}
    \taumax = \tau_{T_0} \left(\frac{T}{T_0}\right)^{T_\mathrm{coeff}} \exp \left[ -\alpha_\tau \left( \frac{T}{T_0} - 1\right)^{\beta_\tau}\right]   
\end{align}
In state-of-the-art simulation tools only \cref{eq:regen_powerT0} and \cref{eq:regen_expSimple} are included.

\paragraph{Surface Recombination Rate}

Surface recombination includes mechanisms that occur on surfaces or interfaces. The primary causes are imperfections or impurities at the transition between two materials. This includes many different surface types and thus also effects, such as semiconductor-oxide or semiconductor-semiconductor and oxides across these. The share of the surface recombination on the overall recombination rate decreases with the thickness of the samples, because the ratio of surface to bulk volume decreases~\cite{klein2010}.

The boundary condition of the diffusion defines the surface recombination (see \cref{eq:regen_surface})~\cite{gulbinas2011,kato2012,kato2020,mori2014}. For the interested reader \citet{mao2023} presented a fantastic review on the causes, characterization methods and possible countermeasures of surface recombination and \citet{gulbinas2011} a theoretical analysis.
\begin{equation}
    \label{eq:regen_surface}
    D \frac{\mathrm{d} \DN(\mathbf{x},t)}{\mathrm{d} \mathbf{x}} = S_0 \DN(\mathbf{x},t)
\end{equation}

Despite the tight connection between diffusion and surface recombination, TCAD tools model the latter by the SRH formalism. The sole difference to the bulk recombination rate is that instead of lifetimes the \textit{surface recombination velocities} $s_{n,p}$ (see Eq.~(\ref{eq:regen_SRH_surface}))~\cite{arafat2012,selberherr1984,schroder2005,buono2012} that depend on the interface trap density $\nitrap$ are used.
\begin{eqnarray}
    \RSRHs &=& \frac{(\nsurf \psurf - \nintr^2)}{(\nsurf+n_{1})/\surfp + (\psurf +p_{1})/\surfn} \label{eq:regen_SRH_surface} \\
    \surfn &=& \sigma_\mathrm{ns} \vthermal \nitrap\\
    \surfp &=& \sigma_\mathrm{ps} \vthermal \nitrap
\end{eqnarray}
TCAD tools expect $\surfn$ and $\surfp$ as inputs. Their dependency on the crystal faces~\cite{kato2007,kato2012,kato2020,kato2024,mori2014} is not included in state-of-the-art simulation tools yet. Some provide the possibility to model a doping dependency but we found no reliable data for 4H-SiC in literature.

The surface recombination velocity depends on the surface quality and the
neighboring material. Therefore, many studies seeked to improve the material quality \cite{chung2009,cheong2003,asada2018,gulbinas2011,hayashi2011,hayashi2014,ichikawa2018,kato2024,kimoto2010a,nonaka2009,boltovets2006} by varying growth mechanisms or by irradiation~\cite{hazdra2017}. Even an elaborate model using trap regions inside the band gap was developed~\cite{buono2010,afanasev1996,afanasev1997}. 

\citet{kato2020} and \citet{klein2010} described the temperature dependency of the surface recombination velocity, which is modeled by \cref{eq:regen_surface_temp}~\cite{fitzgerald1968,yablonovitch1986}. $-\eactsurf$ denotes the band bending near the surface that leads to accumulation of charge carriers of one type at the surface while repelling the other~\cite{yablonovitch1986,kimoto2008}.
\begin{equation}
    \label{eq:regen_surface_temp}
      s_\mathrm{eff}(T) = s_\infty \exp \left(\frac{-\eactsurf}{\kb T}\right)
\end{equation}

\subsubsection{Bimolecular Recombination}

The term $\Rbim$ denotes the recombination rate due to the interaction of two particles, which can be described by \cref{eq:regen_Rbim}~\cite{selberherr1984,lutz2011} with $B$ the bimolecular recombination coefficient~\cite{albanese2010,fang2018}.
\begin{equation}
    \label{eq:regen_Rbim}
    \Rbim = B (np - \nintr^2)
\end{equation}

In literature, bimolecular recombination is sometimes reduced to the radiative band-to-band recombination process emitting a photon~\cite{scajev2010}. However, in the indirect semiconductor 4H-SiC (see \cref{sec:bandgap}) it is less important, because a phonon always has to absorb the momentum of the charge carrier~\cite{scajev2013,booker2015,fang2018,albanese2010,klein2008}. Besides radiative recombination, $\Rbim$ also includes
\begin{enumerate*}[label=(\roman*)]
\item recombinations between donor-acceptor pairs (DAP)~\cite{murata2021,yang2019}
\item the recombination of a charge carrier from the conduction/valence band and an unionized dopant (e.g., e-A)~\cite{murata2021,tanaka2024} and
\item the recombination of excitons (see \cref{sec:bandgap})~\cite{scajev2010}.
\end{enumerate*}

Trap-assisted Auger recombination (TAA)~\cite{fossum1983} is also a bimolecular process but it is handled in differing fashions in the literature. TAA denotes the process when an electron (hole) interacts with a trap (capture resp. emission) and the additional/missing energy is exchanged with a particle of the same kind. In the past, authors included this process in the bimolecular recombination coefficient~\cite{scajev2010,galeckas1997,linnros1998,gao2022a}, the Auger process~\cite{tanaka2023,reshanov2005} and SRH~\cite{fossum1983,landsberg1984,landsberg1992}. The latter is the only option to model TAA in existing TCAD tools, but we were unable to find suitable values for 4H-SiC.

We found additional trap-assisted Auger processes that have to be handled with care: \citet{linnros1998} remarked that the excitonic Auger capture process described by \citet{hangleiter1988} is ``markedly different'' from TAA and should be regarded as an alternative explanation for multi-phonon/SRH recombination. \citet{booker2016} reported a trap-Auger mechanism using a neutral $\mathrm{EH}_{6,7}$ trap that contains two electrons. When a hole recombines with one of the electrons, the excessive energy is transferred to the other electron and ejects it to the conduction band. Finally, \citet{takeshima1981} described a phonon-assisted Auger recombination process.

\subsubsection{Auger Recombination}
\label{sec:auger}

At last, the term $\RAuger$, also called Auger recombination ``(first discovered in atomic systems by Pierre Auger; soft g, please, the gentleman is French not German!)''~\cite{landsberg1992}, denotes a three particle interaction where the excessive resp. missing energy and momentum is taken from/transferred to a third particle (either hole or electron). This process is an intrinsic property of the material~\cite{arafat2012} and dominates for high excessive charge carrier densities. It is described by \cref{eq:regen_Rauger}~\cite{dziewior1977,arvanitopoulos2017,selberherr1984,lutz2011} where $\cn$ denotes the energy transfer to an electron and $\cp$ the transfer to a hole~\cite{galeckas1997}.
\begin{equation}
    \label{eq:regen_Rauger}
    \RAuger = (\cn n + \cp p) (np - \nintr^2)
\end{equation}

We found several approaches to model the temperature
dependency~\cite{grivickas2007,tawara2016} of the
parameters. \citet{galeckas1997} proposed an exponential correlation according
to \cref{eq:regen_auger_temp_sum}.
\begin{equation}
    \label{eq:regen_auger_temp_sum}
    \cn + \cp = \gamma_{30} \exp\left(-\frac{\alpha (T-\SI{300}{\K})}{\kb T}\right)
\end{equation}

\citet{scajev2013} explained their deviating results by the fact that \citet{galeckas1997} did not consider the in-depth profile. For an improved coverage the authors developed the model in \cref{eq:regen_auger_temp_indepth}, with $B_\mathrm{CE}(T)\propto T^{-1.5}$ the Coulomb enhancement coefficient and $a_\mathrm{SC}$ the screening parameter.
\begin{equation}
    \label{eq:regen_auger_temp_indepth}
    C(T,\DN) = \left(C_0 + \frac{B_\mathrm{CE}(T)}{\DN}\right) / \left(1 + \frac{\DN}{a_\mathrm{SC} \times T}\right)^2
\end{equation}  

\citet{scajev2010} mentioned a more simplistic relationship of $C \propto \DN^{-0.3}$ due to the screening of the Coulomb enhancement coefficient, and \citet{tanaka2023} $C \propto \DN^{-0.68}$. The differences may be explainable by the deviating fitting procedures.

In rare cases researchers even reused the formalism to describe the temperature induced changes of the Auger recombination coefficients in Silicon (see \cref{eq:regen_auger_temp_Si})~\cite{lophitis2018,zhang2018}.
\begin{equation}
    \label{eq:regen_auger_temp_Si}
    C_\mathrm{n,p} = \left( A_\mathrm{n,p} + B_\mathrm{n,p} \left(\frac{T}{T_0}\right) + D_\mathrm{n,p} \left(\frac{T}{T_0}\right)^2 \right) \left[ 1 + H_\mathrm{n,p} \exp \left( -\frac{n,p}{N_\mathrm{0n,p}}\right)\right]
\end{equation}

\subsubsection{Analysis}
\label{sec:regen_analysis}

In the sequel we want to analyze the presented equations and extract further useful information. Some reader might have noticed that all recombination rates contain the multiplicative factor shown in \cref{eq:regen_nominator} with $\excessn, \excessp$ the excessive carrier concentrations for electrons resp. holes. Recall that $n=n_0 + \excessn$, $p=p_0 + \excessp$ and $\nintr = n_0p_0$.
\begin{equation}
    \label{eq:regen_nominator}
    (np - \nintr^2)=(n_0 + \excessn) (p_0+ \excessp) - n_0 p_0 = n_0 \excessp + p_0 \excessn + \excessn \excessp
\end{equation}

For a device without traps, i.e., $\excessp = \excessn = \DN$, and for low-level (ll, $\DN \ll n_0, p_0$) resp. high-level (hl, $\DN \gg n_0, p_0$) injections the models can be simplified~\cite{arafat2012,linnros1998,lutz2011,schroder2005,sze2007}, which provides useful insights on how to interpret measurements results~\cite{arafat2012}. After a short calculation we get, due to $R=\DN/\taur$ (see Eq.~(\ref{eq:regen_recomb_lifetime})), for low-level injections the results shown in \cref{eq:regen_ll} and high-levels the results in \cref{eq:regen_hl}.
\begin{subequations}
    \label{eq:regen_ll}
    \begin{eqnarray}
    \tauSRH^\mathrm{ll} &=& \frac{\taup (n_0 +n_1)+ \taun (p_0 + p_1)}{n_0 + p_0} \\
    \taubim^\mathrm{ll} &=& \frac{1}{B (n_0 + p_0)} \\
    \tauAuger^\mathrm{ll} &=& \frac{1}{(\cn n_0 + \cp p_0)(n_0 + p_0)}
    \end{eqnarray}
\end{subequations}
\begin{subequations}
    \label{eq:regen_hl}
    \begin{eqnarray}
    \tauSRH^\mathrm{hl} &=& \taup + \taun \\
    \taubim^\mathrm{hl} &=& \frac{1}{B \DN} \\
    \tauAuger^\mathrm{hl} &=& \frac{1}{(\cn + \cp)\DN^2}
    \end{eqnarray}
\end{subequations}
At high-injection levels $\tauSRH=\taup + \taun$ (also called ambipolar lifetime~\cite{aditya2015,kimoto2014a,hangleiter1987}) and $\tauAuger$ solely provide the sum of electron and hole coefficients. Separate measurements are only possible at low-injection levels and with doped semiconductors, i.e., either $n_0 \gg p_0, p_1, n_1$ or $p_0 \gg n_0, n_1, p_1$. In that case one of the summands can be ignored.

For doped semiconductors we can rewrite the high-level injection results from \cref{eq:regen_hl} as shown in \cref{eq:regen_polynom}~\cite{arafat2012,scajev2010,fang2018,nagaya2020,galeckas1997,albanese2010,danno2007,landsberg1992,lutz2011,linnros1998}, which highlights that the introduced recombination terms actually represent a polynomial approximation of the recombination rate up to degree three in respect to the excess carrier concentration. If no excess charge carriers exist, i.e., $\DN=0$, no recombination is considered ($R=0$). For low excess charge carrier densities the SRH term dominates while for high densities the Auger process is most important. 
\begin{equation}
    \label{eq:regen_polynom}
\begin{split}
    R=\DN \taur^{-1} &= \DN (\tauSRH^{-1} + \taubim^{-1} + \tauAuger^{-1}) \\
&= A\DN + B  \DN^2 + C \DN^3
\end{split}
\end{equation}
%

All these simplifications are only valid for $\DN \geq 0$. If $(np - \nintr^2)<0$ the recombination rate becomes negative, meaning that according to \cref{eq:regen_dN} charge carriers are generated. The corresponding rate of change can be described by using the so-called generation lifetime $\tau_\mathrm{g}$~\cite{chung2008,lutz2011,schroder2005,schroder1982,marinella2010}. Such a reversal is not meaningful for the bimolecular recombination because the respective generation demands incoming photons~\cite{schroder1982}. TCAD tools provide for this purpose dedicated optical generation mechanism, meaning that bimolecular recombination is deactivated for $(np - \nintr^2)<0$.

Impact ionization (see \cref{sec:II}) is often stated as the inverse of Auger recombination~\cite{fossum1982,sze2007,ruff1993}, implying that it also gets deactivated if too few charge carriers are present. However, \citet{selberherr1984} stated that there is a difference: While impact ionization requires high current densities the inverse process of the Auger recombination only requires high charge carrier concentrations with negligible current flow. Despite these facts, $\RAuger$ gets deactivated in TCAD tools when it drops below zero by default, but some allow to explicitly enable it for this case.

\subsubsection{Characterization Methods}
\label{sec:regen_methods}

The recombination lifetime was measured either optically or electrically. Commonly used optical techniques include photoluminescence decay (PLD)~\cite{kordina1996,huh2006,lilja2017}, (transient) (time-resolved) free carrier absorption ((T)(TR-)FCA)~\cite{galeckas1997,grivickas2001,scajev2010,klein2008,linnros1998,scajev2013a,nagaya2020,grivickas2001,scajev2013,suvanam2015,tanaka2023a}, electron beam induced current (EBIC)~\cite{jenny2006}, time-resolved photoinduced absorption (TRPA)~\cite{galeckas1997}, time-resolved photoluminescence (TRPL)~\cite{klein2008,klein2010,mao2023,lilja2013,miyazawa2013,murata2019,storasta2007,storasta2008,tawara2004,tawara2016,tsuchida2016,zhang2003}, time-resolved transient absorption (TRTA), longitudinal optical phonon-plasmon coupling (LOPC)~\cite{meli2022,meli2022a}, transient absorption spectroscopy (TAS)~\cite{fang2018}, four wave mixing (FWM)~\cite{neimontas2006}, low-temperature photoluminescence (LTPL)~\cite{lilja2017,liaugaudas2015}, capacitance transient (C-t)~\cite{cheong2003}, differential transmittivity (DT)~\cite{liaugaudas2015} and (microwave) photoconductance decay (($\mu$)-PCD)~\cite{klein2008,mao2023,chowdhury2015,cui2019,danno2007,erlekampf2021,hayashi2011,hayashi2011a,hayashi2012,hettler2010,hettler2012,hiyoshi2009,ichikawa2012,kaji2015,kato2020,kato2024,kawahara2012,kimoto2008,kimoto2010,kimoto2010a,kimoto2018,miyazawa2010,okuda2013a,okuda2014,okuda2016,saito2016,zhang2023a}. \citet{hirayama2020} described a two-photon absorption (TPA) process to determine the lifetime in a specific depth of the sample.

Electrical measurements included reverse recovery (RR)~\cite{kimoto1999,neudeck1998,udal2000,udal2007}, thyristor turned off gate current (TTOGC)~\cite{agarwal2001}, short-circuit current/open-circuit voltage decay\\(SCCVD/ OCVD)~\cite{albanese2010,reshanov2005,reshanov2009,ivanov1999,ivanov2006b,levinshtein2004,sapienza2020,sozzi2024}, diode current density (DCD)~\cite{dibenedetto2014,hasegawa2017,puzzanghera2016,tian2020}, bipolar transistor emitter current (BTEC)~\cite{domeij2003} and diode forward voltage degradation (DFVD)~\cite{koyama2020}. Also utilized were fittings to measurement (FIT)~\cite{donnarumma2012,maximenko2023,usman2014} or simulations (SIM)~\cite{ivanov2006a}.

The achieved lifetime values depend on the utilized method~\cite{reshanov2005}. It has to be assured that the same quantity is measured, and that injection level and temperature are taken into account~\cite{reshanov2005}. For example, \citet{kato2014} claimed that $\mu$-PCD tends to overestimate the carrier lifetimes in high injection conditions and \citet{tawara2004} that $\mu$-PCD achieves longer lifetimes than the ones by TRPL. Deviations might also results from improper measurement setups: OCVD is limited to low and high injection regions~\cite{albanese2010}. \citet{levinshtein2001a} even argued that RR and OCVD measurement may provide incorrect results~\cite{donnarumma2012}. For more detailed information, e.g., which carrier lifetime is extracted from the decay time for high and low injection by each measurement technique, the interested reader is referred to the dedicated literature~\cite{klein2008,kimoto2014a}.

It is important to consider the impact of the surface recombination $\RSRHs$ as well~\cite{klein2008}, because samples with a thickness in the range of \si{\mm} or even \si{\cm} would be required~\cite{schroder2005} to extract $\RSRHb$ directly~\cite{arafat2012}. In the optimal case $\RSRHs$ is determined separately~\cite{kimoto1999,klein2010,miyazawa2010,neudeck1998}, but that is not always practical~\cite{klein2008}.


\subsection{Results \& Discussion}

In the sequel the results of our analyses will be presented, whereas we will not distinguish the different faces inside the crystal~\cite{kato2020,storasta2008}. We did not include values when solely the effective lifetime was proposed~\cite{werber2007,chung2008,galeckas2001} or if it was not possible to clearly distinguish $\taun$ and $\taup$\cite{hettler2010,hettler2012,scajev2010,meli2022}. We also removed the values by \citet{kato2018} who investigated the recombination at specific crystal stacking faults or dislocations, because the achieved lifetimes were a lot lower than all other values we gathered. We discarded the investigation by \citet{grivickas2001} because no concrete values were mentioned.

\subsubsection{SRH Lifetime}

\begin{table*}[t]
    \setlength{\tabcolsep}{7pt}
    \caption{\label{tab:regen_lte}Electron lifetime results. Column \textit{inj.} denotes the carrier injection level, i.e., low (ll) resp. high (hl), and column \textit{excess} the exact amount. A \textit{y} in column \textit{impr.} highlights that the shown value is the highest lifetime achieved in an optimization process.}
    \resizebox{0.85\linewidth}{!}{%
        \begin{tabular}{l *{8}{c}}
ref. & $\taun$ & dop & conc. &$T$ & inj. & $\DN$ & impr. & method \\ 
 & [\si{\micro\s}] & & [\si{\per\cubic\centi\meter}] & [\si{\K}] && [\si{\per\cubic\centi\meter}] && \\ \hline 
 &&&&&&&& \\[-7pt] 
\reftext{Agar01}\cite{agarwal2001}& \num[exponent-mode=input]{0.6} & p & \num{7e+14} & \num{293} & hl & - & - & TTOGC  \\ 
\reftext{Ivan06a}\cite{ivanov2006a}& \num[exponent-mode=input]{0.066} & p & \num{2e+17} & \num{300} & - & - & - & SIM  \\ 
\reftext{Alba10}\cite{albanese2010}& \num[exponent-mode=input]{0.008} & - & \num{2.04e+17} & \num{300} & - & - & - & OCVD  \\ 
\reftext{Haya11}\cite{hayashi2011}& \num[exponent-mode=input]{1.3} & Al & \num{9e+14} & - & ll & \num{1.5e+14} & y & $\mu$-PCD  \\ 
\reftext{Haya11a}\cite{hayashi2011a}& \num[exponent-mode=input]{1.6} & p & \num{9e+14} & \numrange[exponent-mode=input]{300}{525} & ll & \num{1.5e+14} & y & $\mu$-PCD  \\ 
\reftext{Haya12}\cite{hayashi2012}& \num[exponent-mode=input]{1.7} & p & \num{5.6e+14} & - & ll & \num{1e+15} & y & $\mu$-PCD  \\ 
\reftext{Okud13a}\cite{okuda2013a}& \num[exponent-mode=input]{0.31} & Al & \num{1e+18} & \num{300} & - & \num{9.1e+15} & - & $\mu$-PCD  \\ 
\reftext{Dibe14}\cite{dibenedetto2014}& \num[exponent-mode=input]{0.001} & - & - & - & hl & - & - & DCD  \\ 
\reftext{Okud14}\cite{okuda2014}& \num[exponent-mode=input]{10} & Al & \num{2e+14} & \num{300} & - & \num{3.6e+16} & y & $\mu$-PCD  \\ 
\reftext{Liau15}\cite{liaugaudas2015}& \num[exponent-mode=input]{0.02} & Al & \num{1e+17} & \num{300} & hl & - & - & DT  \\ 
\reftext{Okud16}\cite{okuda2016}& \num[exponent-mode=input]{12} & Al & \num{1e+15} & \num{300} & - & \num{3.6e+14} & y & $\mu$-PCD  \\ 
\reftext{Hase17}\cite{hasegawa2017}& \num[exponent-mode=input]{0.4} & p & \num{8e+14} & - & - & - & - & DCD  \\ 
\reftext{Kato20}\cite{kato2020}& \num[exponent-mode=input]{1.2} & Al & \num{6e+14} & \num{300} & ll & - & y & $\mu$-PCD  \\ 
\reftext{Koya20}\cite{koyama2020}& \num[exponent-mode=input]{0.13} & p & \num{1e+19} & - & - & - & - & DFVD  \\ 
\reftext{Maxi23}\cite{maximenko2023}\footnotemark[1] & \num[exponent-mode=input]{6} & - & - & - & - & - & - & FIT  \\ 
\reftext{Zhan23a}\cite{zhang2023a}& \num[exponent-mode=input]{3.14} & Al & \num{2e+14} & \num{300} & - & - & y & $\mu$-PCD  \\ 
\end{tabular}
        \footnotetext[1]{value fitted to measurements by \citet{kimoto2018}}
    }
\end{table*} 

We identified $16$ investigations of the electron lifetime $\taun$ in 4H-SiC (see \cref{tab:regen_lte}) and $62$ of $\taup$ (see \cref{tab:regen_lth} and \cref{tab:regen_lth_II}). The first scientific reports were published in the late 1990s and up to the present day this is an active research topic. Despite the wide range of values, we still encountered the relation
$\taun = 5 \taup$~\cite{ruff1994,schroeder2006,li2003,shah1998,zhang2010,zhao2003,tiwari2019a,wang1999,johannesson2019,tian2020}, which was originally used for Si and more recently $\taun = \taup$~\cite{levinshtein2001b,bellone2009,nallet1999,pezzimenti2013,rao2022,tamaki2008a,usman2012,khalid2012}. Based on our results the latter seems more reasonable.

\begin{table*}[p]
    \setlength{\tabcolsep}{7pt}
    \caption{\label{tab:regen_lth}Hole lifetime results [1/2]. Column \textit{inj.} denotes the carrier injection level, i.e., low (ll) resp. high (hl), and column \textit{excess} the exact amount. A \textit{y} in column \textit{impr.} highlights that the shown value is the highest lifetime achieved in an optimization process.}
    \resizebox{0.9\linewidth}{!}{%
        \begin{tabular}{l *{8}{c}}
ref. & $\taup$ & dop & conc. &$T$ & inj. & $\DN$ & impr. & method \\ 
 & [\si{\micro\s}] & & [\si{\per\cubic\centi\meter}] & [\si{\K}] && [\si{\per\cubic\centi\meter}] && \\ \hline 
 &&&&&&&& \\[-7pt] 
\reftext{Kord96}\cite{kordina1996}& \num[exponent-mode=input]{2.1} & n & - & \num{300} & - & - & - & PLD  \\ 
\reftext{Gale97}\cite{galeckas1997}& \num[exponent-mode=input]{0.26} & N & \num{5e+15} & - & hl & - & - & TRPA  \\ 
\reftext{Neud98}\cite{neudeck1998}& \num[exponent-mode=input]{0.7} & N & \numrange[exponent-mode=input]{2e16}{4e16} & - & - & - & - & RR  \\ 
\reftext{Gale99a}\cite{galeckas1999a}& \num[exponent-mode=input]{0.5} & N & <\num{1e16} & - & - & - & - & FCA  \\ 
\reftext{Ivan99}\cite{ivanov1999}& \num[exponent-mode=input]{0.6} & n & \num{6e+14} & \num{293} & hl & - & - & OCVD  \\ 
& \num[exponent-mode=input]{3.8} & n & \num{6e+14} & \num{550} & hl & - & - & OCVD  \\ 
\reftext{Kimo99}\cite{kimoto1999}& \num[exponent-mode=input]{0.33} & N & \num{5e+14} & - & - & - & - & RR  \\ 
\reftext{Udal00}\cite{udal2000}& \num[exponent-mode=input]{0.052} & n & \numrange[exponent-mode=input]{0.9e15}{2.6e15} & - & - & - & - & RR  \\ 
\reftext{Cheo03}\cite{cheong2003}& \num[exponent-mode=input]{1} & N & \numrange[exponent-mode=input]{1e16}{1.8e16} & \num{300} & - & - & - & Ct  \\ 
\reftext{Dome03}\cite{domeij2003}& \num[exponent-mode=input]{0.0035} & n & \num{8.6e+15} & - & hl & - & - & BTEC  \\ 
\reftext{Zhan03}\cite{zhang2003}& \num[exponent-mode=input]{0.3} & n & \numrange[exponent-mode=input]{1e14}{2e16} & - & - & - & - & TRPL  \\ 
\reftext{Levi04}\cite{levinshtein2004}& \num[exponent-mode=input]{1.55} & n & \num{3e+14} & \num{293} & hl & - & - & OCVD  \\ 
\reftext{Tawa04}\cite{tawara2004}& \numrange[exponent-mode=input]{0.26e0}{6.8e0} & n & \numrange[exponent-mode=input]{1.8e14}{3.4e15} & \numrange[exponent-mode=input]{300}{500} & - & \numrange[exponent-mode=input]{1.1e15}{4.2e15} & - & TRPL  \\ 
\reftext{Resh05}\cite{reshanov2005}& \num[exponent-mode=input]{1} & n & \num{7e+15} & - & - & - & - & OCVD  \\ 
\reftext{Huh06}\cite{huh2006}& \num[exponent-mode=input]{0.5} & n & \num{1e+14} & - & ll & \num{1e+13} & - & PLD  \\ 
\reftext{Ivan06b}\cite{ivanov2006b}& \num[exponent-mode=input]{3.7} & n & \num{2e+14} & \num{300} & hl & - & - & OCVD  \\ 
\reftext{Jenn06}\cite{jenny2006}& \num[exponent-mode=input]{15.5} & N & \num{5e+15} & - & - & - & - & EBIC  \\ 
\reftext{Neim06}\cite{neimontas2006}& \num[exponent-mode=input]{0.012} & N & \num{1e+16} & - & - & - & - & FWM  \\ 
\reftext{Dann07}\cite{danno2007}& \num[exponent-mode=input]{2.5} & N & \num{1.5e+15} & \num{300} & hl & \numrange[exponent-mode=input]{2e16}{2e17} & y & $\mu$-PCD  \\ 
\reftext{Stor07}\cite{storasta2007}& \num[exponent-mode=input]{0.218} & N & \num{5e+15} & \num{300} & - & - & y & TRPL  \\ 
\reftext{Udal07}\cite{udal2007}& \num[exponent-mode=input]{0.0044} & n & \num{7e+15} & \num{297} & - & - & - & RR  \\ 
\reftext{Kimo08}\cite{kimoto2008}& \num[exponent-mode=input]{8.6} & n & \numrange[exponent-mode=input]{1e15}{2e15} & - & ll & \num{5e+12} & - & $\mu$-PCD  \\ 
\reftext{Stor08}\cite{storasta2008}& \num[exponent-mode=input]{0.99} & n & \num{1e+14} & \num{300} & - & - & y & TRPL  \\ 
\reftext{Hiyo09}\cite{hiyoshi2009}& \num[exponent-mode=input]{1.62} & N & \numrange[exponent-mode=input]{1e15}{5e15} & - & hl & \numrange[exponent-mode=input]{2e16}{2e17} & y & $\mu$-PCD  \\ 
\reftext{Resh09}\cite{reshanov2009}& \num[exponent-mode=input]{2.13} & n & \numrange[exponent-mode=input]{1e15}{1.2e15} & - & hl & - & - & OCVD  \\ 
\reftext{Alba10}\cite{albanese2010}& \num[exponent-mode=input]{1.5e-05} & - & \num{2.21e+17} & \num{300} & - & - & - & OCVD  \\ 
\reftext{Jara10}\cite{jarasiunas2010}& \num[exponent-mode=input]{0.8} & n & \num{4e+14} & \num{300} & - & \numrange[exponent-mode=input]{1e17}{1e19} & - & FCA  \\ 
\reftext{Kimo10}\cite{kimoto2010}& \num[exponent-mode=input]{9.5} & n & \numrange[exponent-mode=input]{0.9e15}{1e15} & - & hl & \numrange[exponent-mode=input]{5e15}{5e16} & y & $\mu$-PCD  \\ 
\reftext{Kimo10a}\cite{kimoto2010a}& \num[exponent-mode=input]{13.1} & N & \num{7e+14} & - & hl & \numrange[exponent-mode=input]{5e15}{5e16} & y & $\mu$-PCD  \\ 
\reftext{Klei10}\cite{klein2010}& >\num{100e0} & N & <\num{1e16} & \num{222} & ll & \num{2e+14} & - & TRPL  \\ 
\reftext{Miya10}\cite{miyazawa2010}& \num[exponent-mode=input]{18.5} & N & \num{7e+13} & - & ll & \num{3e+12} & - & $\mu$-PCD  \\ 
\end{tabular}
    }
\end{table*}     

\begin{table*}[p]
    \setlength{\tabcolsep}{7pt}
    \caption{\label{tab:regen_lth_II}Hole lifetime results [2/2]. Column \textit{inj.} denotes the carrier injection level, i.e., low (ll) resp. high (hl), and column \textit{excess} the exact amount. A \textit{y} in column \textit{impr.} highlights that the shown value is the highest lifetime achieved in an optimization process.}
    \resizebox{0.9\linewidth}{!}{%
        \begin{tabular}{l *{8}{c}}
ref. & $\taup$ & dop & conc. &$T$ & inj. & $\DN$ & impr. & method \\ 
 & [\si{\micro\s}] & & [\si{\per\cubic\centi\meter}] & [\si{\K}] && [\si{\per\cubic\centi\meter}] && \\ \hline 
 &&&&&&&& \\[-7pt] 
\reftext{Haya11a}\cite{hayashi2011a}& \num[exponent-mode=input]{4.6} & n & \num{1.2e+15} & \numrange[exponent-mode=input]{300}{525} & ll & \num{1.5e+14} & y & $\mu$-PCD  \\ 
\reftext{Donn12}\cite{donnarumma2012}\footnotemark[1] & \num[exponent-mode=input]{0.24} & N & \num{1e+16} & - & - & - & - & FIT  \\ 
\reftext{Ichi12}\cite{ichikawa2012}& \num[exponent-mode=input]{33.2} & N & \numrange[exponent-mode=input]{3e14}{8e14} & - & - & \numrange[exponent-mode=input]{1e15}{1e16} & y & $\mu$-PCD  \\ 
\reftext{Kawa12}\cite{kawahara2012}& \num[exponent-mode=input]{6.5} & n & \num{1e+16} & \num{300} & - & \num{1e+16} & y & $\mu$-PCD  \\ 
\reftext{Lilj13}\cite{lilja2013}& \num[exponent-mode=input]{1.6} & N & \num{3e+15} & \num{300} & ll & - & y & TRPL  \\ 
\reftext{Miya13}\cite{miyazawa2013}& \num[exponent-mode=input]{13} & N & \numrange[exponent-mode=input]{2e14}{3e14} & \num{300} & - & - & y & TRPL  \\ 
\reftext{Scaj13}\cite{scajev2013}& \num[exponent-mode=input]{0.55} & n & \num{4e+14} & - & ll & - & - & FCA  \\ 
\reftext{Dibe14}\cite{dibenedetto2014}& \num[exponent-mode=input]{0.01} & - & - & - & hl & - & - & DCD  \\ 
\reftext{Usma14}\cite{usman2014}\footnotemark[2] & \num[exponent-mode=input]{1.3} & n & \num{2e+14} & - & - & - & - & FIT  \\ 
\reftext{Chow15}\cite{chowdhury2015}& \num[exponent-mode=input]{2.5} & n & \num{2.5e+14} & \num{300} & hl & - & - & $\mu$-PCD  \\ 
\reftext{Kaji15}\cite{kaji2015}& \num[exponent-mode=input]{21.6} & N & \num{2e+14} & - & - & \num{1e+14} & y & $\mu$-PCD  \\ 
\reftext{Suva15}\cite{suvanam2015}& \num[exponent-mode=input]{0.35} & N & \num{8e+15} & - & hl & \num{6e+17} & - & FCA  \\ 
\reftext{Puzz16}\cite{puzzanghera2016}& \num[exponent-mode=input]{1} & n & \num{3e+15} & \num{400} & - & - & - & DCD  \\ 
\reftext{Sait16}\cite{saito2016}& \num[exponent-mode=input]{26} & n & \num{1e+14} & \num{300} & - & \num{1.8e+17} & - & $\mu$-PCD  \\ 
\reftext{Stre16}\cite{strelchuk2016}& \num{22.5+-7.5e-3} & n & \num{1.5e+15} & \num{300} & - & - & - & EBIC  \\ 
\reftext{Tawa16}\cite{tawara2016}& \num[exponent-mode=input]{0.3} & N & \num{7.7e+17} & \num{300} & ll & - & - & TRPL  \\ 
\reftext{Tsuc16}\cite{tsuchida2016}& \num[exponent-mode=input]{2.6} & n & \num{1.4e+14} & - & - & - & - & TRPL  \\ 
\reftext{Ayed17}\cite{ayedh2017}& \num[exponent-mode=input]{20} & N & \num{1e+15} & - & - & - & y & -  \\ 
\reftext{Lilj17}\cite{lilja2017}& \num[exponent-mode=input]{3.5} & N & \num{1.3e+15} & \num{300} & ll & - & - & PLD  \\ 
\reftext{Fang18}\cite{fang2018}& \num{11+-3e-3} & N & \num{9.1e+18} & \num{300} & - & - & - & TAS  \\ 
\reftext{Kimo18}\cite{kimoto2018}& \num[exponent-mode=input]{110} & n & \num{1e+14} & - & - & \numrange[exponent-mode=input]{1e14}{1e15} & y & $\mu$-PCD  \\ 
\reftext{Cui19}\cite{cui2019}& \num[exponent-mode=input]{1.05} & N & \num{2e+13} & \num{300} & - & \num{4e+16} & - & $\mu$-PCD  \\ 
\reftext{Mura19}\cite{murata2019}& \num[exponent-mode=input]{9.9} & N & \num{1e+15} & \num{293} & ll & - & - & TRPL  \\ 
\reftext{Kato20}\cite{kato2020}& \num[exponent-mode=input]{0.7} & N & \numrange[exponent-mode=input]{1e15}{1e16} & \num{300} & ll & - & y & $\mu$-PCD  \\ 
\reftext{Naga20}\cite{nagaya2020}& \num[exponent-mode=input]{10} & N & \num{1e+18} & \num{293} & hl & \num{1.5e+18} & - & TR-FCA  \\ 
\reftext{Sapi20}\cite{sapienza2020}& \num[exponent-mode=input]{19} & n & \num{1e+16} & \num{300} & hl & - & - & OCVD  \\ 
\reftext{Erle21}\cite{erlekampf2021}& \num[exponent-mode=input]{5} & N & \numrange[exponent-mode=input]{5e14}{1e15} & - & ll & - & - & $\mu$-PCD  \\ 
\reftext{Mura21}\cite{murata2021}& \num[exponent-mode=input]{0.14} & N & \num{2e+17} & \num{293} & ll & \num{7e+14} & - & TRPL  \\ 
\reftext{Meli22a}\cite{meli2022a}& \num{1.63+-0.18e1} & n & \num{5e+13} & - & hl & \numrange[exponent-mode=input]{5.2e17}{6.6e17} & y & LOPC  \\ 
\reftext{Maxi23}\cite{maximenko2023}\footnotemark[3] & \num[exponent-mode=input]{2} & - & - & - & - & - & - & FIT  \\ 
\reftext{Kato24}\cite{kato2024}& \num[exponent-mode=input]{4.5} & n & \num{1e+15} & \num{300} & - & \numrange[exponent-mode=input]{5e15}{5e16} & - & $\mu$-PCD  \\ 
\reftext{Sozz24}\cite{sozzi2024}& \num[exponent-mode=input]{6.09} & n & \num{1.5e+14} & \num{298} & hl & \num{4e+17} & - & OCVD  \\ 
\end{tabular}
        \footnotetext[1]{value fitted to results by \citet{galeckas1997}}
        \footnotetext[2]{value fitted to forward current measurements by \citet{ryu2012}}
        \footnotetext[3]{value fitted to measurements by \citet{kimoto2018}}   
    }
\end{table*}  

The shown tables represent already a simplification, because many publications include values for multiple operating conditions, i.e., temperature, doping concentration and excess carrier concentration. For example \citet{hayashi2011a} provided measurements for various injection levels and two temperatures (\SI[exponent-mode=input]{300}{\K} and \SI[exponent-mode=input]{525}{\K}). In these cases we only selected the highest lifetime for the table.

The measurements show no difference between $\taun$ and $\taup$, which both decrease with increasing doping density (see \cref{fig:regen_doping}). However, the reported values differ by up to four orders of magnitude for a constant doping concentration. To investigate this circumstance we scaled the maker size with the excess carrier concentration, because for a high level only the sum $\taun+\taup$ is achieved~\cite{huh2006,suvanam2015} (cp. \cref{eq:regen_hl}). Consequently, we expected that the lifetime increases when going from low to high injection level, until it eventually starts to decrease when the bimolecular or Auger recombination become dominant. This effect was explicitly shown in literature~\cite{hayashi2011,hayashi2011a,hayashi2012,kato2024}. In addition, \citet{kimoto2016} stated in 2016 that the injection-level dependence of SRH lifetimes is not known in SiC, calling for further research, \citet{tawara2016} experienced a lifetime decrease eventually approaching a constant value and \citet{scajev2013} concluded that the lifetime is almost injection level independent. Our analysis agrees mostly with the latter, because we were unable to identify any clear tendency in the data. \revision{This includes measurement techniques as all delivered both very low and high lifetimes.}

\begin{figure}[t]
    \centering
    \resizebox{1\linewidth}{!}{%
    \input{figures/regen_doping}
    }
    \caption{\label{fig:regen_doping}Measurements of the minority charge carrier lifetime for the respective doping densities. The mark size indicates the excess carrier concentration. We picked an intermediate size if no data were available. We excluded the results by \citet{albanese2010} for $\taup$ because the value was very low and distorted the plot.
    }
\end{figure}

The lifetime values that were referenced in literature (see \cref{fig:regen_values}) also show a big spread, with values ranging from \SI{15}{\nano\s} to \SI{39.5}{\milli\s}, but similar values for $\taun$ and $\taup$. Compared to other topics investigated in this review lifetimes are rarely referenced, which highlights the impact of the material quality and the necessity for device specific measurements.

\begin{figure}[t]
    \centering
    \resizebox{0.4\linewidth}{!}{%
    \input{figures/regen_values}
    }
    \caption{\label{fig:regen_values}Lifetime values referenced in literature in [\si{\micro\s}].
    }
\end{figure}

\subsubsection{Doping Dependency of SRH Lifetime}

We could solely identify parameter sets for the Scharfetter relation in \cref{eq:regen_scharfetter} (see \cref{tab:regen_scharfetter}) that were fitted to suitable measurements. \citet{liaugaudas2015} conducted their own differential transmittivity measurements and \citet{maximenko2023} fitted to the results of the $\mu$-PCD measurements by \citet{kimoto2018}. These are the only ones that distinguished between electrons and holes. For early publications no 4H-SiC values were available, so \citet{ruff1994} settled for a combination of Silicon based values gathered from various sources. In other cases~\cite{levinshtein2001b,choi2005} we were unable to retrace the origin of the values, although often explicit references were provided.

\begin{table}[t]
    \centering
    \setlength{\tabcolsep}{12pt}
    \renewcommand{\arraystretch}{1.1}
    \caption{\label{tab:regen_scharfetter}Parameters for doping dependency according to the Scharfetter relation in \cref{eq:regen_scharfetter}. \bkgCol{no4HColor} are investigations not focused on 4H-SiC, \bkgCol{fundColor} are fittings to doping variations and \bkgCol{excessColor} fittings to varying excess carrier concentrations (not covered by this model). The last column denotes the source for the fitting data.}
    \resizebox{0.7\linewidth}{!}{%
    \begin{tabular}{l*{2}{c}|*{2}{c}|cc}
&\multicolumn{2}{c|}{electrons} & \multicolumn{2}{c|}{holes} && \\ 
ref. & $\nref$ & $\gamma$ & $\nref$ & $\gamma$ & quantity & fit to \\ 
 & [\si{\per\cubic\centi\meter}] & [1] & [\si{\per\cubic\centi\meter}] & [1] && \\ \hline 
 &&&&&& \\[-7pt] 
\rowcolor{no4HColor} \reftext{Ruff94}\cite{ruff1994}\,\footnotemark[1]& \num{3.000000e+17}& \num{0.3}& \num{3.000000e+17}& \num{0.3}& -- & -- \\ 
\rowcolor{excessColor} \reftext{Nall99}\cite{nallet1999}& \num{1.000000e+16}& \num{1}& \num{1.000000e+16}& \num{1}& excess & \reftext{Gale97}\cite{galeckas1997} \\ 
\reftext{Levi01b}\cite{levinshtein2001b}& \num{7.000000e+17}& \num{1}& \num{7.000000e+17}& \num{1}& -- & -- \\ 
\reftext{Choi05}\cite{choi2005}\,\footnotemark[2]& \num{5.000000e+16}& \num{1}& \num{5.000000e+16}& \num{1}& -- & -- \\ 
\rowcolor{excessColor} \reftext{Donn12}\cite{donnarumma2012}& \num{2e+18}& \num{1.9}& \num{2e+18}& \num{1.9}& excess & \reftext{Gale98}\cite{galeckas1998}\\ 
\rowcolor{excessColor}& \num{4e+18}& \num{1.4}& \num{4e+18}& \num{1.4}& excess & \reftext{Neim06}\cite{neimontas2006}\\ 
\rowcolor{fundColor} \reftext{Liau15}\cite{liaugaudas2015}& \num{5.000000e+18}& \num{1.2}& --& --& doping & -- \\ 
\reftext{Lech21}\cite{lechner2021}\,\footnotemark[3]& \num{7.000000e+16}& \num{1}& \num{7.000000e+16}& \num{1}& -- & -- \\ 
\rowcolor{fundColor} \reftext{Maxi23}\cite{maximenko2023}& \num{5.000000e+15}& \num{0.55}& \num{5.000000e+15}& \num{0.67}& doping & \reftext{Kimo18}\cite{kimoto2018} \\ 
\end{tabular}
        \footnotetext[1]{based on data for Silicon}
        \footnotetext[2]{no data found in provided reference~\cite{harris1995}, according to \citet{albanese2010} based on Silicon}
        \footnotetext[3]{default values from TCAD tool}
    }
 
\end{table} 

We discovered in two cases a mix-up of doping and excess charge carrier concentration during fitting. While both lead to a decrease in lifetime with increasing concentration (as we will show later) the underlying physical processes  are fundamentally different. While more doping results in an increased amount of recombination centers, more excess charge carriers lead to a higher contribution of bimolecular and Auger recombination.
Although the Scharfetter relation only covers the former \citet{nallet1999} and \citet{donnarumma2012} fitted the lifetime versus the excess charge carrier concentration. Despite this discrepancy we included the respective results to paint a more complete picture but highlighted them accordingly.

The fittings predict that $\taup$ starts to decrease for doping concentration $>\SI{e14}{\per\cubic\centi\m}$ and $\taun$ only after a concentration of \SI{e18}{\per\cubic\centi\m} is exceeded (see \cref{fig:regen_scharfetter}). The rate of change is thereby higher for $\taun$. The only exception of this observation is the fitting of $\taun$ by \citet{kimoto2018}, which matches the fittings of $\taup$ from various authors well. To resolve this disagreement and to confirm our hypothesis further measurements would be necessary in the future. The remaining models agree either more to the electron or hole behavior. Solely the Silicon based model (\reftext{Ruff94}\cite{ruff1994}) shows a very slow decrease and thus can not be assumed accurate for 4H-SiC.

\begin{figure}[t]
    \centering
    \resizebox{1\linewidth}{!}{%
    \input{figures/regen_scharfetter}
    }
    \caption{\label{fig:regen_scharfetter}Doping dependency according to the Scharfetter relation in \cref{eq:regen_scharfetter}. In brackets are the values for ($\nref$, $\gamma$). Only one graph is shown if electron and hole parameters are equal (see \cref{tab:regen_scharfetter}). Solid non-opaque lines represent fittings to suitable 4H-SiC measurement data.
    }
\end{figure}

\subsubsection{Temperature Dependency of SRH Lifetime}

We found various fantastic investigations on the temperature dependent lifetime~\cite{kato2020,kato2024}; one even distinguished between surface and bulk lifetime~\cite{klein2010}. Excess charge carriers in the range \SIrange[range-exponents=individual]{1e16}{1e19}{\per\cubic\centi\meter} were investigated~\cite{tanaka2024}, with the conclusion that the temperature behavior depends on the excess carrier concentration~\cite{jarasiunas2010}.

\begin{table}[t]
    \centering
    \setlength{\tabcolsep}{8pt}
    \renewcommand{\arraystretch}{1.1}
    \caption{\label{tab:regen_temperature}Temperature dependency of the
      lifetime. If only one value for $\alpha_\mathrm{n,p}$ is stated no charge
      carrier specification was made in the publication. Column \textit{conf.}
      shows the confidence interval and \textit{equ.} the model.}
    \resizebox{0.75\linewidth}{!}{%
    \begin{tabular}{l rl *{6}{c}}
ref. & $\alpha_\mathrm{n}$ & $\alpha_\mathrm{p}$ & $T_0$& $\eact$ & $C$ & conf. & equ. & method \\ 
 & [1] & [1]& [\si{\K}] & [\si{\electronvolt}] & [1] & [\si{\K}] & \\ \hline 
 &&&&&&&& \\[-7pt] 
\reftext{Kord96}\cite{kordina1996}\,\footnotemark[1] & -- & \num{1.72} & \num{300} & -- & -- & \numrange{300}{500} & (\ref{eq:regen_powerT0}) & PLD \\ 
\reftext{Ivan99}\cite{ivanov1999}\,\footnotemark[3] & -- & -- & -- & \num{0.11} & -- & \numrange{300}{500} & (\ref{eq:regen_actEnergy}) & OCVD \\ 
\reftext{Nall99}\cite{nallet1999}\,\footnotemark[2] & -- & -- & \num{300} & -- & \num{2.55} & -- & (\ref{eq:regen_expSimple}) & -- \\ 
\reftext{Udal00}\cite{udal2000} & -- & \num{2.2} & \num{300} & -- & -- & \numrange{200}{450} & (\ref{eq:regen_powerT0}) & RR \\ 
\reftext{Agar01}\cite{agarwal2001}\,\footnotemark[3] & -- & -- & -- & \num{0.12} & -- & \numrange{300}{500} & (\ref{eq:regen_actEnergy}) & TTOGC \\ 
\reftext{Bala05}\cite{balachandran2005} & \num{5} & -- & \num{300} & -- & -- & -- & (\ref{eq:regen_powerT0}) & -- \\ 
\reftext{Levi05}\cite{levinshtein2005}\,\footnotemark[3] & -- & -- & -- & \num{0.08} & -- & \numrange{300}{550} & (\ref{eq:regen_actEnergy}) & -- \\ 
\reftext{Ivan06a}\cite{ivanov2006a}\,\footnotemark[4] & -- & -- & -- & \num{0.105} & -- & \numrange{300}{500} & (\ref{eq:regen_actEnergy}) & SIM \\ 
\reftext{Udal07}\cite{udal2007} & -- & \num{1.9} & \num{300} & -- & -- & \numrange{300}{700} & (\ref{eq:regen_powerT0}) & RR \\ 
 & -- & \num{4.4} & \num{300} & -- & -- & \numrange{700}{1000} & (\ref{eq:regen_powerT0}) & RR \\ 
\reftext{Scaj13}\cite{scajev2013} & -- & -- & -- & \num{0.125} & -- & \numrange{70}{1000} & (\ref{eq:regen_actEnergy51}) & FCA \\ 
\reftext{Dibe14}\cite{dibenedetto2014} & \num{2.15} & -- & \num{300} & -- & -- & \numrange{250}{500} & (\ref{eq:regen_powerT0}) & DCD \\ 
\reftext{Chow15}\cite{chowdhury2015}\,\footnotemark[5]  & \multicolumn{2}{c}{\num{1.2}} & \num{300} & -- & -- & \numrange{300}{525} & (\ref{eq:regen_powerT0}) & $\mu$-PCD \\ 
\reftext{Rakh20}\cite{rakheja2020} & -- & \num{8} & \num{450} & -- & -- & \numrange{100}{700} & (\ref{eq:regen_power0}) & -- \\ 
 & -- & \num{14} & \num{530} & -- & -- & \numrange{100}{700} & (\ref{eq:regen_power0}) & -- \\ 
\reftext{Sapi20}\cite{sapienza2020}  & \multicolumn{2}{c}{\num{1.5}} & \num{300} & -- & -- & \numrange{300}{450} & (\ref{eq:regen_powerT0}) & OCVD \\ 
\reftext{Tian20}\cite{tian2020} & \num{1.84} & \num{1.84} & \num{300} & -- & -- & \numrange{300}{573} & (\ref{eq:regen_powerT0}) & DCD \\ 
\reftext{Maxi23}\cite{maximenko2023}\,\footnotemark[6] & \num{1.72} & \num{1.72} & \num{300} & -- & -- & -- & (\ref{eq:regen_powerT0}) & FIT \\ 
\reftext{Sozz24}\cite{sozzi2024}  & \multicolumn{2}{c}{\num{1.7}} & \num{298} & -- & -- & \numrange{300}{450} & (\ref{eq:regen_powerT0}) & OCVD \\ 
\end{tabular}
        \footnotetext[1]{fitting provided in citing articles, \citet{sapienza2020} fitted $\alpha=\num{1.9}$}
        \footnotetext[2]{default value in simulation tool, according to \citet{lechner2021} based on the research from \citet{grasser2009}}
        \footnotetext[3]{fitting by \citet{udal2007}}
        \footnotetext[4]{fitting by \citet{tamaki2008}}
        \footnotetext[5]{fitting by \citet{sapienza2020}}
        \footnotetext[6]{fitted to results by \citet{kimoto2018}}
    }
\end{table} 

In the literature predominantly the power law description of \cref{eq:regen_powerT0} was used (see \cref{tab:regen_temperature}). The values for the exponent $\alpha$ thereby commonly vary between $1$ and $2$. \citet{udal2007} stated a fitting in two regions, as the lifetime started to increase faster above \SI{700}{\K} based on charge carrier lifetimes that the authors extracted from reverse recovery measurements by \citet{boltovets2006}. \citet{sapienza2020} achieved $\alpha=1.5$ which is exactly the value expected for Coulomb-attractive charge recombination centers~\cite{udal2007}.

The origin for the value $\alpha_\mathrm{n,p}=5$~\cite{balachandran2005,nawaz2010,usman2014} is unclear. Although we stated \citet{balachandran2005} here we were not able to find a measurement that derived these values. We suspect that these were default values of a TCAD tool but we could not even confirm that. In the same sense we were not able to pinpoint the origin of the value $C=2.55$~~\cite{nallet1999} for \cref{eq:regen_expSimple}. Although \citet{lechner2021} provided a reference~\cite{grasser2009}, we were not able to find anything in there either.

For the exponential based approaches the value of $\eact$ was chosen between \SIrange{110}{125}{\milli\electronvolt} and for \cref{eq:regen_expUpDown} \citet{lechner2021} proposed a variety of parameters. We calculated the average from the single sets and achieved the values shown in \cref{eq:regen_Tcoeff}.
\begin{equation}
    \label{eq:regen_Tcoeff}
    T_\mathrm{coeff} = 0.666,\quad \alpha_\tau = 0.06385,\quad \beta_\tau = 5.716
\end{equation}

A comparison of the single models (see \cref{fig:regen_temperature}) shows a consistent increase of the lifetime with temperature. We added some temperature measurements and scaled them such that the room temperature values are around the arbitrary value of $\SI{1}{\micro\s}$. We had to dismiss the values by \citet{lophitis2018} because the authors only showed the parameters but not the according equations. We also do not show the results by \citet{puzzanghera2016}, who derived solely minor deviations around the room temperature value.

\begin{figure*}[t]
    \centering
    \resizebox{1\linewidth}{!}{%
    \input{figures/regen_temperature}
    }
    \caption{\label{fig:regen_temperature}Temperature Dependency of the SRH lifetime. Models are calibrated to hit \SI{1}{\micro\s} at \SI{300}{\K}. Measurement results are scaled such that \SI{1}{\us} at \SI{296+-4}{\K} is achieved. \textbf{H} denotes high level injection and \textbf{L} low level one.}
\end{figure*}

Some trajectories~\cite{maximenko2023,sozzi2024,tian2020,ivanov2006a,dibenedetto2014,sapienza2020} can be inferred from the other models and are not explicitly shown to improve visibility. For \cref{eq:regen_powerT0} a higher value of $\alpha$ leads to a steeper increase, comparable to a higher value of $\eact$ in \cref{eq:regen_actEnergy}. Different shapes are predicted by \cref{eq:regen_power0}, which shows an increasing steepness at high temperatures, and \cref{eq:regen_expUpDown}, with the eventual decrease of the lifetime at high temperatures.

We want to remind the reader that temperature changes depend on the doping concentration~\cite{murata2021}. In addition, the measured value, e.g., the sum of hole and electron lifetime for high level injection~\cite{chowdhury2015}, can have a differing temperature dependency. \citet{sapienza2020} compared various sources in this respect~\cite{choudhary2021,kordina1996,udal2007}.


\subsubsection{Surface Recombination Velocity}

The surface recombination velocity is, as the name indicates, not a bulk property but depends on many parameters such as crystal faces, i.e., Si-, C-, a- or m-face~\cite{kato2020,mori2014}, interface treatment and the resulting interface traps~\cite{maximenko2010}. For these reasons we will only roughly cover this topic. \citet{mori2014,scajev2010,gulbinas2011} published analyses of the surface recombination velocity for the interested reader.

In general, the surface recombination velocity has to be determined for each sample separately, due to its wide range of dependencies. It scales with temperature~\cite{klein2010,kato2020} and depends on the injection level~\cite{schroder2005,kato2024,scajev2010}, which was explained by a possible band banding near the interface~\cite{kimoto2008}. \revision{In contrast, the type of interface, i.e., a beveled mesa structure or a planar interface, does not seem to have an impact, as in both cases low and high surface velocities were reported.} Suitable values in literature were mainly based on dedicated investigations (see \cref{tab:regen_surface}) or were simply assumed, e.g., $s=\SI{2.2e5}{\centi\meter\per\s}$~~\cite{nonaka2009} and $s=\SI{1e3}{\centi\meter\per\s}$~~\cite{ichikawa2012}.

\begin{table*}[p]
    \centering
    \setlength{\tabcolsep}{6pt}
    \renewcommand{\arraystretch}{1.1}
    \caption{\label{tab:regen_surface}Surface recombination velocity. A \textit{y} in column \textit{impr.} indicates that the velocities represent the end of an optimization process.}
    \resizebox{0.9\linewidth}{!}{%
        \begin{tabular}{l *{6}{c}}
ref. & $s$ & $\surfn$ & $\surfp$ & type\footnotemark[1] & impr. & method \\ 
 & [\si{\centi\meter\per\s}] & [\si{\centi\meter\per\s}]& [\si{\centi\meter\per\s}] &&& \\ \hline 
 &&&&&& \\[-7pt] 
\reftext{Gale97}\cite{galeckas1997} & -- & -- & \num{4e4}/\num{1e4} & S/I & -- & FCA \\ 
\reftext{Neud98}\cite{neudeck1998} & -- & -- & \num{5.00E+04} & M & -- & RR \\ 
\reftext{Gale99a}\cite{galeckas1999a} & -- & -- & \numrange{5e3}{8e3} & S & y & FCA \\ 
\reftext{Kimo99}\cite{kimoto1999} & -- & -- & \num{5.00E+04} & M & -- & CD \\ 
\reftext{Gale01}\cite{galeckas2001} & \numrange{5e3}{5e5} & -- & -- & S & y & TA \\ 
\reftext{Cheo03}\cite{cheong2003} & -- & -- & \numrange{1}{1e3} & S & y & Ct \\ 
\reftext{Huh06}\cite{huh2006} & -- & -- & \num{2.50E+03} & S & -- & TRPL \\ 
\reftext{Ivan06a}\cite{ivanov2006a} & \num{4200} & -- & -- & M & -- & CD \\ 
\reftext{Neim06}\cite{neimontas2006} & -- & -- & \num{4+-1e4} & S & -- & FWM \\ 
\reftext{Griv07}\cite{grivickas2007} & \num{1.00E+04} & -- & -- & S & -- & NFCD \\ 
\reftext{Klei08}\cite{klein2008} & -- & -- & \numrange{5e3}{5e5} & S & y & TFCA \\ 
\reftext{Klei10}\cite{klein2010} & -- & -- & \numrange{400}{6940} & S & -- & TRPL \\ 
\reftext{Scaj10}\cite{scajev2010} & \numrange{2e3}{13e3} & -- & -- & S & -- & FCA \\ 
\reftext{Gulb11}\cite{gulbinas2011} & -- & \numrange{1.5e4}{3e4} & \numrange{1e4}{1e6} & S & y & CPP \\ 
\reftext{Pour11}\cite{pourbagherimahabadi2011} & -- & \num{1.07E+03} & \numrange{1.07e3}{5.7e4} & S & y & FCA \\ 
\reftext{Kato12}\cite{kato2012} & -- & -- & \numrange{1e3}{2e3} & S & y& $\mu$-PCD \\ 
\reftext{Mori14}\cite{mori2014} & -- & -- & \numrange{1500}{7500} & S & y& $\mu$-PCD \\ 
\reftext{Suva15}\cite{suvanam2015} & -- & -- & \numrange{3.5e4}{2e6} & S/I & y & FCA \\ 
\reftext{Asad18}\cite{asada2018} & \numrange{6e5}{1.2e7} & -- & -- & M & y & CD \\ 
\reftext{Ichi18}\cite{ichikawa2018} & -- & \numrange{300}{6e3} & \numrange{200}{2.5e3} & S & y& $\mu$-PCD \\ 
\reftext{Kato20}\cite{kato2020} & -- & \numrange{400}{1950} & \numrange{150}{750} & S & y& $\mu$-PCD \\ 
\reftext{Xian21}\cite{xiang2021} & \numrange{6+-0.7e5}{5.8+-0.5e6} & -- & -- & M & y & CD \\ 
\reftext{Kato24}\cite{kato2024} & -- & -- & \numrange{100}{2500} & S & y& $\mu$-PCD \\ 
\end{tabular}
        \footnotetext[1]{mesa structure (M), surface (S) or interface (I)}
    }
\end{table*}    

To determine the surface recombination velocity complementary measurement methods to the ones mentioned in \cref{sec:regen_methods} were used, for example colinear pump probe (CPP)~\cite{gulbinas2011}, non-equilibrium free-carrier density (NFCD)~\cite{grivickas2007}, the current density (CD)~\cite{asada2018,xiang2021} or reverse recovery (RR)~\cite{kimoto1999}. A common method to extract~\cite{gulbinas2011} and sometimes even eliminate~\cite{miyazawa2010,neudeck1998} the surface recombination was to vary the thickness of the device, but an effective bulk lifetime of more than \SI{1}{\micro\s} can only be measured directly within epitaxial layers thicker than \SI{100}{\micro\meter}~\cite{kimoto2014a}. \citet{kimoto1999,kimoto1999a} varied the radius and extracted $\surfn$ from the change of lifetime versus the perimeter/area ratio.

Due to the numerous growth parameters~\cite{pourbagherimahabadi2011} a wide range of growth optimizations and surface treatments were proposed. While mechanically polishing surfaces leads to an increased roughness (high velocities)~\cite{klein2008}, more recent experiments achieved velocities of $<\SI{200}{\centi\meter\per\s}$, indicating high surface qualities. The amount of publications in the last years also indicates an active research community.

Publications referencing other investigations often combine multiple values to achieve reasonable values, e.g., $s=$\SIrange{1e3}{1e5}{\centi\meter\per\s}~~\cite{ioffe2023,levinshtein2001}, $s=\SI{2e3}{\centi\meter\per\s}$~~\cite{johannesson2019}, $s=\SI{1e4}{\centi\meter\per\s}$~~\cite{liaugaudas2015}, $s=\SI{1e3}{\centi\meter\per\s}$~~\cite{nagaya2020,tanaka2023}, $s=\SI{4e4}{\centi\meter\per\s}$~~\cite{scajev2013} and $\surfn=\surfp=\SI{1e5}{\centi\meter\per\s}$~~\cite{domeij2003}, whereat 4H and 6H are fairly similar in this regard~\cite{galeckas1999a}.


\subsubsection{Bimolecular Recombination}

All studies regarding the bimolecular coefficient $B$ achieved a value of \SI{1.5+-0.5e-12}{\cubic\centi\meter\per\s} (see \cref{tab:regen_bimolecular}) despite the comments that the parameters by \citet{galeckas1997} were just an estimation~\cite{ioffe2023}. Exceptions were the investigation by \citet{neimontas2006}, whose value is one order of magnitude bigger, and \citet{murata2021}, who achieved 3-4 times higher values by also including electron-acceptor recombination. \revision{The results achieved by free carrier absorption (FCA) are very consistent, highlighting the reliability of this technique, but also time-resolved photoluminscence (TRPL) lead to comparable results.} We also found publications~\cite{scajev2013,kato2020,nagaya2020} that only used the radiative component with a value of \SIrange{1e-14}{3e-14}{\cubic\centi\meter\per\s}~\cite{scajev2013a}. The rather high values of $B$ compared to the radiative part was explained by the additionally included trap-assisted Auger~\cite{galeckas1997,scajev2009} and electron-acceptor (e-A) recombination~\cite{murata2021}.

\begin{table}[t]
    \centering
    \setlength{\tabcolsep}{11pt}
    \renewcommand{\arraystretch}{1.1}
    \caption{\label{tab:regen_bimolecular}Fundamental investigations of the bimolecular recombination parameter.}
    \resizebox{0.78\linewidth}{!}{%
        \begin{tabular}{l*{4}{c}}
ref. & $B$ & $T$ & $\DN$ & method \\ 
 & [\si{\cubic\cm\per\s}] & [\si{\K}] &  [\si{\per\cubic\centi\meter}] & \\ \hline 
 &&&& \\[-7pt] 
\reftext{Gale97}\cite{galeckas1997} & \num{1.50E-12} & \num{300} & \numrange[range-exponents=individual]{1.000000e+16}{1.000000e+19} & FCA \\ 
\reftext{Neim06}\cite{neimontas2006} & \num{3+-1e-11} & -- & \numrange[range-exponents=individual]{1.000000e+16}{2.000000e+19} & FWM \\ 
\reftext{Jara10}\cite{jarasiunas2010} & \num{1.2+-0.4e-12} & \num{300} & \numrange[range-exponents=individual]{1.000000e+17}{1.000000e+19} & FCA \\ 
\reftext{Tawa16}\cite{tawara2016} & \num{1.3e-12} & \num{300} & \numrange[range-exponents=individual]{4.000000e+14}{1.000000e+16} & TRPL \\ 
 & \num{0.94e-12} & \num{523} & \numrange[range-exponents=individual]{4.000000e+14}{1.000000e+16} & TRPL \\ 
\reftext{Mura21}\cite{murata2021}\,\footnotemark[1] & \num{5.6e-12} & \num{293} & \num{7.00E+14} & TRPL \\ 
 & \num{4.5e-12} & \num{523} & \num{7.00E+14} & TRPL \\ 
\reftext{Tana23a}\cite{tanaka2023a} & \num{<2e-12} & -- & \numrange[range-exponents=individual]{2.000000e+18}{1.000000e+19} & FCA \\ 
\end{tabular}
        \footnotetext[1]{includes electron acceptor recombination}
    }
\end{table} 

Little data about the temperature dependency of $B$ is available. 
\citet{tawara2016} presented measurements for six different temperatures and concluded that the value stayed constant. If one would, however, ignore the data point at the lowest temperature, a decrease with temperature could be inferred. \citet{scajev2013a} stated that the radiative recombination coefficient doubles in the range of \SIrange{10}{1000}{\K}, which could also lead to an increase of $B$. \citet{jarasiunas2010} proposed an empirical equation for the temperature dependent coefficient (see \cref{eq:regen_bim_temp_jara}), which shows a decrease with increasing temperature. For higher confidence in the results further investigations would be necessary.
\begin{equation}
    \label{eq:regen_bim_temp_jara}
    B(T) = \SI{1.1e-12}{\cubic\centi\m\per\s} + T^{-4} \times \SI{1.7e-4}{\K\tothe{4}\cubic\centi\m\per\s}
\end{equation}

\citet{tanaka2023} noted that for the investigation of excess carrier dependent Auger coefficients, which will be presented in the sequel, the characterization of $B(\DN)$ is essential and thus will be one topic of their future research.


\subsubsection{Auger Recombination}

\citet{galeckas1997} conducted in 1997 the first research on Auger coefficients in 4H-SiC and provided separate values for electrons ($\cn$) and holes ($\cp$). Despite this early availability 4H investigations~\cite{ayalew2004,lechner2021} still used values based on 6H~\cite{ramungul1998} and even based on Silicon~\cite{ruff1994,shah1998,das2015}. \citet{zhao2000} simply assumed $\cn = \cn = \SI{1e-28}{\cm\tothe{6}\per\s}$, which is nearly three orders of magnitude larger than the measured values.

The next fundamental characterization we found after this initial work was conducted one decade later by \citet{grivickas2007}. From then onward new values were published on shorter intervals, up to this day (see \cref{tab:regen_auger}). However, these only proposed combined coefficients or considered just one charge carrier. For the results presented by \citet{tawara2016} we concluded from the description of samples and measurement (low injection and highly n-doped material) that $\cn$ was determined.

\begin{table*}[t]
    \centering
    \setlength{\tabcolsep}{6pt}
    \renewcommand{\arraystretch}{1.1}
    \caption{\label{tab:regen_auger}Auger recombination parameters.}
    \resizebox{0.8\linewidth}{!}{%
        \begin{tabular}{l*{5}{c}}
ref. & $\cn$ & $\cp$ & $T$ & method & $\DN$ \\ 
 & [\si{\cm\tothe{6}\per\s}] & [\si{\cm\tothe{6}\per\s}] & [\si{\K}] &  &  [\si{\per\cubic\centi\meter}] \\ \hline 
 &&&&& \\[-7pt] 
\reftext{Gale97}\cite{galeckas1997} & \num{5+-1e-31} & \num{2+-1e-31} & \num{300} & FCA & \numrange[range-exponents=individual]{2.000000e+18}{2.000000e+19} \\ 
\reftext{Griv07}\cite{grivickas2007} & \num{2.00E-30} & -- & \num{75} & FCA & \num{1.00E+16} \\ 
\reftext{Scaj10}\cite{scajev2010} & \multicolumn{2}{c}{\num{7+-4e-31}} & \num{298} & FCA & \numrange[range-exponents=individual]{1.000000e+18}{3.000000e+18} \\ 
 & \multicolumn{2}{c}{\num{ 0.8+-0.2e-31}} & \num{298} & FCA & \numrange[range-exponents=individual]{1.000000e+19}{1.000000e+20} \\ 
\reftext{Scaj13}\cite{scajev2013} & \multicolumn{2}{c}{\num{5+-1e-31}} & -- & FCA & \num{1.00E+17} \\ 
\reftext{Tawa16}\cite{tawara2016} & \num{1.6e-30} & -- & \num{300} & TRPL & \numrange[range-exponents=individual]{3.000000e+17}{7.000000e+18} \\ 
 & \num{0.44e-30} & -- & \num{523} & TRPL & \numrange[range-exponents=individual]{3.000000e+17}{7.000000e+18} \\ 
\reftext{Tana23}\cite{tanaka2023}\,\footnotemark[1] & \multicolumn{2}{c}{\num{7.4e-19} $\DN^{-0.68}$} & -- & FCA & \numrange[range-exponents=individual]{5.000000e+18}{1.000000e+20} \\ 
\reftext{Tana23a}\cite{tanaka2023a} & \multicolumn{2}{c}{\num{<3e-31}} & -- & FCA & \numrange[range-exponents=individual]{2.000000e+18}{1.000000e+19} \\ 
\end{tabular}
        \footnotetext[1]{added measurements by \citet{scajev2010} to characterization}
    }
\end{table*} 

The achieved values are in the range of \SIrange{1e-31}{11e-31}{\raiseto{6}\centi\meter\per\s}, with the exceptions of \citet{grivickas2007} and \citet{tawara2016}, who derived parameters up to \SI{2e-30}{\raiseto{6}\centi\meter\per\s}. A possible explanation is a dependency on the excess charge carrier density $\DN$, which was first discussed by \citet{scajev2010}, who denoted $C\propto \DN^{-0.3}$. In recent days \citet{tanaka2023} achieved $C\propto \DN^{-0.68}$ and explained the deviation by deviating estimation methods, differences in the samples and a $\DN$-dependency of the bimolecular coefficient $B$. Possible is also a doping material dependency because \citet{murata2021} reported that Auger recombination was not important for an Aluminum doping concentration of \SI{1e19}{\per\cubic\centi\meter} but considerable for a Nitrogen doping in the mid \SI{1e18}{\per\cubic\centi\meter} range. \revision{Free carrier absorption (FCA) again delivered consistent data, but for meaningful statements about time-resolved photoluminescence (TRPL) more data are required.}

The value from \citet{scajev2013} presented in the table corresponds to the temperature-independent unscreened phonon-assisted coefficient. The authors combined the decrease of the Auger coefficient with rising excess carrier concentration, which they attributed to the screening of the carrier-phonon interaction by free carriers, with a temperature dependency (see \cref{eq:regen_auger_temp_indepth}) using the parameters shown in Eq.~(\ref{eq:regen_auger_temp}). We found this kind of description only once again in literature~\cite{gao2022a} but two instances based on the description of Si (see \cref{eq:regen_auger_temp_Si})~\cite{lophitis2018,zhang2018}.  The respective parameters are shown in \cref{tab:auger_temp_Si}, which we encountered also as default value for Si in some TCAD simulation suites.
\begin{subequations}
\label{eq:regen_auger_temp}  
\begin{eqnarray}
    \alpha &=& \SI{0.45}{\milli\eV\per\K} \label{eq:regen_auger_temp_alpha} \\
    B_\mathrm{CE}(T) &=& \num{3.5e-9}\, T^{-3/2}\si{\cubic\centi\meter\per\s} \label{eq:regen_auger_temp_B} \\
    a_\mathrm{SC} &=& \SI{7.8e16}{\per\cubic\centi\meter\K} \label{eq:regen_auger_temp_a}
\end{eqnarray}
\end{subequations}

\begin{table}[t]
    \centering
    \setlength{\tabcolsep}{6pt}
    \caption{\label{tab:auger_temp_Si}Parameters for Auger model shown in \cref{eq:regen_auger_temp_Si}}.
    \resizebox{0.8\linewidth}{!}{%
    \begin{tabular}{l *{7}{c}}
    ref & type & $A$ & $B$ & $D$ & $H$ & $N_0$ & $T_0$ \\
    & & [\si{\raiseto{6}\centi\meter\per\s}] & [\si{\raiseto{6}\centi\meter\per\s}] & [\si{\raiseto{6}\centi\meter\per\s}] && [\si{\per\cubic\centi\meter}] &
    [\si{\K}] \\ \hline
    &&&&&&& \\[-7pt]
    \reftext{Loph18}\cite{lophitis2018} & $\taun$ & \num{6.7e-32} & \num{2.45e-31} & \num{-2.2e-32} & \num{3.47} & \num{e18} & \num{300} \\
    & $\taup$ & \num{7.2e-32} & \num{4.5e-33} & \num{2.63e-32} & \num{8.26} & \num{e18} & \num{300} \\
    \reftext{Zhan18}\cite{zhang2018} & $\taun$ & \num{5e-31} & \num{2.45e-31} & \num{-2.2e-32} & 0 & -- & -- \\
    & $\taup$ & \num{9.9e-32} & \num{4.5e-33} & \num{2.63e-32} & 0 & -- & --
    \end{tabular}
    }
\end{table}

Dedicated 4H-SiC models and measurements predicted a decrease of $C$ with increasing temperature (see \cref{fig:regen_auger_temp}), which confirms the insight that Auger recombination is not significant for power devices operated at high temperatures~\cite{bellone2011}. In contrast, the models developed for Silicon show an increase of $C$.

\begin{figure*}[t]
    \centering
    \resizebox{0.9\linewidth}{!}{%
    \input{figures/regen_auger_temp}
    }
    \caption{\label{fig:regen_auger_temp}Temperature dependency of the Auger recombination coefficients $\cn$ and $\cp$ for $\DN=\SI{2.5e18}{\per\cubic\cm}$.}
\end{figure*}

Finally, we combine all contributions to the lifetime into one plot, showing the lifetime over the excess carrier concentration (see \cref{fig:regen_lifetime}). For the SRH lifetimes we added all measurements that clearly specified the excess carrier concentrations. In this log-log representation changes of the bimolecular and Auger coefficients only cause a horizontal shift while the derivative (-1 for bimolecular; -2 for Auger, except for $\DN$-dependent models) stays constant. The higher the value of $\tauSRH$ the earlier and more pronounced is the impact of the bimolecular recombination. A decrease due to Auger recombination only becomes significant for $\DN > \SI{e18}{\per\cubic\cm}$.

\begin{figure*}[t]
    \centering
    \resizebox{1\linewidth}{!}{%
    \input{figures/regen_lifetime}
    }
    \caption{Charge carrier lifetime considering the contributions of SRH, bimolecular (solid lines) and Auger (long dashed lines). The overall recombination lifetime $\tau_r$ (dashed line) is shown for three values of $A=(\tauSRH)^{-1}$, $B=\SI{3e-11}{\cubic\cm\per\s}$~\cite{neimontas2006} and $C=\SI{5e-31}{\cm\tothe{6}\per\s}$~\cite{galeckas1997} in \cref{eq:regen_polynom}. The marker size scales with the doping concentration. We picked an intermediate size if no data were available.}
    \label{fig:regen_lifetime}
\end{figure*}

At the moment we are unable to explain the lifetime measurements in regions where bimolecular resp. Auger recombination should be the limiting factors.  To explain these deviations we scaled, wherever possible, the marker size with the reported doping concentration, because according to \cref{eq:regen_scharfetter} the lifetime should decrease with increasing doping. However, the data did not show such a behavior. Exceptional in our opinion is also the $\DN$-dependent model of \citet{tanaka2023} that uses $C\propto \DN^{-0.68}$ because the slope in the log-log plot is equal to the bimolecular coefficient $B$. At high concentrations of $\DN$ the model by \citet{scajev2013} even predicts a constant Auger contribution.

\subsubsection{Origin of Parameters}

The doping dependency of 4H-SiC is predominantly modeled by the Silicon based values by \citet{ruff1994} (for a detailed graphical representation see \cref{fig:regen_ref_chain} in Appendix~\ref{sec:refChainRegen}). While some publications are well aware that Silicon data are used~\cite{ayalew2004,buono2010,buono2012} this information seemingly was lost over the years. More recent investigations that were focused on 4H-SiC were almost never referenced. In regard to the temperature dependency the most influential publication is the one by \citet{kordina1996}. In total five fundamental investigations on 4H-SiC were referenced in literature.


For bimolecular and Auger recombination the main source in literature is the publication by \citet{galeckas1997} (see \cref{fig:regen_ref_chain_II} in Appendix~\ref{sec:refChainRegen}). Additional fundamental investigations~\cite{jarasiunas2010,scajev2010,scajev2013,tanaka2023a} are referenced as well. Despite the large amount of references, the coefficients were transferred in almost all cases truthfully. This is extraordinary when we compare it against other topics investigated within this review. All the inconsistencies we encountered for charge carrier recombinations are presented in Appendix~\ref{sec:regen_appendix}.

 
\section{Incomplete Ionization}
\label{sec:incompIon}

To build electronic devices it is indispensable to utilize doping, i.e., to introduce impurity atoms during or even after the growth process. These so-called dopants add energy levels near the conduction (n-type doping) respectively valence band (p-type doping) such that free charge carrier are already available at moderate temperatures. Unintentional impurities, often referred to as traps or defect (centers), distinguish from dopants mainly due to their generally higher ionization energy, but their description in TCAD tools is similar. Consequently, the boundary between traps and dopants is continuous. 

We found several overviews on the most common doping elements and their respective activation energies in 4H-SiC~\cite{atabaev2018,bluet2000,dalibor2003,dhanaraj2010,feng2004a,harris1995,heera2001,kimoto2015,kwasnicki2014,lebedev1999,levinshtein2001,lutz2011,lutz2018,neudeck2001,neudeck2006,pankove2014,pensl2005,persson2005,wang2023,zetterling2002,zhu2008}. In addition, ab initio calculations by \citet{miyata2008} identified Arsenic~\cite{senzaki2000}, Gallium~\cite{troffer1998a} or Antimony as fitting based on their energy level. \citet{krieger2016} investigated group IV elements, \citet{feng2004a,huang2022a} focused on Tantalum and Chromium and \citet{dalibor1997,dalibor2003} on Vanadium. The activation rates~\cite{handy2000} or the impact of hydrogen~\cite{huang2022} prevented so far the deployment of these dopants.

Due to the wide band gap of 4H-SiC the difference between the doping energy level and the conduction resp. valence band, i.e., the dopant activation energy, is large compared to the thermal energy. Consequently, the often used assumption of full ionization is not applicable. Quite the opposite: incomplete ionization has to be considered to accurately predict the amount of free charge carriers and, thus, a realistic conductivity~\cite{xiao1999,donato2018}. \citet{scaburri2011a} and \citet{schoner1994} provided an overview of various physics-based models to describe incomplete ionization.

In this section we will review measurements, models and TCAD parameters used to described the amount of ionized dopants and their respective ionization energies depending on temperature and doping concentration. We focus on the four most common doping species in 4H-SiC~\cite{scaburri2011a,nipoti2018,albanese2010}: Aluminum and Boron for p-type resp. Nitrogen and Phosphorous for n-type doping.  We limit ourselves to simple models that use a single energy level per lattice site (cubic or hexagonal; cp. \cref{sec:sic}), because elaborate descriptions~\cite{pensl1993} would be necessary to cover changes due to the binding type or the dopant location~\cite{bechstedt2004,son2004a,laube2004,torres2022,donato2018}.

Our review revealed that Aluminum and Nitrogen were primarily investigated in the past, while Boron and Phosphorous received less attention. The energy levels introduced by donors are ``shallower'' (less activation energy) than those by acceptors. For the latter we found even deeper energy levels that accompany the doping process and that were occasionally considered the activation energy. For Nitrogen further measurements at high doping densities are required to fit the available doping dependency models.


\subsection{Introduction}

The amount of ionized donors ($\ndp$) and acceptors ($\nam$) can be determined by the Fermi-Dirac distribution~\cite{chen2020a,lutz2011,sze2007,uhnevionak2015,habib2011}, also known as steady-state Gibbs distribution~\cite{donato2018} (see \cref{eq:incompIon_FermiDirac}). $\na$ resp. $\nd$ are the active acceptor resp. donor concentrations, $\ea$ resp. $\ed$ the acceptor resp. donor energy levels and $\efn$ resp. $\efp$ the electron resp. hole Fermi levels.
\begin{equation}\label{eq:incompIon_FermiDirac}
\begin{split}
    \ndp = \frac{\nd}{1 + \gd \exp \left( \frac{\efn - \ed}{\kb T}\right)} \\
    \nam = \frac{\na}{1 + \ga \exp \left( \frac{ \ea - \efp}{\kb T}\right)}
\end{split}
\end{equation}

$\gd$ resp. $\ga$ denote the degeneracy of the energy levels~\cite{blakemore1962}, which was also modeled as $G_\mathrm{A}(T)$ and $G_\mathrm{D}(T)$~\cite{matsuura2004, matsuura2002, schadt1997,troffer1998,pensl2005,pernot2001,scaburri2011a} by adding a temperature dependency. The latter was based on the ionization energy or the energy separation of excited states using varying models, whereat TCAD tools provide at the moment sole the scaling shown in \cref{eq:incompIon_gT}, with $\ded=\ec - \ed$ resp. $\dea = \ea - \ev$ the ionization energies of donors and acceptors relative to the conduction ($\ec$) and valence band ($\ev$). Since the degeneracy factor is still the only parameter here we are not going to discuss this temperature dependency any further.
\begin{equation}
    \label{eq:incompIon_gT}
    G_\mathrm{A,D}(T) = g_\mathrm{A,D}\,\exp\left( \frac{\Delta E_\mathrm{A,D}}{\kb T}\right)
\end{equation}

If solely the Boltzmann statistics are considered, \cref{eq:incompIon_FermiDirac} can be simplified to the expression shown in \cref{eq:incompIon_N_Boltzmann}~\cite{buono2012,hatakeyama2013,scaburri2011a,ayalew2004,zetterling2002} with $\nc$ resp. $\nv$ the effective density of states in the conduction resp. valence band (see \cref{sec:dos}). This representation is often preferred in TCAD simulation tools, because these commonly operate on charge carrier concentrations.

\begin{equation}\label{eq:incompIon_N_Boltzmann}
    \begin{split}
    \ndp = \frac{\nd}{1+\gd\, \frac{n}{n_1}}, \qquad n_1 = \nc \exp \left( -\frac{\ded}{\kb T}\right)\\ 
    \nam = \frac{\na}{1+\ga\, \frac{p}{p_1}}, \qquad p_1 = \nv \exp \left(- \frac{\dea}{\kb T}\right)
    \end{split}
\end{equation}

The carrier concentration can be eliminated completely by using the neutrality equation~\cite{scaburri2011a} shown in \cref{eq:incompIon_neutrality} .
\begin{equation}
    \label{eq:incompIon_neutrality}
    \ndp (\ef) + p(\ef) = \nam (\ef) + n(\ef)   
\end{equation}
If we assume a highly donor doped material ($p(\ef)$ can be neglected) and no compensation ($\nam(\ef)=0$) the expression for $\ndp$ in \cref{eq:incompIon_N_Boltzmann} can be inserted into \cref{eq:incompIon_neutrality} resulting in a quadratic equation whose solution is shown in \cref{eq:incompIon_n_wo_fermi}~\cite{lee2002,zeghdar2019,zeghdar2020,ruff1994,troffer1997,albanese2010,zetterling1998}. For a highly acceptor doped material an according expression is achieved. The calculations of \citet{scaburri2011a} interfered with our analyses, because the author presented on page 22 an expression for $n$ that (i) is not required for the calculation and that (ii) we did not find in the cited publication~\cite{blakemore1962}. We solely achieved the shown result when we calculated $1/n$ and used in one occasion the wrong parameter.
\begin{equation}
    \label{eq:incompIon_n_wo_fermi}
    \ndp = \nd \frac{-1 + \sqrt{1 + 4\,\gd\,\frac{\nd}{\nc}\exp \left( \frac{\ded}{\kb T}\right)}}{2\,\gd\,\frac{\nd}{\nc}\exp \left( \frac{\ded}{\kb T}\right)}
\end{equation}

The ionization energy of a dopant is not constant but can vary due to band gap narrowing effects, which we discussed in \cref{sec:bandgap}, and a doping dependency. The latter will be discussed in the sequel.

\subsubsection{Doping Dependency}

The changing potential energy of charge carriers in the vicinity of ionized atoms, which effectively shielding them~\cite{pearson1949}, was provided as explanation for the doping dependency of the ionization energy. The Pearson-Bardeen~\cite{pearson1949} expression shown in \cref{eq:incompIon_PB} models the decrease of $\Delta E$.
\begin{equation}   
    \Delta E(N) = \Delta E_{0} - \alpha\,N^{1/3}
    \label{eq:incompIon_PB}
\end{equation}
\citet{schoner1994} proposed a more physics based approach shown in \cref{eq:incompIon_PB_schoner} with $f$ being a dimensionless factor that denotes the interaction strength with ionized acceptors and donors. Several other approaches to model the decrease of the ionization energy, e.g., due to interactions with a neighboring ionized dopant of the same kind, a local potential variation due to interaction with charged donors and acceptors~\cite{shklovskii1980} and an extension of the latter in regard to the temperature dependency~\cite{lee1975}, were also discussed~\cite{scaburri2011a,scaburri2011,schoner1994}. These were, however, not applied to 4H-SiC in the literature and are thus not considered in this review.
\begin{equation}
    \label{eq:incompIon_PB_schoner}
    \Delta E(N) = \Delta E_{0} - f \frac{q^2}{4 \pi \es \varepsilon_0} N^{1/3}
\end{equation}

We found some disagreement on what kind of dopants should be used for parameter $N$ in \cref{eq:incompIon_PB}. We found the sum of donors and acceptors $\na+\nd$~\cite{buono2012,hatakeyama2013}, the respective doping concentration ($\na$ or $\nd$)~\cite{arvinte2017,kagamihara2004,matsuura2004,pernot2005,achatz2008,weisse2018} and the ionized dopants~\cite{tanaka2018}. \citet{kajikawa2021} argued that the compensating dopant concentration $\nk$, i.e., donors for the acceptor levels and vice versa, have the bigger impact and should be used instead of the overall donor and acceptor concentrations. This is supported by \citet{scaburri2011a} who used $N=\nd-\nk$ and approaches that proposed to include $\nk$ into the factor $\alpha$~\cite{scaburri2011a,rambach2008,negoro2004}.

\citet{altermatt2006,altermatt2006a} proposed the logistic equation shown in \cref{eq:incompIon_logistic} as an alternative to \cref{eq:incompIon_PB}. For $N=N_\mathrm{E}$, which denotes a reference concentration, the ionization energy dropped to half its initial value $\Delta E_0$. \citet{darmody2019} argued that with \cref{eq:incompIon_PB} it is possible to shift the dopant level into the conduction/valence band and thus ionize all dopants immediately, which is neither physically reasonable nor possible with \cref{eq:incompIon_logistic}. Despite these arguments, the described approach has not yet found its way into the major simulation tools.
\begin{equation}\label{eq:incompIon_logistic}
    \Delta E(N) = \frac{\Delta E_{0}}{1+(N/N_\mathrm{E})^{c}}
\end{equation}

Dopants have differing ionization energies depending on whether they are located in a hexagonal or a cubic lattice site~\cite{ikeda1980} (see \cref{sec:sic}). Consequently \cref{eq:incompIon_N_Boltzmann} has to be adapted to the expression shown in \cref{eq:IncompIon_hex_cubic}~\cite{hatakeyama2013,senzaki2000} with $\Delta\edk$ and $\Delta\edh$ the cubic resp. hexagonal ionization energies for donors and $\Delta\eak$ and $\Delta\eah$ for acceptors.
%
\begin{equation}
    \label{eq:IncompIon_hex_cubic}
    \begin{split}
    \ndp = \frac{\frac{1}{2} \nd}{1+\gd \frac{p}{\nc} \left( \frac{\Delta \edh}{\kb T}\right)} + \frac{\frac{1}{2} \nd}{1+\gd \frac{p}{\nc} \left( \frac{\Delta \edk}{\kb T}\right)} \\    
    \nam = \frac{\frac{1}{2} \na}{1+\ga \frac{p}{\nv} \left( \frac{\Delta \eah}{\kb T}\right)} + \frac{\frac{1}{2} \na}{1+\ga \frac{p}{\nv} \left( \frac{\Delta \eak}{\kb T}\right)}
    \end{split}
\end{equation}
The factors $1/2$ denote that both lattice sites are equally
probable~\cite{ayalew2004}. In TCAD simulations it is often the case that these
separate values are merged to an effective energy level, ending up once again in
a description as shown in
\cref{eq:incompIon_N_Boltzmann}~\cite{lutz2018,ruff1994,bakowski1997,buono2012,bhatnagar2005}. \citet{ladesmartin2000}
and \citet{ayalew2005} reported on the consequences of such simplifications.

\subsubsection{Capture Cross Sections}

For more accurate approximation of dynamic processes around the dopant, i.e., (de)trapping of charge carriers, the electron and hole cross sections are required (see \cref{sec:regen}). The larger the cross section the easier a charge carrier can transition to the dopant energy level. We found multiple descriptions in literature that differ in their temperature scaling~\cite{booker2015}. The multi-phonon capture model is independent of temperature~\cite{landsberg1970,ridley1978,lades1999,schoner1994} and the cascade capture model, which \citet{kaindl1999} credited a better fit, is proportional to $T^{-2}$~~\cite{lax1960,abakumov1991,sridhara1998a,schoner1994} (also [78Aba] in \citet{schoner1994}). \citet{kuznetsov1995} used a scaling with $T^{-3}$. In TCAD tools deviating subsets of these models are supported.

\subsubsection{Methods}

The most commonly method to determine the ionization energy of dopants in
literature is to fit the neutrality equation, i.e., for p-type doping shown in
\cref{eq:IncompIon_neutrality_extract}, to Hall measurements of the conductivity
or charge carrier concentration for varying
temperature~\cite{arvinte2017,asada2016,capano2000,choyke1997d,contreras2016,gotz1993,hitchcock2022,itoh1998,kajikawa2021,kasamakova-kolaklieva2004,kimoto1995}\newline
{}~\cite{koizumi2009,laube2004,lichtenwalner2023,lomakina1973,matsuura1999,nipoti2013,obernhofer2007,parisini2013,pensl2003,pernot2000,pernot2001,pernot2005,rambach2006,rao1999,rao2006,rutsch1998,saks2001,saks2004,schadt1997,schmid2002,schmid2004,schoner1999,sridhara1998a,tanaka2000,terziyska2003,troffer1997,troffer1998,troffer1998a,wagner2002,wang2002,weisse2018}. Due to the anisotropy of the Hall effect it is possible to achieve direction dependent ionization energies~\cite{schadt1997}.
\begin{equation}
    \label{eq:IncompIon_neutrality_extract}
    p + \nk = \frac{\na}{1 + \frac{\ga p}{\nv} \exp \left( \frac{\dea}{\kb T}\right)} 
\end{equation}
Other electrical measurement methods include the fitting to the activation ratio of Hall measurements~\cite{ji2013}, free carrier concentration spectroscopy (FCCS)~\cite{kagamihara2004,matsuura2003,matsuura2004},  (thermal)~\cite{evwaraye1996,smith1998,kawahara2015,beljakowa2010} admittance spectroscopy (AS)~\cite{kaindl1999,kimoto1995,reshanov2005,schadt1997,sridhara1998a,troffer1998,troffer1998a,weisse2018}, electron spin resonance (ESR)~\cite{kisielowski1992}, deep level transient spectroscopy (DLTS)~\cite{kuznetsov1995,schadt1997,zhang2003,kato2022}, thermally stimulated current (TSC)~\cite{mandal2012,muzykov2012} and minority carrier transient spectroscopy (MCTS)~\cite{zhang2003}, which are often combined with Hall measurements for more accurate results. \citet{troffer1998} noted that DLTS is more sensitive but admittance spectroscopy allows to depict time constants below \SI{1}{\micro\s}.

These electrical methods are complemented by optical ones, e.g., (fourier transform infrared) photothermal ionization spectroscopy (PTIS)~\cite{chen2000}, donor-acceptor pair (DAP) luminescence \cite{devaty1997,ikeda1980,ivanov2003,ivanov2005}, free to acceptor (FTA) spectroscopy~\cite{ikeda1980}, infrared absorption (IA)~\cite{gotz1993}, photoluminescence (PL)~\cite{hagen1973,kimoto1995,matsunami1974,suzuki1973,suzuki1977}, time-resolved spectroscopy (TRS)~\cite{hagen1973} or delay measurements (DM)~\cite{hagen1973}. In these cases electrons are empowered and the resulting photon emission is recorded. The latter can be caused by transitions among dopants (traps) or between dopants (traps) and the conduction/valence band~\cite{schadt1997}. Different methods lead to slightly deviating results, even when applied to the same device~\cite{gotz1993}. Also possible are calculations, e.g., Faulkner model (FM) calculations~\cite{ivanov2003a}, density functional theory (DFT)~\cite{lu2021,miyata2008}, effective mass approximation (EMA)~\cite{chen1997}, first principles calculations (FPC)~\cite{huang2022,torres2022,huang2022a} or \textit{ab initio} supercell calculations (AISC)~\cite{son2006}. Finally, some authors defined a value range based on measurements in literature~\cite{ladesmartin2000,albanese2010,ayalew2004}, calculated an average value~\cite{martinez2002}, fitted to existing data~\cite{darmody2019,hatakeyama2013,lechner2021} or to the upper concentration limit of a dopant (UCLF)~\cite{achatz2008}.


\subsection{Results \& Discussion}
    
In the sequel we are going to present measurements and models for the ionization energies $\ded$ resp. $\dea$. To depict reported measurement results we occasionally dropped uncertainties and replaced the sometimes stated free exciton binding energy $\ex$ by \SI{20}{\milli\electronvolt} (see \cref{sec:bandgap}). We had to discard publications that described the samples solely as ``high purity'' or ``unintentional doped''~\cite{choyke1997d,matsunami1974,suzuki1973,suzuki1977} and the data by \citet{laube2002}, which were superseded by a publication of the same authors~\cite{laube2004}. We mark results for the hexagonal lattice site by a trailing \textit{h} and result for the cubic one by a trailing \textit{c}. In the literature the latter is often denoted by the letter ``k'', which, most probably, corresponds to the German word ``kubisch'' for cubic. In fact, some of the literature is written in German~\cite{troffer1998,schadt1997}, posing a considerable barrier for the international community.

In the theoretical analysis of the previous section we showed that the incomplete ionization is dominated by two parameters: the degeneracy factor and the ionization energy. For the former commonly the values $\ga=4$ (spin up and spin down plus two valence bands) and $\gd=2$ (spin up and down)~\cite{xiao1999,schmid2004,ayalew2004,bakowski1997,bellone2011,buono2012,chen2020a,habib2011,hatakeyama2013,khalid2012,ladesmartin2000,lechner2021,li2003,liu2021,lophitis2018,lutz2018,pezzimenti2009,rao1999,rao2022,schadt1997,schoner1994,song2012,sze2007,tian2020,zeghdar2019,zeghdar2020,zhang2010} were used. We also found $\gd=6$~\cite{troffer1998}, a spin degeneracy of $\gd=4$ for Phosphorous~\cite{scaburri2011a,laube2002}, $\ga=\gd=3$~\cite{balachandran2005,nawaz2010}, $\ga=2$~\cite{persson1998}, $\ga=\gd =2$~\cite{lv2023} and $g_\mathrm{k}=6$ resp. $g_\mathrm{h}=2$ for the cubic resp. hexagonal site of Nitrogen~\cite{pernot2000}. 

Many fundamental studies and measurements of incomplete ionization were published over more than five decades. The acquired data do not converge on single values but shows deviations of up to \SI{+-20}{\percent}. This spread did not improve with high-quality samples: Even as we limited our analysis to the last two decades we achieved the same spread. The fact that values for hexagonal and cubic lattice sites are often used without appropriate notation adds to this uncertainty and leads to confusions and errors. These are most striking for Boron, where the deep level is approximately twice the shallow one.

\subsubsection{Ionization Energy of Al}

For Aluminum we found measurements and models in a time span of almost three decades (see \cref{fig:IncompIon_Al}), which qualitatively agree on a value of \SI{225+-25}{\milli\eV} at low doping. This value is almost tenfold the thermal voltage at room temperature ($\kb \times \SI{300}{\K} = \SI{26}{\milli\eV}$). Most models show a noticeable change in the energy level for doping concentrations bigger than $\SI{e18}{\per\cubic\centi\m}$.

\cref{eq:incompIon_PB} was almost exclusively used to approximate the measurements. \citet{achatz2008} determined a critical aluminum concentration of $\SI{8.7e20}{\per\cubic\centi\m}$ for the doping-induced metal-insulator transition to set $\dea=0$ (\citet{kimoto2014a} predicted for the solubility limit \SI{1e21}{\per\cubic\centi\m}). In contrast \citet{darmody2019} used the logistic equation in \cref{eq:incompIon_logistic} that shows a slower decrease of $\dea$ below \SI{100}{\milli\electronvolt}. Aluminum is the only material where we encountered this model with the parameters shown in \cref{eq:incompIon_logistic_params}.
\begin{equation}
    \label{eq:incompIon_logistic_params}
    \Delta E_0 = \SI{214.86}{\meV}\qquad,\qquad N_\mathrm{E} = \SI{8.12e19}{\per\cubic\centi\meter}\qquad,\qquad c=0.632
\end{equation}

\begin{figure*}[t]
    \centering
    \resizebox{\textwidth}{!}{%
    \input{figures/IncompIon_Al}
    }
    \caption{\label{fig:IncompIon_Al}Ionization energy of Aluminum. Marks refer to measurements and lines to fittings. The letter \textit{d} after the reference indicates a deep level whose origin is still discussed (see text).}
\end{figure*}

To properly describe the measurements of low-doped and compensated devices a second energy level denoted with letter \textit{d} in the figure is required~\cite{pernot2005}. The origin of this deep level is still discussed in literature: \citet{matsuura2003,matsuura2004} were not able to provide any explanation, \citet{weisse2018} suspected excited states of the aluminum ground state and \citet{smith1998,pernot2005} described them as the cubic lattice site, which contradicts, however, several other investigations~\cite{smith1998,ivanov2005,pernot2005}. Although \citet{smith1998} stated that for higher concentrations only the hexagonal site is measured we show the data as they were published. An exception are the values by \citet{saks2001}: The authors did not specify the lattice site but \citet{pernot2005} later denoted them as cubic.

The values of $\Delta E_0$ for the fittings according to \cref{eq:incompIon_PB} are within \SI{220+-20}{\milli\eV}, with the exception of \citet{schoner1994} and \citet{koizumi2009}, who proposed slightly higher values (see \cref{tab:IncompIon_PB_acc}). \revision{We were unable to relay this discrepancy to the utilized Hall measurement method, because similar investigations ended up with lower ionization energies.}  For $\alpha$ the values agree upon \SI{3.2+-1.7e-5}{\milli\eV\centi\m}. This is confirmed by a statistical analysis (see \cref{fig:IncompIon_Al_stat}), which reveals that \SI{75}{\percent} of all proposed values for $\alpha$ are below \SI{3.2e-5}{\milli\eV\centi\m}. \revision{More recent research suggest a higher value, which we retraced to missing measurements at doping concentrations $\geq \SI{e20}{\per\cubic\centi\m}$. Although older models only utilize one or two measurement points in this regime as well, they can be considered more accurate in this regard. \citet{achatz2008} even added the doping concentration limit to achieve a value of $\alpha=\SI{2.23e-5}{\milli\eV\centi\m}$.}

\begin{table*}[t]
    \centering
    \setlength{\tabcolsep}{13pt}
    \renewcommand{\arraystretch}{1}
    \caption{\label{tab:IncompIon_PB_acc}Changing ionization energy of acceptors with doping (\cref{eq:incompIon_PB}). Column N denotes the interpretation of factor $N$: active dopants (dop), all dopants (tot), compensating ones (comp) or simply a fitting (fit). The site denotes besides hexagonal and cubic also a combined effective energy level (eff) and the deep level for Aluminum (deep).}
    \resizebox{0.75\textwidth}{!}{%
    \begin{tabular}{lc*{2}{l} *{2}{c}}
ref. & method & N & site & $\Delta E$ & $\alpha$ \\ 
 & & & & [\si{\milli\electronvolt}] & [\si{\milli\electronvolt \cm}]\\ \hline 
\\[-14pt] \multicolumn{5}{l}{\textbf{Acceptor}} \\ 
\reftext{Tama08a}\cite{tamaki2008a} & -- & dop & -- & \num{191} & \num{3e-5}\\ 
\\ \multicolumn{5}{l}{\textbf{Aluminum}} \\ 
\reftext{Scho94}\cite{schoner1994} & Hall & dop & -- & \num{241.6+-4.8} & \num{2.35e-5}\\ 
\reftext{Mats04}\cite{matsuura2004} & FCCS & dop & -- & \num{220} & \num{1.9e-5}\\ 
 &  &  & deep & \num{413} & \num{2.07e-4}\\ 
\reftext{Pern05}\cite{pernot2005} & Hall & dop & -- & \num{205} & \num{1.7e-5}\\ 
\reftext{Acha08}\cite{achatz2008} & UCLF & dop & -- & \num{220} & \num{2.32e-5}\\ 
\reftext{Koiz09}\cite{koizumi2009} & Hall & dop & -- & \num{265} & \num{3.6e-5}\\ 
\reftext{Buon12}\cite{buono2012} & -- & tot & eff & \num{210} & \num{3.1e-5}\\ 
\reftext{Arvi17}\cite{arvinte2017} & Hall & dop & -- & \num{230+-10} & \num{2.8+-0.3e-5}\\ 
\reftext{Weis18}\cite{weisse2018} & Hall AS & dop & -- & \num{210} & \num{3e-5}\\ 
\reftext{Kaji21}\cite{kajikawa2021} & HC & comp & -- & \num{220} & \num{4.7e-5}\\ 
\reftext{Lech21}\cite{lechner2021} & FIT & tot & -- & \num{230} & \num{1.8e-5}\\ 
\\ \multicolumn{5}{l}{\textbf{Boron}} \\ 
\reftext{Lech21}\cite{lechner2021} & FIT & tot & -- & \num{345} & \num{3.1e-5}\\ 
t.w. & FIT & fit & -- & \num{311} & \num{1.41E-05}\\ 
\end{tabular}
    }
\end{table*}

\begin{figure}[t]
    \centering
    \resizebox{0.75\textwidth}{!}{%
    \input{figures/IncompIon_Al_stat}
    }
    \caption{\label{fig:IncompIon_Al_stat}Statistical analysis of Aluminum doping dependency models. Shown are the 0th, 25th, 50th, 75th and 100th quartile. The mean value is added in numerical form.}
\end{figure}

\subsubsection{Ionization Energy of B}

We found only a few measurements for Boron in the range of \SIrange{285}{332}{\milli\eV} that date back to the last millennium (see \cref{fig:IncompIon_B}). We suspect the main cause in the deep D-center~\cite{troffer1998,torres2022,darmody2019,gali1999} that comes with a Boron doping. It introduces an efficient recombination center with an ionization energy of \SIrange{495}{630}{\meV}~~\cite{kuznetsov1995,sridhara1998a,troffer1998,zhang2003,ikeda1980,devaty1997}. \citet{deak2003} calculated that the shallower Boron level corresponds to a hydrogen assisted incorporation at the Silicon site while the deeper is located at a carbon vacancy~\cite{bockstedte2004}. The lack of hydrogen during an implantation leads to more deep levels but in chemical vapor deposition (CVD) growth more shallow ones are observed. Nevertheless, in certain publications~\cite{ikeda1980,ioffe2023,greulich-weber1997,devaty1997,fan2014,kwasnicki2014,madelung1991,suzuki1977} and simulation tools the D-center serves as the Boron ionization energy.

\begin{figure*}[t]
    \centering
    \resizebox{\textwidth}{!}{%
    \input{figures/IncompIon_B}
    }
    \caption{\label{fig:IncompIon_B}Ionization energy of Boron. Marks refer to measurements and lines to fittings. We added a fitting to all available data points (t.w.).}
\end{figure*}

Due to the fact that only a single fitting according to \cref{eq:incompIon_PB} was available for Boron~\cite{lechner2021} we used all the available data to generate an additional one (t.w.). Both fittings show a decrease of the ionization energies for $\na > \SI{e17}{\per\cubic\centi\m}$ and a solubility limit at or above \SI{e20}{\per\cubic\centi\m}. This is in contradiction to the value of \SI{2e19}{\per\cubic\centi\m} reported by \citet{kimoto2014a}.

\subsubsection{Ionization Energy of N}

For the n-type donor Nitrogen almost all publications distinguish between cubic and hexagonal site (see \cref{fig:IncompIon_N}), but one can observe a discrepancy in the data, especially towards higher doping concentrations. Due to the high solubility limit (\SI{1e19}{\per\cubic\centi\m} for annealing at \SI{1700}{\K}, up to \SI{3e20}{\per\cubic\centi\m} for annealing at \SI{2500}{\K}~\cite{bockstedte2004,kimoto2019,ivanov2005,madelung1991,pankove2014}, \SI{2e21}{\per\cubic\centi\m}~~\cite{kimoto2014a}) these results might still be realistic, but further investigations at doping concentrations beyond \SI{e19}{\per\cubic\centi\m} would be required.

\begin{figure*}[t]
    \centering
    \resizebox{\textwidth}{!}{%
    \input{figures/IncompIon_N}
    }
    \caption{\label{fig:IncompIon_N}Ionization energy of Nitrogen. Marks refer to measurements and lines to fittings. Dashed lines represent the maximum value of $\alpha$ for \citet{schoner1994}.}
\end{figure*}

The measurements still show that the ionization energy of the cubic lattice site lies within a range of \SIrange{80}{130}{\milli\eV}, which is higher than the energy for the hexagonal lattice site in a range of \SIrange{30}{80}{\milli\eV} at low to moderate doping concentrations. \citet{gorban1987} used this circumstance to predict $\Delta \edk$ based on $\Delta \edh=\SI{66}{\milli\eV}$~\cite{ikeda1980}.

For Nitrogen only fittings according to \cref{eq:incompIon_PB} are available. \citet{kagamihara2004} provided a fitting for both hexagonal and cubic lattice site, which were combined by \citet{hatakeyama2013} to an effective ionization energy model. \citet{buono2012} also proposed an effective level but the respective values are surprisingly low. This can be explained by the selection of $\Delta E_0=\SI{65}{\meV}$, which was determined by \citet{bakowski1997} based on the values from \citet{gotz1993}. The latter extracted this ionization level value for a doping concentration of \SI{1e17}{\per\cubic\centi\meter} resp. \SI{1e18}{\per\cubic\centi\meter}, where a deviation from the low-doping levels has to be expected (cp. \cref{fig:IncompIon_N}). In \cref{eq:incompIon_PB}, however, $\Delta E_0$ denotes the ionization energy at zero doping causing the predicted values to be too low.

\revision{For the cubic lattice site only two fittings exist, whereat the one by \citet{kagamihara2004} is closer to the majority of measurement results. Nevertheless, it is also not capable to cover $\Delta\edk > \SI{100}{\milli\eV}$ for $\nd > \SI{3e17}{\per\cubic\cm}$. We investigated these measurements in detail but were not able to identify anything that would cause an increase in ionization energy. For the hexagonal lattice site the model by \citet{kagamihara2004} again fits the measurements best.}

The detailed model parameters are shown in \cref{tab:IncompIon_PB_don}. \citet{schoner1994} conducted multiple fittings using different models, whose ionization energies were all within \num{53.3+-10.5} for the hexagonal and \num{99.6+-8.3} for the cubic lattice sites. One approximation used \cref{eq:incompIon_PB_schoner} with $\es=9.8$ and $f=\num{1.12+-0.88}$ for cubic lattice sites and $f=\num{0.96+-0.7}$ for hexagonal ones. The values for $\alpha$ deviate by up to a factor of three, resulting in shallower and steeper approximations (cp. \cref{fig:IncompIon_N}). Compared to Aluminum the values of $\alpha$ are higher, indicating a more rapid decline with increasing doping concentration.

\begin{table*}[t]
    \centering
    \setlength{\tabcolsep}{13pt}
    \renewcommand{\arraystretch}{1}
    \caption{\label{tab:IncompIon_PB_don}Changing ionization energy of donors with doping (\cref{eq:incompIon_PB}). Column N denotes the interpretation of factor $N$: active dopants (dop), all dopants (tot), compensating ones (comp) or simply a fitting (fit). The site denotes besides hexagonal and cubic also a combined effective energy level (eff).}
    \resizebox{0.75\textwidth}{!}{%
    \begin{tabular}{lc*{2}{l} *{2}{c}}
ref. & method & N & site & $\Delta E$ & $\alpha$ \\ 
 & & & & [\si{\milli\electronvolt}] & [\si{\milli\electronvolt \cm}]\\ \hline 
\\[-14pt] \multicolumn{5}{l}{\textbf{Donor}} \\ 
\reftext{Tama08a}\cite{tamaki2008a} & -- & dop & -- & \num{66} & \num{1.9e-5}\\ 
\\ \multicolumn{5}{l}{\textbf{Nitrogen}} \\ 
\reftext{Scho94}\cite{schoner1994} & Hall & dop & hex & \num{53.3+-10.5} & \num{ 1.65+-1.29e-5}\\ 
 &  &  & cubic & \num{99.6+-8.3} & \num{1.41+-1.03e-5}\\ 
\reftext{Kaga04}\cite{kagamihara2004} & FCCS & dop & hex & \num{70.9} & \num{3.38e-5}\\ 
 &  &  & cubic & \num{123.7} & \num{4.65e-5}\\ 
\reftext{Buon12}\cite{buono2012} & -- & tot & eff & \num{65} & \num{3.1e-5}\\ 
\reftext{Hata13}\cite{hatakeyama2013} & FIT & tot & eff & \num{105} & \num{4.26e-5}\\ 
\reftext{Lech21}\cite{lechner2021} & FIT & tot & -- & \num{52.5} & \num{3.38e-5}\\ 
\\ \multicolumn{5}{l}{\textbf{Phosphorous}} \\ 
t.w. & FIT & fit & hex & \num{57} & \num{9.54E-06}\\ 
 &  &  & cubic & \num{113} & \num{1.26E-05}\\ 
\end{tabular}
    }
\end{table*}

\subsubsection{Ionization Energy of P}

Compared to Nitrogen less measurements are available for Phosphorous (see \cref{fig:IncompIon_P}), all roughly two decades old. Again, hexagonal and cubic lattice site were always distinguished, but this time the values are much more concise. The cubic lattice site showed again the higher values around \SI{100+-20}{\milli\eV} while the hexagonal ionization energy was determined within \SI{55+-5}{\milli\eV}. These values are by a factor two to six smaller than the energies we found for the p-type doping. Consequently, more free electrons can be expected for an equivalent n-type doping.

\begin{figure*}[t!]
    \centering
    \resizebox{\textwidth}{!}{%
    \input{figures/IncompIon_P}
    }
    \caption{\label{fig:IncompIon_P}Ionization energy of Phosphorous.}
\end{figure*}

Due to the lack of fittings according to \cref{eq:incompIon_PB} we used the
available data to generate one ourselves. For the cubic site we excluded the
results by \citet{wang2002} as the measured values had to be considered outliers
in respect to the remaining data. The achieved models agree quite well with the solubility limits we found in literature. (\SI{6e18}{\per\cubic\centi\m} for annealing at \SI{1700}{\K}, up to \SI{2e20}{\per\cubic\centi\m} for annealing at \SI{2500}{\K}~\cite{bockstedte2004,kimoto2019,ivanov2005,madelung1991,pankove2014}, \SI{1e21}{\per\cubic\centi\m}~~\cite{kimoto2014a}). The exact parameters are shown in \cref{tab:IncompIon_PB_don}.

\subsubsection{Capture Cross Sections}

For 4H-SiC various models with varying temperature dependencies were published at the end of the last century (see \cref{tab:IncompIon_CS}). Since then we only found data for Aluminum, whereat for n-type doping only one investigation by \citet{kaindl1999} could be acquired.

\begin{table}[p]
    \centering
    \setlength{\tabcolsep}{13pt}
    \renewcommand{\arraystretch}{1.1}
    \caption{\label{tab:IncompIon_CS}Dopants and their respective cross sections with the charge carriers in the energetically closer band (conduction or valence). Different temperature dependencies are indicated in the column $T^\alpha$.}
    \resizebox{0.85\textwidth}{!}{%
    \begin{tabular}{lc*{2}{l}*{2}{c}}
ref. & method & site & $T^\alpha$ & $\Delta E$ & $\sigma$  \\ 
 & & & & [\si{\milli\electronvolt}] & [\si{\square\cm}] \\ \hline 
\\[-14pt] \multicolumn{5}{l}{\textbf{Nitrogen}} \\ 
\reftext{Kain99}\cite{kaindl1999} & AS & cubic & $T^0$ & \num{77} & \num{7.92e-15}\\ 
 &  &  & $T^{-2}$ & \num{90} & \num{3.57e-10}\\ 
\\ \multicolumn{5}{l}{\textbf{Aluminum}} \\ 
\reftext{Kuzn95}\cite{kuznetsov1995} & DLTS &  & $T^{-3}$ & \num{229} & \num{8e-13}\\ 
\reftext{Scha97}\cite{schadt1997} & Hall AS DLTS &  & $T^0$ & \numrange{164}{179} & \numrange{2.2e-13}{7.6e-13}\\ 
 &  &  & $T^{-2}$ & \numrange{189}{202} & \numrange{1.7e-12}{5.6e-12}\\ 
\reftext{Kain99}\cite{kaindl1999} & AS &  & $T^0$ & \num{189} & \num{2.58e-13}\\ 
 &  &  & $T^{-2}$ & \num{208} & \num{2.57e-8}\\ 
\reftext{Resh05}\cite{reshanov2005} & AS &  & $T^0$ & \num{185} & \num{1e-14}\\ 
 &  &  & $T^{-2}$ & \num{210} & \num{1e-13}\\ 
\reftext{Belj10}\cite{beljakowa2010} & AS &  & $T^0$ & \num{200} & \num{1e-12}\\ 
\reftext{Mand12}\cite{mandal2012} & TSC &  & $T^0$ & \num{220} & \num{1e-16}\\ 
\reftext{Kawa15}\cite{kawahara2015} & AS &  & $T^0$ & \num{190} & \num{1.4e-13}\\ 
\reftext{Kato22}\cite{kato2022} & DLTS &  & $T^0$ & \numrange{120}{170} & \numrange{1e-17}{100e-17}\\ 
\\ \multicolumn{5}{l}{\textbf{Boron}} \\ 
\reftext{Srid98a}\cite{sridhara1998a} & Hall AS &  & $T^0$ & \numrange{259}{262} & --\\ 
 &  &  & $T^{-2}$ & \numrange{284}{295} & --\\ 
\reftext{Trof98}\cite{troffer1998} & Hall AS &  & $T^0$ & \num{292} & \num{6e-15}\\ 
 &  &  & $T^{-2}$ & \num{314} & \num{5e-14}\\ 
\reftext{Kain99}\cite{kaindl1999} & AS &  & $T^0$ & \num{312} & \num{2.1e-14}\\ 
 &  &  & $T^{-2}$ & \num{375} & \num{9.69e-9}\\ 
\reftext{Zhan03}\cite{zhang2003} & DLTS MCTS &  & $T^0$ & \numrange{230}{280} & \numrange{2e-14}{30e-14}\\ 
\end{tabular}
    }
\end{table}

The ionization energy in column $\Delta E$ shows large variations, even among the same temperature dependency models. This is also the case for the cross section $\sigma$, which refers to the charge carrier in the energetically closer band (conduction or valence) because the interaction with the other band is significantly lower~\cite{kaindl1999,troffer1998}. Differences of more than five orders of magnitude were encountered.

\subsubsection{Values in Literature}

Ionization energies collected from overviews or TCAD simulations (see \cref{fig:IncompIon_TCAD}) show a wide range of values for each dopant. These are, however, well within the boundaries of the fundamental investigations we identified. An exception are the values of the D-center for Boron, that was specified in eleven publications as the ionization energy. Interestingly, we were unable to identify clear favorites for each dopant. Instead, several values are utilized equally often. Only for Al the value $\dea=\SI{191}{\milli\eV}$ has a slight edge.

\begin{figure*}[p]
    \centering
    \resizebox{0.95\textwidth}{!}{%
    \input{figures/IncompIon_TCAD}
    }
    \caption{\label{fig:IncompIon_TCAD}Energy levels for each investigated dopant. \bkgCol{fundColor} are fundamental investigations and \bkgCol{no4HColor} research not focused on 4H. \textit{h}, \textit{c} and \textit{e} denote hexagonal, cubic and effective lattice sites respectively.}
\end{figure*}

In some cases the dopant was not clearly specified: Instead only the acceptor and donor energy levels were stated (see \cref{fig:IncompIon_undef}). While the acceptor values clearly correspond to Aluminum the n-type values could belong to both Nitrogen and Phosphorous.

\begin{figure*}[t]
    \centering
    \resizebox{0.45\textwidth}{!}{%
    \input{figures/IncompIon_undef}
    }
    \caption{\label{fig:IncompIon_undef}Energy levels used in literature for acceptors and donors without specification of dopant.}
\end{figure*}


\subsubsection{Origin of Values}

Due to missing references it is often impossible to retrace where and how the used ionization energies were determined. Nevertheless, we tried to create reference chains for n-type (for a detailed graphical representation see \cref{fig:IncompIon_ref_chainN} in Appendix~\ref{sec:refChainIncompIon}) and p-type (see \cref{fig:IncompIon_ref_chainP} in Appendix~\ref{sec:refChainIncompIon}) doping and inferred missing connections based on the used values. On the bright sight, we rarely encountered 6H values~\cite{huh2006,rakheja2020}, although they seem to match 4H ones~\cite{arvanitopoulos2017,donato2018,lophitis2018}.


The ionization energy for Nitrogen was often referenced from the publications by \citet{ikeda1980}, \citet{gotz1993} and \citet{ivanov2005} while the energy for Phosphorous goes regularly back to the investigations by \citet{capano2000} and \citet{ivanov2005}. The latter reported $\Delta\edk=\SI{60.7}{\milli\eV}$ and $\Delta\edh=\SI{120+-20}{\milli\eV}$, i.e., $\Delta\edk<\Delta\edh$, which contradicts every other measurement we found. In fact, all papers~\cite{kimoto2014a,kimoto2015,nipoti2018,scaburri2011} that referenced \citet{ivanov2005} changed the values to $\Delta\edh=\SI{60.7}{\milli\eV}$ and $\Delta\edk=\SI{120+-20}{\milli\eV}$ so we assume a typographical issue in the paper.

Twelve fundamental investigations were referenced for n-type ionization energies, which is a lot more compared to other topics we studied in this review. The concrete values were, in general, transferred correctly, but in some cases the information regarding cubic or hexagonal lattice site was dropped. \citet{ikeda1980}, \citet{troffer1997}, \citet{ladesmartin2000}, \citet{ivanov2005} and \citet{koizumi2009} represent the main sources for the ionization energies of Aluminum and Boron, whereat overall thirteen different fundamental investigations were at least once referenced.


For the parameters to model the doping dependency according to
\cref{eq:incompIon_PB} we only found two (n-type) respectively three (p-type)
fundamental investigations that were referenced (see
\cref{fig:IncompIon_ref_chain_alpha} in Appendix~\ref{sec:refChainIncompIon}).
\citet{song2012} introduced an
unfortunate typographical error by changing the acceptor ionization energy from \SI{191}{\milli\eV}~\cite{tamaki2008a} to \SI{19}{\milli\eV}. In terms of charge carrier densities this results in a huge difference. The incorrect value was later referenced by \citet{khalid2012}, which shows the importance of referencing the original research results. A comprehensive listing of all inconsistencies for incomplete ionization can be found in Appendix~\ref{sec:incompIon_appendix}.



\section{Mobility}
\label{sec:mobility}

Charge carrier mobilities ($\mun$ for electrons; $\mup$ for holes) of a material determine its conductivity $\sigma$ (see \cref{eq:mob_conductivity}; $n$ and $p$ are the free electron/hole concentrations), which influences all transient processes. Surfaces, inversion channels in MOS structures and the bulk~\cite{lechner2021} may have different mobilities, whereat we will investigate in this review solely the latter. We, thereby, extend the analysis of measurements by \citet{darmody2019} and the summaries of mobility models published by \citet{neilainglesias2012}, \citet{stefanakis2014} and \citet{tian2020}.
\begin{equation}
    \label{eq:mob_conductivity}
    \sigma = e (n \mun + p \mup)
\end{equation}

Overall, our analyses reveal many investigations targeted toward the mobility. The change with doping concentration is well covered, however, for holes a large spread of values is apparent. More problematic is the temperature dependency, because only few models predict a decrease of the mobility at low temperatures, as observed in measurements. A better agreement is visible in the high-field regime, but only few values of the hole saturation velocity exist, calling for further investigations. Except for one publication, all indicate a decrease of the velocity with temperature. For carrier-carrier scattering only a single publication could be identified in literature.


\subsection{Introduction}

Charge carriers in 4H-SiC accelerate along an electric field $F$ until they are ``scattered'', i.e., they drastically change their velocity and/or direction by interacting with other particles. Theoretical analyses investigated the individual contributions in detail~\cite{koizumi2009,pernot2000,pernot2001,pernot2005,adachi2005,neilainglesias2012,selberherr1984,meyer2000}. In this context, the mobility defines the average time between two scattering events~\cite{ladesmartin2000} and thus provides a link between charge carriers and scattering processes~\cite{cheng2022}. The most prominent processes for 4H-SiC are
\begin{enumerate*}[label=(\roman*)]
    \item phonon, i.e., acoustic phonon, (non-)polar optical phonon,  zero and first order optical intervalley phonon,
    \item defect, i.e., ionized and neutral impurity, and
    \item carrier-carrier scattering~\cite{adachi2005,lv2004,pernot2005,ishikawa2021,koizumi2009,kaji2015}.
\end{enumerate*}
An overall mobility is derived by combining the individual contributions using the Matthiessen rule~\cite{adachi2005,selberherr1984,kimoto2014a} (see \cref{eq:mob_matthiessen}), such that the lowest value dominates.
\begin{equation}
    \label{eq:mob_matthiessen}
    \frac{1}{\mu} = \sum_i \frac{1}{\mu_i}\ .
\end{equation}

Due to the various scattering processes, the mobility depends on several material parameters such as doping concentration~\cite{das2015}, degree of compensation~\cite{lutz2011}, spatial direction~\cite{iwata2000,iwata2001,harima1995}, temperature~\cite{das2015} and whether majority or minority charge carriers~\cite{adachi2005} are described. For holes even the separate valence bands (heavy-hole, light-hole, split-off; see \cref{sec:dos})~\cite{adachi2005,pernot2005,zhao2000} have to be considered.

In the past years, hopping conduction~\cite{efros1979}, also denoted as nearest-neighbor-hopping (NNH)~\cite{matsuura2018,matsuura2020a} or variable-range-hopping (VRH)~\cite{matsuura2018}, was investigated in 4H-SiC. This process denotes the tunneling of charge carriers bound to a dopant from one impurity to the next, possibly alternated by conventional drifting phases in the bands. This effect was described at temperatures below \SI{100}{\K}~\cite{kajikawa2021} and at very high doping concentrations~\cite{kim2024,matsuura2018}, whereat \citet{darmody2019} calculated the critical limit to $N_\mathrm{crit} \approx \SI{1e20}{\per\cubic\centi\meter}$. Since this is a relatively new effect the available information are very limited, so we will not further consider it in this review.

In TCAD tools high detailed descriptions are not convenient~\cite{koizumi2009}. Instead, empirical models describe the mobility in the low- and high-field region, which we will elaborate in the sequel.


\subsubsection{Low-Field Mobility}

At low electric fields the carrier velocity $v$ and the electric field strength $F$ are directly related via the mobility $\mu$ (see \cref{eq:mob_vmuF})~\cite{roschke2001}. The latter depends on phonon and impurity scattering~\cite{schroeder2006}, meaning that both temperature and doping concentration have to be included in the same model.
\begin{equation}
    \label{eq:mob_vmuF}
    v = \mu(N,T) F
\end{equation}

Each dopant in 4H-SiC represents a coulomb scattering center~\cite{baliga2019} that causes the charge carrier mobility to decrease. \citet{caughey1967} mathematically described this effect by \cref{eq:mobility_CT}. $\mumin$ can be interpreted as the mobility for very high doping where impurity scattering is dominant~\cite{stefanakis2014,neilainglesias2012} and $\mumax$ the highest possible mobility at low doping, i.e., when lattice (phonon) scattering is dominant~\cite{lechner2021,stefanakis2014,neilainglesias2012}. $\nref$ denotes the doping concentration where the mobility is exactly in between those values~\cite{roschke2001,lutz2011} and $\delta$ how quickly the transition from one to the other occurs. \citet{arora1982} later simplified the model by replacing the expression $\mumax - \mumin$ by $\mu_0$~\cite{huang1998}.
\begin{equation}
    \label{eq:mobility_CT}
    \mu(N) = \mumin + \frac{\mumax - \mumin}{1+(N/\nref)^\delta}
\end{equation}

There is some disagreement in literature (discussed by \citet{vasilevskiy2017}) whether $N$ denotes all dopants~\cite{ schaffer1994,bakowski1997,huang1998,mickevicius1998,linewih2003,ayalew2004,mcnutt2004,perez-tomas2006,cha2008,koizumi2009,buono2012,hatakeyama2013,stefanakis2014,liaugaudas2015,sharma2015,arvanitopoulos2017,naugarhiya2017,vasilevskiy2017,tanaka2018,kimoto2019,ishikawa2021,huang2022b,ishikawa2023,ishikawa2024} or just the ionized ones~\cite{codreanu2000,hatakeyama2003,li2003,lv2004,negoro2004,zhang2010,megherbi2018,tanaka2018,biondo2012}. An argument to use all dopants was that this model is just a fit~\cite{tian2020,stefanakis2014} or that also scattering on neutral dopants decreases the mobility~\cite{koizumi2009,hatakeyama2013}. \citet{roschke2001} stated that using something different than the absolute doping for $N$ did not improve the results but led to convergence issues.

\citet{masetti1983} extended the model by an additive term that leads to a further reduction of the mobility at high doping densities (see \cref{eq:mobility_masetti})~\cite{lombardi1988}. We found only one instance~\cite{zhang2018} where this model was used; other publications referencing this model~\cite{lechner2021,nallet1999} chose $\mu_1=0$ leading to the description in \cref{eq:mobility_CT}.
\begin{equation}
    \label{eq:mobility_masetti}
    \mu(N) = \mumin + \frac{\mumax - \mu_\mathrm{min2}}{1+(N/\nref)^\delta} -\frac{\mu_1}{1 + (N_\mathrm{ref2}/N)^\kappa}
\end{equation}

The mobility scaling with temperature depends on the dominant scattering mechanism~\cite{koizumi2009,ishikawa2021}, whereat each has its own temperature dependency~\cite{kimoto2019}. Acoustic phonon scattering showed a decrease $\propto T^{-1.5}$~~\cite{rambach2006,rambach2008,matsunami2004a,ladesmartin2000,schaffer1994,schoner1994} and optical-mode phonons $\propto T^{-2.5}$~~\cite{ladesmartin2000}. A reduction $\propto T^{-2.6}$ was attributed to nonpolar optical phonon scattering~\cite{koizumi2009,matsunami2004a}, but \citet{adachi2005} stated a dependency according to $T^{-1.5}$ for this process. Ionized impurities scale with $T^{3/2}$~~\cite{rambach2006,rambach2008,ishikawa2021,liaugaudas2015,schoner1994}, neutral impurities with $T^0$~~\cite{ishikawa2021,adachi2005,liaugaudas2015,schoner1994}, Coulomb scattering with $T^1$ and phonon scattering with $T^{-1}$~~\cite{berens2021}.

Phonon scattering dominates in hot samples because more phonons are generated~\cite{das2015,baliga2019,gotz1993}, and (un)ionized impurity scattering at low temperatures~\cite{mnatsakanov2002,kimoto2019}. The exact amount of ``low temperature'' in this context increases with the doping concentration, i.e., more dopants lead to an earlier impact of impurity scattering. According to the temperature dependencies mentioned before, the mobility thus rises in cold samples with temperature before it decreases in phonon-dominated temperature regimes~\cite{koizumi2009}.

These dependencies are added to \cref{eq:mobility_CT} by scaling the single parameters with temperature (see Eq.~(\ref{eq:mob_temp_scaling}))~\cite{albanese2010,hatakeyama2013}. We listed all scaling parameters we found in literature, although some of them are redundant ($\gNNref = -\delta \gref$). In most publications only a subset of these parameters is used~\cite{brosselard2004a}, whereat differing origins were stated. Referenced were, among others, the publications by \citet{mohammad1993} and \citet{sotoodeh2000} but also the description \textit{Masetti model}~\cite{das2015,li2003} was used, although the second additive term in \cref{eq:mobility_masetti} was missing.
\begin{eqnarray}
    \label{eq:mob_temp_scaling}
    \theta &=& \theta_{300} \left( \frac{T}{300} \right)^{\zeta}\\
    (\theta,\zeta) &\in& [(\mumin,\gmin), (\mumax,\gmax), (\mu_0,\gamma_0), (\nref,\gref), (\delta,\gdelta), ((N/\nref)^\delta,\gNNref)]
\end{eqnarray}  

The temperature-dependent mobility can be properly described by the approach in Eq.~(\ref{eq:mob_temp_scaling}) if the parameters are carefully determined, e.g., as done by \citet{hatakeyama2013}. In most cases (details follow in \cref{sec:mob_results}), however, the decrease of $\mu$ at low temperatures is not properly covered making these approximations only adequate for high temperatures~\cite{dixit2023}. As a workaround \citet{schroeder2006} defined on page $668$ equations below/above $\SI{200}{\K}$ with deviating temperature scalings, but only provided parameters for Si.

We found various additional models to describe the temperature and doping dependencies of the mobility in 4H-SiC. \citet{rambach2008}~\cite{rambach2006,rambach2008} proposed a more inclusive approach based on the doping dependent parameters $\delta(\na)$, $\nref(\na)$, $\gmin(\na)$ and $N=p(T,\na,\nk)$. In another model for $T>\SI{250}{\K}$ the exponential parameter $\zeta$ had the same structure as \cref{eq:mobility_CT} (see Eq.~(\ref{eq:mob_mup}) and Eq.~(\ref{eq:mob_bp}))~\cite{kagamihara2004,matsuura2004}.
\begin{eqnarray}
    \label{eq:mob_mup}
    \mu(T,N) &=& \mu(300,N) \left(\frac{T}{300}\right)^{\beta(N)} \\
    \label{eq:mob_bp} \beta(N) &=&  \bmin + \frac{\bmax - \bmin}{1+(N/\Np)^\eta}
\end{eqnarray}

\citet{lavia2005} used the phenomenological function shown in \cref{eq:temp_pheno} to model the increasing mobility due to ionized impurity scattering at low temperature ($T^{3/2}$) and an adjustable decrease for high temperatures. This description was further simplified by \citet{mitchel2007}, who just used $\mu(T) = A\, T^{-n}$.
\begin{equation}
    \label{eq:temp_pheno}
    \mu(T) = \left( \frac{A}{T^{3/2}} + \frac{B}{T^{-n}} \right) ^{-1}
\end{equation}

\citet{uhnevionak2015} found discrepancies between simulated and measured MOSFET currents with varying temperature. To better match the bulk-phonon scattering, the author split $\mumax$ into two additive parts with separate temperature scaling factors (see \cref{eq:mobility_sum_max}).
\begin{equation}
    \label{eq:mobility_sum_max}
    \mumax \left(\frac{T}{300}\right)^{\gmax} = \mu_\mathrm{max1} \left(\frac{T}{300}\right)^{\gamma_\mathrm{max1}} + \mu_\mathrm{max2} \left(\frac{T}{300}\right)^{\gamma_\mathrm{max2}}
\end{equation}

\citet{mnatsakanov2002} tackled the challenge of doping dependent temperature coefficients by separating impurity and lattice scattering and scaling each independently with temperature. Starting from \cref{eq:mobility_CT} this led to the expressions shown in Eq.~(\ref{eq:mobility_mnat_mu}) and Eq.~(\ref{eq:mobility_mnat_B}) with $\gI$ the temperature scaling factor of the impurity scattering contribution. For $T=T_0$ Eq.~(\ref{eq:mobility_mnat_mu}) collapses to \cref{eq:mobility_CT}. The authors claimed that this model is able to simultaneously describe the temperature behavior for low and very high doping.
%
\begin{eqnarray}
    \label{eq:mobility_mnat_mu}
    \mu(N,T) &=& \mumax(T_0) \frac{B(N) \left(\frac{T}{T_0}\right)^{\gI}}{1+B(N) \left(\frac{T}{T_0}\right)^{\gI+\gmax}}\\
    \label{eq:mobility_mnat_B}
    B(N) &=& \left. \left[ \frac{\mumin + \mumax \left(\frac{\nref}{N} \right)^{\delta}}{\mumax - \mumin} \right] \right\vert_{T=T_0}
\end{eqnarray}
    
Finally, \citet{klaassen1992} proposed a unified description of majority and minority charge carriers including screening effects, also called Philips model, which was recently used in conjunction with 4H-SiC~\cite{dixit2023}. Although this model is already included in state of the art simulation frameworks we did not consider it in our review, because we could not find explicit parameters for 4H-SiC.


\subsubsection{High-Field Mobility}

At high electric fields the charge carrier velocity approaches a maximum value, the so-called \textit{saturation velocity}. Higher optical phonon scattering~\cite{cha2008,das2015,lee2002} or elastic and nonelastic scattering rates owing to the increase in carrier energy~\cite{hatakeyama2013} explain this behavior. According to \cref{eq:mob_vmuF} the velocity is proportional to the field strength, meaning that the mobility has to decrease. \citet{caughey1967} modeled this field dependency according to \cref{eq:mob_high}: $\mulow$ denotes the low field mobility as described in the previous section and $\vsat$ the saturation velocity (\citet{kimoto2014a} calls the latter sound velocity). In some cases an additive factor $\alpha$ was introduced in various spots of this equation~\cite{arvanitopoulos2017,lophitis2018,biondo2012}, but it was always set to $0$, rendering it irrelevant. \citet{chen2012} used the hydrodynamic version of this equation, which we are not going to discuss in this review.
\begin{equation}
    \label{eq:mob_high}
    \mu = \frac{\mulow}{\left[ 1+ \left( \frac{\mulow F}{\vsat}\right)^\beta\right]^{\frac{1}{\beta}}}
\end{equation}

The high field mobility is temperature dependent~\cite{adachi2005,baliga2019,banerjee2021} so \citet{canali1975} suggested to scale the parameters $\beta$ and $\vsat$ with temperature as was shown for the low-field case in Eq.~(\ref{eq:mob_temp_scaling})~\cite{ladesmartin2000,hatakeyama2013}. This model is thus often called Canali model~\cite{jin2024,arvanitopoulos2019}. \citet{quay2000} proposed a very similar model of the form $\vsat / [ (1-A) + A(T/300)]$ for semiconductors in general, but we did not find a single application for 4H-SiC yet. In some occasions~\cite{albanese2010,bertilsson2004a} the temperature dependency of $\vsat$ was modeled by the approach commonly used for Si (see \cref{eq:mob_vsat})~\cite{jacoboni1977,wang1999}. 
\begin{equation}
    \label{eq:mob_vsat}
   \vsat = \frac{\vmax}{1+d \exp\left(\frac{T}{600}\right)}\ .
\end{equation}

\citet{roschke2001} proposed an exponential and \citet{bertilsson2004a} a linear dependency of $\beta$ with temperature. For a more concise presentation we merged both in \cref{eq:mob_beta}.
\begin{equation}
    \label{eq:mob_beta}
    \beta = \beta_0 + a\,\exp\left(\frac{T-\Tref}{b}\right) + c\, T
\end{equation}

Monte-Carlo simulations~\cite{rakheja2020,akturk2009,bellotti1999,bellotti2000,bertilsson2004,bertilsson2004a,hjelm2003,lv2004,mickevicius1998,neilainglesias2012,nilsson2000,nilsson1996} indicated a velocity peak, i.e., a negative differential mobility at fields near \SI{1e6}{\V/\cm}~\cite{roschke2001}. \citet{foutz1998} proposed a new model for wide band gap materials that covered this \textit{overshoot} (see \cref{eq:mob_overshoot})~\cite{lv2004}. This approach was used for GaN~\cite{polyakov2001} and lately also for 4H-SiC~\cite{lv2004} but is not yet available in TCAD tools. A simplified version ($\alpha = -\infty$) was denoted as transferred-electron model~\cite{turin2005,selberherr1984}.
\begin{equation}
  \label{eq:mob_overshoot}
 v(F) = \frac{\mu_0 F + \mu_1 F (F/F_0)^\alpha + \vsat (F/F_1)^\beta}{1 + (F/F_0)^\alpha + (F/F_1)^\beta} 
\end{equation}

Finally, \citet{baliga2019} used the representation shown in \cref{eq:mob_mu_simple} to describe the high field mobility. The parameters from \cref{eq:mobility_CT} can be achieved by using $\vsat=A$ and $\mulow=\vsat/B^{1/\beta}$.
\begin{equation}
    \label{eq:mob_mu_simple}
    \mu = \frac{A}{[B+F^{\beta}]^{1/\beta}}
\end{equation}


\subsubsection{Carrier-Carrier Scattering}

This scattering process denotes interactions among the same type of charge carriers, e.g., electron-electron~\cite{adachi2005}, or between electrons and holes~\cite{lutz2011}, which decreases the mobility at high injection levels~\cite{baliga2019,dorkel1981,tian2020}. In the literature it was either assumed negligible~\cite{werber2008} or described by the Conwell-Weisskopf equation (see \cref{eq:mob_conwell_weisskopf})~\cite{choo1972,ayalew2004,onoda2007,tripathi2019}, which was also used in the past to describe silicon. For the latter, \citet{lutz2011} mentioned the possibility to cover carrier-carrier-scattering by including the free charge carriers in $N$ of \cref{eq:mobility_CT} but we did not encounter this approach for 4H-SiC.
\begin{equation}
    \label{eq:mob_conwell_weisskopf}
    \muccs = \frac{D \left( \frac{T}{T_0} \right)^{\frac{3}{2}}}{\sqrt{np}} \left[ \ln \left( 1 + F \left(\frac{T}{T_0}\right)^2 (pn)^{-\frac{1}{3}} \right) \right]^{-1}
\end{equation}

TCAD tools combine the carrier-carrier scattering and the low/high field mobility by utilizing the Matthiessen rule (see \cref{eq:mob_matthiessen}).


\subsubsection{Hall Scattering Factor}

The mobility can be experimentally extracted from conductivity measurements, delivering the so-called conductivity or drift mobility (see \cref{eq:mob_conductivity})~\cite{adachi2005}. The main issue with this method is that the charge carriers concentrations have to be known as well. Instead, the simpler Hall measurements are often preferred that, however, deliver a slightly different mobility, the Hall mobility $\muH$~\cite{arvinte2017}. In general, it is said that the Hall mobility is easier to measure, while the conductivity mobility is easier to calculate~\cite{adachi2005}.

The ratio of Hall and conductivity mobility is called Hall factor $\rH$ (see Eq.~(\ref{eq:mob_r}))~\cite{adachi2005,gotz1993,ishikawa2021,iwata2001,pensl2003,schaffer1994}. To calculate $\rH$ for holes one has to consider both light and heavy holes~\cite{parisini2013,pernot2005}. Th Hall factor must not be confused with the Hall coefficient $\RH$ (see Eq.(\ref{eq:mob_RH})), which is the value actually derived in Hall measurements. It depends on the Hall charge carrier count $\nH$~\cite{tanaka2018,asada2016,contreras2016,koizumi2009,rutsch1998} and is used to connect conductivity $\sigma$ and mobility (see Eq.~(\ref{eq:mob_muH}))~\cite{baliga2019,pensl1993,schadt1997}. For further theoretical analyses and overviews on Hall mobilities and measurements the interested reader is referred to the dedicated literature~\cite{pernot2001,neilainglesias2012,schmid2004,schroder2005}.
\begin{eqnarray}
    \label{eq:mob_r}
    \rH &=& \frac{\muH}{\muc} = \frac{n}{\nH}\\
    \label{eq:mob_RH}
    \RH &=& \frac{1}{\nH e} = \frac{\rH}{n e}\\
    \label{eq:mob_muH}
    \muH &=& \sigma \RH = \frac{1}{e \nH \rho}
\end{eqnarray}

Consequently, one can not use the Hall mobility directly in \cref{eq:mob_conductivity} to calculate the conductivity of a material. To properly handle values that were proposed in the literature it is, therefore, mandatory to consider the utilized characterization method.

\revision{It was shown that the Hall scattering factors for electrons and holes depend on the the magnetic field~\cite{rutsch1998,rutsch2000,schmid2004}, the temperature~\cite{nipoti2013,pernot2005,schmid2004,parisini2013,pensl2003,pensl2005,schadt1997,tanaka2018,habib2011,koizumi2009,iwata2001}, the anisotropy of the effective masses~\cite{iwata2001} and the doping concentration~\cite{tanaka2018,habib2011,asada2016}. The impact of compensation~\cite{koizumi2009} as well as the anisotropy~\cite{rutsch1998,rutsch2000} is still under investigation.  }

\subsubsection{Methods}

Several approaches to determine the mobility were proposed in the literature. These include simulations such as Monte Carlo~\cite{akturk2009,bertilsson2001,bertilsson2004,hjelm2003,joshi1995,mickevicius1998,nilsson1996,nilsson2000}, full band monte carlo (FBMC)~\cite{cheng2020,tanaka2024}, empirical pseudo potentials (EPM)~\cite{bellotti1999,bellotti2000,hjelm2003}, monte carlo particle (MCP)~\cite{zhao2000}, non equilibrium statistical ensemble formalism (NESEF)~\cite{vasconcelos2019}, linear augmented plane wave (LAPW)~\cite{martinez2002}, density functional theory (DFT)~\cite{tanaka2018}, extraction from the diffusion coefficient (DIFF)~\cite{neimontas2006}, conductivity tensor calculations (COTE)~\cite{iwata2000,iwata2000a,iwata2001,kinoshita1998}, general calculations~\cite{pernot2000,pernot2001,pernot2005} and fitting (FIT)~\cite{cheng2022,hatakeyama2013,jin2024,kagamihara2004,kajikawa2021,ladesmartin2000,lv2004,mnatsakanov2001,roschke2001}. \citet{nilsson2001} provided a comparison among several Monte-Carlo models.

Measurements include collected charge (CCh)~\cite{belas2022}, nanosecond pulsed conductance (NPC)~\cite{ardaravicius2005,cha2005,khan1998,khan2000b,khan2000}, resistance measurements (RES)~\cite{izzo2008,lavia2005}, Schottky barrier diode I-V fitting (SBD-IV)~\cite{jaiswal2024}, Raman scattering (Raman)~\cite{harima1995,nakashima1997,sun2011a}, low temperature photoluminescence (LTPL)~\cite{liaugaudas2015}, spectroscopic ellipsometry (SE)~\cite{mainali2024}, optical detection of cyclotron resonance (ODCR)~\cite{meyer2000,son1995}, diode I-V (DIV)~\cite{werber2007,jaiswal2024}, bipolar transistor I-V (BIV)~\cite{sankin2002} and Hall\newline measurements~\cite{arvinte2017,barrett1967,burk1998,choyke1997d,contreras2006,contreras2016,fujihara2017,galeckas2002,gotz1993,hatakeyama2003,ishikawa2021,ishikawa2023,ishikawa2024,itoh1994,kasamakova-kolaklieva2004,kimoto2001,koizumi2009,laube2004,lomakina1972,matsunami1997,matsunami2004a,matsunami2020,matsuura2004,negoro2004,nipoti2013,pensl2005,rambach2006,schadt1994a,schadt1997,schaffer1994,schoner1994,schoner1999,siergiej1999,terziyska2003,vasilevskiy2017}. The majority of publications provided\newline Hall~\cite{schaffer1994,schadt1997,burk1998,kinoshita1998,schoner1999,siergiej1999,iwata2000,iwata2000a,pernot2000,pernot2001,hatakeyama2003,terziyska2003,kasamakova-kolaklieva2004,laube2004,negoro2004,pensl2005,pernot2005,rambach2006,koizumi2009,contreras2016,fujihara2017,vasilevskiy2017,ishikawa2021,ishikawa2023,ishikawa2024} and only few conductivity mobilities~\cite{harima1995,nakashima1997,mickevicius1998,meyer2000,cheng2020}.

It is also possible to calculate the mobility from the charge carrier lifetime
$\tau$ (cp. \cref{sec:regen}) and the conductivity mass $\mcond$ (cp. \cref{sec:dos}) according to \cref{eq:mob_lifetime_mass}~\cite{tiwald1999,harima1995,ishikawa2021,ishikawa2024,kordina1995,tanaka2018}. \citet{neimontas2006} calculated it from the diffusion coefficient using the Einstein relationship $\mu=De/\kb T$.
\begin{equation}
    \label{eq:mob_lifetime_mass}
    \mu = \frac{e \tau}{\mcond\,m_0}
\end{equation}


\subsection{Results \& Discussion}
\label{sec:mob_results}

In the sequel we present the gathered results on mobility in 4H-SiC. We had to  exclude publications that solely focused on channel or inversion layer mobilities~\cite{bader2020,huang1998,noguchi2017,tanaka2024a,weitzel1994,cha2005,choi2005,zhang2008,jin2024,friedrichs2009a,poggi2010}, that did not clearly specify the SiC polytype~\cite{mantooth2015,persson2005,schroeder2006,trew1991,murray2002,ferry1975,vuillod1995}, that
dealt with mobility variations induced by irradiation defects~\cite{vobecky2015} or that did not clearly specify the used equations, making a unique mapping of the parameter values impossible~\cite{bhatnagar2005}. We also did not consider \citet{wright1996} that was superseded by \citet{wright1998}.


\subsubsection{Hall Scattering Factor}

\revision{ We discussed in the last section that the results of Hall measurements have to be scaled by the Hall scattering factor $\rH$ to obtain the conductivity mobility. In early investigations $\rH$ was assumed approximately unity~\cite{ladesmartin2000,roschke2001,kimoto2001,darmody2019,noguchi2017,harima1995} due to the lack of reliable data~\cite{gotz1993}. Later it was shown that this is only the case for high magnetic fields~\cite{schmid2004}. For all other cases described in the previous section we found variations of $\rH$ predominantly in the range of $[0.5,1.5]$ for holes~\cite{asada2016,schmid2004,pensl2003,pensl2005,tanaka2018} and in a smaller range of $[0.9,1.2]$ for electrons~\cite{schadt1997,rutsch1998,rutsch2000,schmid2004,iwata2001}.

  For the temperature variation of the electron Hall scattering factor, which at first decreases before rising again above \SI{150}{\K}, we found few investigations~\cite{iwata2001,schmid2004,rutsch2000,rutsch1998}. More data is available for holes: Parallel to the c-axis $\rH$ shows a steady decrease with temperature. \citet{pernot2005} fitted the results by \citet{pensl2003} (see Eq.~(\ref{eq:mob_rH_fit})) with a rapid decline from $1.2$ at \SI{300}{\K} to approx. $0.5$ at \SI{800}{\K}. Perpendicular to the c-axis $\rH$ rises with temperature~\cite{koizumi2009,iwata2001,ishikawa2024}.  }
\begin{equation}
    \label{eq:mob_rH_fit}
    \begin{split}
    \rH = \num{1.74823} &- \num{6.22e-3}\,T + \num{1.36729e-5}\,T^2 \\
    &- \num{1.44837e-8}\,T^3 + \num{5.86498e-12}\,T^4
    \end{split}
\end{equation}

\citet{tanaka2018} obtained for holes an analytic equation of the temperature and doping dependent variations (see \cref{eq:mob_rH_tanaka}). For this purpose the authors fitted the Hall and conductivity mobility with the model shown in \cref{eq:mobility_CT}. A division of these fits then led to the shown expression, which predicted a steady decrease with rising temperature.
\begin{equation}
    \label{eq:mob_rH_tanaka}
    \rH = 1.16 \left( \frac{T}{\SI{300}{\K}} \right)^{-0.9} \frac{
        1 + \left( \frac{T}{\SI{300}{\K}} \right)^{-1.5} \left( \frac{\na}{\SI{1e19}{\per\cubic\centi\meter}} \right)^{0.7}
    }{
        1 + \left( \frac{T}{\SI{300}{\K}} \right)^{-1.8} \left( \frac{\na}{\SI{3e18}{\per\cubic\centi\meter}} \right)^{0.6}
    }
\end{equation}

\revision{ In summary, there exists no simple and unique translation between Hall and conductivity mobility. Consequently, we are going to present the values as they were published and denote their meaning with $H$ (Hall) resp. $C$ (conductivity) whenever possible, whereas the absence of both indicates that we were unable to classify the results.

  At last, we are going to discuss the results by \citet{darmody2019}, who investigated both the Hall and conductivity mobility, in greater detail because the achieved results of $\rH>5$ for holes contradict commonly agreed values. In that paper, the authors introduced the ratio $\muc/\muH$, which is the inverse of $\rH$, as the ratio of the free charge carriers and the doping concentration. This was based on the assumption that all dopants contribute to the conductivity mobility, i.e., are ionized, which is not the case in 4H-SiC (cp. \cref{sec:incompIon}). For this purpose the authors defined the Hall scattering coefficient equal to $1$. The conductivity mobility values shown in the publication were much lower than the Hall ones, which resulted from the fact that the authors gathered the former from resistivity measurements, assuming that the amount of charge carriers was equal to the doping concentration. A division of the mobility values led in some case to $\rH>5$, which is more than three times the biggest value we encountered in the remaining literature.  }


\subsubsection{Low-Field Mobility}

The simplest way to describe the mobility is to use a constant value~\cite{beyer2011}. Although this approach is also applicable in TCAD tools it mainly serves the purpose of a rough and quick comparison among different materials. Consequently, often no information about doping concentration and temperature are provided. We found a wide range of values for the individual spatial directions and also combined to an effective mobility (see \cref{fig:mobility_values}), which reflects range of dependencies, e.g., doping concentration, temperature and field strength. Nevertheless, it becomes clear that the mobility of electrons is almost one order of magnitude higher than the mobility of holes.

\begin{figure*}[t]
    \centering
    \resizebox{0.99\linewidth}{!}{%
    \input{figures/mobility_values}
    }
    \caption{\label{fig:mobility_values}Constant mobility values for 4H-SiC in directions relative to the c-axis.}
\end{figure*}

The reference chain (a detailed graphical representation is shown in \cref{fig:mobility_ref_chain} in Appendix~\ref{sec:refChainMobility}) shows that these values are not derived from a main source but from small clusters. This indicates that the authors assumed the mobilities as common knowledge, without the need to provide references.


In the data an anisotropy of the mobility in regard to the spatial direction is visible. Some authors argued that it is low enough such that just considering the mobility in the base plane is a reasonable approximation~\cite{albanese2010}. However, \citet{hatakeyama2003} showed that not even within the basal plane the mobility is constant.  The anisotropy of the mobility is most probably caused by the anisotropy of the effective masses~\cite{ishikawa2023,hatakeyama2013,iwata2000,iwata2001,vasconcelos2019} (cp. \cref{sec:dos}) and thus depends on additional parameters such as the temperature~\cite{ishikawa2021}.

We found $14$ investigations on the electron but only three on the hole mobility ratio, i.e., $\mu^\perp/\mu^\parallel$, whose results agree well (see \cref{tab:mobility_uRatio}). All but one fundamental investigation predicted for electrons a higher mobility parallel to the c-axis with $\munperp/\munpara=\num{0.83+-0.07}$. Solely \citet{harima1995} stated the exact opposite, i.e., $1/0.83=1.2$. Also various predominantly qualitative overview papers used $\munperp>\munpara$~\cite{chow1996,chow1997,chow2004,chow2017,cole2004,dhanaraj2010,dmitriev2000,elasser2002,klein2009}, while others predicted no anisotropy at all~\cite{wijesundara2011,sriram1997,neudeck2001,neudeck2006,mukherjee2023}. \revision{Due to the similarity of the values we concluded that the methods are all equally reliable.}

\begin{table}[t]
    \centering
    \setlength{\tabcolsep}{6pt}
    \renewcommand{\arraystretch}{1.1}
    \caption{\label{tab:mobility_uRatio}Fundamental investigations on the ratio of mobilities perpendicular ($\perp$) and parallel ($\parallel$) to the c-axis for electrons and holes.}
    \resizebox{0.58\linewidth}{!}{%
        \begin{tabular}{l|*{5}{c}}
ref. & $\munperp/\munpara$ & $\mupperp/\muppara$ & $T$ & type\footnotemark[1] & method \\ 
 & [1] & [1] & [\si{\K}] && \\ \hline 
 &&&&& \\[-7pt] 
\reftext{Scha94a}\cite{schadt1994a} & \num{0.83+-0.03} & -- & 300 & H & Hall\\ 
\reftext{Scha94}\cite{schaffer1994} & \num{0.83} & -- & 300 & H & Hall\\ 
\reftext{Hari95}\cite{harima1995} & \num{1.2+-0.3} & -- & -- & C & Raman\\ 
\reftext{Josh95}\cite{joshi1995} & \num{0.75} & -- & 300 & C & MC\\ 
\reftext{Son95}\cite{son1995} & \num{0.7} & -- & 2R6 & C & ODCR\\ 
\reftext{Nils96}\cite{nilsson1996} & \num{0.82} & -- & 300 & C & MC\\ 
\reftext{Choy97d}\cite{choyke1997d} & \num{0.86} & -- & 300 & H & Hall\\ 
\reftext{Mats97}\cite{matsunami1997} & $(\num{1.2})^{-1}$ & -- & -- & H & Hall\\ 
\reftext{Mick98}\cite{mickevicius1998} & \num{0.79} & -- & 300 & C & MC\\ 
\reftext{Bell99}\cite{bellotti1999} & -- & \num{1.33} & -- & C & EPM\\ 
\reftext{Iwat00}\cite{iwata2000} & \num{0.7} & -- & -- & H & COTE\\ 
\reftext{Bert01}\cite{bertilsson2001} & \num{0.85} & -- & -- & C & MC\\ 
\reftext{Hata03}\cite{hatakeyama2003} & \num{0.83} & \num{1.15} & 300 & H & Hall\\ 
\reftext{Chen20}\cite{cheng2020} & $(\num{1.75})^{-1}$ - $(\num{1.25})^{-1}$\footnotemark[2] & -- & -- & C & FBMC\\ 
\reftext{Ishi21}\cite{ishikawa2021} & $(\num{1.17})^{-1}$ & -- & 300 & H & Hall\\ 
\reftext{Ishi24}\cite{ishikawa2024} & -- & $(\num{0.9})^{-1}$ & 300 & H & Hall\\ 
\end{tabular}
        \footnotetext[1]{type of mobility: Hall (H), conductivity (C)}
        \footnotetext[2]{ratio is lower for smaller doping concentrations and lower voltages}
    } 
\end{table}

\citet{iwata2000} investigated the mobility in varying in-plane directions, revealing a maximum anisotropy factor for electrons of $0.7$. \citet{cheng2020} stated that previous publications~\cite{schaffer1994} did not specify the exact perpendicular direction. Thus their values for $\munpara/\munperp$ range between $1.25$ and $1.75$, compared to the previously achieved $1.2$~\cite{schaffer1994}.  For holes all investigations predicted a ratio bigger than one with the most common values of \num{1.15+-0.04}. For electric fields $<\SI{50}{\kilo\V\per\cm}$ \citet{vasconcelos2019} calculated a ratio $> 10$, which drops below six for fields $\geq \SI{200}{\kilo\V\per\cm}$.

\citet{ishikawa2021,ishikawa2023} utilized the resistivity ratio to achieve $\mu^\perp/\mu^\parallel$ because in this way the Hall scattering factor cancels. We also found several investigations on temperature and doping concentration dependencies~\cite{ishikawa2021,ishikawa2024,joshi1995,lomakina1973,martinez2002,mickevicius1998,nilsson1996,schadt1994a,schadt1997,schaffer1994}.  \citet{schadt1994a} reported an increasing ratio (up to unity, see \cref{fig:mobility_ratio_temp}) with rising temperature but \citet{schaffer1994} stated that the value of $0.83$ did not change over temperature in homogeneous samples. Only for a wafer with an additional deep donor level a comparable tendency was observed in the range $[0.83,1]$ when varying the temperature within \SIrange{100}{600}{\K}. \citet{joshi1995} showed a slight decrease from $0.78$ to $0.73$ within \SIrange{100}{650}{\K}, \citet{nilsson1996} various values with no clear tendency in $[0.81,0.84]$ within \SIrange{100}{500}{\K} and \citet{mickevicius1998} as well as \citet{ishikawa2021} a slight decrease with rising temperature. The ratio of hole mobilities was reported constant with temperature~\cite{ishikawa2024}.

\begin{figure}[t]
    \centering
    \resizebox{0.99\linewidth}{!}{%
    \input{figures/mobility_ratio_temp}
    }
    \caption{\label{fig:mobility_ratio_temp}Ratio of electron mobilities in the direction perpendicular ($\perp$) and parallel ($\parallel$) to the c-axis versus temperature. The entry ``\reftext{Scha94}~\cite{schaffer1994} DD'' denotes the measurements including deep donor levels (see text).}
\end{figure}

Despite these various dependencies, which render constant values not well suited~\cite{bertilsson2004}, almost exclusively constant factors are used in literature. In accordance with the fundamental investigations shown earlier, the majority agrees upon $\munperp < \munpara$ and $\mupperp > \muppara$ (see \cref{fig:mobility_ratio}).

\begin{figure}[t]
    \centering
    \resizebox{0.86\linewidth}{!}{%
    \input{figures/mobility_ratio}
    }
    \caption{\label{fig:mobility_ratio}Constant values for the ratio of mobilities
      perpendicular ($\perp$) and parallel ($\parallel$) to the c-axis for
      electrons and holes.}
\end{figure}

\paragraph{Doping Dependency}

The low-field mobility in regard to the doping concentration was heavily
investigated~\cite{schaffer1994,schadt1997,burk1998,kinoshita1998,schoner1999,siergiej1999,iwata2000,iwata2000a,pernot2000,pernot2001,hatakeyama2003,terziyska2003,kasamakova-kolaklieva2004,laube2004,negoro2004,pensl2005,pernot2005}\newline
{}~\cite{rambach2006,koizumi2009,contreras2016,fujihara2017,vasilevskiy2017,ishikawa2021,ishikawa2023,ishikawa2024,harima1995,nakashima1997,mickevicius1998,meyer2000,sun2011a,cheng2020,tiwald1999,rutsch2000},
which makes it impossible to show all gathered results in this review. Instead we focus on the models that were developed based on these data.


We found $26$ investigations focused on the doping dependency of electrons in 4H-SiC based on \cref{eq:mobility_CT} (see \cref{tab:mobility_CT_elec}) over a time span of three decades. The maximum mobility varies within a range of \SIrange{800}{1240}{\square\centi\meter\per\V\per\s} while the reference doping concentration $\nref$ varies only slightly in the low \SI{e17}{\per\cubic\centi\m} range. We also highlighted the deviating interpretations of the parameter $N$ in the table, which underlines the large discrepancy in literature. The overview shows that different models share some of their values, which suggests that these were created by combining others or extending them by additional parameters. For $\nref$ the values \SI{1.94e17}{\per\cubic\centi\m} and \SI{2e17}{\per\cubic\centi\m} are used in $16$ out of $28$ models and for $\delta$, \num{0.76} and \num{0.61} are used in $11$ out of $28$. The most prominent value for $\mumin$ is \SI{40}{\square\centi\m\per\V\per\s} in $11$ and \SI{0}{\square\centi\m\per\V\per\s} in $7$ out of $28$ while for $\mumax$ \SIrange{947}{954}{\square\centi\m\per\V\per\s} was chosen in $10$ out of $28$ investigations.

\revision{Hall measurements provided a wide range of mobilities, indicating a high dependency on the sample quality. For the diode I-V approach (DIV) the highest mobility values were achieved for Schottky~\cite{morisette2001} or Merged-Pin Schottky~\cite{sharma2015} diodes. Pure pin diodes delivered lower mobilities, whereat $\mumax=\SI{200}{\square\centi\m\per\V\per\s}$ by \citet{pezzimenti2008} was remarkably low.}

\begin{table*}[p]
    \centering
    \setlength{\tabcolsep}{6pt}
    \renewcommand{\arraystretch}{1.1}
    \caption{\label{tab:mobility_CT_elec}Parameters for the Caughey-Thomas model in \cref{eq:mobility_CT} used to describe the electron mobility.}
    \resizebox{0.99\linewidth}{!}{%
    \begin{tabular}{l|*{15}{c}}
ref. & dir & $\mumin$ & $\mumax$ & $\mu_0$ & $\nref$ & $\delta$ & $\gmin$ & $\gmax$ & $\gamma_{0}$ & $\gref$ & $\gdelta$ & $\gNNref$ & N\footnotemark[1] & K\footnotemark[2] & method \\ 
 && [\si{\square\centi\m\per\V\per\s}] & [\si{\square\centi\m\per\V\per\s}] & [\si{\square\centi\m\per\V\per\s}] & [\si{\per\cubic\centi\m}] & [1] & [1] & [1] & [1] & [1] & [1] & [1] &&&\\ \hline 
 &&&&&&&&&&&&&&& \\[-7pt] 
\rowcolor{no4HColor} \reftext{Ruff94}\cite{ruff1994} & -- & \num{20} & -- & \num{380} & \num{4.50E+17} & \num{0.45} & -- & -- & \num{-3} & -- & -- & -- & -- & -- & --\\ 
\reftext{Scha94}\cite{schaffer1994} & $\perp$ & -- & \num{947} & -- & \num{1.94E+17} & \num{0.61} & -- & \num{-2.15}\footnotemark[3] & -- & -- & -- & -- & S & H & Hall\\ 
\reftext{Mick98}\cite{mickevicius1998} & -- & -- & \num{1071} & -- & \num{1.94E+17} & \num{0.4} & -- & -- & -- & -- & -- & -- & S & C & MC\\ 
\reftext{Rosc98}\cite{roschke1998}\footnotemark[4] & -- & \num{40} & \num{800} & -- & \num{4.00E+17} & \num{0.44} & -- & -- & -- & -- & -- & -- & D & -- & FIT\\ 
\reftext{Shah98}\cite{shah1998} & -- & -- & \num{947} & -- & \num{1.94E+17} & \num{0.61} & -- & -- & -- & -- & -- & -- & -- & -- & --\\ 
\reftext{Wrig98}\cite{wright1998}\footnotemark[5] & -- & \num{88} & -- & \num{970} & \num{1.43E+17} & \num{1} & \num{-0.57} & -- & \num{-2.7} & \num{2.55} & -- & -- & I & -- & --\\ 
\reftext{Levi01a}\cite{levinshtein2001a}\footnotemark[6] & -- & \num{79.8} & \num{855} & -- & \num{2.00E+17} & \num{1} & -- & -- & -- & -- & -- & -- & D & -- & FIT\\ 
\reftext{Mnat01}\cite{mnatsakanov2001}\footnotemark[7] & -- & \num{30} & \num{880} & -- & \num{2.00E+17} & \num{0.67} & -- & -- & -- & -- & -- & -- & D & C & FIT\\ 
\reftext{Mori01}\cite{morisette2001} & $\parallel$ & -- & \num{1141} & -- & \num{1.94E+17} & \num{0.61} & -- & -- & -- & -- & -- & -- & S & C & DIV\\ 
\reftext{Rosc01}\cite{roschke2001}\footnotemark[8] & $\perp$ & \num{40} & \num{950} & -- & \num{2.00E+17} & \num{0.76} & \num{-0.5} & \num{-2.4} & -- & \num{1} & -- & -- & D & -- & FIT\\ 
\reftext{Hata03}\cite{hatakeyama2003} & -- & -- & \num{954} & -- & \num{1.28E+17} & \num{0.61} & -- & -- & -- & -- & -- & -- & I & H & Hall\\ 
\reftext{Kaga04}\cite{kagamihara2004}\footnotemark[9] & -- & -- & \num{977} & -- & \num{1.17E+17} & \num{0.49} & -- & -- & -- & -- & -- & -- & S & H & Hall\\ 
\reftext{Adac05}\cite{adachi2005} & -- & -- & \num{1400} & -- & \num{1.00E+17} & \num{0.5} & -- & -- & -- & -- & -- & -- & I & -- & --\\ 
\reftext{Bala05}\cite{balachandran2005} & -- & \num{40} & \num{950} & -- & \num{2.00E+17} & \num{0.73} & -- & \num{-2.4} & -- & -- & -- & \num{-0.76} & -- & -- & --\\ 
\reftext{Werb07}\cite{werber2007} & -- & \num{33} & \num{771} & -- & \num{2.00E+17} & \num{0.76} & -- & -- & -- & -- & -- & -- & D & C & DIV\\ 
\reftext{Cha08}\cite{cha2008} & -- & -- & \num{950} & -- & \num{1.90E+17} & \num{0.6} & \num{1} & \num{-2.15} & -- & -- & -- & \num{0.05} & D & -- & --\\ 
\reftext{Pezz08}\cite{pezzimenti2008} & -- & \num{6} & \num{200} & -- & \num{1.00E+16} & \num{1} & -- & -- & -- & -- & -- & -- & -- & C & DIV\\ 
\reftext{Habi11}\cite{habib2011}\footnotemark[10] & -- & \num{40} & -- & \num{910} & \num{2.00E+17} & \num{0.76} & \num{-1.538} & -- & \num{-2.397} & \num{0.75} & \num{0.722} & -- & S & -- & FIT\\ 
\reftext{Hata13}\cite{hatakeyama2013}\footnotemark[11] & -- & \num{5} & \num{1010} & -- & \num{1.25E+17} & \num{0.65} & \num{-0.57} & \num{-2.6} & -- & \num{2.4} & \num{-0.146} & -- & S & -- & FIT\\ 
\reftext{Stef14}\cite{stefanakis2014}\footnotemark[12] & $\perp$ & \num{28} & \num{950} & -- & \num{1.94E+17} & \num{0.61}\footnotemark[13] & -- & \num{-2.4} & -- & -- & -- & \num{0.73} & I & -- & FIT\\ 
\reftext{Shar15}\cite{sharma2015} & $\parallel$ & \num{40} & \num{947} & -- & \num{1.94E+17} & \num{0.61} & \num{-0.5} & \num{-2.9} & -- & -- & -- & \num{2.4} & S & C & DIV\\ 
\reftext{Arva17}\cite{arvanitopoulos2017} & -- & \num{40} & \num{950} & -- & \num{1.94E+17} & \num{0.61} & \num{-1.536} & \num{-2.4} & -- & -- & -- & -- & D & -- & --\\ 
\reftext{Vasi17}\cite{vasilevskiy2017} & -- & \num{20} & \num{950} & -- & \num{2.00E+17} & \num{0.8} & -- & -- & -- & -- & -- & -- & D & H & Hall\\ 
\reftext{Ishi21}\cite{ishikawa2021} & $\perp$ & \num{40} & -- & \num{970} & \num{2.4e17} & \num{0.7} & -- & -- & \num{-2.58}\footnotemark[14] & -- & -- & -- & D & H & Hall\\ 
 & $\parallel$ & \num{60} & -- & \num{1120} & \num{2.3e17} & \num{0.74} & -- & -- & \num{-2.67}\footnotemark[14] & -- & -- & -- & D & H & Hall\\ 
\reftext{Lech21}\cite{lechner2021} & $\parallel$ & \num{15.48} & -- & \num{1415.4} & \num{1.80E+17} & \num{0.560723} & -- & -- & \num{-2.5} & -- & -- & -- & -- & -- & --\\ 
\reftext{Rao22}\cite{rao2022} & -- & \num{40} & \num{950} & -- & \num{2.00E+17} & \num{0.76} & \num{-0.5} & \num{-2.15} & -- & -- & -- & \num{-0.76} & I & -- & --\\ 
\reftext{Ishi23}\cite{ishikawa2023} & $\perp$ & \num{40} & -- & \num{1000} & \num{2.2e17} & \num{0.68} & \num{-0.7} & -- & \num{-2.9} & -- & -- & \num{-2.5} & D & H & Hall\\ 
 & $\parallel$ & \num{20} & -- & \num{1240} & \num{2e17} & \num{0.64} & \num{0.3} & -- & \num{-3.2} & -- & -- & \num{-2.7} & D & H & Hall
\end{tabular}
\footnotetext[1]{meaning of N: doping (D), ionized (I), sum of all dopants (S), intrinsic (N)}
\footnotetext[2]{type of mobility: Hall (H), conductivity (C)}
\footnotetext[3]{$T>\SI{250}{\K}$ in \hkl<11-20> direction ($\gmax=-2.4$ towards \hkl<0001>). $T<\SI{250}{\K}$: $\gmax=-1.18$ resp. $-1.2$}
\footnotetext[4]{fitted to~\cite{harris1995}}
\footnotetext[5]{parameters referenced from Si based investigation~\cite{arora1982} but not found there}
\footnotetext[6]{fitted to~\cite{lindefelt1998,schaffer1994}}
\footnotetext[7]{fitted to~\cite{schaffer1994} using dedicated Hall scattering factor~\cite{rutsch1998}}
\footnotetext[8]{fitted to~\cite{carterjr.1999,schaffer1994,nakashima1996,kinoshita1998,mickevicius1998,joshi1995}}
\footnotetext[9]{fitted to~\cite{schaffer1994,pernot2000,matsunami1997,gotz1993,carterjr.1999,schoner1999}}
\footnotetext[10]{fitted to~\cite{mnatsakanov2002,nakashima1996,schaffer1994,carterjr.1999}}
\footnotetext[11]{fitted to~\cite{matsunami1997}}
\footnotetext[12]{fitted to~\cite{kinoshita1998,pernot2005,koizumi2009,nawaz2010}}
\footnotetext[13]{In the paper $\delta=-0.61$ was stated, which did not match the shown plots.}
\footnotetext[14]{$\nd=\SI{2.1e15}{\per\cubic\centi\m}$, increases with doping concentration}
    }
\end{table*} 

The models agree qualitatively (see \cref{fig:mobility_CT_elec}). Exceptions are the ones by \citet{ruff1994}, which was fitted to 6H, \citet{pezzimenti2008}, who predicted a very low maximum mobility, and \citet{zhang2018}, who used the model in \cref{eq:mobility_masetti} with the parameters in \cref{tab:mobility_masetti_param} to predict the decrease in mobility at higher doping densities. In the plot an additional reduction at doping densities around $\SI{e20}{\per\cubic\centi\m}$ is visible. Since the authors did not provide $\mumax$ we picked $\SI{950}{\square\centi\meter\per\V\per\s}$. The remaining descriptions mainly differ in the value of $\mumax$.

\begin{figure*}[t]
    \centering
    \resizebox{1\linewidth}{!}{%
    \input{figures/mobility_CT_doping_elec}
    }
    \caption{\label{fig:mobility_CT_elec}Electron mobility approximations using the Caughey-Thomas model in \cref{eq:mobility_CT} at $T=\SI{300}{\K}$. The models are only shown in the region used for characterization.}
\end{figure*}

\begin{table*}[t]
    \centering
    \setlength{\tabcolsep}{6pt}
    \renewcommand{\arraystretch}{1.1}
    \caption{\label{tab:mobility_masetti_param}Doping dependent mobility parameters for the model in \cref{eq:mobility_masetti}.}
    \resizebox{0.75\linewidth}{!}{%
    \begin{tabular}{c|*{8}{c}}
        ref & mob. & $\mumin$ & $\mu_\mathrm{min2}$ & $\nref$ & $\delta$ & $\mu_1$ & $N_\mathrm{ref2}$ & $\kappa$ \\ 
        && [\si{\square\centi\meter\per\V\per\s}] & [\si{\square\centi\meter\per\V\per\s}] & [\si{\per\cubic\centi\m}] & [1] & [\si{\square\centi\meter\per\V\per\s}] & [\si{\per\cubic\centi\m}] & [1] \\ \hline
        \reftext{Zhan18}\cite{zhang2018} & $\mun$ & \num{88} & \num{0} & \num{5e18} & \num{1} & \num{43.4} & \num{3.43e20} & \num{2} \\
        & $\mup$ & \num{44} & \num{0} & \num{5e19} & \num{1} & \num{29} & \num{6.1e20} & \num{2}
    \end{tabular}
    }
\end{table*} 

We also found a more direct modeling of the mobility as shown in \cref{eq:mob_direct}~\cite{baliga2006,baliga2019}. According to the authors these expressions are, however, based on \citet{ruff1994} which only contains values on 6H-SiC. For the sake of comparison we added it to \cref{fig:mobility_CT_elec} as well.
\begin{equation}
\begin{split}  
    \label{eq:mob_direct}
    \mun &= \frac{\num{4.05e13} + 20\,\nd^{0.61}}{\num{3.55e10} + \nd^{0.61}}\\
    \mup &= \frac{\num{4.05e13} + 10\,\na^{0.65}}{\num{3.3e11} + \na^{0.65}}
\end{split}    
\end{equation} 

Similarly, \citet{pernot2000} fitted the Hall mobility to the free electron density $n$ as shown in \cref{eq:mob_pernot}. In the plot we set $n=N$.
\begin{equation}
    \label{eq:mob_pernot}
    \mun = \num{-39000} + \num{7436}\,\log(n) -\num{450.5}\,\log^2(n) + \num{8.81}\,\log^3(n)
\end{equation}

For the hole mobility less models were proposed with a larger spread in the respective parameters (see \cref{tab:mobility_CT_hole}). Qualitatively, the decrease of the hole mobility, compared to the electron one, starts at higher doping densities (see \cref{fig:mobility_CT_hole}), i.e., the maximum mobility can be maintained longer. In contrast to electrons we do not see a reuse of model parameters here, which might be an explanation for the big discrepancies: The models differ in the absolute mobility values at low and high doping concentrations, at which concentration and even at which rate the transition occurs. An outlier is the model by \citet{pezzimenti2008}, showing a very low mobility. \revision{Once again the pin diodes provided rather low mobility results, but in contrast to electrons occasionally also Hall measurements delivered similar values.}

\begin{table*}[t]
    \centering
    \setlength{\tabcolsep}{6pt}
    \renewcommand{\arraystretch}{1.1}
    \caption{\label{tab:mobility_CT_hole}Parameters for the Caughey-Thomas model in \cref{eq:mobility_CT} used to describe the hole mobility.}
    \resizebox{0.99\linewidth}{!}{%
    \begin{tabular}{l|*{15}{c}}
ref. & dir & $\mumin$ & $\mumax$ & $\mu_0$ & $\nref$ & $\delta$ & $\gmin$ & $\gmax$ & $\gamma_{0}$ & $\gref$ & $\gdelta$ & $\gNNref$ & N\footnotemark[1] & K\footnotemark[2] & method \\ 
 && [\si{\square\centi\m\per\V\per\s}] & [\si{\square\centi\m\per\V\per\s}] & [\si{\square\centi\m\per\V\per\s}] & [\si{\per\cubic\centi\m}] & [1] & [1] & [1] & [1] & [1] & [1] & [1] &&&\\ \hline 
 &&&&&&&&&&&&&&& \\[-7pt] 
\rowcolor{no4HColor} \reftext{Ruff94}\cite{ruff1994} & -- & \num{5} & -- & \num{70} & \num{1.00E+19} & \num{0.5} & -- & -- & \num{-3} & -- & -- & -- & -- & -- & --\\ 
\reftext{Scha94}\cite{schaffer1994} & $\perp$ & \num{15.9} & \num{124} & -- & \num{1.76E+19} & \num{0.34} & -- & -- & -- & -- & -- & -- & S & H & Hall\\ 
\reftext{Shah98}\cite{shah1998} & -- & \num{25.9} & \num{128.1} & -- & \num{1.00E+19} & \num{0.24} & -- & -- & -- & -- & -- & -- & -- & -- & --\\ 
\reftext{Wrig98}\cite{wright1998}\footnotemark[3] & -- & \num{74} & -- & \num{43} & \num{1.43E+17} & \num{1} & \num{-0.57} & -- & \num{-2.7} & \num{2.55} & -- & -- & I & -- & --\\ 
\reftext{Levi01a}\cite{levinshtein2001a}\footnotemark[4] & -- & \num{65} & \num{114} & -- & \num{5.00E+17} & \num{1} & -- & -- & -- & -- & -- & -- & D & -- & FIT\\ 
\reftext{Mnat01}\cite{mnatsakanov2001}\footnotemark[5] & -- & \num{33} & \num{117} & -- & \num{1.00E+19} & \num{0.5} & -- & -- & -- & -- & -- & -- & D & C & FIT\\ 
\reftext{Hata03}\cite{hatakeyama2003} & -- & \num{15.9} & \num{120} & -- & \num{1.80E+18} & \num{0.65} & -- & -- & -- & -- & -- & -- & I & H & Hall\\ 
\reftext{Mats04}\cite{matsuura2004} & -- & \num{37.6} & \num{106} & -- & \num{2.97E+18} & \num{0.356} & -- & -- & -- & -- & -- & -- & S & H & Hall\\ 
\reftext{Bala05}\cite{balachandran2005} & -- & \num{53.3} & \num{105.4} & -- & \num{2.20E+18} & \num{0.7} & -- & \num{-2.1} & -- & -- & -- & -- & -- & -- & --\\ 
\reftext{Werb07}\cite{werber2007} & -- & \num{10} & -- & \num{81} & \num{1.00E+19} & \num{0.5} & -- & -- & -- & -- & -- & -- & D & C & DIV\\ 
\reftext{Cha08}\cite{cha2008} & -- & \num{16} & \num{140} & -- & \num{1.70E+19} & \num{0.34} & \num{-1.6} & \num{-2.14} & -- & -- & -- & \num{0.17} & D & -- & --\\ 
\reftext{Pezz08}\cite{pezzimenti2008} & -- & \num{2} & \num{20} & -- & \num{1.00E+16} & \num{1} & -- & -- & -- & -- & -- & -- & -- & C & DIV\\ 
\reftext{Koiz09}\cite{koizumi2009}\footnotemark[6] & -- & -- & \num{114.1} & -- & \num{5.38E+18} & \num{0.66} & -- & \num{-2.72} & -- & -- & \num{-0.35} & \num{2.44} & D & H & Hall\\ 
\reftext{Habi11}\cite{habib2011}\footnotemark[7] & -- & \num{40} & -- & \num{82} & \num{6.30E+18} & \num{0.55} & \num{-1.538} & -- & \num{-2.2397} & \num{0.75} & \num{0.722} & -- & S & -- & FIT\\ 
\reftext{Hata13}\cite{hatakeyama2013}\footnotemark[8] & -- & -- & \num{113.5} & -- & \num{2.40E+18} & \num{0.69} & \num{-0.57} & \num{-2.6} & -- & \num{2.9} & \num{-0.2} & -- & S & -- & FIT\\ 
\reftext{Liau15}\cite{liaugaudas2015} & -- & -- & \num{75} & -- & \num{2.00E+19} & \num{0.7} & -- & -- & -- & -- & -- & -- & D & C & LTPL \\ 
\reftext{Shar15}\cite{sharma2015} & $\parallel$ & \num{15.9} & \num{124} & -- & \num{1.76E+19} & \num{0.34} & \num{-0.5} & \num{-2.9} & -- & -- & -- & \num{2.3} & S & C & DIV\\ 
\reftext{Tana18}\cite{tanaka2018} & -- & -- & \num{110} & -- & \num{3e18} & \num{0.6} & -- & \num{-3} & -- & -- & -- & \num{-1.8} & D & H & DFT\\ 
 & -- & -- & \num{95} & -- & \num{1e19} & \num{0.7} & -- & \num{-2.1} & -- & -- & -- & \num{-1.5} & D & C & DFT\\ 
\reftext{Lech21}\cite{lechner2021} & -- & \num{2.529} & -- & \num{469.42607} & \num{1.28E+19} & \num{0.332645} & -- & -- & \num{-2} & -- & -- & -- & -- & -- & --\\ 
\reftext{Rao22}\cite{rao2022} & -- & \num{15.9} & \num{125} & -- & \num{1.76E+19} & \num{0.76} & \num{-0.5} & \num{-2.15} & -- & -- & -- & \num{-0.76} & I & -- & --\\ 
\reftext{Ishi24}\cite{ishikawa2024} & $\perp$ & \num{20} & -- & \num{74} & \num{6.2e18} & \num{0.72} & \num{-2.2} & -- & \num{-2.3} & -- & -- & \num{-0.9} & D & H & Hall\\ 
 & $\parallel$ & \num{20} & -- & \num{63} & \num{6.4e18} & \num{0.83} & \num{-2.2} & -- & \num{-2.3} & -- & -- & \num{-0.9} & D & H & Hall
\end{tabular}
\footnotetext[1]{meaning of N: doping (D), ionized (I), sum of all dopants (S), intrinsic (N)}
\footnotetext[2]{type of mobility: Hall (H), conductivity (C)}
\footnotetext[3]{parameters referenced from Si based investigation~\cite{arora1982} but not found there}
\footnotetext[4]{fitted to~\cite{lindefelt1998,schaffer1994}}
\footnotetext[5]{fitted to~\cite{schaffer1994} using dedicated Hall scattering factor~\cite{rutsch1998}}
\footnotetext[6]{fitting done by \citet{stefanakis2014}}
\footnotetext[7]{fitted to~\cite{mnatsakanov2002,nakashima1996,schaffer1994,carterjr.1999}}
\footnotetext[8]{fitted to~\cite{matsunami1997}}
    }
\end{table*} 

\citet{negoro2004} pointed out that the value $\mumin=\SI{16}{\square\centi\meter\per\V\per\s}$ proposed by \citet{hatakeyama2003} was too high. Similarly \citet{stefanakis2014} argued that \citet{schaffer1994} overestimated the hole mobility at high doping values, although the latest studies indicated a value larger than zero. Exceptional in this regard is the large value of \citet{wright1998} that even exceeded some values for $\mumax$, which vary in the range of \SIrange{75}{140}{\square\centi\meter\per\V\per\s}.

\begin{figure*}[t]
    \centering
    \resizebox{1\linewidth}{!}{%
    \input{figures/mobility_CT_doping_hole}
    }
    \caption{\label{fig:mobility_CT_hole}Hole mobility approximations by the Caughey-Thomas model in \cref{eq:mobility_CT} at $T=\SI{300}{\K}$. The models are only shown in the region used for characterization.}
\end{figure*}

We also added the model by \citet{zhang2018} to \cref{fig:mobility_CT_hole}, who utilized \cref{eq:mobility_masetti} with the parameters in \cref{tab:mobility_masetti_param}. Since the authors did not define the maximum mobility we picked $\mumax=\SI{80}{\square\centi\meter\per\V\per\s}$. This model shows the most delayed transition from high to low mobility in literature. In addition, we fitted the doping dependent parameters by \citet{rambach2008} using a polynomial of degree two (see \cref{eq:mob_rambach_dop_dep}). Although this fit solely used values in a doping range of \SIrange{0.5e19}{5e19}{\per\cubic\centi\m} we extended the model in the plot, showing that it is also accurate well beyond these borders. Thereby we picked $\mumin=\SI{15.9}{\square\centi\meter\per\V\per\s}$ and $\mumax=\SI{124}{\square\centi\meter\per\V\per\s}$.
\begin{equation}
    \label{eq:mob_rambach_dop_dep}
    \begin{split}
        \gmin(\na)=\gmax(\na) &= -1.24 + \num{5e-20}\, \na - \num{4e-40}\, \na^2 \\
        \delta(\na) &= 1.14 + \num{9.4e-21}\, \na + \num{5.55e-40}\, \na^2 \\
        \nref(\na) &= \num{1.2e18} + \num{5.4e-2}\, \na + \num{2e-21}\, \na^2
    \end{split}
\end{equation}

Some authors also investigated the impact of compensation on the mobility. \citet{pernot2005} fitted the Hall mobility using the logarithmic doping concentration for non-compensated (see Eq.~(\ref{eq:mob_pernot_noncomp})) and weakly-compensated (see Eq.~(\ref{eq:mob_pernot_weakcomp})) devices. In the graphical representation, where we used $\na=N$, we only show the former case, since the latter results in negative mobility values.
%
\begin{eqnarray}
\text{non-compensated: } \sigma R_H(\SI{292}{\K}) &=& \num{2964.3} - \num{648.72}\log(\na) + \num{53.393} \log^2(\na) \nonumber \\
    && - \num{1.8717} \log^3(\na) + \num{0.002296} \log^4(\na) \label{eq:mob_pernot_noncomp}\\
\text{weakly-compensated: }  \sigma R_H(\SI{292}{\K}) &=& \num{24617} - \num{5982} \log(\na) + \num{536.12} \log^2(\na) \nonumber \\
    && - \num{21.151} \log^3(\na) + \num{0.30937} \log^4(\na) \label{eq:mob_pernot_weakcomp}
\end{eqnarray}


\paragraph{Temperature Dependency}

In our review we encountered numerous investigations of the
Hall~\cite{parisini2013,arvinte2017,contreras2006,contreras2016,dixit2023,gotz1993,habib2011,ishikawa2021,ishikawa2023,ishikawa2024,itoh1994,iwata2000,iwata2000a,iwata2001,kagamihara2004,kasamakova-kolaklieva2004,kimoto1997,kimoto2001,kinoshita1998,koizumi2009,laube2004,lomakina1972,matsunami1997,matsunami2004a}\newline
{}\cite{matsunami2020,matsuura2004,mnatsakanov2002,neimontas2006,nipoti2013,pensl2005,pernot2000,pernot2001,pernot2005,rambach2006,rutsch2000,schadt1997,schaffer1994,schoner1994,schoner1999,tanaka2018,terziyska2003} and conductivity mobility~\cite{izzo2008,joshi1995,lavia2005,liaugaudas2015,martinez2002,meyer2000} variations with changing temperature. 

The temperature scaling parameters of electrons in \cref{eq:mobility_CT} (see \cref{tab:mobility_CT_elec}) showed a consistent picture. We found $16$ fits, which all agree on an increase of $\mumin$, $\mumax$ and $\mu_0$ for decreasing temperatures. An exception is $\mumin$ by \citet{ishikawa2023} parallel to the c-axis, but in this case the remaining temperature dependencies ensure the previously described behavior (see \cref{fig:mobility_CT_temp_elec}). The values for $\gdelta$ and $\gNNref$ are less consistent; even positive and negative values were proposed. Nevertheless, due to the complicated interactions discussing parameters in isolation is not meaningful. Instead it is favorable to investigate the complete model at once.

\begin{figure*}[t]
    \centering
    \resizebox{1\linewidth}{!}{%
    \input{figures/mobility_CT_temp_elec}
    }
    \caption{\label{fig:mobility_CT_temp_elec}Electron mobility models for varying temperature according to \cref{eq:mobility_CT} for a doping of \SI{4e16}{\per\cubic\centi\meter}.}
\end{figure*}

For an in-detail comparison we plot all models for $N=\SI{4e16}{\per\cubic\centi\meter}$ (see \cref{fig:mobility_CT_temp_elec}). As we discussed earlier, the mobility values change with doping concentration, so this plot serves as a qualitative comparison of the temperature dependency. In this regard the predictions by \cref{eq:temp_pheno} are an exception, because the mobility always approaches zero for $T \to 0$, independent of the doping concentration.

For low temperatures ($< \SI{200}{\kelvin}$) almost all models agree on high
mobilities (\SI{1000}{\square\centi\m\per\V\per\s}) that further increase with
decreasing temperature. This includes the description according to
Eq.~(\ref{eq:mob_mup}), whose parameters are shown in
\cref{tab:mobility_temp_beta}. However, such a tendency contradicts the increasing impact of impurity scattering, which is only covered by three models (four if we include the one by \citet{hatakeyama2013} that is only plotted for $T>\SI{300}{\K}$): \citet{lavia2005} used \cref{eq:temp_pheno} with the parameters shown in \cref{eq:mob_temp_lavia}, whereat we picked suitable values for $A$ and $B$. \citet{izzo2008}, who also used this mode, derived $n=3.06$.
\begin{equation}
    \label{eq:mob_temp_lavia}
    n=3\ ,\qquad A=1\ ,\qquad B=\num{1e-11}
\end{equation}

\begin{table*}[t]
    \centering
    \setlength{\tabcolsep}{8pt}
    \renewcommand{\arraystretch}{1.1}
    \caption{\label{tab:mobility_temp_beta}Parameters for the model in Eq.~(\ref{eq:mob_mup}) and Eq.~(\ref{eq:mob_bp}).}
    \resizebox{0.75\linewidth}{!}{%
    \begin{tabular}{c|*{7}{c}}
        ref & mob. & $\bmin$ & $\bmax$ & $\Np$ & $\eta$ & K &  method \\ 
        && [1] & [1] & [\si{\per\cubic\centi\m}] & [1] && \\ \hline
        \reftext{Kaga04}\cite{kagamihara2004} & $\mun$ & \num{1.54} & \num{2.62} & \num{1.14e17} & \num{1.35} & -- & FIT \\
        \reftext{Mats04}\cite{matsuura2004} & $\mup$ & \num{2.51} & \num{3.04} & \num{8.64e17} & \num{0.456} & H & Hall
    \end{tabular}
    }
\end{table*} 

For the model in Eq.~(\ref{eq:mobility_mnat_mu}) we used the parameters provided by \citet{neimontas2006} (see \cref{tab:mobility_mnat_param}), whereat \citet{mnatsakanov2001} already proposed the values $\gmax=2.6$ and $\gI=0.5$ earlier. The predicted mobility values are lower than in other investigations, because the model described the electron mobility in heavily p-doped 4H-SiC, i.e., the minority carrier mobility.
\begin{table*}[t]
    \centering
    \setlength{\tabcolsep}{6pt}
    \renewcommand{\arraystretch}{1.1}
    \caption{\label{tab:mobility_mnat_param}Parameters for the model in Eq.~(\ref{eq:mobility_mnat_mu}) and Eq.~(\ref{eq:mobility_mnat_B}).}
    \resizebox{0.8\linewidth}{!}{%
    \begin{tabular}{c|*{8}{c}}
        ref & mob. & $\mumin$ & $\mumax$ & $\nref$ & $\delta$ & $\gmax$ & $\gI$ \\ 
        && [\si{\square\centi\meter\per\V\per\s}] & [\si{\square\centi\meter\per\V\per\s}] & [\si{\per\cubic\centi\m}] & [1] & [1] & [1] \\ \hline
        \reftext{Neim06}\cite{neimontas2006} & $\mun$ & \num{100} & \num{320} & \num{2e17} & \num{0.67} & \num{2.6} & \num{0.5}
    \end{tabular}
    }
\end{table*} 

For the model proposed by \citet{uhnevionak2015} (cp. \cref{eq:mobility_sum_max}) we used the parameters shown in \cref{eq:mob_temp_uhne} and the remaining ones according to \citet{roschke2001}. The increase of the mobility at low temperatures is thereby not automatically caused by splitting the maximum mobility but relies on a careful selection of all parameters.
\begin{equation}
    \label{eq:mob_temp_uhne}
    \begin{split}
    \mu_\mathrm{max1}=\SI{500}{\square\centi\m\per\V\per\s},\qquad \mu_\mathrm{max2}=\SI{450}{\square\centi\m\per\V\per\s},\qquad \gamma_\mathrm{max1} = -11.6\\
    \gamma_\mathrm{max2} = -2.74,\qquad \gNNref=-12.5
    \end{split}
\end{equation}

At $T>\SI{300}{\K}$ all models predict a decreasing mobility. Some observed a dependency of $T^{-2.1}$ to $T^{-2.5}$ instead of the expected $T^{-1.5}$~~\cite{matsunami1997,matsunami2004a,matsunami2020}. The temperature dependency even showed anisotropy. \citet{schaffer1994} proposed a temperature scaling parallel to the c-axis of $T^{-2.4}$ and perpendicular of $T^{-2.15}$ above \SI{200}{\K}. Below that temperature they got $T^{-1.2}$ (parallel) and $T^{-1.18}$ (perpendicular). \citet{ishikawa2021} investigated the temperature dependency using a multiplicative factor $T^{-\beta}$ for three differing doping concentrations, observing an absolute decrease of $\beta$ with increasing doping concentration and an anisotropy~\cite{ishikawa2024}. In \cref{fig:mobility_CT_temp_elec} we chose $\beta=-2.58$ perp. to the c-axis and $\beta=-2.67$ parallel to it which corresponds to measurements at a doping concentration of $\nd = \SI{2.1e15}{\per\cubic\centi\m}$. \citet{kagamihara2004} stated that according to theoretical consideration the temperature parameters are $1.5$ for low temperatures and $2.6$ for high ones, which is close to $\bmin=1.54$ and $\bmax=2.62$ in their fit. \citet{mitchel2007} used $\propto T^{-1.8}$, which is close to the expected $T^{-1.5}$ due to phonon scattering. \citet{buono2012} pointed out that $\gmax=-2.15$ for both charge carriers is smaller than the ideal factor of $-1.5$ that is expected from lattice scattering. This was believed to be due to non-polar optical-phonon scattering. \citet{ladesmartin2000} achieved $-1.8 \geq \gmax \geq -2.2$, which lies between acoustical-mode phonon ($-1.5$) and optical-mode phonon ($-2.5$) scattering. Thus, the author assumed that other scattering factors contributed less to limit the mobility.

We also show the simplified models by \citet{cheng2022}, who fitted the expression shown in Eq.~(\ref{eq:mob_temp_cheng}) to the measurements by \citet{schaffer1994}, and \citet{baliga2006}, who fitted Eq.~(\ref{eq:mob_temp_baliga_elec}) to data by \citet{koizumi2009}.
\begin{eqnarray}  
    \label{eq:mob_temp_cheng}
    \mun(T) &=& 5422\,\exp\left(-\frac{T}{128}\right) + 95 \\
    \label{eq:mob_temp_baliga_elec}
    \mun(T) &=& 1140\,\left(\frac{T}{300}\right)^{-2.7}
\end{eqnarray}

The investigations of hole mobilities show the same tendencies for $\gmin$, $\gmax$ and $\gamma_0$ and inconsistencies for $\gdelta$ and $\gNNref$ that we observed for electrons. Although overall less investigations are available, the amount of temperature dependency studies is comparable ($13$ for holes, $16$ for electrons). The main difference is that only one model, i.e., the one by \citet{hatakeyama2013}, was proposed that covers the decreasing mobility at low temperatures (see \cref{fig:mobility_CT_temp_hole}). For increasing temperature all models agree upon a continuous decrease of the mobility. The parameters for the model by \citet{matsuura2004} are shown in \cref{tab:mobility_temp_beta} and for the model by \citet{baliga2006} we used the parameters from \cref{eq:mob_temp_baliga_hole}.
\begin{equation}
    \label{eq:mob_temp_baliga_hole}
    \mup(T) = 120\,\exp\left(\frac{T}{300}\right)^{-3.4}
\end{equation}

\begin{figure*}
    \centering
    \resizebox{1\linewidth}{!}{%
    \input{figures/mobility_CT_temp_hole}
    }
    \caption{\label{fig:mobility_CT_temp_hole}Hole mobility models for varying temperature according to \cref{eq:mobility_CT} for a doping of \SI{4e16}{\per\cubic\centi\meter}.}
\end{figure*}


\paragraph{Analysis}

It is common in literature to mix the parameters of different models, e.g., using the mobility values from one and the temperature scaling from another. We regularly encountered~\cite{ayalew2004,lee2002,chen2015,arvanitopoulos2017,arvanitopoulos2019} a combination of the models by \citet{schaffer1994} and \citet{roschke2001} (see \cref{fig:mobility_ref_chain_low_elec} in Appendix~\ref{sec:refChainMobility}). These are also the most influential publications, whereat newer studies were barely adopted in literature. On the positive side almost all references go back to one of the fundamental studies and almost exclusively 4H values are used. The only exception are the values by \citet{ruff1993} and \citet{ruff1994} which were referenced three times.


For holes the majority of the values goes back to a single publication by \citet{schaffer1994} (see \cref{fig:mobility_ref_chain_low_hole} in Appendix~\ref{sec:refChainMobility}), whereat, again, newer values were not widely adopted in the community. We are also unsure about the temperature scaling of the hole mobility in the publications citing \citet{schaffer1994}, because the latter stated $\gmax=-2.15$ only for electrons. There were also multiple instances~\cite{zhao2003,zhang2009,zhang2010,wang1998,wang1999,chen2015} where $\mumax$ was interpreted as $\mu_0$, i.e., without subtracting $\mumin$.


We were unable to retrace some values~\cite{balachandran2005,nawaz2010,usman2014,lophitis2018} back to any scientific publication. In other occasions references were provided but the presented values could not be found therein~\cite{lechner2021,lophitis2018}. Some publications~\cite{denapoli2022,kimoto2014a,rakheja2020} even present multiple models that contradict each other.

At last we want to discuss two previous reviews on mobility models, as we encountered problematic parameters. We begin with the analysis by \citet{stefanakis2014} who investigated six models and fitted a seventh to measurement results for holes. The model denoted as ``Reggio Calabria Uni.'' was cited from \citet{pezzimenti2013} but goes back to \citet{schaffer1994} for holes and \citet{roschke2001} for electrons. In addition, the shown parameters for this model contain some flaws: $\delta$ of the electrons should be $0.76$ instead of $0.34$, while $\gNNref$ of the holes should be $-0.34$ instead of $-0.76$. A similar confusion occurred for the values of the model denoted as ``Nawaz''~\cite{nawaz2010}, which used for the electrons $\delta=0.73$ instead of the stated $0.34$ and $\gNNref=-0.76$ instead of $0.73$. For holes $\mumin$ should be \SI{53.3}{\square\centi\meter\per\V\per\s} instead of the stated \SI{15.3}{\square\centi\meter\per\V\per\s}. The authors also fitted the measurements presented by \citet{koizumi2009} but extracted $\nref$ at a temperature of \SI{400}{\K} instead of \SI{300}{\K}. The value $\gNNref=2.44$ further inferred a decrease of $\nref$ with rising temperature, but the measurements showed the opposite. Finally, there is a typographical error in $\delta$ of the proposed model for electrons, which should be $0.61$ instead of the proposed $-0.61$.

The second overview paper was published by \citet{tian2020}. For the values cited from \citet{nawaz2010} $\delta$ of the electrons should be $0.73$ instead of the stated $0.34$ and $\gNNref=-0.76$ instead of $0.73$. For holes $\mumin$ should be \SI{53.3}{\square\centi\meter\per\V\per\s} instead of \SI{15.3}{\square\centi\meter\per\V\per\s}. Surprisingly, these are the same discrepancies that we discovered for the review discussed in the last paragraph. Despite these variations, the values from \citet{nawaz2010} achieved the best results and was chosen for the simulations by the authors. For an other approach, the maximum mobility for electrons was changed from \SI{950}{\square\centi\meter\per\V\per\s}~\cite{buono2012} to \SI{947}{\square\centi\meter\per\V\per\s}. The model proposed by \citet{megherbi2018} was extended by a temperature scaling, which matched an earlier publication by the same authors~\cite{megherbi2015} with deviations for $\gNNref$ ($0$ instead of $-0.76$ for electrons and $-0.34$ for holes).

The models called ``Bakowski'' by \citet{stefanakis2014} and ``Gustaffson'' by \citet{tian2020} share the same values, which were primarily adopted from \citet{schaffer1994}. However, the origin of $\mumin=\SI{88}{\square\cm\per\V\per\s}$ for electrons and $\mumin=\SI{74}{\square\cm\per\V\per\s}$ for holes is indeterminate. Both reviews denoted \citet{bakowski1997} as the model source, but we were unable to locate it there. In fact, \citet{wright1996,wright1998} first proposed the stated values in the 1990s and provided as reference the publication by \citet{arora1982}. In that investigation of silicon the authors stated for the electron mobility $\mumin=\SI{88.3}{\square\cm\per\V\per\s}$ at $\SI{300}{\K}$ and $\mumin=\SI{73.78}{\square\cm\per\V\per\s}$ at $\SI{400}{\K}$, which seemingly got adapted by \citet{wright1998}. A comprehensive listing of all inconsistencies for the mobility can be found in Appendix~\ref{sec:mobility_appendix}.


\subsubsection{High-Field Mobility}

Early publications on SiC had to rely on silicon parameters for the high-field mobility~\cite{ruff1994}, but starting from the year 1995 we identified $20$ investigations of the electric field dependency of electron and hole velocities~\cite{akturk2009,mickevicius1998,khan2000,hjelm2003,ardaravicius2005,belas2022,bellotti1999,bellotti2000,bertilsson2004a,cha2005,khan1998,khan2000b,kimoto2019,nilsson2000,zhao2000,cheng2020}. For this purpose, simulations and measurements were used to the same extent. The achieved parameters for electrons (see \cref{tab:mobility_vsat_elec}) reveal saturation velocities in the range from a few \SI{e6}{\V\per\centi\m} to a few \SI{e7}{\V\per\centi\m}. \citet{hjelm2001} investigated the saturation velocity for varying field angle in respect to the steps in the interface, which reduced with increasing angle.

\revision{The bipolar transistor I-V (BIV) measurement method provided the lowest mobilities. At least a factor of two higher values in the range of \SIrange{8e6}{2e7}{\centi\m\per\s} were achieved by using diode I-V (DIV). More consistent are the results by Monte-Carlo simulations (MC) and nanosecond pulsed conductance measurements (NPC), whose results all exceed \SI{1e7}{\centi\m\per\s}.}

\begin{table*}[t]
    \setlength{\tabcolsep}{8pt}
    \renewcommand{\arraystretch}{1.1}
    \caption{\label{tab:mobility_vsat_elec}High field mobility parameters in \cref{eq:mob_high} for electrons.}
    \resizebox{0.95\linewidth}{!}{%
    \begin{tabular}{l|*{9}{c}}
ref. & $\vs$ & $\vsperp$ & $\vspara$ & $\beta$ & $\gsat$ & $\gbeta$ & $T$ & K\footnotemark[1] & method \\ 
 & [\si{\centi\m\per\s}] & [\si{\centi\m\per\s}] & [\si{\centi\m\per\s}] & [1] & [1] & [1] & [\si{\K}] &&\\ \hline 
 &&&&&&&&& \\[-7pt] 
\rowcolor{no4HColor} \reftext{Ruff94}\cite{ruff1994}\footnotemark[2]  & \num{2.00E+07} & -- & -- & \num{2} & -- & -- & -- & -- & --\\ 
\reftext{Josh95}\cite{joshi1995}  & -- & \num{2.10E+07} & \num{2.70E+07} & -- & -- & -- & \num{300} & C & MC\\ 
\reftext{Nils96}\cite{nilsson1996}  & -- & \num{2.10E+07} & \num{1.80E+07} & -- & -- & -- & \num{300} & C & MC\\ 
\reftext{Khan98}\cite{khan1998}  & -- & \num{2.08E+07}\footnotemark[3] & -- & \num{0.825} & -- & -- & \num{300} & C & NPC\\ 
\reftext{Mick98}\cite{mickevicius1998}  & -- & \num{2.00E+07} & \num{2.50E+07} & \num{1} & -- & -- & \num{300} & C & MC\\ 
\reftext{Khan00}\cite{khan2000}  & -- & \num{2.2e7} & -- & \num{1.2} & -- & -- & \num{296} & C & NPC\\ 
  & -- & \num{1.6e7} & -- & \num{2.2} & -- & -- & \num{593} & C & NPC\\ 
\reftext{Lade00}\cite{ladesmartin2000}\footnotemark[4]  & -- & \num{2.20E+07} & -- & \num{1.2} & \num{-0.44} & \num{1} & -- & -- & FIT\\ 
\reftext{Nils00}\cite{nilsson2000}  & -- & \num{2.26E+07} & \num{1.64E+07} & -- & -- & -- & -- & C & MC\\ 
\reftext{Sank00}\cite{sankin2000}  & \num{3.30E+06} & -- & -- & -- & -- & -- & -- & C & BIV\\ 
\reftext{Vass00}\cite{vassilevski2000}  & -- & -- & \num{8e6} & -- & -- & -- & \num{300} & C & DIV\\ 
  & -- & -- & \num{7.5e6} & -- & -- & -- & \num{460} & C & DIV\\ 
\reftext{Zhao00}\cite{zhao2000}  & -- & -- & \num{1.83E+07} & -- & -- & -- & -- & C & MCP\\ 
\reftext{Bert01}\cite{bertilsson2001}  & -- & \num{2.10E+07} & \num{1.70E+07} & 0.84/1.1\footnotemark[5] & -- & -- & -- & C & MC\\ 
\reftext{Rosc01}\cite{roschke2001}\footnotemark[6]  & \num{2.40E+07} & -- & -- & \num{0.85} & --\footnotemark[7] & --\footnotemark[7] & \num{300} & -- & FIT\\ 
\reftext{Hjel03}\cite{hjelm2003}  & -- & \num{2.12E+07} & \num{1.58E+07} & -- & -- & -- & -- & C & EPM\\ 
\reftext{Bert04a}\cite{bertilsson2004a}\footnotemark[8]  & -- & \num{2.00E+07} & \num{1.70E+07} & 0.9/1.1\footnotemark[5] & --\footnotemark[7] & --\footnotemark[7] & \num{300} & -- & FIT\\ 
\reftext{Arda05}\cite{ardaravicius2005}  & -- & \num{1.40E+07} & -- & -- & -- & -- & \num{293} & C & NPC\\ 
\reftext{Aktu09}\cite{akturk2009}  & \num{1.60E+07} & -- & -- & -- & -- & -- & \num{300} & C & MC DFT-DOS\\ 
\reftext{Donn09}\cite{donnarumma2009}  & -- & \num{1.85e7} & -- & -- & -- & -- & \num{296} & C & MC\\ 
  & -- & \num{1.5e7} & -- & -- & -- & -- & \num{593} & C & MC\\ 
\reftext{Sun10}\cite{sun2010}  & \num{1.80E+07} & -- & -- & -- & -- & -- & \num{300} & C & MC\\ 
\reftext{Hata13}\cite{hatakeyama2013}\footnotemark[9]  & -- & \num{2.20E+07} & -- & \num{1.2} & \num{-0.46} & \num{0.88} & -- & -- & FIT\\ 
\reftext{Das15}\cite{das2015}  & \num{2.00E+07} & -- & -- & \num{1} & \num{0.87} & \num{0.66} & -- & -- & --\\ 
\reftext{Bela22}\cite{belas2022}  & \num{8.70E+06} & -- & -- & \num{2} & -- & -- & -- & C & CCh\\ 
\reftext{Jais24}\cite{jaiswal2024}  & \num{2.00E+07} & -- & -- & -- & -- & -- & -- & C & DIV\\ 
\reftext{Tana24}\cite{tanaka2024}  & -- & \num{1.40E+07} & -- & -- & -- & -- & \num{300} & -- & --
\end{tabular}
\footnotetext[1]{type of mobility: Hall (H), conductivity (C)}
\footnotetext[2]{$\beta$ taken from Silicon}
\footnotetext[3]{Value extrapolated. Highest measured velocity was \SI{1.5e7}{\centi\m\per\s}.}
\footnotetext[4]{fitted to~\cite{khan2000}}
\footnotetext[5]{values of $\beta$ $\perp/\parallel$ to c-axis}
\footnotetext[6]{fitted to~\cite{khan2000}}
\footnotetext[7]{temperature scaling according to \cref{eq:mob_vsat}}
\footnotetext[8]{fitted to~\cite{nilsson1996}}
\footnotetext[9]{fitted to~\cite{khan2000,ladesmartin2000}}
    }
\end{table*} 

A statistical interpretation shows that the values parallel to the c-axis and those without direction information have a similar mean value (see \cref{fig:mobility_vsat_stat_elec}) but also a considerable uncertainty. For the direction perpendicular to the c-axis the results agree better and indicate a higher velocity. Exceptions are the investigations by \citet{mickevicius1998} and \citet{joshi1995}, whose parallel values were larger than their perpendicular ones. \citet{hatakeyama2004} estimated the anisotropy based on impact ionization coefficients, which are also higher for the perpendicular direction (see \cref{sec:incompIon}) and \citet{hjelm2001} explained it by the anisotropy in the band structure. \citet{hatakeyama2013} obtained a ratio of $\vspara/\vsperp=0.6$.

\begin{figure*}[t]
    \centering
    \resizebox{0.86\linewidth}{!}{%
    \input{figures/mobility_vsat_stat_elec}
    }
    \caption{\label{fig:mobility_vsat_stat_elec}Statistical evaluation of electron saturation velocity. Shown are the 0th, 25th, 50th, 75th and 100th quartile. The mean value is added in numerical form.}
\end{figure*}

The hole saturation velocity is lower than the electron one but $\vsperp > \vspara$ is still satisfied (see \cref{tab:mobility_vsat_hole}). \citet{hatakeyama2013} calculated a ratio of $\vspara/\vsperp=0.8$. The key difference to electrons is the amount of conducted investigations. We only found six publications focusing on holes starting in the year 2000, whereat the latest was published a decade ago. In consequence, the value of the hole saturation velocity is often just assumed~\cite{arvanitopoulos2017,kimoto2014a} or set equal to the electron one~\cite{ruff1994,pezzimenti2013,wang1999,roschke1998,arvanitopoulos2017,rakheja2020}. \citet{vasconcelos2019} calculated the field dependency but only up to $F=\SI{200}{\kilo\V\per\cm}$, where they obtained $v_\perp \approx \SI{5e6}{\centi\m\per\s}$ and $v_\parallel \approx \SI{1e6}{\centi\m\per\s}$. A fit of \citet{guo2025} to these values predicted $\vs=\SI{1e7}{\centi\m\per\s}$.

\begin{table*}[t]
    \setlength{\tabcolsep}{8pt}
    \renewcommand{\arraystretch}{1.1}
    \caption{\label{tab:mobility_vsat_hole}High field mobility parameters in \cref{eq:mob_high} for holes.}
    \resizebox{0.95\linewidth}{!}{%
    \begin{tabular}{l|*{9}{c}}
ref. & $\vs$ & $\vsperp$ & $\vspara$ & $\beta$ & $\gsat$ & $\gbeta$ & $T$ & K\footnotemark[1] & method \\ 
 & [\si{\centi\m\per\s}] & [\si{\centi\m\per\s}] & [\si{\centi\m\per\s}] & [1] & [1] & [1] & [\si{\K}] &&\\ \hline 
 &&&&&&&&& \\[-7pt] 
\rowcolor{no4HColor} \reftext{Ruff94}\cite{ruff1994}\footnotemark[2]  & \num{2.00E+07} & -- & -- & -- & -- & -- & -- & -- & --\\ 
\reftext{Nils00}\cite{nilsson2000}  & -- & \num{1.10E+07} & \num{6.50E+06} & -- & -- & -- & -- & C & MC\\ 
\reftext{Zhao00}\cite{zhao2000}  & -- & -- & \num{8.60E+06} & -- & -- & -- & -- & C & MCP\\ 
\reftext{Hjel03}\cite{hjelm2003}  & -- & \num{1.08E+07} & \num{7.30E+06} & -- & -- & -- & -- & C & EPM\\ 
\reftext{Aktu09}\cite{akturk2009}  & \num{1.00E+07} & -- & -- & -- & -- & -- & -- & C & MC DFT-DOS\\ 
\reftext{Kimo14a}\cite{kimoto2014a}\footnotemark[3]  & \num{1.30E+07} & -- & -- & -- & -- & -- & -- & -- & --\\ 
\reftext{Das15}\cite{das2015}  & \num{2.00E+07} & -- & -- & \num{1.213} & \num{0.52} & \num{0.17} & -- & -- & --\\ 
\reftext{Guo25}\cite{guo2025}\footnotemark[4]  & \num{1.00E+07} & -- & -- & \num{1.2} & -- & -- & -- & -- & FIT
\end{tabular}
\footnotetext[1]{type of mobility: Hall (H), conductivity (C)}
\footnotetext[2]{$\vsat$ set equal to electron saturation velocity}
\footnotetext[3]{values estimated}
\footnotetext[4]{fitted to~\cite{vasconcelos2019,tanaka2024b,khan2000}}
    }
\end{table*}

We mentioned in the introduction that Monte-Carlo simulations revealed a maximum in the charge carrier velocity, followed by a decrease with increasing field\newline strengths~\cite{akturk2009,bellotti1999,bellotti2000,bertilsson2004,bertilsson2004a,hjelm2003,lv2004,mickevicius1998,neilainglesias2012,nilsson2000,nilsson1996,tanaka2024}. \citet{lv2004} used the model introduced in \cref{eq:mob_overshoot} with the parameter shown in \cref{eq:mob_vsat_lv2004} to model this \enquote{overshoot}~\cite{neilainglesias2012,lv2004,rakheja2020}.
\begin{equation}
    \begin{split}
    \label{eq:mob_vsat_lv2004}
    \mu_0 &= 0.17 \mu_1 \text{  ,  } \alpha = -1.95 \text{  ,  } \beta = 3\\
    F_0 &= \SI{3.05e4}{\V\per\centi\m}\text{  ,  } F_1 = \SI{2.8e5}{\V\per\centi\m}\text{  ,  } \vmax = \SI{4.8e7}{\centi\m\per\s}
    \end{split}
\end{equation} 

Interestingly, none of the publications that presented these decreasing velocities used the term ``overshoot''. Instead, the peak value, i.e., the maximum, was denoted as the saturation velocity. The question remains whether the decrease in velocity is real or just a simulation artifact, also because this effect has yet to be seen in experiments~\cite{rakheja2020}. \citet{nilsson1996} explained the pronounced peak in the velocity due to band bending at the zone boundaries, which decreases the energy gradient there and \citet{tanaka2022} named Bloch oscillations as the main reason~\cite{tanaka2024}. \citet{mickevicius1998} stated that, within their simplified analytical band model, the velocity decreased at high electric fields due to conduction band non-parabolicity at higher energies. The authors then showed on 3C-SiC that impact ionization cools the electrons down, which effectively increased their velocity, resulting in a constant velocity at high fields. However, later studies that also took impact ionization into account still observed a velocity decrease~\cite{nilsson2000,hjelm2001}. In the majority of cases the decrease is simply not commented, although the velocity can decrease up to one order of magnitude~\cite{zhao2000,bertilsson2001,sun2010,bertilsson2004a}. Exceptional are the results presented by \citet{akturk2009}, whose electron velocity stabilized at \SI{7e6}{\centi\m\per\s} after passing the peak value of \SI{1.6e7}{\centi\m\per\s}. The predicted hole velocity passes a saddle point, i.e., it further increased for high fields.

The models show a good agreement for varying field strengths (see \cref{fig:mobility_vsat_field_elec}). The first deviations among the model are visible already at a few \si{\kV\per\cm}, whereat considerable deviations from the low field mobility are detectable around \SI{10}{\kV\per\cm}~\cite{baliga2019}. This is significantly smaller than the \SI{200}{\kV\per\cm} proposed by \citet{lophitis2018}. We were unable to recreate the plots shown by \citet{lv2004} with the provided parameters (see \cref{eq:mob_vsat_lv2004}). In the paper a continuously decreasing derivative of $v$ is visible with increasing field strength, with a peak at $F=E_1$ and a value of $v=\SI{1e7}{\cm\per\s}$ at $F=\SI{2e6}{\V\per\cm}$. We do not explicitly show the results for holes as we only found a single model in literature.

\begin{figure*}[t]
    \centering
    \resizebox{1\linewidth}{!}{%
    \input{figures/mobility_vsat_field_elec}
    }
    \caption{\label{fig:mobility_vsat_field_elec}Electron carrier velocity with varying field for $\mulow=\SI{400}{\square\centi\meter\per\V\per\s}$. The models are only plotted in the range used for the characterization.}
\end{figure*}


The temperature dependency of the high-field velocity was measured by \citet{khan1998}~\cite{khan1998,khan2000} and investigated by simulations~\cite{nilsson1996,bertilsson2004,bertilsson2004a}. With increasing temperature the velocity decreases (see \cref{fig:mobility_vsat_temp_elec}). Later these results were numerically fitted~\cite{ladesmartin2000,hatakeyama2013}. \citet{ladesmartin2000} used a linear fit for the exponent $\beta$ meaning that the value published by \citet{khan2000} for \SI{620}{\K} could not be perfectly matched ($2.2$ vs. $2.4$). The values published by \citet{das2015} predict a steep increase of the velocity with temperature also for holes. Since this is the only fit for holes we found, further comments about its accuracy are impossible. We did not include the results by \citet{hatakeyama2013} in the plot for improved readability due to a high similarity with \citet{ladesmartin2000}.

\begin{figure*}[t]
    \centering
    \resizebox{0.86\linewidth}{!}{%
    \input{figures/mobility_vsat_temp_elec}
    }
    \caption{\label{fig:mobility_vsat_temp_elec}Temperature dependency of electron velocity for $F=\SI{e6}{\V\per\cm}$ and $\mulow=\SI{450}{\square\centi\meter\per\V\per\s}$. The dashed line shows the model by \citet{roschke2001} with $d=0.8$ and $\vmax=\SI{5.56e7}{\cm\per\s}$ in \cref{eq:mob_vsat} (see text).}
\end{figure*}

\citet{roschke2001} and \citet{bertilsson2004a} used the models in \cref{eq:mob_vsat} and \cref{eq:mob_beta} with the parameters shown in \cref{tab:mobility_jacoboni}. Multiple investigations~\cite{albanese2010,bellone2014,neilainglesias2012,wang1999} later reused these values. For \cref{eq:mob_vsat} we regularly found $d=0.8$~\cite{wang1999,ruff1994,selberherr1984}, which we could trace back to the publication by \citet{jacoboni1977}. Another popular value, $d=0.6$~\cite{roschke2001,neilainglesias2012,lv2004}, was referenced from an early edition of \citet{sze2007}, but the version we used in this review also stated $d=0.8$. This indicates that $d=0.6$ is outdated.

\begin{table}
    \centering
    \setlength{\tabcolsep}{8pt}
    \caption{\label{tab:mobility_jacoboni}Model parameters for temperature dependent carrier velocity in \cref{eq:mob_vsat} and \cref{eq:mob_beta}.}
    \begin{tabular}{c|cccccccc}
        ref. & dir & $\vmax$ & $d$ &$\beta_0$ & $\Tref$ & $a$ & $b$ & $c$\\ 
        &&\ [\si{\centi\m\per\s}] & [1] & [1] & [\si{\K}] & [1] & [\si{\K}] & [\si{\per\K}] \\ \hline
        \reftext{Rosc01}\cite{roschke2001}\footnotemark[1] & - & \num{4.77e7} & \num{0.6} & \num{0.816} & \num{327} & \num{4.27e-2} & \num{98.4} & 0 \\
        \reftext{Bert04a}\cite{bertilsson2004a}\footnotemark[2] & $\perp$ & \num{2.77e7} & \num{0.23} & \num{0.6} & $0$ & $0$ & $\infty$ & \num{e-3} \\
        & $\parallel$ & \num{2.55e7} & \num{0.3} & \num{1.01} & $0$ & $0$ & $\infty$ & \num{3e-4}
    \end{tabular}
    \footnotetext[1]{fitted to~\cite{khan2000,nilsson1996}}
    \footnotetext[2]{fitted to simulations by \citet{nilsson1996}}
\end{table}

In simple overviews the electron saturation velocity is dominantly denoted as \SI{2e7}{\cm\per\s} (see \cref{fig:mobility_vsat_values_elec}). This popular value was already reported by \citet{v.muench1977a} for 6H, as pointed out by \citet{khan1998}, and then reused for 4H. Three investigations~\cite{mickevicius1998,das2015,jaiswal2024} achieved the same value also for 4H. Slightly lower/higher values are also available but are cited not nearly as prominently.

\begin{figure*}[t]
    \centering
    \resizebox{1\linewidth}{!}{%
    \input{figures/mobility_vsat_values_elec}
    }
    \caption{\label{fig:mobility_vsat_values_elec}Electron saturation velocity values used in literature. \bkgCol{fundColor} are fundamental investigations and \bkgCol{no4HColor} research not focused on 4H.}
\end{figure*}

For holes very similar values were proposed (see \cref{fig:mobility_vsat_values_hole}) but, again, in less amount. The fact that the most popular value is equal to electrons indicates that this value was reused for holes as well. Considering the high impact ionization coefficient of holes and its importance for a wide range of devices, more data will be required for the high-field behavior of holes in the future.

\begin{figure*}[t]
    \centering
    \resizebox{0.8\linewidth}{!}{%
    \input{figures/mobility_vsat_values_hole}
    }
    \caption{\label{fig:mobility_vsat_values_hole}Hole saturation velocity values used in literature. \bkgCol{fundColor} are fundamental investigations and \bkgCol{no4HColor} research not focused on 4H.}
\end{figure*}

The reference chain for electrons (for a detailed graphical representation see \cref{fig:mobility_vsat_ref_chain_elec} in Appendix~\ref{sec:refChainMobility}) confirms our earlier intuition. The value $\vsat = \SI{2e7}{\cm\per\s}$ could not be traced back to any 4H based publication. Instead, it is mentioned often without proper references. The analysis shows, similar to the low-field mobility, that small clusters are formed, especially for the most prominent values. This is a sign that the respective values are commonly accepted within the community. On the bright side eight fundamental investigations were referenced at least once.


The reference chain for holes (see \cref{fig:mobility_vsat_ref_chain_hole} in Appendix~\ref{sec:refChainMobility}) paints a similar picture. Small clusters for the most prominent values are formed, but also four out of the seven fundamental investigations directly referenced. In addition, we see for many entries blank lines, which indicates that the cited publication did not state any values. This shows that values were misused, e.g., by using the electron saturation velocity for holes~\cite{rakheja2020}.



\subsubsection{Carrier-Carrier Scattering}

The only investigation of carrier-carrier scattering in 4H-SiC we found was conducted by \citet{ladesmartin2000}, who started from the parameters of \cref{eq:mob_conwell_weisskopf} for silicon and scaled them until the results fit to 4H-SiC measurements. In this fashion the values shown in \cref{eq:mob_ccs_params} were achieved, which are already reused at various occasions~\cite{ayalew2004,tripathi2019,schroeder2006}
\begin{equation}
    \label{eq:mob_ccs_params}
    D = \SI{6.9e20}{\per\centi\m\per\V\per\s}\text{  ,  }F = \SI{7.452e13}{\per\square\centi\m}
\end{equation}

We found further sources that were not suited for this review: \citet{onoda2007} proposed values for 6H and  \citet{bhatnagar2005} separate values for electrons and holes but used an equation that was developed by \citet{dorkel1981} for impurity scattering. Finally, \citet{lechner2021} referenced the values by \citet{fletcher1957}, which were fitted for silicon.

\section{Conclusion}
\label{sec:conclusion}


The models and parameters we found for each of the covered properties were propagated but also changed and misinterpreted within the community. We were able to show, among others, that (i) the majority of permittivity values are based on a single publication from the year 1970 that focused on a different SiC polytype, (ii) the increase of the hole density-of-states mass by \SI{100}{\percent} between $0$ and \SI{300}{\K} is rarely considered, (iii) the most common band gap values at room temperature are based on a lower energy gap at zero Kelvin, (iv) impact ionization data perpendicular to the c-axis, where electrons and holes behave similarly, are almost non-existent, and (v) reliable data on the hole saturation velocity and velocity overshoots at high electric fields are missing.

Despite our extensive analyses this work is unable to provide a common material parameter set that matches every device. Instead we experienced in our daily business that each physical object is unique, i.e., simulations always have to be calibrated. However, we are confident that the provided overview improves the initial parameter values and decreases the required effort. The achieved results are not limited to TCAD simulation but applicable to any task that requires knowledge of the physical properties of 4H-SiC, e.g., in material science.

For future research we envision a critical evaluation of the current knowledge base within the scientific community. In detail, further characterizations (measurements, calculations, simulations) will be required to obliterate the shortcomings we discovered.


\begin{acknowledgments}
This work was supported by the Austrian Research Promotion Agency (FFG) in the
project RadHardDetSim (895291).

The authors would like to thank Christa Bay from the library at TU Wien, who
helped to acquire countless publications.
\end{acknowledgments}

\section*{Author Declarations}

\subsection*{Conflict of Interest}
\noindent The authors have no conflicts to disclose.

\subsection*{Author Contributions}

\noindent
\textbf{J{\"u}rgen Burin:} Conceptualization (lead), Data curation (lead), Investigation (lead), Methodology (lead), Resources (lead), Validation (lead), Visualization (lead), Writing - original draft (lead), Writing - review \& editing (lead).
\textbf{Philipp Gaggl:} Conceptualization (supporting), Data curation (equal), Resources (equal), Validation (supporting), Visualization (supporting), Writing - original draft (equal).
\textbf{Simon Waid:} Conceptualization (supporting), Methodology (supporting), Visualization (supporting), Writing - review \& editing (equal).
\textbf{Andreas Gsponer:} Conceptualization (supporting), Methodology (supporting), Validation (supporting), Visualization (supporting), Writing - review \& editing (equal).
\textbf{Thomas Bergauer:} Funding acquisition (lead), Project administration (lead).

\section*{Data Availability Statement}
The data, evaluation scripts and figures of this study are openly available in
the repository \href{https://gitlab.com/dd-hephy/HiBPM/review_4HSiC}{Data of 4H
  SiC TCAD Parameter Review}~\cite{burin2025}.

\FloatBarrier

\appendix
\section{\label{sec:appendix}Inaccuracies}

In this section we list inaccuracies we encountered throughout our review. The purpose of this listing is not to blame any of the authors or suggest any wrongdoing. Instead, we think that mistakes are unavoidable (there are definitely a lot of them also in this review) and that it is important to highlight them in order to prevent them from spreading. If not further specified the described actions were done by the authors of the respective publication.


\subsection{\label{sec:perm_appendix}Permittivity}

\begin{appendixTable}

\appE{Acha17}{acharyya2017} & We could not find the value $\es=8.5884$ in the referenced paper~\cite{ioffe2023}. Even using the available values and calculating the effective relative permittivity did not yield the desired result. \\

\appE{Arpa06}{arpatzanis2006} & Changed $\espa=10.03$~\cite{patrick1970} to $\es=10$. \\

\appE{Arva17}{arvanitopoulos2017} & Changed $\espe=9.66$~\cite{patrick1970} to $\es=9.66$. \\

\appE{Chen94}{chen1994} & The second column in Table~II denotes 4H but is labeled 2H. \\

\appE{Choi05}{choi2005} & Changed $\espe=9.66$~\cite{harris1995} to $\es=9.7$. \\

\appE{Darm19}{darmody2019} & Changed $\espe=9.76$~\cite{kimoto2014a} to $\es=9.76$. \\

\appE{Egil99}{egilsson1999} & Changed $\espe=9.66$~\cite{patrick1970} to $\es=9.7$. \\

\appE{Elah17}{elahipanah2017} & Changed $\es=9.94$~\cite{kimoto2014a} to $\es=10$. \\

\appE{Hari98}{harima1998} & Changed $\eipa=6.78$~\cite{harima1995} to $\ei=6.8$. \\

\appE{Harr95}{harris1995} & Changed $\espa=10.03$~\cite{patrick1970} to $\espa=10.3$. \\

\appE{Huan98}{huang1998} & Changed $\es=9.66$~\cite{casady1996} to $\es=9.7$. \\


\appE{Kimo19}{kimoto2019} & Used $\espa=10.32$ instead of $\espa=9.98$ specified in the referenced publication~\cite{ninomiya1994}. \\

\appE{Klah20}{klahold2020} & Changed $\espa=10.03$~\cite{patrick1970} to $\es=10$. \\

\appE{Kova20}{kovalchuk2020} & Changed $\espe=9.66$~\cite{patrick1970} to $\es=9.67$. \\

\appE{Micc19}{miccoli2019} & Changed $\espe=9.66$~\cite{patrick1970} to $\es=9.66$. \\

\appE{Neil12}{neilainglesias2012} &  The used $\es=9.72$ is actually a 3C value~\cite{patrick1970}. \\

\appE{Neud01}{neudeck2001} & Changed $\espe=9.66$~\cite{harris1995} to $\es=9.7$. \\

\appE{Neud06}{neudeck2006} & Changed $\espe=9.66$~\cite{harris1995} to $\es=9.7$. \\

\appE{Ozpi04}{ozpineci2004} & Changed $\es=10.007$~\cite{agarwal2004} to $\es=10.1$. \\

\appE{Pear23}{pearton2023} & Changed $\espe=9.66$~\cite{harris1995} to $\es=9.7$. \\

\appE{Ryba17}{rybalka2017} & Changed $\espe=9.76$~\cite{kimoto2014a} to $\es=9.7$. \\

\appE{Scho94}{schoner1994} & The specified reference [77Pan] could not be found in the reference list. The similar reference [75Pan]~\cite{pantelides1975} did also not contain the desired values. \\

\appE{Torp01}{torpo2001} & Used $\es=6.7$, which is, however, a value for the high-frequency one. It is unclear whether the wrong value was picked from the reference or the textual description is flawed. \\

\appE{Trip19}{tripathi2019} & Changed $\espe=9.66$~\cite{patrick1970} to $\es=9.66$. \\

\appE{Wije11}{wijesundara2011} & Changed $\es=9.7$~\cite{neudeck2006} to $\es=10$. \\

\appE{Yosh18}{yoshioka2018} & Changed $\espe=9.66$~\cite{patrick1970} to $\espe=9.7$. \\

\appE{Zatk21}{zatko2021} & Changed $\espe=9.66$~\cite{patrick1970} to $\es=9.66$. \\

\appE{Zipp11}{zippelius2011} & Changed $\espe=9.76$~\cite{troffer1998} to $\es=9.76$. \\

\end{appendixTable}


\subsection{\label{sec:dos_appendix}DOS Mass}

\begin{appendixTable}
   
\appE{Arpa06}{arpatzanis2006} & Changed $\mde=0.19$~\cite{gotz1993} to $0.2$. \\


\appE{Fang05}{fang2005} & Refernced $\mde=0.4\,m_0$~\cite{look1983} but that publication is focused on GaAs and we could not find the respective values. \\

\appE{Flor03}{flores2003} & Present values for the hole effective masses but the labels $m_{\textrm{M}\Gamma}$ and $m_\textrm{ML}$ refer to electron effective masses. \\

\appE{Gale98}{galeckas1997} & Changed $\mhgm = 4.23$~\cite{kackell1994} to $4.2$.  \\

\appE{Harr95}{harris1995} & Changed $\mdeperp=0.21$~\cite{lomakina1973} to $0.24$, $\mdeperp=0.176$~\cite{gotz1993} to $0.18$ and $\mdepara=0.224$~\cite{gotz1993} to $\mdepara=0.22$. \\

\appE{Hemm97}{hemmingsson1997} & Calculated $\mde = 0.37$ but it is actually $0.39$. \\

\appE{Itoh95}{itoh1995} & Changed $\mde=0.19$~\cite{gotz1993} to $0.2$. \\

\appE{Kim24}{kim2024} & Changed $\mde=0.39$~\cite{negoro2004a} to $0.2$. \\

\appE{Lind98}{lindefelt1998} & We could not retrace the values for the hole masses in the referenced publication~\cite{persson1998}. \\

\appE{Penn01}{pennington2001} & Only used the transversal masses to calculate the DOS masses. \\

\appE{Penn04}{pennington2004} & Changed $\memk=0.28$~\cite{lambrecht1995} to $0.29$ and $\meml=0.31$~\cite{lambrecht1995} to $0.33$. Only used the transversal masses to calculate the DOS masses. \\

\appE{Pens93}{pensl1993} & In the definition $\mde = (\mdeperp \mdepara )^{(1/3)}$ the square for $\mdeperp$ is missing. Changed $\mdeperp=0.176$~\cite{gotz1993} to $0.17$ and $\mdepara=0.224$~\cite{gotz1993} to $0.22$. \\

\appE{Pere06}{perez-tomas2006} & The origin of the used values is unclear, because the cited publication~\cite{harris1995} corresponds to hole effective masses of 3C and 6H. \\

\appE{Pern05}{pernot2005} & Calculated $\mdh = 2.66$ but our calculations resulted in $\mdh=2.64$. \\

\appE{Pers98a}{persson1998a} & We were not able to confirm $\mdh = 0.88$ based on the referenced data~\cite{persson1997}. Instead we achieved $\mdh= 0.78$. Similarly, instead of $\mde=0.45$ we got $\mde=0.44$. \\


\appE{Resh05}{reshanov2005} & Changed $\memg =0.57$\cite{persson1997} to $0.58$. \\

\appE{Rodr21}{rodrigues2021} & Changed $\mdeperp=0.176$, $\mdepara=0.224$~\cite{gotz1993} to $\mdeperp=0.18$, $\mdepara=0.22$.\\

\appE{Scha94a}{schadt1994a} & Changed $\mdeperp=0.176$~\cite{gotz1993} to $0.17$ and $\mdepara=0.224$~\cite{gotz1993} to $0.22$. \\

\appE{Scho94}{schoner1994} & Changed $\mdeperp=0.176$~\cite{gotz1993} to $0.18$ and $\mdepara=0.224$~\cite{gotz1993} to $0.22$. Assumed $\mdh=1$, $\ga=4$ and $\gd=2$.\\

\appE{Sozz19}{sozzi2019} & Changed $\mde = 0.394$~\cite{wellenhofer1997} to $0.4$. \\

\appE{Yang19}{yang2019} & Calculated $\mde=0.36$, but we achieved $\mde=0.37$. 

\end{appendixTable}


\subsection{\label{sec:bandgap_appendix}Band Gap}

\begin{appendixTable}

\appE{Alba10}{albanese2010} & In the doping dependency factor $\Delta E_\mathrm{ga}$ the exponent for the prefactor \num{1.57e-2} should be $1/4$ instead of $1/3$ and the factor \num{1.54e-2} should be \num{1.54e-3}. In addition, factor \num{1.7e-2} in $\Delta E_\mathrm{gd}$ is actually \num{1.17e-2}. \\

\appE{Back94}{backes1994} & Stated that the band gap energies by \citet{choyke1964} suffer from an inaccuracy in the order of \SI{0.015}{\eV}. This value, however, refers to the non-measured values. The correct inaccuracy is \SI{0.003}{\eV}. \\

\appE{Bade20}{bader2020} & Changed $\eg=\SI{3.26}{\eV}$~\cite{chow2017} to $\eg=\SI{3.3}{\eV}$. \\

\appE{Baie19}{baierhofer2019} & The value $\eg=\SI{3.268}{\eV}$ does not match $\eg=\SI{3.26}{\eV}$~\cite{kimoto2015} and $\egx=\SI{3.263}{\eV}$~\cite{fan2014} from the provided references. \\

\appE{Bako97}{bakowski1997} & The temperature dependency $\alpha =
\SI{-3.3e-3}{\eV\per\K}$ is stated, but it should be $\alpha=\SI{-3.3e-4}{\eV\per\K}$~\cite{choyke1957}. \\

\appE{Bane21}{banerjee2021} & Used the value $\eg(300)=\SI{3.23}{\eV}$~\cite{ioffe2023} but interpreted it as $\eg(0)$. Consequently the authors end up with $\eg(300)=\SI{3.1934}{\eV}$. \\

\appE{Bech04}{bechstedt2004} & Changed $\eg=\SI{3.265}{\eV}$~\cite{haberstroh1994} to $\eg=\SI{3.27}{\eV}$. \\

\appE{Bell99}{bellotti1999} & Changed $\egx=\SI{3.263}{\eV}$~\cite{yoshida1995} to $\eg=\SI{3.26}{\eV}$. \\

\appE{Bere21}{berens2021} & Changed $\egx=\SI{3.265}{\eV}$~\cite{kimoto2014a} to $\eg=\SI{3.26}{\eV}$.\\

\appE{Bion12}{biondo2012} & Changed $\egx=\SI{3.263}{\eV}$~\cite{yoshida1995} to $\eg=\SI{3.26}{\eV}$. \\

\appE{Buon12}{buono2012} & Eq.~(2.41) denotes $\Delta E_v$ instead of $\Delta E_c$. The respective parameters need to be positive and the exponent of the first summand is $1/4$ not $1/3$. \\


\appE{Cama08}{camassel2008} & Changed $\egx$~\cite{choyke1964} to $\eg$. \\

\appE{Capa22}{capan2022} & Changed $\egx=\SI{3.265}{\eV}$~\cite{kimoto2014a} to $\eg=\SI{3.26}{\eV}$. \\

\appE{Casa96}{casady1996} & Changed $\egx=\SI{3.265}{\eV}$~\cite{haberstroh1994} to $\eg=\SI{3.26}{\eV}$. \\

\appE{Cha08}{cha2008} & States $\eg(300)=\SI{3.25}{\eV}$ but the respective reference~\cite{nallet1999} denotes it as the band gap at \SI{0}{\K}. \\

\appE{Choi05}{choi2005} & Changed $\egx=\SI{3.263}{\eV}$~\cite{yoshida1995} to $\eg=\SI{3.26}{\eV}$. \\

\appE{Dena22}{denapoli2022} & According to the presented equation for the temperature induced band gap narrowing $\alpha$ has to be negative to fit the
description in the text. \\
& Changed $\beta = \num{1.8e3}$~\cite{kimoto2014a} to \num{1.3e3}.  \\
& Changed $\egx=\SI{3.265}{\eV}$~\cite{kimoto2014a} to $\eg=\SI{3.26}{\eV}$. \\

\appE{Dhan20}{dhanaraj2010} & Changed $\eg=\SI{3.26}{\eV}$~\cite{chow2000} to $\eg=\SI{3.3}{\eV}$. \\

\appE{Donn12}{donnarumma2012} & Lists a range of values for $\eg(300)$ but in one~\cite{harima1995} of the provided references only $\eg(4)$ is available. The value $\eg(300)=\SI{3.03}{\eV}$ was taken from \citet{sandeep2011} instead of \citet{galeckas1998}, as stated in the paper. The origin of the remaining value $\eg(4)=\SI{3.28}{\eV}$ could not be retraced. \\

\appE{Egil99}{egilsson1999} & Changed exciton band gap energy $\egx$ from \SI{3.266}{\eV}~\cite{kordina1995} to $\SI{3.265}{\eV}$. \\

\appE{Elah17}{elahipanah2017} & Changed $\egx=\SI{3.265}{\eV}$~\cite{kimoto2014a} to $\eg=\SI{3.2}{\eV}$. \\

\appE{Feng04a}{feng2004a} & Changed $\egx=\SI{3.265}{\eV}$~\cite{choyke1969} to $\egx=\SI{3.26}{\eV}$. \\

\appE{Gale97}{galeckas1997} & Changed $\egx=\SI{3.265}{\eV}$~\cite{itoh1996} to $\eg=\SI{3.275}{\eV}$. \\

\appE{Gale02}{galeckas2002} & According to the provided definition of the
temperature dependency using $\mathrm{d}\eg / \mathrm{d}T$ the paramete should
be \SI{2.4e-4}{\electronvolt\per\K} instead of \SI{-2.4e-4}{\electronvolt\per\K}. Otherwise the band gap would increase with increasing temperature. \\

\appE{Griv07}{grivickas2007} & The values from \citet{itoh1996} were displayed in a plot but got shifted by \SI{2}{\milli\eV} to compensate for differing band gap energies at \SI{0}{\K}. \\
& It is unclear how the phonon dispersion $\Delta=0.29$ was derived, as it leads
to $p=3.59$. The stated $p=2.9$ would, in contrast, result in $\Delta=0.37$. \\

\appE{Huan98}{huang1998} & Changed $\egx=\SI{3.265}{\eV}$~\cite{haberstroh1994} to $\eg=\SI{3.26}{\eV}$. \\

\appE{Hudg03}{hudgins2003} & It is unclear to us how $\eg(300)=\SI{3.25}{\eV}$ was derived based upon $\egx(4.2)=\SI{3.265}{\eV}$~\cite{marshall1974}. \\

\appE{Ioff23}{ioffe2023} & All Lindefelt parameters have negative values,
however, only the ones impacting the conduction band should be $<0$. The
exponent $1/4$ with parameters $\anv$ and $\apc$ were changed to $1/3$. \\
& The authors state that the values from \citet{dubrovskii1977} denote the exciton band gap energy $\egx$ instead of $\eg$. \\

\appE{Joha19}{johannesson2019} & Changed $\egx=\SI{3.265}{\eV}$~\cite{ladesmartin2000} to $\eg=\SI{3.26}{\eV}$. \\

\appE{Khal12}{khalid2012} & Referenced an investigation on GaN~\cite{baik2003}
that only contains the model. The origin of the shown values is for us unclear. \\
& Proposed $\alpha = \num{-2.206e-2}$, which leads to an increasing band gap
with temperature. This contradicts the presented reference~\cite{baik2003}. \\

\appE{Kimo19}{kimoto2019} & It is unclear how the value $\eg(0)=\SI{3.292}{\eV}$ was derived. \\

\appE{Kohl03}{kohlscheen2003} & Changed $\egx=\SI{3.263}{\eV}$~\cite{yoshida1995} to $\eg=\SI{3.26}{\eV}$. \\

\appE{Kwas14}{kwasnicki2014} & Changed $\egx$~\cite{choyke1964a} to $\eg$. \\

\appE{Lade00}{ladesmartin2000} & The values $\ex=\SI{40}{\milli\electronvolt}$ was picked as average value from \SIrange{10}{80}{\milli\electronvolt}~\cite{devaty1997}, which we could, however, not retrace. The only value that is stated for 4H by \citet{devaty1997} is $\ex=\SI{20}{\milli\electronvolt}$~\cite{dubrovskii1975}. \\

\appE{Lebe99}{lebedev1999} & Changed $\eg=\SI{3.26}{\eV}$~\cite{matsunami1997} to $\eg=\SI{3.2}{\eV}$. \\

\appE{Lech21}{lechner2021} & Changed $\egx=\SI{3.265}{\eV}$~\cite{ladesmartin2000} to $\eg=\SI{3.265}{\eV}$. \\

\appE{Levc11}{levcenco2011} & Changed $\egx=\SI{3.267}{\eV}$~\cite{grivickas2007} to $\eg=\SI{3.267}{\eV}$. \\

 

\appE{Made91}{madelung1991} & Changed $\eg$~\cite{dubrovskii1977} to $\egx$. \\

\appE{Maxi23}{maximenko2023} & The value $\eg=\SI{3.23}{\eV}$ could not be found in the provided references. We achieved this value only by using the 2H band gap from \citet{persson1997} and scaling it with temperature. \\

\appE{Mcnu04}{mcnutt2004} & Stated $\alpha = \SI{3.3e-3}{\eV\per\K}$, which
should be $\alpha=\SI{3.3e-4}{\eV\per\K}$. \\

\appE{Megh18a}{megherbi2018a} & Changed $\eg=\SI{3.26}{\eV}$~\cite{li2003} to $\eg=\SI{3.2}{\eV}$. \\

\appE{Mill00}{miller2000} & For reasonable results an exponent of $-1$ had to be added to the term $\left( \exp(\theta_\mathrm{E}/T) -1 \right)$ in Eq.~(2). \\

\appE{Ozpi04}{ozpineci2004} & Changed $\eg=\SI{3.25}{\eV}$~\cite{agarwal1999} to $\eg=\SI{3.26}{\eV}$.\\ 

\appE{Pear23}{pearton2023} & Changed $\egx=\SI{3.263}{\eV}$~\cite{yoshida1995} to $\eg=\SI{3.26}{\eV}$. \\

\appE{Pers97}{persson1997} & The stated value $\eg=\SI{3.29}{\eV}$ appeared in the referenced publication~\cite{harris1995} only in the context of 2H- and 3C-SiC. \\

\appE{Pers99}{persson1999} & Changed $\eg=\SI{3.285}{\eV}$~\cite{freitas1995} to $\SI{3.29}{\eV}$. \\

\appE{Rayn10}{raynaud2010} & Changed $\eg$~\cite{yoshida1995} to $\egx$. \\

\appE{Resc18}{rescher2018} & Changed $\egx=\SI{3.265}{\eV}$~\cite{kimoto2014a} to $\eg=\SI{3.26}{\eV}$. \\

\appE{Resh05}{reshanov2005} & We could not find the stated value $\eg=\SI{3.26}{\eV}$ in the referenced publication~\cite{persson1997}. \\

\appE{Scho94}{schoner1994} & We did not find the specified reference [77Pan]
from the reference list. The similar reference [75Pan]~\cite{pantelides1975} did also not contain the desired values. \\

\appE{Sole19}{soler2019} & Changed $\eg=\SI{3.25}{\eV}$~\cite{pearton2018} to $\eg=\SI{3.2}{\eV}$. \\

\appE{Son12}{son2012} & We were unable to locate the specified values in the provided reference~\cite{casady1996} but the same values were used by \citet{tamaki2008a} four years earlier. \\

\appE{Stef14}{stefanakis2014} & In the Lindefelt model the dopants should be $\ndp$ for the n-type and $\nam$ for the p-type semiconductor. The exponents for $\anv$ and $\apv$ should be $1/4$ instead of $1/3$. \\ 
& In Table~1 the band gap for the ``Vienna Uni.'' model is $\egx=\SI{3.265}{\eV}$ at \SI{0}{\K}~\cite{ayalew2004} instead of the stated $\eg=\SI{3.265}{\eV}$ at \SI{300}{\K}. \\
& In the Slotboom model the logarithmic term in the quadratic root was only applied to the fraction $N/N_0$ and not to the term $+C$ as well.\\
&The parameters for the Pssler (sic!) model are referenced from \citet{grivickas2007} and not the specified \citet{passler1999}. For the latter $\egx=\SI{3.267}{\eV}$ got changed to $\eg=\SI{3.27}{\eV}$. \\

\appE{Tama08a}{tamaki2008a} & We could not find the shown values in the provided reference~\cite{casady1996}. \\

\appE{Tann07}{tanner2007} & Changed $\egx=\SI{3.263}{\eV}$~\cite{yoshida1995} to $\eg=\SI{3.26}{\eV}$. \\

\appE{Trof98}{troffer1998} & Changed $\egx=\SI{3.263}{\eV}$~\cite{choyke1969} to $\eg=\SI{3.265}{\eV}$. \\

\appE{Uhne15}{uhnevionak2015} & Changed $\egx=\SI{3.263}{\eV}$~\cite{yoshida1995} to $\eg=\SI{3.26}{\eV}$. \\

\appE{Usma14}{usman2014} & Changed band gap energy $\eg$ from \SI{3.26}{\eV}~\cite{nawaz2010} to $\SI{3.24}{\eV}$. \\

\appE{Wrig98}{wright1998} & The second term in the temperature dependent band gap, i.e., $T^2/(T+\beta)$ should be preceeded by a minus sign. \\

\appE{Yosh18}{yoshioka2018} & Changed $\egx=\SI{3.265}{\eV}$~\cite{patrick1965} to $\eg=\SI{3.26}{\eV}$. \\

\appE{Zipp11}{zippelius2011} & Changed $\egx=\SI{3.263}{\eV}$~\cite{yoshida1995} to $\eg=\SI{3.26}{\eV}$.

\end{appendixTable}



\subsection{\label{sec:II_appendix}Impact Ionization}

\begin{appendixTable}

\appE{Aktu08}{akturk2008} & We were unable to confirm the electron parameters for Model 3. In fact the used values $\aelec=\SI{2.5e5}{\per\cm}$ and $E_c^n=\SI{1.84e7}{\V\per\cm}$ match very well the impact ionization parameters for holes in 6H-SiC by \citet{raghunathan1997}, i.e., $\ahole=\SI{2.5+-0.1e6}{\per\cm}$ and $\bhole=\SI{1.48+-0.1e7}{\V\per\cm}$, if we consider a small typographical error for the latter.\\
& Switched parameters $\gamma_n$ and $\gamma_p$ for model 2~\cite{ng2003}. \\

\appE{Arva19}{arvanitopoulos2019} & Turned coefficients parallel to c-axis from \citet{niwa2015} to perpendicular ones. \\

\appE{Ayal04}{ayalew2004} & Seemingly changed parameter $b$ for the electrons in 6H from \SI{2.58e6}{\V\per\cm}~\cite{bakowski1997} to \SI{2.58e7}{\V\per\cm}. This value presumably then got the default value in simulation tools and got reused~\cite{tripathi2019}.\\

\appE{Bako97}{bakowski1997} & Defined $\alpha_\perp = \beta/3.5$, which we assume should be $\alpha/3.5$. \\

\appE{Bane21}{banerjee2021} & In Table 1 on page 159 the range of the field for 4H-SiC should be \SIrange{10e7}{100e7}{\V\per\m}. \\

\appE{Bert04}{bertilsson2004} & Changed $\bp$ ${}_\perp$ from \SI{2.5e7}{\V\per\centi\m} to \SI{1.8e7}{\V\per\centi\m}. \\

\appE{Bion12}{biondo2012} & In Eq.~(21) the temperature scaling in the exponent
is flawed. Instead of $1+DT-300$ it should be $1+D(T-300)$. In Eq.~(21) $B$ and $C$ should be switched. \\
& We changed $m$ from $2$ to $1$ and $n$ from $1$ to $0$ to fit the results shown in the paper. \\

\appE{Bros12}{brosselard2004} & The value $\belec=\SI{1.67e5}{\V\per\cm}$ leads to unrealistic results. We changed it to $\belec=\SI{1.67e7}{\V\per\cm}$. \\

\appE{Buon12}{buono2012} & Changed $\aelec=\SI{2.1e7}{\per\cm}$~\cite{hatakeyama2004} to $\aelec=\SI{2.1e8}{\per\cm}$. \\

\appE{Chea21}{cheang2021} & $a$ and $b$ were presumably stated in \si{\per\meter} and \si{\mega\volt\per\meter} in the paper, although they are denoted as \si{\per\centi\meter} and \si{\mega\volt\per\centi\meter}. \\

\appE{Gree12}{green2012} & Stated that for electrons  and
$F>\SI{2.5}{\mega\volt\per\cm}$ the parameters proposed by \citet{ng2003} ($a =
\SI{1.98e6}{\per\cm}$, $b=\SI{9.46}{\mega\V\per\cm}$, $m=\num{1.42}$) were used,
but the presented values slightly differ ($a = \SI{1.878e6}{\per\cm}$, $b=\SI{9.134}{\mega\V\per\cm}$, $m=\num{1.459}$). \\
& The exponent of the Ocuto-Crowell models in Eqs.~(8) to (10) have to be pulled inside the squared brackets. \\
& We changed $\ahole=\SI{6e4}{\per\cm}$ to $\SI{6e6}{\per\cm}$ and
$\bhole=\SI{1.387e6}{\V\per\cm}$ to \SI{1.387e7}{\V\per\cm} to fit the shown plots. \\

\appE{Hata09}{hatakeyama2009} & Changed
$\aelec=\SI{2.1e7}{\per\cm}$~\cite{hatakeyama2004} to $\aelec=\SI{2.1e8}{\per\cm}$ and $\ahole=\SI{2.96e7}{\per\cm}$~\cite{hatakeyama2004} to $\ahole=\SI{2.96e8}{\per\cm}$. \\

\appE{Khal12}{khalid2012} & Eqs.~(19) and (20) both denote $a_n$. While the
references for the parameters in Eq.~(19) are
clear~\cite{hatakeyama2004,biondo2012}, the sources for $a$ and $b$ in Eq.~(20)
could not be retraced by us. \\
& Changed $\ahole=\SI{29.6e6}{\per\cm}$~\cite{hatakeyama2004} to
$\SI{29e6}{\per\cm}$ and $\bhole=\SI{16e6}{\V\per\cm}$~\cite{hatakeyama2004} to $\SI{14e6}{\V\per\cm}$. \\

\appE{Kimo19}{kimoto2019} & Changed parallel $\ahole=\SI{3.14e6}{\per\cm}$ to
$\ahole=\SI{3.12e6}{\per\cm}$. \\

\appE{Kons97}{konstantinov1997} & The factor $3\,\ephonon$ of \cref{eq:II_TH} is not shown for $\beta$ in the paper. We assume a typographical error as the division sign is visible. \\

\appE{Loh09}{loh2009} & The exponent $c$ in Eq.~(2) has to be pulled inside the exponential function. \\
& We changed $\bhole=\num{8.9e6} - \num{4.95e3}\, T$ to $\bhole=\num{8.9e6} + \num{4.95e3}\, T$ to better match the results in the paper. \\

\appE{Megh15}{megherbi2015} & Changed parallel $\ahole=\SI{3.41e8}{\per\cm}$ to $\ahole=\SI{2.41e8}{\per\cm}$. \\

\appE{Nall99}{nallet1999} & Used for the electron and hole impact ionization coefficient the 6H parameters for holes published by \citet{raghunathan1997}. \\

\appE{Nall00}{nallet2000} & The values for $a$ and $b$ could not be found in the provided reference~\cite{nallet1999}. \\

\appE{Niwa15}{niwa2015} & Once $\ahole$ is specified as $\SI{3.12e6}{\per\cm}$ and once as $\SI{3.14e6}{\per\cm}$. \\

\appE{Pezz13}{pezzimenti2013} & The used values $\aelec=\SI{2.5e5}{\per\cm}$ and $n=\SI{1.84e7}{\V\per\cm}$ match the impact ionization parameters for holes in 6H-SiC by \citet{raghunathan1997}, i.e., $\ahole=\SI{2.5+-0.1e6}{\per\cm}$ and $\bhole=\SI{1.48+-0.1e7}{\V\per\cm}$, if we consider a small typographical error for the latter. These are the same errors as done by \citet{akturk2008}, although no direct connection could be found between these publications. \\

\appE{Rayn09}{raynaud2009} & Referred to intermediate results by \citet{konstantinov1997} for a comparison. \\

\appE{Rayn10}{raynaud2010} & Turned $\aelec=\SI{0.408e6}{\per\cm}$~\cite{brosselard2004} to \SI{0.41e6}{\per\cm}. \\

\appE{Sher00}{sheridan2000} & Fitted to \citet{konstantinov1997}, but only the holes. Interestingly the achieved results are equal to those presented in~\cite{ruff1994}, whereat these were based on 6H-SiC measurements. \\

\appE{Stef21}{stefanakis2021} & The hole parameters presented for ``Nguyen''~\cite{nguyen2012} ($a_\mathrm{p}=\SI{4.0e7}{\per\cm}$, $b_\mathrm{p}=\SI{1.89e7}{\V\per\cm}$) could not be retraced. \\
& The parameter $b_\mathrm{p}$ for ``Loh Power''~\cite{loh2008} should be $\SI{0.35e7}{\V\per\cm}$ instead of $\SI{0.035e7}{\V\per\cm}$. \\
& $c_\mathrm{n}$ and $c_\mathrm{p}$ for entry ``Akturk Power'' were transferred correctly from the cited publication~\cite{akturk2008} but there the values got switched. Thus, it should be $c_\mathrm{n}=1.42$ and $c_\mathrm{p}=1.06$~\cite{ng2003}. Consequently, the fitting for the hole impact ionization coefficient is more than one order of magnitude lower than the other models. \\
& The value $a_\mathrm{p}=\SI{8.5e6}{\per\cm}$ for entry ``Raghunathan'' could not be found in the provided reference~\cite{raghunathan1997}. The only explanation we found was that the authors misinterpreted \SI{3.5+-0.5e6}{\per\cm} with \SI{3.5+-5e6}{\per\cm}. \\

\appE{Stei23}{steinmann2023} & In Fig. 6 the high temperature plots should refer to \SI{470}{\K} instead of \SI{470}{\degreeCelsius} for a good fit of the describing model. \\
& For the fitting to \citet{niwa2015} we had to change
$\delec=\SI{-0.72e-3}{\per\K}$ to $\delec=\SI{-0.72e-6}{\per\K}$ to achieve reasonable results. \\

\appE{Trip19}{tripathi2019} & It is safe to assume that the value of $\belec$ is
an order of magnitude too high and stems from a 6H-SiC sample (see the analysis of \citet{ayalew2004} in this section). \\

\appE{Wang22a}{wang2022a} & The used parameters match the values by \citet{loh2008} with the exception that the exponent $m$ in the Ocuto-Crowell model was not considered. This causes an implicit change of $m=\num{1.37}$ for electrons and $m=\num{1.09}$ to $\num{1}$. \\

\appE{Zhao19}{zhao2019} & The proposed $\chole = \SI{6.19e-3}{\per\K}$ and $\dhole = \SI{1.15e-3}{\per\K}$ cause $\beta$ to increase in the range \SIrange{150}{300}{\K}. This was not observed with other models. \\
    
\end{appendixTable}


\subsection{\label{sec:regen_appendix}Charge Carrier Recombination}

\begin{appendixTable}
\appE{Adit15}{aditya2015} & The referenced lifetime values could not found in provided references. \\

\appE{Alba10}{albanese2010} & Used $\nintr$ instead of $n_1$ and $p_1$ in the SRH lifetime model. \\

\appE{Arva17}{arvanitopoulos2017} & The value of $\taup=\SI{0.5}{\micro\s}$ could not be found in the provided reference~\cite{aditya2015}, where $\taup=\SI{0.6}{\micro\s}$ was proposed. The used $\taun=\SI{1}{\nano\s}$ also does not match $\taun=\SI{2.5}{\micro\s}$ in the reference. \\

\appE{Ayal04}{ayalew2004} & Used the Auger coefficients from 6H~\cite{ramungul1998}. \\

\appE{Bell11}{bellone2011} & Changed $\taun=\taup=\SI{15}{\nano\s}$~\cite{albanese2010} in the bipolar transistor base to $\taun=\taup=\SI{16}{\nano\s}$. \\

\appE{Bion12}{biondo2012} & The expression for $n_1$ and $p_1$ in Eq.~(18) are flawed. $n_i^2$ should be $n_i$ and the exponent for $n_1$ is actually negative. \\
& Used the sum $\cn + \cp$~\cite{galeckas1997} as $\cn$ and $\cp$. \\

\appE{Choi05}{choi2005} & The provided Scharfetter parameters could not found in the provided references~\cite{harris1995}. \\

\appE{Das15}{das2015} & The Auger coefficient $\cp$ was taken from \citet{ruff1994} which is based on Silicon values. \\

\appE{Gao22a}{gao2022a} & In the temperature dependent Auger recombination coefficient the first part of the product scales with $T^{-1.5}=T^{-3/2}$~~\cite{scajev2013} but a value of $T^{-2/3}$ was used instead.\\

\appE{Kaka20}{kakarla2020} & For the temperature dependency of the SRH lifetime values determined for Si~\cite{schenk1992} were used. \\

\appE{Khal12}{khalid2012} & Used the sum $\cn + \cp$~\cite{galeckas1997} as $\cn$ and $\cp$. \\

\appE{Lech21}{lechner2021} & Used 6H values for Auger coefficients~\cite{ayalew2004}. \\

\appE{Levi01b}{levinshtein2001b} & Changed $B=
\SI{5e-12}{\cubic\cm\per\s}$~\cite{galeckas1997} to $B=
\SI{1e-12}{\cubic\cm\per\s}$. \\ 
& Turned $\taun = \SI{600}{\nano\s}$~\cite{agarwal2001} into $\taun = \taup = \SI{300}{\nano\s}$. Since the measurements were done under high injection levels this might be reasonable. \\

\appE{Liu21}{liu2021} & Change $\gamma=0.3$~\cite{usman2014} to $\gamma=1$. \\

\appE{Megh15}{megherbi2015} & Changed $\taup = \SI{12}{\nano\s}$~\cite{donnarumma2012} to $\taun=\SI{10}{\nano\s}$. \\

\appE{Megh18a}{megherbi2018a} & Referenced a very high value of $\nref=\SI{1e30}{\per\cubic\centi\m}$~~\cite{bakowski1997,landsberg1984}. \\

\appE{Nall99}{nallet1999} & Proposed value of $\taun=\taup = \SI{50}{\nano\s}$ could not be found in the stated reference~\cite{galeckas1997}.\\
& The temperature dependency is presented in a ambiguous fashion, because it is not clear that the term $\left( \frac{T}{300} - 1\right)$ belongs to \textit{Coeff} in the exponential. \\

\appE{Rao22}{rao2022} & The term $\nintr^2$ in the definition of $n_1$ and $p_1$ should be $\nintr$. \\

\appE{Scho94}{schoner1994} & Denoted in Eqs.~(5.21) and (5.22) a decrease of the SRH lifetime with increasing temperature, which may go back to an error in the sign of the exponential (is negative, should be positive). \\

\appE{Usma14}{usman2014} & Changed the Auger coefficients by one order of magnitude~\cite{galeckas1997} to $\cn=\SI{5e-32}{\cm\tothe{6}\per\s}$ and $\cp=\SI{2e-32}{\cm\tothe{6}\per\s}$. \\
& In the text a value of the Scharfetter parameter $\alpha=1.72$ was stated but in the overall parameter listing $\alpha=5$. \\

\appE{Tama08}{tamaki2008} & Changed the activation energy from $\eact= \SI{0.11}{\eV}$~\cite{ivanov2006a} to $\eact= \SI{0.105}{\eV}$. \\

\appE{Zegh20}{zeghdar2020} & The value $\cn=\SI{5e31}{\cm\tothe{6}\per\s}$ was provided for the Auger coefficient but it should be $\SI{5e-31}{\cm\tothe{6}\per\s}$. \\

\appE{Zhan18}{zhang2018} & Unit of Auger coefficients stated as \si{\per\cubic\cm\per\s} but should be \si{\cm\tothe{6}\per\s}.

\end{appendixTable}


\subsection{\label{sec:incompIon_appendix}Incomplete Ionization}

\begin{appendixTable}

\appE{Arva17}{arvanitopoulos2017} & Referenced a Nitrogen ionization energy of \SI{71}{\milli\eV}~\cite{lebedev1999} but only \SI{81}{\milli\eV} for 6H-SiC was provided. \\


\appE{Darm19}{darmody2019} & Equation~(18), i.e., $\frac{p}{N_A} = \beta \frac{p^v}{N_A} + (1-\beta)$ could not be retraced given the expressions $p = p^v + p^i$ (Eq.~(14)) and $p^i = (1-\beta)N_A^0$ (Eq.~(16)).\\

\appE{Dona18}{donato2018} & We were unable to confirm the used ionization energies for Phosphorous (hexagonal: \SI{55}{\milli\eV}, cubic: \SI{102}{\milli\eV}). \\
& For Nitrogen the ionization energies (hexagonal: \SI{70}{\milli\eV}, cubic: \SI{120}{\milli\eV}) are presumably based on the values by \citet{ikeda1980} (hexagonal: \SI{66}{\milli\eV}, cubic: \SI{124}{\milli\eV}). \\

\appE{Feng04a}{feng2004a} & Changed for Nitrogen the ionization energy on the cubic lattice site from \SI{91.8}{\milli\eV}~\cite{gotz1993} to \SI{92}{\milli\eV} and on the hexagonal site from \SI{52.1}{\milli\eV}~\cite{gotz1993} to \SI{52}{\milli\eV}. \\

\appE{Huh06}{huh2006} & Used 6H values~\cite{storasta2002} and changed the ionization energy for Boron from \SI{270}{\milli\eV} to \SI{300}{\milli\eV}. \\

\appE{Ivan05}{ivanov2005} & The values for cubic and hexagonal site of Phosphorous were switched. These incorrect results were later copied~\cite{janzen2008} and implicitly corrected~\cite{scaburri2011a}.\\

\appE{Khal12}{khalid2012} & Changed the value of the donor energy level from \SI{67}{\milli\eV}~\cite{song2012} to \SI{650}{\milli\eV}.\\
& The ionized acceptors are denoted as $N_D^-$ instead of $\nam$. \\

\appE{Kuzn95}{kuznetsov1995} & In the conclusion the cross section of Aluminum
was stated as $\num{8e-3}(300/T)^3\si{\cm\squared}$ but the first term should be
$\num{8e-13}$ as noted earlier in the paper. \\

\appE{Lebe99}{lebedev1999} & Changed the ionization energy of Nitrogen on the cubic lattice site from \SI{91.8}{\milli\eV}~\cite{gotz1993} to \SI{92}{\milli\eV} and on the hexagonal site from \SI{52.1}{\milli\eV}~\cite{gotz1993} to \SI{52}{\milli\eV}. \\

\appE{Lech21}{lechner2021} & The used values for Aluminum seem to go back to \citet{arvinte2017} but the parameter $\alpha$ to describe the doping dependency was \SI{1.8e-5}{\milli\eV \cm} instead of \SI{2.8e-5}{\milli\eV \cm}. \\

\appE{Levi01}{levinshtein2001} & Changed the ionization energy of Aluminum from \SI{191}{\milli\eV}~\cite{ikeda1980} to \SI{190}{\milli\eV} and for Boron from \SI{647}{\milli\eV}~\cite{ikeda1980} to \SI{650}{\milli\eV}. \\

\appE{Lu21}{lu2021} & Changed the ionization energy of Aluminum from \SI{230}{\milli\eV}~\cite{lebedev1999} to \SI{220}{\milli\eV}. \\

\appE{Maxi23}{maximenko2023} & In equations~(6a) and (6b), corresponding to \cref{eq:incompIon_FermiDirac}, $\nc$ and $\nv$ were added as multiplicative factor to the exponential in the denominator. \\

\appE{Nipo18}{nipoti2018} & Presumably used the ionization energies by \citet{ivanov2005} but changed the value for the hexagonal lattice site from \SI{197.9}{\milli\eV} to \SI{198}{\milli\eV} and the value for the cubic from \SI{201.3}{\milli\eV} to \SI{210}{\milli\eV}. \\
& Changed the ionization energy for Phosphorous on the hexagonal site from \SI{60.7}{\milli\eV}~\cite{ivanov2005} to \SI{60}{\milli\eV}. \\

\appE{Pank14}{pankove2014} & Referenced \SI{90}{\milli\eV} for the ionization energy of the cubic lattice site of Nitrogen from \citet{hagen1973}, but there only \SI{90}{\milli\eV} for the hexagonal or \SI{130}{\milli\eV} for the cubic lattice site were proposed. \\

\appE{Pens93}{pensl1993} & Changed the ionization energy of Nitrogen on the cubic lattice site from \SI{91.8}{\milli\eV}~\cite{gotz1993} to \SI{91.4}{\milli\eV} and on the hexagonal site from \SI{52.1}{\milli\eV}~\cite{gotz1993} to \SI{51.8}{\milli\eV}, which can be explained by the fact that the reference was not yet published when the values were taken. \\

\appE{Pers05}{persson2005} & Changed the ionization energy of Nitrogen on the cubic lattice site from \SI{91.8}{\milli\eV}~\cite{gotz1993} to \SI{92}{\milli\eV} and on the hexagonal site from \SI{52.1}{\milli\eV}~\cite{gotz1993} to \SI{52}{\milli\eV}. \\

\appE{Rakh20}{rakheja2020} & Referenced a 6H source~\cite{sullivan2008} for the energy levels of Boron, Vanadium and Nitrogen but the values could not be found there. \\

\appE{Scab11a}{scaburri2011a} & In the derivation of \cref{eq:incompIon_n_wo_fermi} on page 22 an expression for $n$~\cite{blakemore1962} was presented that (i) is not required for the calculation and (ii) we were unable to retrace in the original publication. We
solely achieved the shown result if we calculated $1/n$ and used in one occasion the wrong parameter. \\

\appE{Song12}{song2012} & Changed the donor ionization energy from \SI{66}{\milli\eV}~\cite{tamaki2008a} to \SI{67}{\milli\eV} and the acceptor one from \SI{191}{\milli\eV}~\cite{tamaki2008a} to \SI{19}{\milli\eV}. \\

\appE{Sozz19}{sozzi2019} & Interpreted the maximum ionization energy of Nitrogen on the hexagonal lattice site of \SI{50}{\milli\eV}~\cite{kimoto1995} as the effective value. \\

\appE{Tian20}{tian2020} & Used the ionization energy \SI{201.3}{\milli\eV}~\cite{ivanov2005} of Aluminum on a cubic lattice site as the effective value. \\

\appE{Yang19}{yang2019} & Changed the ionization energy of Nitrogen on a hexagonal lattice site from \SI{61.4+-0.5}{\milli\eV}~\cite{ivanov2003a} to an effective value of \SI{61}{\milli\eV}. \\

\appE{Yosh18}{yoshioka2018} & Changed the ionization energy of Aluminum from \SI{191}{\milli\eV}~\cite{ikeda1980} to \SI{190}{\milli\eV}. \\

\appE{Zett02}{zetterling2002} & Changed the ionization energy of Nitrogen on the cubic lattice site from \SI{91.8}{\milli\eV}~\cite{gotz1993} to \SI{92}{\milli\eV} and on the hexagonal site from \SI{52.1}{\milli\eV}~\cite{gotz1993} to \SI{50}{\milli\eV}. \\

\appE{Zhan18}{zhang2018} & Used in equations~(13) and (14), corresponding to \cref{eq:incompIon_FermiDirac}, instead of the donor ($\ed$) and acceptor ($\ea$) energy level the conduction and valence band energies. \\

\appE{Zhu08}{zhu2008} & Changed the ionization energy of Nitrogen on the cubic lattice site from \SI{91.8}{\milli\eV}~\cite{gotz1993} to \SI{92}{\milli\eV} and on the hexagonal site from \SI{52.1}{\milli\eV}~\cite{gotz1993} to \SI{50}{\milli\eV}. \\
& Changed the ionization energy of Phosphorous on the hexagonal lattice site from \SI{53}{\milli\eV}~\cite{capano2000} to \SI{54}{\milli\eV}. \\

\end{appendixTable}


\subsection{\label{sec:mobility_appendix}Mobility}

\begin{appendixTable}

\appE{Aktu09}{akturk2009} & Changed the following electron saturation velocities: $\SI{2.12e7}{\centi\m\per\s}$~\cite{hjelm2001} to $\SI{2e7}{\centi\m\per\s}$, $\SI{1.58e7}{\centi\m\per\s}$~\cite{hjelm2001} to $\SI{1.6e7}{\centi\m\per\s}$, $\SI{2.5e7}{\centi\m\per\s}$~\cite{mickevicius1998} to $\SI{2.4e7}{\centi\m\per\s}$ and $\SI{1.83e7}{\centi\m\per\s}$~\cite{zhao2000} to $\SI{2e7}{\centi\m\per\s}$. \\

\appE{Alba10}{albanese2010} & Changed $\mumax$ of holes from $\SI{124}{\square\centi\meter\per\V\per\s}$~\cite{schaffer1994} to $\SI{125}{\square\centi\meter\per\V\per\s}$.\\

\appE{Arda05}{ardaravicius2005} & Changed the electron low field mobility from $\SI{720}{\square\centi\meter\per\V\per\s}$~\cite{masri2002} to $\SI{730}{\square\centi\meter\per\V\per\s}$.\\

\appE{Arva17}{arvanitopoulos2017} 
& Changed $\gsat$ for electrons from $-0.44$~\cite{ladesmartin2000} to $0.44$. \\

\appE{Arva19}{arvanitopoulos2019} & Rounded $\mumin$ for holes from $\SI{15.9}{\square\centi\meter\per\V\per\s}$~\cite{schaffer1994} to $\SI{16}{\square\centi\meter\per\V\per\s}$. \\


\appE{Ayal04}{ayalew2004} & Changed $\mumax$ of holes from $\SI{124}{\square\centi\meter\per\V\per\s}$~\cite{schaffer1994} to $\SI{125}{\square\centi\meter\per\V\per\s}$.\\

\appE{Bane21}{banerjee2021} & In equation (10), describing the carrier velocity at high fields, the exponent $1/\kappa$ should be only applied to the denominator and not also to the dominator. \\

\appE{Bela22}{belas2022} & The unit of the saturation velocity is stated as \si{\square\centi\meter\per\V\per\s} instead of \si{\centi\m\per\s}. \\

\appE{Bell11}{bellone2011} & Changed $\mumax$ of holes from $\SI{124}{\square\centi\meter\per\V\per\s}$~\cite{schaffer1994} to $\SI{125}{\square\centi\meter\per\V\per\s}$.\\

\appE{Bhat05}{bhatnagar2005} & The saturation velocity is stated as \SI{2e17}{\centi\m\per\s} instead of \SI{2e7}{\centi\m\per\s}. \\
& Only the parameters are shown but not the corresponding equations. Therefore, it is not possible to uniquely identify the correct values. \\

\appE{Buon12}{buono2012} & Dismissed the leading $\mumin$ from \cref{eq:mobility_CT}. \\
& Changed $\mumax$ of electrons from $\SI{947}{\square\centi\meter\per\V\per\s}$~\cite{schaffer1994} to $\SI{950}{\square\centi\meter\per\V\per\s}$.\\

\appE{Capa22}{capan2022} & Rounded the hole mobility from $\SI{118}{\square\centi\meter\per\V\per\s}$~\cite{kimoto2014a} to $\SI{120}{\square\centi\meter\per\V\per\s}$. \\

\appE{Chen22}{cheng2022} & Referenced the electron mobility parallel to the c-axis from \cite{ladesmartin2000} but there it is actually stated perpendicular to the c-axis. \\


\appE{Das15}{das2015} & In equation (4), describing the carrier velocity at high fields, the exponent $1/\beta 1$ should be only applied to the denominator and not also to the dominator. \\

\appE{Elah17}{elahipanah2017} & $\vsat$ was stated as \SI{2}{\centi\m\per\s}. We assume that \SI{2e7}{\centi\m\per\s} was intended. \\

\appE{Gotz93}{gotz1993} & Took saturation velocity from literature~\cite{powell1989}, where the investigated material is only specified as $\alpha$-SiC. So it is not clear whether 4H-SiC was meant.\\

\appE{Huan98}{huang1998} & Referenced values for $\beta$ from \citet{ruff1994}, who used Silicon values only. \\

\appE{Joha19}{johannesson2019} & We were unable to locate the stated maximum hole mobility $\mumax=\SI{20}{\square\centi\meter\per\V\per\s}$ in the specified reference. We assumed a typographical mistake and it should be \SI{120}{\square\centi\meter\per\V\per\s} (rounded from \SI{124}{\square\centi\meter\per\V\per\s}~\cite{schaffer1994}). \\
& Changed $\mumax$ of electrons from $\SI{947}{\square\centi\meter\per\V\per\s}$~\cite{schaffer1994} to $\SI{940}{\square\centi\meter\per\V\per\s}$. \\

\appE{Josh95}{joshi1995} & In the text the value of the saturation velocity is provided in \si{\m\per\s} but as unit \si{\centi\m\per\s} was specified. \\
& Changed $\mumax$ of electrons from $\SI{947}{\square\centi\meter\per\V\per\s}$~\cite{schaffer1994} to $\SI{940}{\square\centi\meter\per\V\per\s}$. \\

\appE{Kimo97}{kimoto1997} & Turned $\mun=\SI{720}{\square\centi\meter\per\V\per\s}$~\cite{itoh1994} to $\mun=\SI{724}{\square\centi\meter\per\V\per\s}$. \\

\appE{Lade00}{ladesmartin2000} & For parameter $F$ of the carrier-carrier scattering the wrong unit was specified. It should be \si{\per\square\centi\m} instead of \si{\centi\m\tothe{2/3}}. \\

\appE{Lang22}{langpoklakpam2022} & In the referenced publication~\cite{wijesundara2011} the electron mobility was denoted as $\SI{900}{\square\centi\meter\per\V\per\s}$ not $\SI{800}{\square\centi\meter\per\V\per\s}$.\\

\appE{Lech21}{lechner2021} & We were unable to locate the used mobility values in the stated reference~\cite{ayalew2004}. \\

\appE{Liu21}{liu2021} & Changed $\gmax$ for electrons from $-2.9$~\cite{tamaki2008a} to $-2.8$. \\
& Turned $\mumax$ for holes from \SI{124}{\square\centi\meter\per\V\per\s}~\cite{tamaki2008a} to $\SI{125}{\square\centi\meter\per\V\per\s}$ and the reference doping concentration $\nref$ from \SI{1.76e19}{\per\cubic\centi\m}~\cite{tamaki2008a} to \SI{1.76e17}{\per\cubic\centi\m}. \\
& Used as hole saturation velocity the electron saturation velocity~\cite{zhang2008}. \\
& In the high field equations in Table 1 (cp. \cref{eq:mob_high}) $\beta$ is defined over the whole sum in the denominator and the exponent outside the brackets should be $-1/\beta$ instead of $-\beta$. \\

\appE{Lv04}{lv2004} & We were not able to recreate the high field velocity curves shown in Fig. 2 of the paper with the presented parameters. \\

\appE{Mcnu04}{mcnutt2004} & Turned $\mumax$ for holes from \SI{108.1}{\square\centi\meter\per\V\per\s}~\cite{bakowski1997}
to $\SI{108.9}{\square\centi\meter\per\V\per\s}$. \\

\appE{Megh18}{megherbi2018} & Turned $\mumax$ for holes from \SI{124}{\square\centi\meter\per\V\per\s}~\cite{li2003} to $\SI{125}{\square\centi\meter\per\V\per\s}$. \\


\appE{Neud01}{neudeck2001} & We could not find the stated electron mobility parallel to the c-axis (\SI{800}{\square\centi\meter\per\V\per\s}) in the mentioned reference~\cite{pensl1998}. \\

\appE{Ostl24}{ostling2024} & $\vsat$ was stated as \SI{2}{\centi\m\per\s}. We assumed that \SI{2e7}{\centi\m\per\s} was intended. \\

\appE{Palm97}{palmour1997} & Changed $\mumax$ of electrons from $\SI{947}{\square\centi\meter\per\V\per\s}$~\cite{schaffer1994} to $\SI{950}{\square\centi\meter\per\V\per\s}$. \\

\appE{Pere06}{perez-tomas2006} & Changed $\gmax=-2.4$~\cite{roschke2001} for electrons to $-2$. \\

\appE{Pezz08}{pezzimenti2008} & The unit of the velocity saturation was stated in \si{\square\cm\per\s} but should be \si{\cm\per\s}. \\

\appE{Powe02}{powell2002} & The unit of the saturation velocity was stated in \si{\m\per\s} but the value indicates \si{\centi\m\per\s}. \\

\appE{Rakh20}{rakheja2020} & Changed $\mumax$ of electrons from $\SI{947}{\square\centi\meter\per\V\per\s}$~\cite{ladesmartin2000} to $\SI{950}{\square\centi\meter\per\V\per\s}$. \\
& Changed $\mumax$ of holes from $\SI{124}{\square\centi\meter\per\V\per\s}$~\cite{schaffer1994} to $\SI{120}{\square\centi\meter\per\V\per\s}$. \\
& The velocity saturation for holes was extracted from \cite{ladesmartin2000} but there this value denotes the electron saturation velocity. \\

\appE{Sand11}{sandeep2011} & It is reasonable to assume that in Table~II the columns for 4H and 6H got exchanged, especially because of the deviating band gap values. \\

\appE{Shah98}{shah1998} & References \citet{schaffer1994} regarding the hole mobility but all parameters are different. \\

\appE{Shar15}{sharma2015} & In the Caughey-Thomas equation (cp. \cref{eq:mobility_CT}) $\mumin$ also shows up in the denominator. \\

\appE{Sole19}{soler2019} & We were unable to locate the presented values in the provided reference~\cite{pearton2018}. \\

\appE{Stef14}{stefanakis2014} & See detailed analysis in \cref{sec:mobility}. \\

\appE{Tama08a}{tamaki2008a} & The charge carrier mobility was referenced from \citet{ruff1994} but the values fit the models by \citet{schaffer1994} resp. \citet{roschke2001} better. \\

\appE{Tian20}{tian2020} & See detailed analysis in \cref{sec:mobility}. \\

\appE{Tila07}{tilak2007} & We could not find $\gmax=2.4$ in the provided reference~\cite{mickevicius1998}. Also the value contradicts many others. \\

\appE{Trip19}{tripathi2019} & Corrected the wrong value of $\delta=-0.61$~\cite{stefanakis2014} to $\delta=0.61$. \\
& The value of $\nref=\SI{1e16}{\per\cubic\centi\meter}$ could not be found in the reference~\cite{stefanakis2014}. Overall, using this parameter set is risky, as pointed out in \cref{sec:mobility}. \\

\appE{Vasc19}{vasconcelos2019} & In the overview table 2 the minimum instead of the maximum mobility was used from \citet{arvanitopoulos2019}. The value for \citet{contreras2006} should be $\SI{45.2}{\square\centi\meter\per\V\per\s}$ instead of $\SI{37}{\square\centi\meter\per\V\per\s}$. For \citet{stefanakis2014} a value was stated \SI{116.1}{\square\centi\meter\per\V\per\s}, but the authors used $\mumax=\SI{114}{\square\centi\meter\per\V\per\s}$. \\

\appE{Wang98}{wang1998} & Changed $\gmax=-2.15$~\cite{schaffer1994} to $\gmax=-2$. \\

\appE{Wang99}{wang1999} & In the description of the high-field mobility (Eq.~(4)) only the saturation velocity was squared instead of the whole fraction $F \mun/\vsat$ (cp. \cref{eq:mob_high}). \\
& Changed $\gmax=-2.15$~\cite{schaffer1994} to $\gmax=-2$. \\

\appE{Wrig98}{wright1998} & Used for carrier-carrier scattering an equation that was proposed to describe the impurity scattering in Silicon by \citet{dorkel1981}. \\

\appE{Zegh19}{zeghdar2019} & Changed $\mumax$ of holes from $\SI{124}{\square\centi\meter\per\V\per\s}$~\cite{schaffer1994} to $\SI{125}{\square\centi\meter\per\V\per\s}$.\\

\appE{Zegh20}{zeghdar2020} & All temperature dependency parameters $\gamma$ are positive, which leads to $\mu(0)=0$ and a continuously increasing mobility with temperature. \\

\appE{Zhan18}{zhang2018} & In Eq.~(2) we presumed that $C_S$ should be replaced by $T$ and $N_i$ by $T_0$. \\
& In Eq.~(4) $\beta$ is defined by using $\beta$. \\

\appE{Zhou16}{zhou2016} & Rounded $\delta=0.76$~\cite{roschke2001} to $1$. \\

\end{appendixTable}


\section{\label{sec:appendixRefChain}Reference Chains}

\revision{In this section we present the detailed reference chains for each parameter. These analyses show which parameter values are predominantly used in literature, how they might have changed over time and which publications determined the currently used values. The amount of references can also be used as a measure for the reliability of the proposed parameters values.}

\clearpage
\subsection{\label{sec:refChainPermittivity}Permittivity}

\begin{figure*}[ht!]
    \centering
    \resizebox{0.86\textwidth}{!}{%
    \input{figures/permittivity_ref_chain}
    }
    \caption{\label{fig:permittivity_ref_chain}Reference chains for the permittivity. A single value denotes $\es$ respectively $\ei$ without direction. \bkgCol{no4HColor} are investigations not focused on 4H-SiC and \bkgCol{fundColor} novel analyses on 4H-SiC.}
\end{figure*}

\FloatBarrier

\clearpage
\subsection{\label{sec:refChainDOS}Density-of-States Mass}

\begin{figure*}[ht!]
    \centering
    \resizebox{0.97\linewidth}{!}{%
    \input{figures/dos_ref_chain_h}
    }
    \caption{\label{fig:dos_ref_chain_h}Reference chain for hole DOS mass. `M'
      denotes multiple values. \bkgCol{no4HColor}~were not focused on 4H-SiC,
      \bkgCol{fundColor}~indicates novel analyses on 4H-SiC and \bkgCol{guessColor}~shows connections we inferred based on the used data.}
\end{figure*}

\begin{figure*}[p]
    \centering
    \resizebox{0.99\linewidth}{!}{%
    \input{figures/dos_ref_chain_e}
    }
    \caption{\label{fig:dos_ref_chain_e}Reference chain for electron DOS
      mass. `M' denotes multiple values. \bkgCol{no4HColor}~were not focused on
      4H-SiC, \bkgCol{fundColor}~indicates novel analyses on 4H-SiC and \bkgCol{guessColor}~shows connections we inferred based on the used data.}
\end{figure*}  

\FloatBarrier

\clearpage
\subsection{\label{sec:refChainBandgap}Band Gap}

\begin{figure*}[ht!]
    \centering
    \resizebox{0.5\linewidth}{!}{%
    \input{figures/bandgap_ref_chain_Ex}
    }
    \caption{\label{fig:bandgap_ref_chain_Ex}Reference chain of the free exciton binding energy $\ex$. \bkgCol{no4HColor} indicates values that were not determined for 4H-SiC, \bkgCol{fundColor} fundamental investigations and \bkgCol{guessColor} connections predicted from the used values.}
\end{figure*}

\begin{figure*}[p]
    \centering
    \resizebox{0.98\linewidth}{!}{%
    \input{figures/bandgap_ref_chain_bandgap}
    }
    \caption{\label{fig:bandgap_ref_chain_bandgap}Reference chain for band gap energies. \bkgCol{fundColor} are fundamental investigations, \bkgCol{calcColor} calculations and \bkgCol{guessColor} connections predicted from the used values.}
\end{figure*}

\begin{figure*}[t]
    \centering
    \resizebox{0.98\linewidth}{!}{%
    \input{figures/bandgap_ref_chain_temp}
    }
    \caption{\label{fig:bandgap_ref_chain_temp}Reference chain for the temperature dependency of the band gap. \bkgCol{no4HColor} indicates values that were not determined for 4H-SiC, \bkgCol{fundColor} are fundamental investigations and \bkgCol{guessColor} connections predicted from the used values.}
\end{figure*}

\begin{figure*}[th]
    \centering
    \resizebox{0.95\linewidth}{!}{%
    \input{figures/bandgap_ref_chain_doping}
    }
    \caption{\label{fig:bandgap_ref_chain_doping}Reference chain for the doping dependency of the band gap. If solely a single value is shown for parameter $B$ then the publication only provided $B_{x\mathrm{c}}+B_{x\mathrm{v}}$. \bkgCol{fundColor} are fundamental investigations and \bkgCol{guessColor} connections predicted from the used values.}
\end{figure*}

\FloatBarrier

\clearpage
\subsection{\label{sec:refChainII}Impact Ionization}

\begin{figure*}[h!]
    \centering
    \resizebox{0.97\linewidth}{!}{%
    \input{figures/II_ref_chain}
    }
    \caption{\label{fig:II_ref_chain} Reference chain for impact ionization parameters. \bkgCol{no4HColor} are not focused on 4H-SiC, while \bkgCol{fundColor} are novel analyses on 4H-SiC, and \bkgCol{guessColor} indicates an educated guess on the reference based on the given values in cases the source is not explicitly stated in the publication.}
\end{figure*}

\FloatBarrier

\clearpage
\subsection{\label{sec:refChainRegen}Charge Carrier Recombination}

\begin{figure*}[h!]
    \centering
    \resizebox{0.99\linewidth}{!}{%
    \input{figures/regen_ref_chain}
    }
    \caption{\label{fig:regen_ref_chain}Reference chain for doping and temperature dependency of the SRH lifetime. \bkgCol{fundColor} are fundamental investigations, \bkgCol{no4HColor} research not focused on 4H and \bkgCol{guessColor} connections predicted from the used values. }    
\end{figure*}

\begin{figure*}[t]
    \centering
    \resizebox{0.99\linewidth}{!}{%
    \input{figures/regen_ref_chain_II}
    }
    \caption{\label{fig:regen_ref_chain_II}Reference chain for bimolecular and Auger charge carrier recombination. \bkgCol{fundColor} are fundamental investigations, \bkgCol{no4HColor} research not focused on 4H and \bkgCol{guessColor} connections predicted from the used values. }
\end{figure*}

\FloatBarrier

\clearpage
\subsection{\label{sec:refChainIncompIon}Incomplete Ionization}

\begin{figure*}[h!]
    \centering
    \resizebox{0.95\textwidth}{!}{%
    \input{figures/IncompIon_ref_chainN}
    }
    \caption{\label{fig:IncompIon_ref_chainN}Reference chain for n-type doping. \bkgCol{fundColor} are fundamental investigations, \bkgCol{no4HColor} research not focused on 4H and \bkgCol{guessColor} references we inferred based on the used data.}
\end{figure*}

\begin{figure*}[t]
    \centering
    \resizebox{0.95\textwidth}{!}{%
    \input{figures/IncompIon_ref_chainP}
    }
    \caption{\label{fig:IncompIon_ref_chainP}Reference chain for p-type doping. \bkgCol{fundColor} are fundamental investigations, \bkgCol{no4HColor} research not focused on 4H and \bkgCol{guessColor} references we inferred based on the used data.}
\end{figure*}

\begin{figure*}[t]
    \centering
    \resizebox{0.75\textwidth}{!}{%
    \input{figures/IncompIon_ref_chain_alpha}
    }
    \caption{\label{fig:IncompIon_ref_chain_alpha}Reference chain for doping dependency modeling. \bkgCol{fundColor} are fundamental investigations, \bkgCol{no4HColor} research not focused on 4H and \bkgCol{guessColor} references we inferred based on the used data.}
\end{figure*}

\FloatBarrier

\clearpage
\subsection{\label{sec:refChainMobility}Mobility}

\begin{figure*}[h!]
    \centering
    \resizebox{0.99\linewidth}{!}{%
    \input{figures/mobility_ref_chain_low_elec}
    }
    \caption{\label{fig:mobility_ref_chain_low_elec}Reference chain for low field electron mobility models. \bkgCol{fundColor} are fundamental investigations, \bkgCol{no4HColor} research not focused on 4H and \bkgCol{guessColor} connections predicted from the used values.}
\end{figure*}

\begin{figure*}[p]
    \centering
    \resizebox{0.99\linewidth}{!}{%
    \input{figures/mobility_ref_chain}
    }
    \caption{\label{fig:mobility_ref_chain}Reference chain for constant mobility
      values in different spatial directions. If only a single value is shown
      the value denotes the effective value. \bkgCol{fundColor} are fundamental investigations, \bkgCol{no4HColor} research not focused on 4H and \bkgCol{guessColor} references we inferred based on the used data.}
\end{figure*}

\begin{figure*}[t]
    \centering
    \resizebox{0.99\linewidth}{!}{%
    \input{figures/mobility_ref_chain_low_hole}
    }
    \caption{\label{fig:mobility_ref_chain_low_hole}Reference chain for low field hole mobility models. \bkgCol{fundColor} are fundamental investigations, \bkgCol{no4HColor} research not focused on 4H and \bkgCol{guessColor} connections implied by the values used in the publication.}
\end{figure*}

\begin{figure*}[t]
    \centering
    \resizebox{0.99\linewidth}{!}{%
    \input{figures/mobility_vsat_ref_chain_elec}
    }
    \caption{\label{fig:mobility_vsat_ref_chain_elec}High-field mobility reference chain for electrons. \bkgCol{fundColor} are fundamental investigations, \bkgCol{no4HColor} research not focused on 4H and \bkgCol{guessColor} connections implied by the values used in the publication.}
\end{figure*}

\begin{figure*}[t]
    \centering
    \resizebox{0.79\linewidth}{!}{%
    \input{figures/mobility_vsat_ref_chain_hole}
    }
    \caption{\label{fig:mobility_vsat_ref_chain_hole}High-field mobility
      reference chain for holes. \bkgCol{fundColor} are fundamental investigations, \bkgCol{no4HColor} research not focused on 4H and \bkgCol{guessColor} connections implied by the values used in the publication.}
\end{figure*}

\FloatBarrier

%

\end{document}